\documentclass[english,11pt]{article}
\usepackage[T1]{fontenc}
\usepackage[utf8]{inputenc}
\usepackage[pdftex]{graphicx}
\usepackage{verbatim}
\usepackage[width=\textwidth]{caption}
\usepackage{amsmath,bbm,amssymb,amsopn,mathrsfs,mathtools,bm}
\usepackage{a4wide}
\usepackage[english]{babel}
\usepackage[]{float}
\usepackage[]{placeins}
\usepackage{flafter}
\usepackage[]{longtable}
\usepackage{verbatim}
\usepackage{dsfont}
\usepackage{color}
\usepackage{array}
\usepackage{natbib}
\usepackage{xr}
\usepackage{marginnote}
\usepackage{subfig}

\newtheorem{thm}{Theorem}[section]

\newtheorem{lem}{Lemma}[section]
\newtheorem{prop}[thm]{Proposition}
\newtheorem{hyp}{Assumption}

\newcommand{\convP}{\stackrel{\mathbb{P}}{\longrightarrow}}
\newcommand{\convD}{\stackrel{d}{\longrightarrow}}
\newcommand{\convDboot}{\stackrel{d^*}{\longrightarrow}}
\newcommand{\convAS}{\stackrel{a.s.}{\longrightarrow}}
\newcommand{\E}{\mathbb E}
\newcommand{\G}{\mathbb G}
\newcommand{\V}{\mathbb V}
\newcommand{\R}{\mathbb R}
\newcommand{\N}{\mathbb N}

\newcommand{\Supp}{\text{Supp}}
\newcommand{\indep}{\perp \!\!\! \perp}
\newcommand{\eps}{\varepsilon}

\renewcommand{\j}{\bm{j}}

\newcommand{\un}{\boldsymbol{1}}
\newcommand{\deux}{\boldsymbol{2}_i}
\newcommand{\deu}{\boldsymbol{2}}
\newcommand{\C}{\bm{C}}
\newcommand{\Cov}{\mathbb{C}ov}
\newcommand{\e}{\bm{e}}

\renewcommand{\c}{\bm{c}}
\newcommand{\zero}{\boldsymbol{0}}
\newcommand{\Id}{\text{Id}}

\newcolumntype{C}[1]{>{\centering\arraybackslash}p{#1}}

\newenvironment{tablenotes}[1][Notes]{\begin{tabular}{p{\linewidth}}\footnotesize{\itshape#1: }}{\end{tabular}}

\newcolumntype{C}[1]{>{\centering\let\newline\\\arraybackslash\hspace{0pt}}m{#1}}

\DeclarePairedDelimiter\abs{\lvert}{\rvert}%
\DeclareMathOperator*{\argmin}{arg\,min}


\date{August 3, 2018}

\begin{document}
	
\title{Asymptotic results under multiway clustering\thanks{We would like to thank Stéphane Bonhomme, Clément de Chaisemartin, Isabelle Méjean, seminar participants at Bristol, CREST and Yale and attendees of the 2018 IAAE conference for helpful comments.}}

\author{Laurent Davezies\thanks{CREST, laurent.davezies@ensae.fr}
\and Xavier D'Haultf\oe uille
\thanks{CREST. xavier.dhaultfoeuille@ensae.fr}
\and Yannick Guyonvarch
\thanks{CREST. yannick.guyonvarch@ensae.fr}}

\maketitle

\bigskip
\begin{abstract}
If multiway cluster-robust standard errors are used routinely in applied economics, surprisingly few theoretical results justify this practice. This paper aims to fill this gap. We first prove, under nearly the same conditions as with i.i.d. data, the weak convergence of empirical processes under multiway clustering. This result implies central limit theorems for sample averages but is also key for showing the asymptotic normality of nonlinear estimators such as GMM estimators. We then establish consistency of various asymptotic variance estimators, including that of \cite{cameron2011} but also a new estimator that is positive by construction. Next, we show the general consistency, for linear and nonlinear estimators, of the pigeonhole bootstrap, a resampling scheme adapted to multiway clustering. Monte Carlo simulations suggest that inference based on our two preferred methods may be accurate even with very few clusters, and significantly improve upon inference based on \cite{cameron2011}.

\medskip
\textbf{Keywords:} Multiway clustering, Empirical processes, Cluster-robust standard errors, Pigeonhole bootstrap, GMM.

\medskip
\textbf{JEL codes:} C13, C15, C21, C23.

\end{abstract}

%\end{frontmatter}

\newpage

\section{Introduction}

%!TEX root = /Users/xdhaultfoeuille/Dropbox/Cluster/paper/main.tex
Taking into account dependence between observations is crucial for making correct inference. Common shocks tend to correlate observations positively, leading to overly optimistic inference when ignored \citep{bertrand2004}. As a result, estimation of standard errors robust to clustering has become pervasive in applied economics. In particular, following the very influential work of \cite{cameron2011},\footnote{According to the Web of Science and Google Scholar, \cite{cameron2011} is the most cited paper in econometrics since 2009.} empirical studies now routinely report standard errors accounting for multiway clustering. Perhaps surprisingly however, econometric theory has lagged behind this practice. \cite{cameron2011} conduct simulations suggesting the validity of their method but neither prove that their variance estimators are consistent, nor that estimators of parameters of interest are themselves asymptotically normal. And more generally, there are still very few theoretical results under multiway clustering.

\medskip
The goal of this paper is to fill this gap, by developing general tools for inference on linear but also nonlinear estimators with multiway clustering. We consider for that purpose a fairly general set-up including as particular cases one- and two-way clustering. To understand the key underlying restrictions, let us consider the example of two-way clustering where the first dimension is the sector of activity and  the second is the area of residence, e.g. counties or states. Such an example would be appropriate when studying for instance individual wages. We index the two dimensions respectively by $j_1\in \{1,...,C_1\}$ and $j_2\in\{1,...,C_2\}$. We call a cell any pair $(j_1,j_2)$, corresponding therefore to a specific sector of activity and area of residence. %We then let $N_{j_1,j_2}$ denote the number of units in cell $(j_1,j_2)$, with possibly $N_{j_1,j_2}=0$. 
Then two units in different cells $(j_1,j_2)$ and $(j'_1,j'_2)$ are assumed independent whenever $j_1\neq j'_1$ and $j_2\neq j'_2$. Otherwise they may be dependent in an unrestricted way. The idea behind is that units sharing at least one cluster may be affected by common shocks, e.g. sectorial shocks or local shocks in the previous example.

\medskip
Following most of the literature, we consider an asymptotic framework where $\underline{C}=\min(C_1,C_2)$ tends to infinity.\footnote{A growing strand of the literature on one-way clustering has also considered fixed-$\underline{C}$ asymptotics, with cluster sizes tending to infinity. We refer in particular to \cite{donald2007inference}, \cite{ibragimov2010},  \cite{bester2011inference}, \cite{ibragimov2016} and \cite{canay2018}. To our knowledge, no paper has considered such a set-up with multiway clustering yet.} In particular, we allow for random, possibly unbounded, cell sizes. Cell sizes may also be correlated with the data themselves. These features are important to account for cluster heterogeneity, following the terminology of \cite{carter2017}. Given this set-up, our first contribution is a general weak convergence result on the empirical process. To our knowledge, this weak convergence result is new, even under one-way clustering. When considering processes indexed by a finite class of functions, it is equivalent to a simple multivariate central limit theorem (CLT) on sample averages. But when considering infinite classes of functions, this result is also key for proving asymptotic normality of nonlinear estimators like GMM estimators or smooth functionals of the empirical cumulative distribution function (cdf). Also, up to moment restrictions that have to be slightly adapted, our conditions on the class of functions indexing the empirical process are the same as with i.i.d. data. This means that results on, e.g., GMM estimators already established for i.i.d. data can be extended directly to multiway clustering. 

\medskip
Then, we prove the consistency of three asymptotic variance estimators, including that suggested by \cite{cameron2011}.\footnote{\cite{mackinnon2017} also prove the consistency of the estimator of \cite{cameron2011} under two-way clustering, but under restrictions that may not hold in practice, as we argue below.} If this latter estimator is asymptotically valid, it has the drawback of being possibly negative in practice. We develop another simple estimator that is also consistent and avoids this drawback. Our Monte Carlo simulations suggest that this estimator may perform significantly better than that suggested by \cite{cameron2011} when $\underline{C}$ is small.

\medskip
Next, we prove the asymptotic validity of a general bootstrap scheme adapted to multiway clustering, called the pigeonhole bootstrap. This resampling scheme differs from the usual multinomial bootstrap by explicitly taking into account the particular dependence structure implied by multiway clustering. The idea is to sample independently each dimensions of clustering and to select cells (with possible repetitions) that are at the intersection of selected clusters in various dimensions. This bootstrap was suggested by \cite{mccullagh2000resampling} and studied by \cite{owen2007pigeonhole} but to our knowledge, no weak convergence result has been obtained on it yet, even for sample averages. Again, we prove a general weak convergence result on the pigeonhole bootstrap process. This result implies the validity of the pigeonhole bootstrap for sample averages but also for GMM or smooth functionals of the cdf.\footnote{Similarly to the usual multinomial bootstrap but contrary to, e.g., the wild bootstrap, this bootstrap has the advantage of being universal. Namely, as a resampling scheme, it can be applied in the same way irrespective of the estimation procedure.} Monte Carlo simulations suggest that the pigeonhole bootstrap may work very well even with $\underline{C}$ as small as 5 (resp. 3) under two-way (resp. three-way) clustering.

\medskip
As in the i.i.d. setting, weak convergence of the empirical process relies on two main ingredients: a multivariate CLT and the asymptotic equicontinuity of the process. To prove the multivariate CLT, we use the Aldous-Hoover representation for exchangeable arrays \citep{Aldous1981,Aldous83,Hoover1979,kallenberg05} and techniques related to U-statistics, in particular H\'ajek projections. % \citep[see, e.g.][Chapters 11.3 and 12]{vanderVaart2000}. 
For the asymptotic equicontinuity, a key step, as in the i.i.d. setting, is to symmetrize the initial process. We do so by generalizing the standard symmetrization lemma \citep[see for instance Lemma 2.3.1 in][]{vanderVaartWellner1996}, using again the Aldous-Hoover representation and an adaptation to our framework of arguments used in \cite{ArconesGine1993}. The same kind of strategy is used conditional on the data and combined with ergodicity arguments to establish the consistency of the pigeonhole bootstrap process.

\medskip
The literature on clustering is vast but has mostly focused on linear models under one-way clustering, following the seminal papers of \cite{pfeffermann1981}, \citeauthor{moulton1986} (\citeyear{moulton1986}, \citeyear{moulton1990}) \cite{liang1986} and \cite{arellano1987}. Without being exhaustive, we also refer to \cite{hansen2007}, \cite{cameron2008bootstrap}, \cite{carter2017}, \cite{mackinnon2017wild} and \cite{hansen2017} for more recent contributions. 

\medskip
The only papers we are aware of considering multiway clustering are the recent works of \cite{menzel2017} and \cite{mackinnon2017}. \cite{menzel2017} focuses on sample averages. Contrary to us, he studies inference both with and without asymptotically normality. He also shows that refinements in asymptotic approximations are possible using the wild bootstrap. \cite{mackinnon2017} focus on linear regressions with two-way clustering. For such models, they show asymptotic normality and the consistency of the variance estimator of \cite{cameron2011}. They also show the validity of a certain wild boostrap in this context. 

\medskip
Compared to these papers, our contributions are the following. First, our empirical process result allows us to consider nonlinear estimators. To our knowledge, we are thus the first to show the asymptotic normality of general GMM estimators with multiway clustering. Second, we propose for linear and nonlinear models a new variance estimator that is always positive, very simple to compute and that seems to perform better in practice than that of \cite{cameron2011}. Third, we show the general validity of the pigeonhole bootstrap with multiway clustering. Finally, even in linear models, we obtain our results under different conditions from those in \cite{menzel2017} and \cite{mackinnon2017}. Contrary to \cite{menzel2017}, we do not impose cell sizes equal to one, or i.i.d. units within cells. \cite{mackinnon2017} assume, through their Assumption 3, that $\overline{N}$, the average of the cell sizes, satisfies $\overline{N} \underline{C}^{2/(2+\lambda)} \rightarrow 0$ for some $\lambda>0$. In other words, the vast majority of cells has to become empty as $\underline{C}$ tends to infinity. This condition may not hold in applications. In contrast, while our framework allows for empty cells, it implies that $\overline{N}$ converges in probability to a positive constant. 

 %We impose no restrictions on $N_{j_1,j_2}$, except that its distribution is assumed independent of $\underline{C}$.  

\medskip
The paper is organized as follows. Section \ref{sec:setup} describes the assumptions we impose on the data generating process and the parameters of interest we consider afterwards. Section \ref{sec:gen_results} provides our main results on the convergence of the empirical process and the pigeonhole bootstrap empirical process. Section \ref{sec:appli} discusses applications of these results to linear and nonlinear estimators. In particular, we show therein the consistency of various asymptotic variance estimators, and asymptotic normality of GMM and smooth functionals of the cdf. We also show the consistency of the pigeonhole bootstrap for inference on such estimators. Section \ref{sec:MC} explores through simulations the finite-sample properties of inference based on asymptotic normality or the pigeonhole bootstrap. Section \ref{sec:conclu} concludes. The appendix gathers extensions, additional details on simulations and all the proofs of our results.

\section{The set up}
\label{sec:setup}
%!TEX root = /Users/xdhaultfoeuille/Dropbox/Cluster/paper/main.tex

In this section, we define and discuss the restrictions we impose on the data generating process, and the parameters of interest. We suppose to have $k$ non-nested partitions of the population, which correspond to the different dimensions of clustering. We denote the index of the first dimension of clustering (e.g. sector of activity) by $j_1$, the second (e.g. area of residence) by $j_2$ etc. Hereafter, the intersection of $k$ given clusters in the different dimensions (e.g., the second sector of activity and the third area of residence if $(j_1,j_2)=(2,3)$) is called a cell. Cells are indexed by the $k$-tuple $\j=(j_1,...,j_k)$ for $j_i=1,...,C_i$, where  $C_i$ denotes the number of clusters in the sample for dimension $i$. With $k=2$, cells may be seen as matrix entries where the dimensions of clustering would be rows and columns. With $k>2$, cells correspond to the entries of a multidimensional array.  We let $\j\geq \j'$ to mean that $j_i\geq j'_i$ for all $i=1,...,k$. In the following, we let $\un=(1,...,1)$ and $\C=(C_1,...,C_k)$. The number of observations within each cell is denoted by $N_{\j}$. The random vector corresponding to unit $\ell=1,...,N_{\j}$ in cell $\j$ (with $\un\leq \j\leq \C$) is then denoted $Y_{\ell,\j}$, with $Y_{\ell,\j}\in \mathcal{Y} \subset \R^l$.  

\medskip
The key assumptions of this $k$-way clustering are the following. First, the sequences $(N_{\j},(Y_{\ell,\j})_{\ell \geq 1})$ are identically distributed, but not necessarily independent across $\j$, since two cells with at least one common cluster may face common shocks. Second, $(N_{\j}, (Y_{\ell,\j})_{\ell\geq 1})$
 and $(N_{\j'}, (Y_{\ell,\j'})_{\ell\geq 1})$ are independent if $j_i\neq j'_i$ for all $i=1,...,k$. Third, we consider a sample $(Y_{1,\j},...,Y_{N_{\j},\j})_{\un\leq \j\leq \C}$ where $\underline{C}=\min_{i\in\{1,...,k\}}C_i$ tends to infinity. Assumption \ref{as:dgp} formalizes all these conditions.

\begin{hyp}\label{as:dgp}
~
	\begin{enumerate}
%		\item $(U_{j_1}^1)_{j_1\geq 1}$,..., $(U_{j_k}^k)_{j_k\geq 1}$ are mutually independent random sequences;
		\item The array $(N_{\j}, (Y_{\ell,\j})_{\ell\geq 1})_{\j\geq \un}$ is separately exchangeable. Namely, for any $(\pi_1,...,\pi_k)$ $k$-tuple of permutations of $\mathbb{N}$,
		$$(N_{\j},(Y_{\ell,\j})_{\ell\geq 1})_{\j\geq \un}\overset{d}{=}(N_{\pi_1(j_1),...,\pi_k(j_k)},(Y_{\ell,\pi_1(j_1),...,\pi_k(j_k)})_{\ell\geq 1})_{\j\geq \un}.$$
		\item For any $\c\geq \un$, $(N_{\j}, (Y_{\ell,\j})_{\ell\geq 1})_{\un \leq \j \leq \c}$ is independent of $(N_{\j'}, (Y_{\ell,\j'})_{\ell\geq 1})_{\j'\geq \c+\un}$. 
		\item $E(N_{\un})>0$.
		\item The econometrician observes $(N_{\j},(Y_{\ell,\j})_{1\leq \ell \leq N_{\j}})_{\un \leq \j \leq \C}$, with $\underline{C}\rightarrow \infty$ and for all $i=1...k$, $\underline{C}/C_i\rightarrow \lambda_i\geq 0$.
%		\item For every $\j$, the $(Y_{\ell,\j})_{\ell\geq 1}$ are identically distributed.
	\end{enumerate}
\end{hyp}

To better understand Assumptions \ref{as:dgp}.1 and \ref{as:dgp}.2, consider first for simplicity two-way clustering with $N_{\j}=1$ almost surely. The data can then be depicted as follows.

\begin{figure}[H]%C{2.0cm}
$$\begin{array}{|c|c|c|c|c|} \hline
	&1&2& \cdots &C_2\\
	\hline
1& Y_{1,(1,1)} & Y_{1,(1,2)} & \cdots & Y_{1, (1,C_2)} \\	
\hline
2& Y_{1,(2,1)}& Y_{1,(2,2)}& \cdots & Y_{1,(2,C_2)} \\
\hline
\vdots & \vdots & \vdots & \vdots & \vdots \\
\hline
C_1& Y_{1,(C_1,1)} & Y_{1,(C_1,2)}& \cdots & Y_{1,(C_1,C_2)} \\
\hline
\end{array}$$
\end{figure}

Assumption \ref{as:dgp}.1 imposes for instance that $(Y_{1,(1,1)}, Y_{1,(1,2)})$ has the same distribution as $(Y_{1,(2,1)}$, $Y_{1,(2,2)})$. More generally, data of all rows, or data of all columns, are assumed to have the same distribution. Another way to state this is that the DGP is invariant by a relabelling of each dimension of clustering. This assumption is natural in many settings, with the notable exception of time series. Importantly, Assumption \ref{as:dgp}.1 does not impose that $(Y_{1,(1,1)},Y_{1,(1,2)})$ has the same distribution as $(Y_{1,(1,1)}, Y_{1,(2,2)})$. This would indeed amount to neglecting possible dependence within a specific row.  

\medskip
Assumption \ref{as:dgp}.2 imposes that any two blocks on the diagonal that do not overlap are independent. In particular $Y_{1,(1,1)}$ and $Y_{1,(2,2)}$ are assumed independent, contrary to, e.g., $Y_{1,(1,1)}$ and $Y_{1,(1,2)}$.\footnote{To be precise, Assumption \ref{as:dgp}.2 remains silent on the joint distribution of cells sharing at least one cluster: they may or may not be independent. Thus, i.i.d. sampling of cells is compatible with Assumption \ref{as:dgp}.} When combined with Assumption \ref{as:dgp}.1, it also implies that cells sharing no rows and columns are mutually independent, since they have the same distribution as cells on the diagonal, which themselves are mutually independent (by applying repeatedly Assumption \ref{as:dgp}.2).

%We impose no restrictions on $N_{j_1,j_2}$, except that its distribution is assumed independent of $\underline{C}$.

\medskip
Let us come back to the general case with possibly $N_{\j}\neq 1$. Assumption \ref{as:dgp}.2 does not impose any restriction on the distribution of $(N_{\j}, (Y_{\ell,\j})_{\ell\geq 1})$. Hence, the dependence between $N_{\j}$ and the $(Y_{\ell,\j})_{\ell\geq 1}$, and the dependence between the $(Y_{\ell,\j})_{\ell\geq 1}$ within cell $\j$, are left unrestricted. This implies for instance that conditional on $N_{\j}$, the correlation between $Y_{\ell,\j}$ and $Y_{\ell',j}$ may vary with $N_{\j}$. In this sense, we allow for cluster heterogeneity, as defined by \cite{carter2017}. Also, $Y_{\ell,\j}$ may have a different distribution from $Y_{\ell',\j}$, for $\ell\neq \ell'$. 

\medskip
Assumption \ref{as:dgp}.3 only excludes arrays that are almost surely empty. Assumption \ref{as:dgp}.4 states that only  the $N_{\j}$ first units in each cell $\j$ are observed. It also specifies our asymptotic framework, in which all dimensions of the array grow large. The condition that $\underline{C}/C_i$ tends to $\lambda_i\geq 0$ is very mild since it allows for different rates of convergence along the different dimensions of clustering.  

\medskip
Whereas the data generating process is defined at the cell level, parameters of interest are virtually always defined at the unit level. To see this, consider again the example of wages. When considering average wages, one usually focuses on units (e.g., individuals) rather than cells, which means that the parameter of interest satisfies
\begin{equation}
\theta_0 =\frac{\E\left(\sum_{\ell=1}^{N_{\un}}Y_{\ell,\un}\right)}{\E(N_{\un})}.	
	\label{eq:expectation}
\end{equation}
This definition differs from $\E\left(Y_{\ell,\un}\right)$ or its symmetrized version $$\theta_{0,w}=\E\left(\frac{1}{N_{\un}} \sum_{\ell=1}^{N_{\un}}Y_{\ell,\un}\right),$$ 
which may seem more natural.\footnote{Note though that the three parameters coincide when $N_{\un}=1$ or when $N_{\un}$ is independent of $(Y_{\ell,\un})_{\ell \geq \un}$} But $\theta_0$ is actually the right parameter of interest if one wants to weight equally each individual, rather than weighting equally each cell (e.g., each sector $\times$ area of residence). In the latter case, we would put more weight on individuals lying in small cells. The plug-in estimator of $\theta_0$ corresponds to the empirical mean at the individual level:
\begin{equation}
\widehat{\theta}= \frac{\sum_{\un \leq \j \leq \C}\sum_{\ell=1}^{N_{\j}}Y_{\ell,\j}}{\sum_{\un \leq \j \leq \C}N_{\j}}= \frac{\frac{1}{\Pi_C}\sum_{\un \leq \j \leq \C} \sum_{\ell=1}^{N_{\j}}Y_{\ell,\j}}{\frac{1}{\Pi_C}\sum_{\un \leq \j \leq \C} N_{\j}},	
	\label{eq:sample_avg}
\end{equation}
where $\Pi_C=\prod_{i=1}^k C_i$ denotes the total number of cells. We study inference on $\theta_0$ based on $\widehat{\theta}$ in Section \ref{sub:sample_averages} below.

\medskip
More generally, we consider parameters of interest that depend on the unit-level distribution of $Y$, defined by 
\begin{equation}
F_Y(y) = \frac{\E\left(\sum_{\ell=1}^{N_{\un}}\mathds{1}\{Y_{\ell,\un}\leq y\}\right)}{\E(N_{\un})}.	
	\label{eq:def_F}
\end{equation}
This implies that the median of wages at the individual level is defined by $\theta_0=F_Y^{-1}(1/2)$. We consider smooth functionals of $F_Y$ in Section \ref{sub:nonlinear_functionals_of_the_distribution} below.

\medskip
Finally, we consider in Section \ref{sub:gmm_estimators} moment restrictions at the unit level, rather than at the cell level. Namely, we consider a parameter of interest  $\theta_0\in \Theta$ satisfying 
\begin{equation}
\E\left(\sum_{\ell=1}^{N_{\un}}m(Y_{\ell,\un},\theta_0)\right)=0,	
	\label{eq:GMM}
\end{equation}
for a vector-valued function $m(y,\theta)$. The average parameter defined by \eqref{eq:expectation} is a particular case of \eqref{eq:GMM}, with $m(Y_{\ell,\un},\theta)=Y_{\ell,\un}-\theta$. This GMM framework also encompasses linear models and pseudo maximum likelihood estimators of nonlinear models such as logit or probit models. These latter estimators are covered by taking $m$ as the score of the model. Such estimators are not usual maximum likelihood estimators, since they ignore potential correlations between observations within cells, and between cells sharing at least one cluster. 

\medskip
We establish below that in the three cases above, namely expectations, smooth functionals of $F_Y$ and GMM, the corresponding estimators are asymptotically normal. We also develop valid inference on the corresponding estimands. To establish such results, we first study in Section \ref{sec:gen_results} the asymptotic behavior of empirical processes and their bootstrap counterparts. %These theorical results imply also new practical recommendations.

\section{Weak convergence results}
\label{sec:gen_results}

\subsection{Empirical processes}%Weak convergence of empirical processes with clustering}
\label{sec:emp_process}
Let $\mathcal{F}$ denote a class of real-valued functions. %such that $\E\left[\left(\sum_{\ell=1}^{N_{\un}}f(Y_{\ell,\un})\right)^2\right]<\infty$ for any $f\in \mathcal{F}$. 
In this section, we study the empirical process $\mathbb{G}_{C}$ defined on $\mathcal{F}$ by
$$\mathbb{G}_{C}f=\sqrt{\underline{C}}\left\{\frac{1}{\Pi_C}\sum_{\un \leq \j \leq \C}\sum_{\ell=1}^{N_{\j}} f(Y_{\ell,\j}) - \E\left[\sum_{\ell=1}^{N_{\un}}f(Y_{1,\un})\right]\right\}.$$
Specifically, we prove that under restrictions on $\mathcal{F}$,  $\mathbb{G}_{C}$ converges weakly to a Gaussian process as  $\underline{C}$ tends to infinity. While we  refer to, e.g., \cite{vanderVaartWellner1996} for a formal definition of weak convergence of empirical processes, we recall that this result is stronger than  pointwise asymptotic normality of $\mathbb{G}_{C}f$. Our result below will therefore entail central limit theorems for means of the form
$$\frac{1}{\Pi_C} \sum_{\un \leq \j \leq \C}\sum_{\ell=1}^{N_{\j}} f(Y_{\ell,\j}),$$
and therefore, by the delta method (considering $f(y)=y$ and $f(y)=1$), for sample averages defined by \eqref{eq:sample_avg}. But such a result is not sufficient for the asymptotic normality of, e.g.,  smooth functionals of the empirical cdf or GMM estimators. Convergence of the whole process, on the other hand, allows one to establish such results. To establish this convergence, we cannot apply standard results on the empirical process for two reasons. First, the different cells are potentially dependent rather than i.i.d. Second, even if they were i.i.d., we do not consider the usual empirical process at the cell-level, because the class of functions is defined at the unit level and we sum over a random number of units within each cell.

%Second, even if this were the case, we do not consider the usual empirical process over cells, because we also sum over a random number of units within each cell.

\medskip
Before giving our main asymptotic result on $\mathbb{G}_{C}$, we introduce additional notation related to a generic class $\mathcal{G}$. An envelope of $\mathcal{G}$ is a measurable function $G$ satisfying $G(u)\geq\sup_{f\in \mathcal{G}}|f(u)|$. For any $\varepsilon>0$ and any norm $||.||$ on a space containing $\mathcal{G}$, $N(\varepsilon,\mathcal{G},||.||)$ denotes the minimal number of $||.||$-closed balls of radius $\varepsilon$ with centers in $\mathcal{G}$ needed to cover $\mathcal{G}$.\footnote{With a slight abuse of language, we use here the term norms in lieu of seminorms. For instance, Assumption \ref{as:vc} involves seminorms rather than norms. Also, we use $\|.\|$ for (semi)norms on functions and $|.|$ for norms on finite-dimensional objects. Specifically, for any vector $b$, $|b|$ denotes the Euclidean norm of $b$; and for any matrix $A$, $|A|$ denotes the Frobenius norm of $A$.} The norms we consider hereafter are $\|f\|_{\mu,r}=(\int |f|^rd\mu)^{1/r}$ for any $r\geq 1$ and probability measure $\mu$. Finally, a class of measurable functions $\mathcal{G}$ is pointwise measurable if there exists a countable subclass $\mathcal{H}\subset \mathcal{G}$ such that elements of $\mathcal{G}$ are pointwise limit of elements of $\mathcal{H}$.

\medskip
We consider the following standard assumptions on the class $\mathcal{F}$ indexing $\mathbb{G}_C$.

\begin{hyp}\label{as:measurability}
$\mathcal{F}$ is a pointwise measurable class of functions.
\end{hyp}

\begin{hyp}\label{as:vc}
The class $\mathcal{F}$ admits an envelope $F$ with either:\\
- $\E\left[\left(\sum_{\ell=1}^{N_{\un}} F\left(Y_{\ell,\un}\right)\right)^2\right]<+\infty$ and $\mathcal{F}$ is finite; \\
- or $\E\left[N_{\un}^2\right]<+\infty$, $\E\left[N_{\un}\sum_{\ell=1}^{N_{\un}} F\left(Y_{\ell,\un}\right)^2\right]<+\infty$ and
\begin{equation*}
    \int_0^{+\infty}\sup_Q\sqrt{\log N\left(\varepsilon||F||_{Q,2},\mathcal{F},||.||_{Q,2}\right)}d\varepsilon<+\infty,
\end{equation*}
where the supremum is taken over the set of probability measures with finite support on $\mathcal{Y}$.
\end{hyp}

Assumption \ref{as:measurability} is not necessary but usually imposed \cite[see, e.g.][]{cherno2014,Kato2016} to avoid measurability issues and the use of outer expectations. For further discussion about these classes, we refer to \citeauthor{Kosorok2006} (\citeyear{Kosorok2006}, pp.137-140). Assumption \ref{as:vc} imposes a condition on what is usually referred to as the uniform entropy integral, see, e.g., \cite{vanderVaartWellner1996}. Finiteness of the uniform entropy integral is satisfied by any VC-type class of functions \citep[see][for a definition]{cherno2014}, or by the convex hull of such classes under some restrictions. These conditions are nearly the same as those used with i.i.d. data. The only difference lies in the moment conditions. When $\mathcal{F}$ is finite, we require a second moment condition that is the exact analog of the moment condition for usual central limit theorems. When $\mathcal{F}$ is infinite, on the other hand, we require the slightly stronger condition $\E\left[N_{\un}^2\right]<+\infty$ and $\E\left[\left(N_1 \sum_{\ell=1}^{N_{\un}} F^2\left(Y_{\ell,\un}\right)\right)\right]<+\infty$. Note however that the two conditions are equivalent whenever $N_{\un}$ is bounded.

\begin{thm} \label{thm:unifTCL}
Suppose that Assumptions~\ref{as:dgp}-\ref{as:vc} hold. Then the process $\mathbb{G}_C$ converges weakly to a centered Gaussian process $\mathbb{G}$ on $\mathcal{F}$ as $\underline{C}$ tends to infinity. Moreover, the covariance kernel $K$ of $\mathbb{G}$ satisfies:
$$K(f_1,f_2) = \sum_{i=1}^{k}\lambda_i \Cov\left(\sum_{\ell=1}^{N_{\un}}f_1(Y_{\ell,\un}), \sum_{\ell=1}^{N_{\deux}}f_2(Y_{\ell,\deux})\right),$$
where $\deux$ is the $k$-tuple with $2$ in each entry but 1 in entry $i$.
\end{thm}

Theorem \ref{thm:unifTCL} shows the weak convergence of $\mathbb{G}_C$ towards a centered gaussian process $\mathbb{G}$, and gives the form of the covariance kernel of $\mathbb{G}$. The result holds under Assumption \ref{as:vc}, but this is not the only possible restriction on the class of functions. In Appendix \ref{app:smoothness_class}, we show the same result under smoothness restrictions on $\mathcal{F}$ instead of Assumption \ref{as:vc}.

\medskip
Let us summarize the proof of Theorem \ref{thm:unifTCL}. Weak convergence of $\G_C$ holds under two main conditions. First, $(\G_Cf_1,...,\G_C f_m)$ should be asymptotically normal for any $(f_1,...,f_m)$ in $\mathcal{F}$ and any $m\geq 1$. Second, one should establish asymptotic equicontinuity. Regarding finite-dimensional convergence, we proceed in several steps. To simplify the discussion, we consider here the case where $m=1$ and two-way clustering. We first exploit the Aldous-Hoover representation \citep[][for the almost-sure version]{Aldous1981,Hoover1979,kallenberg05}, which extends de Finetti's theorem to separately exchangeable random sequences. This result ensures the existence of mutually independent variables  $(U_{j_1, 0}, U_{0,j_2}, U_{\j})_{j_1\geq 1, j_2\geq 1, \j\geq \un}$ such that for all $\j$,
\begin{equation}
(N_{\j}, (Y_{\ell, j})_{\ell\geq 1})= \tau(U_{j_1, 0},U_{0,j_2}, U_{\j}).	
	\label{eq:representation}
\end{equation}
The variable $U_{j_1,0}$ (resp. $U_{0, j_i}$) may be seen as a shock specific to cluster 1 (resp. 2), while $U_{\j}$ can be interpreted as a shock specific to Cell $\j$.\footnote{With more than two dimensions of clustering, the representation is similar but we have to include shocks specific to each subset of $(j_1,...,j_k)$. For instance, with $k=3$, we have also to consider shocks such as $U_{j_1,j_2,0}$.} %Our representation result is based on the Aldous-Hoover representation theorem , which itself extends de Finetti's theorem to separately exchangeable random sequences. This representation has been exploited by quoted authors to establish some ergodicity results ensuring the convergence in $L^1$ and almost-surely\citep[see also][for a recent reference]{kallenberg05}. %We can improve the Aldous-Hoover representation in our context, using the fact that cells sharing no common clusters are independent.

\medskip
In the second step, we consider the H\'ajek projection of $\mathbb{G}_C f_1$ on the set $\mathscr{S}$ of random variables depending only on the marginal cluster specific factors, namely
$$\mathscr{S}=\left\{\sum_{j_1=1}^{C_1} g_{j_1,0}(U_{j_1,0}) + \sum_{j_2=1}^{C_2} g_{0,j_2}(U_{0,j_2}), \; g_{j_1,0}\in L^2(U_{j_1,0}), g_{0,j_2}\in L^2(U_{0,j_2})\right\}.$$
We prove that $\mathbb{G}_Cf_1$ gets close, in a $L^2$ sense, to its H\'ajek projection as $\underline{C}\rightarrow\infty$.  Asymptotic normality then follows by a simple CLT on the H\'ajek projection.

\medskip
To complete the proof of the theorem, we have to establish asymptotic equicontinuity. Roughly speaking, this means that whenever $f_1$ and $f_2$ are close to each other, $\G_Cf_1-\G_Cf_2$ is close to zero \citep[see, e.g.,][Section 2.1.2, for a formal definition]{vanderVaartWellner1996}. For that purpose, we prove a symmetrization lemma similar to Lemma 2.3.1 in \cite{vanderVaartWellner1996}. To do so, we adapt arguments used in the proofs of Theorem 3.1 in \cite{ArconesGine1993} where independent copies of random variables are introduced to control U-statistics. Following this idea, we introduce independent copies of the $(U_{\j})_{\j>0}$ that come from the Aldous-Hoover representation. By the symmetrization lemma, we can then bound fluctuations of $\G_C$ by a function of the entropy of the class $$\widetilde{\mathcal{F}}=\left\{g(n,y_1,...,y_n)=\sum_{i=1}^{n}f(y_i): n\in \mathbb{N}, (y_1,...,y_n) \in \mathcal{Y}^{n}; f\in \mathcal{F} \right\}.$$
Note that this class is related to, but different from $\mathcal{F}$. We have defined the class of function $\mathcal{F}$ at the unit (e.g., individual) level because parameters of interest are defined at this level. But the stochastic model in Assumption \ref{as:dgp} is stated at the cell level, which explains why, intuively, we need to control the complexity of $\widetilde{\mathcal{F}}$. We show that this is possible under Assumption \ref{as:vc}. By what precedes, this implies the asymptotic equicontinuity of $\G_C$.

%Assumption \ref{as:vc} imposes some restrictions on the statistical model at the unit level and allows us to derive consistent estimators. But as with one-way clustering, the stochastic restrictions used in Assumption \ref{as:dgp} are stated at the cell level and allowed to derive the asymptotic variance of the estimator. So, to show the asymptotic equicontinuity, we link the complexity of functional classes defined at unit level with the complexity of their aggregation at cell-level to control the weak limit of the process.

%We then use Assumption \ref{as:vc} to control this entropy and prove asymptotic equicontinuity. Note that Assumption \ref{as:vc} imposes some restrictions on the statistical model at the unit level and allows us to derive consistent estimators. But as with one-way clustering, the stochastic restrictions used in Assumption \ref{as:dgp} are stated at the cell level and allowed to derive the asymptotic variance of the estimator. So, to show the asymptotic equicontinuity, we link the complexity of functional classes defined at unit level with the complexity of their aggregation at cell-level to control the weak limit of the process.

\medskip
We now comment on the asymptotic kernel $K$ of $\G_C$. For simplicity, let $f_1(y)=f_2(y)=y$ and define $S_{\j}=\sum_{\ell=1}^{N_{\un}} Y_{\ell,\j}$. Theorem \ref{thm:unifTCL} implies that the asymptotic variance of $\sum_{\un\leq \j\leq \C} S_{\j}/\Pi_C$ is  
\begin{equation}
\sum_{i=1}^k \lambda_i \Cov\left(S_{\un}, S_{\deux}\right).	
	\label{eq:formula_variance}
\end{equation}
This formula may seem surprising, because it is not obvious at first glance that it is positive. But it turns out that under Assumption \ref{as:dgp}, each covariance term is positive. Specifically, and considering for simplicity $k=2$, we establish in the proof of Theorem \ref{thm:unifTCL} that
$$\Cov\left(S_{\un}, S_{\bm{2}_1}\right) = \V\left(\E\left(\sum_{\ell=1}^{N_{\un}}f(Y_{\ell,\un})\bigg| U_{\un, 0}\right)\right),\; \Cov\left(S_{\un}, S_{\bm{2}_2}\right) = \V\left(\E\left(\sum_{\ell=1}^{N_{\un}}f(Y_{\ell,\un})\bigg| U_{0, \un}\right)\right),$$
where $U_{\un, 0}$ and $U_{0,\un}$ appear in the representation \eqref{eq:representation}.

\medskip
Now, let us give some intuitions on \eqref{eq:formula_variance}. This formula involves cells sharing exactly one common cluster, namely cluster 1 in dimension $i$. To better understand why only such terms appear, consider
$$\V\left(\frac{\sqrt{\underline{C}}}{\Pi_C}\sum_{\un \leq \j \leq \C} S_{\j}\right).$$
This variance is complicated because of the particular dependence structure due to multiway clustering. To simplify it, we can write it as the sum of covariances between cells sharing no common cluster, cells sharing one common cluster... and finally the covariances of cells with themselves. The number of pairs of cells sharing no common cluster is $\Pi_C \times \prod_{i=1}^k (C_i-1)$, which is of the order $\Pi^2_C$ as $\underline{C}$ tends to infinity. The number of pairs of cells sharing one common cluster is $\Pi_C \sum_{i=1}^k \prod_{j\neq i} (C_j-1)$, which is smaller than $k\Pi^2_C/\underline{C}$. Hence, the number of such pairs of cells is negligible compared to the number of pairs of cells sharing no common cluster. Similarly, we can prove that the number of cells sharing more than one common cluster is negligible compared with the number of cells sharing one common cluster. Hence, intuitively, the variance will be equivalent to the sum of only covariances between cells sharing either no or just one common cluster. But by independence, the covariance between cells sharing no common cluster is actually zero. Hence, at the end of the day, we only get covariances between cells sharing just one common cluster.

\subsection{Pigeonhole bootstrap processes}
\label{sub:boot}
%!TEX root = /Users/xdhaultfoeuille/Dropbox/Cluster/paper/main.tex
\label{sec:boot}

We now consider the bootstrap counterpart of the weak convergence result in Theorem \ref{thm:unifTCL}. Bootstrap offers several advantages over usual inference based on asymptotic normality. First, it avoids the computation of theoretical formulas of asymptotic variances, which can be difficult with, e.g., multistep estimators. Second, it often exhibits a better behavior than normal approximations in finite samples. Still, a consistent bootstrap scheme in our clustering setting needs to reproduce the dependence between cells. We consider for that purpose the ``pigeonhole bootstrap'', suggested by \cite{mccullagh2000resampling} and studied, in the case of the sample mean and for particular models, by \cite{owen2007pigeonhole}. We are, however, not aware of any result concerning the asymptotic validity of the pigeonhole bootstrap for inference. Theorem \ref{thm:boot_unif} below aims to fill this gap. 

\medskip
We first recall the principle of the pigeonhole bootstrap:
\begin{enumerate}
\item For each $i\in\{1,...,k\}$, $C_i$ elements are sampled with replacement and equal probability in the set $\{1,...,C_i\}$. For each $j_i$ in this set, let $W^i_{j_i}$ denote the number of times $j_i$ is selected this way. 
\item Cell $\j=(j_1,...,j_k)$ is then selected $W_{\j}=\prod_{i=1}^k W^i_{j_i}$ times in the bootstrap sample. 
\end{enumerate} 

By construction, any bootstrap sample consists of exactly $\Pi_C$ cells. Also, dependence between cells sharing cluster $i$ is achieved through the term $W^i_{j_i}$. Actually, one can check that conditional on the data $(N_{\j},(Y_{\ell,\j})_{\ell\geq 1})_{\un \leq \j \leq \C}$, the bootstrap weights $(W_{\j})_{\un\leq \j\leq \C}$ satisfy the first condition in Assumption \ref{as:dgp}, and the second asymptotically.\footnote{As in the i.i.d. setting where bootstrap weights are asymptotically independent, the weights of cells sharing no common cluster become independent as $\underline{C}\rightarrow\infty$.} This suggests that the pigeonhole bootstrap could be asymptotically valid.

\medskip
We now consider the bootstrap counterpart of the empirical process $\G_C$. For any $f\in\mathcal{F}$, let us define
$$\mathbb{G}^{\ast}_C(f)=\frac{\sqrt{\underline{C}}}{\Pi_C}\sum_{\un \leq \j\leq \C} \left(W^{\C}_{\j}-1\right) \sum_{\ell=1}^{N_{\j}}f(Y_{\ell,\j}).$$

The asymptotic validity of the pigeonhole bootstrap amounts to showing that conditional on the data $\{N_{\j},(Y_{\ell,\j})_{\ell\geq 1}\}_{\j\geq \un}$, $\mathbb{G}_C^{\ast}$ converges weakly in probability to the process $\mathbb{G}$ defined in Theorem \ref{thm:unifTCL}. %Note that we study the behavior of $\mathbb{G}_C^{\ast}$ conditional on the data because the variations we rely on to draw bootstrap-based inference are only due to the randomness of the weights $(W_{\j})_{\un\leq \j\leq \C}$. 
As discussed in, e.g., \citeauthor{vanderVaartWellner1996} (1996, Chapter 3.6), conditional weak convergence in probability amounts to proving
\begin{equation}
\sup_{h\in \text{BL}_1} \left|\E\left(h(\mathbb{G}_C^{\ast})|\{N_{\j},(Y_{\ell,\j})_{\ell\geq 1}\}_{\j\geq \un}\right)-\E\left(h(\mathbb{G})\right)\right|\convP 0,	
	\label{eq:conv_boot_proc}
\end{equation}
where $\text{BL}_1$ is the set of bounded and Lipschitz functions from $\ell^{\infty}(\mathcal{F})$ to $\mathbb{R}$. 

\begin{thm}
Suppose that Assumptions \ref{as:dgp}-\ref{as:vc} hold. Then $\mathbb{G}^{\ast}_{C}$ converges weakly to $\mathbb{G}$ in probability, namely \eqref{eq:conv_boot_proc} holds.
\label{thm:boot_unif}
\end{thm}

As we shall see below, this theorem ensures the asymptotic validity of the pigeonhole bootstrap not only for sample means, but also for smooth functionals of the empirical cdf and GMM estimators. The proof of Theorem \ref{thm:boot_unif} follows the same lines as that of Theorem \ref{thm:unifTCL} to get weak convergence of the bootstrap process conditionally on the original data. To ensure
the unconditional boostrap consistency we also use some ergodicity arguments \citep{kallenberg05} and we prove Lindeberg-Feller conditions for some statistics defined on the exchangeable array.

\section{Applications}

\label{sec:appli}

\subsection{Simple averages and linear models} % (fold)
\label{sub:sample_averages}

%!TEX root = /Users/xdhaultfoeuille/Dropbox/Cluster/paper/main.tex
As before, let $S_{\j}=\sum_{\ell=1}^{N_{\j}} Y_{\ell,\j}$. We first investigate here how inference can be conducted on $\theta_0=E(S_{\un})$ based on the plug-in estimator
\begin{equation}
\widehat{\theta}=\frac{1}{\Pi_C}\sum_{\un \leq \j \leq \C}S_{\j}.	
	\label{eq:def_theta_hat}
\end{equation}
We focus first on $\widehat{\theta}$ for simplicity but show at the end of the section how our reasoning extends to sample averages as defined by \eqref{eq:sample_avg} and linear models. Provided that $E(S_{\j}^2)<+\infty$, we have, by Theorem \ref{thm:unifTCL},
\begin{equation}
\sqrt{\underline{C}}\left(\widehat{\theta}-\theta_0\right) \convD \mathcal{N}\left\{0,\sum_{i=1}^{k}\lambda_i \Cov\left(S_{\un}, S_{\deux}\right)\right\}.	
	\label{eq:TCL_simple}
\end{equation}
A first strategy to make inference on $\theta_0$ is therefore to use the normal approximation and a consistent estimator of the asymptotic variance. A second strategy is to rely on the pigeonhole bootstrap.\footnote{A third strategy is to rely on other bootstrap schemes. We refer to \cite{menzel2017} for the construction and analysis of a wild bootstrap procedure for sample averages on such clustered data.} 

\medskip
First, let us consider inference based on asymptotic normality. The asymptotic variance $V$ depends on $\lambda_i$ and $\Cov\left(S_{\un}, S_{\deux}\right)=\E\left((S_{\un}-\theta_0)(S_{\deux}-\theta_0)\right)$, for $i=1,...,k$.  $\lambda_i$ can simply be approximated by $\underline{C}/C_i$. Regarding the covariance term, observe that $\un$ and $\deux$ share exactly one cluster. It is then natural to consider the estimator
$$\widehat\Cov\left(S_{\un}, S_{\deux}\right) = \frac{1}{C_i\prod_{s\neq i}C_s(C_s-1)}\sum_{(\j,\j')\in\mathcal{A}_i}\left(S_{\j}-\widehat{\theta}\right)\left(S_{\j'}-\widehat{\theta}\right)',$$
where $\mathcal{A}_i:=\left\{(\j,\j'):j_i=j_i',\quad j_s\neq j_s'\quad\forall s\neq i\right\}$. This estimator is the average of cross products between clusters sharing just one common cluster, the denominator $C_i\prod_{s\neq i}C_s(C_s-1)$ corresponding to the number of such pairs. This leads to the following estimator for $V$:
\begin{equation}
	\label{eq:Vch2}
\widehat{V}_2 = \sum_{i=1}^k \frac{\underline{C}}{C_i} \frac{1}{C_i\prod_{s\neq i}C_s(C_s-1)}\sum_{(\j,\j')\in\mathcal{A}_i}\left(S_{\j}-\widehat{\theta}\right)\left(S_{\j'}-\widehat{\theta}\right)'.	
\end{equation}
We will show that $\widehat{V}_2$ is consistent for $V$. A major drawback of this estimator, however, is that it is not necessarily positive. Also, $V$ is the variance of a H\'ajek projection, as explained above. As such, it is likely to underestimate $\V(\sqrt{\underline{C}}\widehat{\theta})$. Because $\widehat\Cov\left(S_{\un}, S_{\deux}\right)$ itself slightly underestimates $\Cov\left(S_{\un}, S_{\deux}\right)$ ($\E[\widehat\Cov\left(S_{\un}, S_{\deux}\right)] = \Cov\left(S_{\un}, S_{\deux}\right) - \V(\widehat{\theta})$), we can expect the corresponding confidence regions to undercover in practice. This intuition is confirmed in our simulations below.

\medskip
To avoid these issues, we suggest to simply add to $\widehat{V}_2$ pairs sharing more than one cluster. Specifically, we consider
\begin{align}
\widehat{V}_1 & = \sum_{i=1}^k\frac{\underline{C}}{C_i} \frac{1}{C_i\prod_{s\neq i}C_s^2}\sum_{(\j,\j'): \j_i=\j'_i} \left(S_{\j}-\widehat{\theta}\right) \left(S_{\j'}-\widehat{\theta}\right)' \nonumber \\
& = \sum_{i=1}^k\frac{\underline{C}}{C_i} \frac{1}{C_i}\sum_{j'_i=1}^{C_i} \left(\frac{1}{\prod_{s\neq i}C_s}\sum_{\j:j_i=j'_i} S_{\j}-\widehat{\theta}\right)
\left(\frac{1}{\prod_{s\neq i}C_s}\sum_{\j:j_i=j'_i} S_{\j}-\widehat{\theta}\right)'. \label{eq:V1_positive}
\end{align}
From an asymptotic point of view, the additional terms in $\widehat{V}_1$ correspond to pairs sharing more than one cluster. We show in the proof of Proposition \ref{prop:as_var} below that such terms are negligible, implying that $\widehat{V}_1$ is consistent, just as $\widehat{V}_2$. In finite samples, on the other hand, $\widehat{V}_1$ has several advantages over $\widehat{V}_2$. First, $\widehat{V}_1$ is positive, as \eqref{eq:V1_positive} shows. Also, it is likely to overestimate $V$, but this may somewhat compensate the fact that $V$ itself underestimates $\V(\sqrt{\underline{C}}\widehat{\theta})$. And indeed, in the simulations considered in Section \ref{sec:MC} below, inference is more accurate when using  $\widehat{V}_1$ rather than $\widehat{V}_2$, in particular when $\underline{C}$ is small. A last advantage is computational. Equation \eqref{eq:CGM} shows that we can compute this estimator using variance estimators of $\widehat{\theta}$ assuming only one-way clustering along dimensions $i\in\{1,...,k\}$, and then summing these different variances. $\widehat{V}_2$, on the other hand, cannot be obtained as easily. For all these reasons, we recommend using  $\widehat{V}_1$ rather than $\widehat{V}_2$ in practice.

\medskip
We now compare our two estimators with that proposed by \cite{cameron2011}. Their estimator relies on a reformulation of $\V(\widehat{\theta})$. For any $m\in\{1,...,k\}$ and $1\leq i_1<...<i_m\leq k$, let $\mathcal{B}_{i_1,...,i_m}=\left\{(\j,\j'):j_{i_1}=j'_{i_1},...,j_{i_m}=j'_{i_m}\right\}$. Then
\begin{align*}
\V(\widehat{\theta}) & = \frac{1}{\Pi_C^2} \sum_{(\j, \j') \in \cup_{i=1}^k \mathcal{B}_i} \Cov\left(S_{\j}, S_{\j'}\right) \\
& = \sum_{m=1}^k (-1)^m \sum_{1\leq i_1 < ... < i_m \leq k} \frac{1}{\Pi_C^2} \sum_{(\j, \j')\in \mathcal{B}_{i_1,...,i_m}} \Cov\left(S_{\j}, S_{\j'}\right).
\end{align*}
The first line follows because if $(\j,\j')$ share no cluster, $\Cov\left(S_{\j}, S_{\j'}\right)=0$. The second line follows from the inclusion-exclusion principle. This leads to the following estimator for the asymptotic variance of $\widehat{\theta}$
\begin{equation}
\widehat{V}_{\text{cgm}} = \underline{C} \sum_{m=1}^k  (-1)^m \sum_{1\leq i_1 < ... < i_m \leq k} \frac{1}{\Pi^2_C} \sum_{(\j, \j')\in \mathcal{B}_{i_1,...,i_m}} \left(S_{\j}-\widehat{\theta}\right) \left(S_{\j'}-\widehat{\theta}\right)'.	
	\label{eq:CGM}
\end{equation}
We can consider various finite sample adjustments where $1/(\Pi_C)^2$ is replaced by $c_{i_1,...,i_m}/(\Pi_C)^2$, with $c_{i_1,...,i_m}$ tending to one as $\underline{C}$ tends to infinity. We refer to \cite{cameron2011} for more details. As with $\widehat{V}_1$, the appeal of Formula \eqref{eq:CGM} is that we can compute this estimator using variance estimators of $\widehat{\theta}$ assuming only one-way clustering along dimensions $(i_1,...,i_m)$, for all $1\leq i_1 < ... < i_m \leq k$. The estimator $\widehat{V}_{\text{cgm}}$ is still slightly more complicated to compute than $\widehat{V}_1$, as the latter only requires the computation of one-way clustering variances along dimensions $i=1,...,k$.

\medskip
To further understand the links and differences between $\widehat{V}_1$ and $\widehat{V}_{\text{cgm}}$, it is instructive to consider the case $k=2$. Then the formulas simplify to
\begin{align}
	\widehat{V}_1 & = \underline{C} \sum_{i_1=1}^2 \frac{1}{\Pi^2_C} \sum_{(\j, \j')\in \mathcal{B}_{i_1}} \left(S_{\j}-\widehat{\theta}\right) \left(S_{\j'}-\widehat{\theta}\right)', \nonumber \\
	\widehat{V}_{\text{cgm}} & = \widehat{V}_1 - \frac{\underline{C}}{\Pi^2_C} \sum_{\un \leq \j \leq \C} \left(S_{\j}-\widehat{\theta}\right) \left(S_{\j}-\widehat{\theta}\right)'.\label{eq:compar_V1_Vcgm}
\end{align}
In other words, $\widehat{V}_1$ estimates $V$ by counting twice the pairs $(\j,\j')$ sharing two clusters, or equivalently the pairs $(\j,\j)$, $\un \leq \j\leq \C$. $\widehat{V}_{\text{cgm}}$ counts such pairs only once, whence the correction in \eqref{eq:compar_V1_Vcgm}. The cost of this correction is that $\widehat{V}_{\text{cgm}}$ is not always positive. Finally, note that there are only $\Pi_C$ pairs $(\j,\j)$. Thus, the second term in \eqref{eq:compar_V1_Vcgm} is of order $\underline{C}/\Pi_C$ and tends to 0 as $\underline{C}$ tends to infinity. We can therefore expect $\widehat{V}_1$ and $\widehat{V}_{\text{cgm}}$ to be asymptotically equivalent.

\medskip
Finally, for any $k\in \{1,2,\text{cgm}\}$ and $\alpha\in (0,1)$, we consider confidence regions $R_k^{1-\alpha}$ for $\theta_0$ defined by
$$R_k^{1-\alpha}= \left\{\theta: \, \underline{C}(\theta-\widehat{\theta})' \widehat{V}_k^{-1} (\theta-\widehat{\theta}) \leq \chi^2_L(1-\alpha)\right\},$$
where $\chi^2_L(1-\alpha)$ is the quantile of order $1-\alpha$ of a $\chi^2_L$ distribution. Proposition \ref{prop:as_var} shows that $\widehat{V}_1$, $\widehat{V}_2$ and $\widehat{V}_{\text{cgm}}$ are all consistent estimators of $V$, implying that the confidence regions are also asymptotically valid, as long as $V$ is positive definite.

\begin{prop}
\label{prop:as_var}
Suppose that Assumption \ref{as:dgp} holds and $\mathbb{E}\left[S_{\un}S_{\un}'\right]<+\infty$. Then $\widehat{V}_1$, $\widehat{V}_2$ and $\widehat{V}_{\text{cgm}}$ are consistent for $V$. Moreover, if $V$ is positive definite, we have, for any $k\in \{1,2,\text{cgm}\}$ and $\alpha\in (0,1)$,
$$\lim_{\underline{C}\rightarrow+\infty} \mathbb{P}\left(R_k^{1-\alpha}\ni\theta_0\right)=1-\alpha$$
\end{prop}

\medskip
The condition that $V$ is positive definite basically states that at least one of the dimension of clustering matters, in the sense that for at least one $i\in\{1,...,k\}$, $\Cov(S_{\un},S_{\deux})$ is positive definite. Note that under Assumption \ref{as:dgp}, $\Cov(S_{\un},S_{\deux})$ is necessarily positive; but it may not be positive definite. For instance, consider two-way clustering and $S_{\j}=U_{j_1,0}+U_{0,j_2}+U_{\j}\in \R$, where the $(U_{j_1,0})_{j_1}$, $(U_{0,j_2})_{j_2}$ and $(U_{\j})_{\j}$ are all mutually independent. Then $\Cov(S_{\un},S_{\deux})>0$ if and only if $\V(U_{j_1,0})+\V(U_{0,j_2})>0$. As discussed in \cite{menzel2017}, $\Cov(S_{\un},S_{\deux})$ may not be positive definite even when  $S_{\un}$ and $S_{\deux}$ are dependent. This is the case if we modify the example above by assuming instead
\begin{equation}
S_{\j}=(U_{j_1,0} - \E(U_{j_1,0})) (U_{0,j_2} - \E(U_{0,j_2}))+U_{\j}.	
	\label{eq:non_normal_example}
\end{equation}
If $V$ is not positive definite, standard tests and confidence regions % , which rely on inverses of estimators of $V$,
are not valid in general. When $V=0$, $\widehat{\theta}$ actually converges at a rate faster than $1/\sqrt{\underline{C}}$ and its asymptotic distribution may be non-normal. This is the case for instance if  \eqref{eq:non_normal_example} holds. We refer to \cite{menzel2017}, Example 1.6, for more details.

\medskip
We now turn to the pigeonhole bootstrap. Let
$$\widehat{\theta}^* = \frac{1}{\Pi_C} \sum_{\un\leq \j\leq \C} W_{\j} S_{\j},$$
where $W_{\j}$ is defined in Section \ref{sub:boot} above. Then let $q_{1-\alpha}^*$ denote the quantile of order $1-\alpha$ of the distribution of $|\widehat{\theta}^*-\widehat{\theta}|$ conditional on the data.  We consider the confidence region $R_{\text{boot}}^{1-\alpha}$ for $\theta_0$ defined by
$$R_{\text{boot}}^{1-\alpha}= \left\{\theta: \, |\widehat{\theta}-\theta|\leq q_{1-\alpha}^*\right\}.$$
The asymptotic validity of $R_{\text{boot}}^{1-\alpha}$ is an immediate consequence of Theorem \ref{thm:boot_unif}.

\begin{prop}
\label{prop:boot_mean}	
Suppose that Assumption \ref{as:dgp} holds, $\mathbb{E}\left[S_{\un}S_{\un}'\right]<+\infty$ and $V$ is positive definite. Then  $$\lim_{n\rightarrow\infty} \mathbb{P}\left(R_{\text{boot}}^{1-\alpha} \ni \theta_0\right) = 1-\alpha.$$
\end{prop}

When  $\theta_0\in \R$, an alternative, popular confidence region is the percentile bootstrap. This amounts to considering $[q_{\alpha/2}(\widehat{\theta}^*), q_{1-\alpha/2}(\widehat{\theta}^*)]$. This interval is also valid asymptotically, since the asymptotic distribution of $\widehat{\theta}-\theta_0$ is normal, and therefore symmetric.

\medskip
We now discuss how Propositions \ref{prop:as_var} and \ref{prop:boot_mean} extend to other parameters of interest. First, let us consider $\theta_0=E(S_{\un})/E(N_{\un})$, as in Section \ref{sec:setup}. Assume that $\E(S_{\un}^2)<\infty$ and $\E(N_{\un}^2)<\infty$. By Theorem \ref{thm:unifTCL} applied to $\mathcal{F}=\{\Id,1\}$ and the delta method, we have 
\begin{equation}
\sqrt{\underline{C}}\left(\frac{\sum_{\un\leq \j\leq \C}S_{\j}}{\sum_{\un\leq \j\leq \C} N_{\j}}-\theta_0\right) = \frac{\sqrt{\underline{C}}}{\Pi_C}\sum_{\un\leq \j\leq \C} T_{\j} +o_p(1).	
	\label{eq:lin_ratio}
\end{equation}
where $T_{\j} = (S_{\j} - N_{\j} \theta_0)/\E(N_{\un})$. We can then estimate the asymptotic variance of the sample average as previously, by simply replacing $S_{\j}-\widehat{\theta}$ in \eqref{eq:Vch2}, \eqref{eq:V1_positive} and \eqref{eq:CGM} by 
$$\widehat{T}_{\j} = \frac{S_{\j} - N_{\j} \widehat{\theta}}{\frac{1}{\Pi_C} \sum_{\un \leq \j\leq \C} N_{\j}}.$$
Consistency follows as in the proof of Proposition \ref{prop:as_var}, using consistency of $\widehat{\theta}$ and $\sum_{\un \leq \j\leq \C} N_{\j}/\Pi_C$. The pigeonhole bootstrap is also valid for $\E(S_{\un})/\E(N_{\un})$ by applying the simple delta method for the bootstrap, see e.g. Theorem 23.5 in \cite{vanderVaart2000}. More generally, Propositions \ref{prop:as_var} and \ref{prop:boot_mean} extend to parameters of the form $g(\theta_{01},...,\theta_{0R})$, where  $\theta_{0r}=\E(\sum_{\ell=1}^{N_{\un}} q_r(N_{\un},Y_{\ell,\un}))$ ($r=1,...,R$), provided that $g$ is continuously differentiable at $(\theta_{01},...,\theta_{0R})$.

\medskip
Finally, let us consider linear models. Then $Y_{\ell,\j}=(\tilde{Y}_{\ell,\j},X'_{\ell,\j})'$, with $\tilde{Y}_{\ell,\j}$ the outcome variable and $X_{\ell,\j}$ a  vector of covariates. Then the parameter of interest $\theta_0$ and its estimator satisfy 
\begin{align}
\theta_0 & = \E\left[\sum_{\ell=1}^{N_{\un}} X_{\ell,\un} X_{\ell,\un}' \right]^{-1} \E\left[\sum_{\ell=1}^{N_{\un}}  X_{\ell,\un} \tilde{Y}_{\ell,\un} \right] \label{eq:def_theta_lin} \\
\widehat{\theta} & = \left(\frac{1}{\Pi_C}\sum_{\un \leq \j \leq \C} X_{\ell,\j} X_{\ell,\j}'  \right)^{-1} \left(\frac{1}{\Pi_C} \sum_{\ell=1}^{N_{\un}} \sum_{\un \leq \j \leq \C} X_{\ell,\j} \tilde{Y}_{\ell,\j}\right). \label{eq:def_thetahat_lin}
\end{align}
We first show that $\widehat{\theta}$ is asymptotically normal and characterize its asymptotic variance. Hereafter, we define $u_{\ell,\j}=\widetilde{Y}_{\ell,\j}-X'_{\ell,\j}\theta_0$.

\begin{prop}\label{prop:linear}
	Suppose that Assumption \ref{as:dgp} holds with $Y_{\ell,\j}=(\tilde{Y}_{\ell,\j},X'_{\ell,\j})'$. Suppose also
$\E\left((\sum_{\ell=1}^{N_{\un}}|Y_{\ell,\un}|^2)^2\right)<+\infty$ and $\E\left(\sum_{\ell=1}^{N_{\un}}X_{\ell,\un} X'_{\ell,\un}\right)$ non-singular. Let $\theta_0$ and $\widehat{\theta}$ be defined by \eqref{eq:def_theta_lin} and \eqref{eq:def_thetahat_lin}. Then
$$\sqrt{\underline{C}}\left(\widehat{\theta}-\theta_0\right)\convD\mathcal{N}(0,V),$$
with $V=J^{-1}HJ^{-1}$, $J=\E\left(\sum_{\ell=1}^{N_{\un}}X_{\ell,\un} X'_{\ell,\un}\right)$ and
$$H=\sum_{i=1}^k\lambda_i \E\left[\left(\sum_{\ell=1}^{N_{\un}}X_{\ell,\un} u_{\ell,\un}\right)\left(\sum_{\ell=1}^{N_{\deux}}u_{\ell,\deux} X'_{\ell,\deux} \right)\right].$$
\end{prop}

Next, we show similar results as in Propositions \ref{prop:as_var} and \ref{prop:boot_mean}. For conciseness, we focus  on an estimator of $V$ similar to $\widehat{V}_1$ rather than $\widehat{V}_2$ and $\widehat{V}_{\text{cgm}}$. Let $\widehat{J}=\frac{1}{\Pi_C}\sum_{\un \leq \j \leq \C}\sum_{\ell=1}^{N_{\j}}X_{\ell,\j} X'_{\ell,\j}$, $\widehat{u}_{\ell,\j}=\widetilde{Y}_{\ell,\j}-X'_{\ell,\j}\widehat{\theta}$ and
$$\widehat{H}=\sum_{i=1}^k\frac{\underline{C}}{C_i} \frac{1}{C_i}\sum_{j'_i=1}^{C_i} \left(\frac{1}{\prod_{s\neq i}C_s}\sum_{\j:j_i=j'_i} \sum_{\ell=1}^{N_{\j}}X_{\ell,\j}\widehat{u}_{\ell,\j}\right)
\left(\frac{1}{\prod_{s\neq i}C_s}\sum_{\j:j_i=j'_i} \sum_{\ell=1}^{N_{\j}} \widehat{u}_{\ell,\j} X'_{\ell,\j}\right)$$
Then let $\widehat{V}=\widehat{J}^{-1}\widehat{H}\widehat{J}^{-1}$. 

\begin{prop}\label{prop:linear2}
	Under the assumptions of Proposition \ref{prop:linear}, $\widehat{V} \convP V$. If $V$ is positive definite, inference based on either asymptotic normality and $\widehat{V}$, or the pigeonhole bootstrap, is valid.
\end{prop}

Propositions \ref{prop:linear} and \ref{prop:linear2} complement the results of \cite{mackinnon2017} by showing asymptotic normality and the validity of two inference methods without assuming that the average of the cell sizes $\overline{N}$ satisfies $\overline{N} \underline{C}^{2/(2+\lambda)} \rightarrow 0$ for some $\lambda>0$. Proposition \ref{prop:linear2} also shows the consistency of a new, positive, variance estimator and the asymptotic validity of the pigeonhole bootstrap in this context of linear models.

% subsection sample_averages (end)

\subsection{Nonlinear functionals of the distribution} % (fold)
\label{sub:nonlinear_functionals_of_the_distribution}

%!TEX root = /Users/xdhaultfoeuille/Dropbox/Cluster/paper/main.tex
Simple central limit theorems and the usual delta method are not sufficient to yield the asymptotic normality of estimators such as the sample median. We now show how the results in Section \ref{sec:gen_results} can be applied to such smooth, nonlinear functionals of the empirical distribution. Let $F_Y$ be defined as in \eqref{eq:def_F} and let $\theta_0=g(F_Y)$. To take examples related to income inequalities (so that here the support of $Y$ is $\R^+$), we may consider for instance quantiles, interquantile ratios and poverty rates, for which we have respectively $g(F_Y)=F_Y^{-1}(\tau)$ for any $\tau\in (0,1)$, $g(F_Y)=F_Y^{-1}(\tau)/F_Y^{-1}(1-\tau)$ for $q\in (1/2,1)$ and $g(F_Y)=F_Y(\alpha F_Y^{-1}(\beta))$ for $(\alpha,\beta)\in (0,1)^2$. Other examples include the Kaplan-Meier functional \citep[see, e.g., Example 20.15 in][]{vanderVaart2000} or the nonlinear difference-in-difference estimand of \cite{Athey06},  for which $\theta_0=\int y dF_1(y) - \int \left[F_2^{-1}\circ F_3(y)\right] dF_4(y)$, where $(F_1,...,F_4)$ are the cdf's of $Y$ on four distinct subpopulations..

\medskip
We consider the plug-in estimator $\widehat{\theta}=g(\widehat{F_Y})$ of $\theta_0$, with
\begin{equation}
\widehat{F_Y}(y) = \frac{\sum_{\un\leq \j\leq \C}\sum_{\ell=1}^{N_{\j}}\mathds{1}\{Y_{\ell,\j}\leq y\}}{\sum_{\un\leq \j\leq \C} N_{\j}}.
	\label{eq:def_Fhat}
\end{equation}
To state the smoothness condition on $g$, we need additional notation and definitions. Let $\mathbb{D}$ denote a subset of the set of all cumulative distribution functions on $\R^\ell$ and suppose that $g:\mathbb{D}\mapsto \R^r$. We consider for simplicity here vector-valued functions $g$, but could easily extend our result below to functions taking values in normed spaces. We say that $g$ is Hadamard differentiable at $F_Y$ tangentially to $\mathbb{D}_0$ if there exists a continuous, linear map $g'_{F_Y}:\mathbb{D}\mapsto \R^r$ such that for every $(h_t)_{t\in \R^+}$ such that $h_t\rightarrow h\in\mathbb{D}_0$ as $t\downarrow 0$,
$$\lim_{t\downarrow 0} \left|\frac{g(F_Y+th_t) - g(F_Y)}{t} - g'_{F_Y}(h)\right|=0.$$

Proposition \ref{prop:functional_dm} shows that if $g$ is Hadamard differentiable at $F_Y$, $g(\widehat{F_Y})$ will be asymptotically normal. We also consider confidence regions based on the bootstrap. As before, we let $R_{1-\alpha}^{\text{boot}}= \left\{\theta: \, |\theta-\widehat{\theta}|\leq q_{1-\alpha}^*\right\}$, where $q_{1-\alpha}^*$ denotes the quantile of order $1-\alpha$ of the distribution of $|\widehat{\theta}^*-\widehat{\theta}|$ conditional on the data.

\begin{prop}
	\label{prop:functional_dm}
	Suppose that $\theta_0=g(F_Y)$ and $\widehat{\theta}=g(\widehat{F_Y})$, where $F_Y$ and $\widehat{F_Y}$ are defined respectively by \eqref{eq:def_F} and \eqref{eq:def_Fhat} and $g$ is Hadamard differentiable at $F_Y$ tangentially to $\mathbb{D}_0$.	Suppose also that Assumption \ref{as:dgp} holds and $\E(N_{1,\un}^2)<+\infty$. Then:
\begin{enumerate}
	\item $\sqrt{\underline{C}}(\widehat{F_Y}-F_Y)$ converges weakly, as a process indexed by $y$, to a Gaussian process $\mathbb{G}$ with kernel $K$ satisfying	
\begin{equation}
K(y_1,y_2)=\frac{1}{\E(N_{1,\un})^2} \sum_{i=1}^{k}\lambda_i \Cov\left(\sum_{\ell=1}^{N_{\un}}\mathds{1}\{Y_{\ell,\un}\leq y_1\}, \sum_{\ell=1}^{N_{\deux}}\mathds{1}\{Y_{\ell,\deux}\leq y_2\}\right).
\label{eq:kernel_had_diff}
\end{equation}
	\item If $\mathbb{G} \in \mathbb{D}_0$ with probability one,
$$\sqrt{\underline{C}} \left(\widehat{\theta}-\theta_0\right) \rightarrow \mathcal{N}(0,\V(g'_{F_Y}(\mathbb{G}))).$$
	\item If $\mathbb{G} \in \mathbb{D}_0$ with probability one and $\V(g'_{F_Y}(\mathbb{G}))$ is positive definite, $$\lim_{n\rightarrow\infty} \mathbb{P}\left(R_{1-\alpha}^{\text{boot}} \ni \theta_0\right) = 1-\alpha.$$
\end{enumerate}
\end{prop}

The first part follows from Theorem \ref{thm:unifTCL}, and a linearization of the ratio akin to \eqref{eq:lin_ratio}. The second follows from the first part and the functional delta method \citep[see, e.g.,][Theorem 20.8]{vanderVaart2000}. The third part is a direct consequence of Theorem \ref{thm:boot_unif} and the functional delta method for the bootstrap \citep[see, e.g.,][Theorem 3.9.11]{vanderVaartWellner1996}. 

\medskip
As an illustration, let us consider the example of a quantile, $\theta_0=F_Y^{-1}(\tau)$ for some $\tau\in (0,1)$. Suppose that $F_Y$ is differentiable at $\theta_0$. Then the function $g(F_Y)=F_Y^{-1}(\tau)$ is Hadamard differentiable at $F_Y$, tangentially to the set of functions that are continuous at $\theta_0$ \citep[see, e.g.][Lemma 21.3]{vanderVaart2000}. Moreover, we prove in Appendix \ref{app:example_quantile} that if  $\E[N_{\un}^{2+\zeta}]<+\infty$ for some $\zeta>0$, $\mathbb{G}$ is almost surely continuous at $\theta_0$. Hence, Proposition \ref{prop:functional_dm} ensures that $\widehat{\theta}$ is asymptotically normal. Moreover, by Point 3 of the proposition, inference based on the bootstrap is valid, as long as its asymptotic variance is strictly positive. 

\medskip
The third part ensures the consistency of the pigeonhole bootstrap. in principle, one could also use the normal approximation and a consistent estimator of $\V(g'_{F_Y}(\mathbb{G}))$ to make inference on $\theta_0$. $g'_{F_Y}(\mathbb{G})$ is a linear functional of $\mathbb{G}$, so the same ideas as in Section \ref{sub:sample_averages} above can be applied. This linear functional may however depend on complicated functions of $F_Y$ that must be estimated. For instance,  when $g(F)=F^{-1}(\tau)$, $g'_{F_Y}(\mathbb{G})$ depends on the derivative of $F_Y$ taken at $F^{-1}(\tau)$. As a result, additional restrictions may be necessary to achieve the consistency of the variance estimator. We do not explore this avenue further here, as it depends very much on the functional $g$, but consider explicitly this approach in the following section on GMM.

% subsection nonlinear_functionals_of_the_distribution (end)

\subsection{GMM estimators} % (fold)
\label{sub:gmm_estimators}

%!TEX root = /Users/xdhaultfoeuille/Dropbox/Cluster/paper/main.tex
Finally, we consider parameters defined by the moment restrictions \eqref{eq:GMM}, with possibly nonsmooth moments. We suppose that $m(y,\theta)\in \R^L$, with $m(y,\theta)=(m_1(y,\theta),...,m_L(y,\theta))'$. We show the asymptotic normality of $\widehat{\theta}$ under the following condition.

\begin{hyp}
	~
\begin{enumerate}
    \item $\theta_0$ belongs to the interior of $\Theta$, a compact subset of $\mathbb{R}^p$.
    \item $\E\left[\sum_{\ell=1}^{N_{\un}}m(Y_{\ell,\un},\theta)\right]=0$ if and only if $\theta=\theta_0$.
    \item For any $\theta\in \Theta$ we have $\lim_{\theta'\rightarrow \theta}\E\left[\left|\sum_{\ell=1}^{N_{\un}}m(Y_{\ell,\un},\theta')-\sum_{\ell=1}^{N_{\un}}m(Y_{\ell,\un},\theta)\right|^2\right]=0$.
    \item $\theta\mapsto \E\left(\sum_{\ell=1}^{N_{\un}}m(Y_{\ell,\un},\theta)\right)$ is differentiable at $\theta_0$ with a jacobian matrix $J$ of rank $p$.
    \item For all $s=1,...,L$ the class $\mathcal{F}_s=\{y\mapsto m_s(y,\theta):\theta \in \Theta\}$ fulfills Assumptions \ref{as:measurability}-\ref{as:vc}.
\item $\widehat{\Xi}$ is a sequence of random symmetric matrices of size $L$ tending in probability to $\Xi$, which is positive definite.
\end{enumerate}
\label{as:GMM}
\end{hyp}

Assumptions \ref{as:GMM}.1, \ref{as:GMM}.2 and \ref{as:GMM}.6 are standard. Assumption \ref{as:GMM}.3, combined with \ref{as:GMM}.1 and \ref{as:GMM}.2, ensures that the minimum of $$\theta\mapsto \E\left(\sum_{\ell=1}^{N_{\un}}m(Y_{\ell,\un},\theta)\right)'\Xi \, \E\left(\sum_{\ell=1}^{N_{\un}}m(Y_{\ell,\un},\theta)\right)$$
is well-separated on $\Theta$, thus ruling out possible inconsistency of the GMM estimator. Note that Assumption \ref{as:GMM}.3 is weaker than the standard continuity assumption of $\theta \mapsto m(Y_{\ell,\un},\theta)$, which may fail if  $m$ includes for instance indicator functions. Assumption \ref{as:GMM}.4 is standard in GMM with nonsmooth moments where $\theta \mapsto m(Y_{\ell,\un},\theta)$ is not differentiable, see e.g. Condition (ii) in Theorem 7.2 of  \cite{newey1994large}. Finally, by Theorem \ref{thm:unifTCL}, Assumption \ref{as:GMM}.5 ensures the stochastic equicontinuity condition  \citep[e.g., Condition (v) in Theorem 7.2 of][]{newey1994large}, which together with \ref{as:GMM}.4, is key to obtain $\sqrt{\underline{C}}$-asymptotic normality of $\widehat{\theta}$. 

\medskip
To illustrate that Assumption \ref{as:GMM} can handle nonsmooth moments, let us consider the example of quantile IV regressions. Let $Y_{\ell, \j}=(W_{\ell, \j}, X'_{\ell, \j}, Z'_{\ell, \j})'$, where $W_{\ell, \j}\in \R$ denotes the outcome variable, $X_{\ell, \j}\in \R^p$ denotes the explanatory, potentially endogenous variable and $Z_{\ell, \j}\in \R^L$ denotes the set of instruments ($Z_{\ell, \j}$ may include some components of $X_{\ell, \j}$). The moment functions are then
$$m(Y_{\ell,\un},\theta)=Z_{\ell, \j}\left(\tau - \mathds{1}\{W_{\ell, \j}- X'_{\ell, \j}\theta\leq 0\}\right).$$
Let us assume for simplicity that the $(Y_{\ell, \un})_{\ell\geq 1}$ are identically distributed. We show in Appendix \ref{sub:quantile_IV} that Assumptions \ref{as:GMM}.3-\ref{as:GMM}.5 hold if, basically, $\E[N_{\un}^2 |Z_{1,\un}|^2]<+\infty$, $X$ is in a compact set, the conditional cdf $F_{W_{1,\un}|X_{1,\un},Z_{1,\un}}(\cdot|X_{1,\un},Z_{1,\un})$ is continuous everywhere and admits a bounded derivative $f_{W_{1,\un}|X_{1,\un},Z_{1,\un}}(\cdot|X_{1,\un},Z_{1,\un})$ in a neighborhood of $X_{1,\un}'\theta_0$ and the rank of $\E\left[N_{\un} X_{1,\un} Z'_{1,\un}f_{W_{1,\un}|X_{1,\un},Z_{1,\un}}(X'_{\ell, \j}\theta_0|X_{1,\un},Z_{1,\un})\right]$ 
is equal to $p$.\footnote{For the exact conditions, see Assumption \ref{as:quantile_IV} in Appendix \ref{sub:quantile_IV}.}  

%Let $\mathbb{H}_C(\theta)$ be defined by %the random element of $\ell^{\infty}(\Theta)$ defined by
%$$\mathbb{H}_C(\theta)=\frac{\sqrt{\underline{C}}}{\Pi_C}\sum_{\un \leq \j \leq \C} \left[\sum_{\ell=1}^{N_{\j}}m(Y_{\ell,\j},\theta)-\E\left(\sum_{\ell=1}^{N_{\un}}m(Y_{\ell,\un},\theta)\right)\right].$$

\begin{thm}
Suppose that Assumptions \ref{as:dgp} and \ref{as:GMM} hold.  Then $\widehat{\theta}$ is well-defined with probability approaching one and
%$$\sqrt{\underline{C}}\left(\widehat{\theta}-\theta_0\right)=-(J' \Xi J)^{-1}J'\Xi \mathbb{H}_C(\theta_0)+o_p(1),$$
$$\sqrt{\underline{C}}\left(\widehat{\theta}-\theta_0\right)\convD\mathcal{N}\left(0,V_0\right),$$
where $V_0 = (J'\Xi J)^{-1} J' \Xi  H \Xi J (J'\Xi J)^{-1}$ and
$$H = \sum_{i=1}^k \lambda_i  \E\left[\left(\sum_{\ell=1}^{N_{\un}}m(Y_{\ell,\un},\theta_0)\right)\left(\sum_{\ell=1}^{N_{\deux}}m(Y_{\ell,\deux},\theta_0)\right)'\right].$$
\label{thm:AN_GMM}
\end{thm}

To our knowledge, Theorem \ref{thm:AN_GMM} is the first result on the asymptotic normality of GMM estimators under multiway clustering. Theorem \ref{thm:AN_GMM} gives also the expression of the asymptotic variance $V_0$. This matrix takes the usual form, except that the matrix $H$, which would simply be $E[m(Y_{\ell,\un},\theta_0)m(Y_{\ell,\un},\theta_0)']$ without clustering, takes a more complicated form here. This form is in line with our result on the covariance kernel of the empirical process considered above.
%$$\mathbb{H}_C(\theta)=\frac{\sqrt{\underline{C}}}{\Pi_C}\sum_{\un \leq \j \leq \C} \left[\sum_{\ell=1}^{N_{\j}}m(Y_{\ell,\j},\theta)-\E\left(\sum_{\ell=1}^{N_{\un}}m(Y_{\ell,\un},\theta)\right)\right].$$

\medskip
We now turn to inference on $\theta_0$. As for sample averages, we consider inference based on asymptotic normality and a consistent estimator of $V_0$, or the pigeonhole bootstrap. To ensure the consistency of our estimator of $V_0$, we impose the following additional regularity condition.

\begin{hyp}
	~
\begin{enumerate}
\item The jacobian matrix $J$ of $\theta\mapsto\E\left(\sum_{\ell=1}^{N_{\un}}m(Y_{\ell,\un},\theta)\right)$ at $\theta_0$ admits the following representation
$$J=\E\left(\sum_{\ell=1}^{N_{\un}}d(Y_{\ell,\un},\theta)\right)$$
  for some matrix-valued function $d(.,.)=\left(d_{r,s}(.,.)\right)_{1\leq r\leq p,1\leq s\leq L}$.
        \item For all $(r,s)\in\{1,...,p\}\times\{1,...,L\}$, the class $\mathcal{F}_{r,s}=\{y\mapsto d_{r,s}(y,\theta):\theta \in \Theta\}$ fulfills Assumption \ref{as:measurability} and admits an envelope $F_{r,s}$ such that $E[\sum_{\ell=1}^{N_{\un}}F_{r,s}(Y_{\ell, \un})]<+\infty$,  and for any $\varepsilon>0$, $\sup_{Q} N(\varepsilon ||.||_{Q,1},\mathcal{F}_{r,s},||.||_{Q,1})<\infty$ where the supremum is taken over the set of probability measures with finite support on $\mathcal{Y}$.		
		\item $\lim_{\theta'\to\theta_0}\mathbb{E}\left[\sum_{\ell=1}^{N_{\un}}d(Y_{\ell,\un},\theta')\right]=\mathbb{E}\left[\sum_{\ell=1}^{N_{\un}}d(Y_{\ell,\un},\theta_0)\right]$.
     \item For every $i=1,...,k$, $$\lim_{\theta'\to\theta_0}\mathbb{E}\left[\sum_{\ell=1}^{N_{\un}}\sum_{\ell=1}^{N_{\deux}}m(Y_{\ell,\un},\theta')m(Y_{\ell,\deux},\theta')\right]=\mathbb{E}\left[\sum_{\ell=1}^{N_{\un}}\sum_{\ell=1}^{N_{\deux}}m(Y_{\ell,\un},\theta_0)m(Y_{\ell,\deux},\theta_0)\right].$$
        \end{enumerate}
\label{as:varGMM}
\end{hyp}

Assumption \ref{as:varGMM}.1 refines Assumption \ref{as:GMM}.4 by imposing some structure on the Jacobian matrix of $\theta\mapsto\E\left(\sum_{\ell=1}^{N_{\un}}m(Y_{\ell,\un},\theta)\right)$. In Assumption \ref{as:varGMM}.2, the condition on the classes  are of Glivenko-Cantelli type, and weaker than Assumption \ref{as:vc}. The continuity conditions in Points 3 and 4 are similar to Assumption \ref{as:GMM}.3, but are imposed on different functions related to $J$ and $H$ rather than on the moment conditions themselves.

\medskip
We now define our estimator of $V_0$, which is based on estimators of $J$ and $H$. Given Assumption \ref{as:varGMM}.1, $\widehat{J}$ is the simple plug-in estimator
$$\widehat{J}=\frac{1}{\Pi_C}\sum_{\un\leq \j \leq \C} \sum_{\ell=1}^{N_{\j}}d(Y_{\ell,\j},\widehat{\theta}).$$
To estimate $H$, we adapt our previous estimator $\widehat{V}_1$ to this context by considering 
\begin{align*}
	&\widehat{H}=\sum_{i=1}^k\frac{\underline{C}}{C_i} \frac{1}{C_i}\sum_{j'_i=1}^{C_i} \left(\frac{1}{\prod_{s\neq i}C_s}\sum_{\j:j_i=j'_i} \sum_{\ell=1}^{N_{\j}}m\left(Y_{\ell,\j},\widehat{\theta}\right)\right)
\left(\frac{1}{\prod_{s\neq i}C_s}\sum_{\j:j_i=j'_i} \sum_{\ell=1}^{N_{\j}}m\left(Y_{\ell,\j},\widehat{\theta}\right) \right)'.
\end{align*}
Our variance estimator is then $\widehat{V}=(\widehat{J}'\widehat{\Xi}\widehat{J})^{-1}\widehat{J}'\widehat{\Xi} \widehat{H}\widehat{\Xi} \widehat{J}(\widehat{J}'\widehat{\Xi}\widehat{J})^{-1}$.

\begin{thm}
\label{thm:var_gmm}
Assume that Assumptions \ref{as:dgp} and \ref{as:GMM} hold and $H$ is positive definite. Then:
\begin{enumerate}
	\item If Assumption \ref{as:varGMM} holds as well, $\widehat{V}\convP V_0$ and confidence regions and tests on $\theta_0$ based on asymptotic normality and $\widehat{V}$ are asymptotically valid.
	\item Confidence regions and tests on $\theta_0$ based on the pigeonhole bootstrap are asymptotically valid.
\end{enumerate}
\end{thm}

Note that the pigeonhole bootstrap does not require any additional condition, above that ensuring the $\sqrt{\underline{C}}$-asymptotic normality of the GMM estimator and the fact that $H$ is positive definite. Hence, Theorem \ref{thm:var_gmm} implies for instance that under the conditions displayed above, the pigeonhole bootstrap is valid for quantile IV regressions under multiway clustering.

% subsection gmm_estimators (end)

\section{Monte Carlo Simulations}
\label{sec:MC}
%!TEX root = /Users/xdhaultfoeuille/Dropbox/Cluster/paper/main.tex
We now investigate the finite sample properties of the different inference strategies we have considered. We study the coverage rate of confidence intervals based on either asymptotic normality or the pigeonhole bootstrap. We consider the following different cases:
\begin{enumerate}
	\item ``two-way, Gaussian'': our baseline scenario is a two-way balanced design ($C_1=C_2=\underline{C}$) with one observation per cell ($N_{\j}=1$). Each $Y_{1,\j}$ is drawn in a standard Gaussian distribution, but the variance due to cell shocks only represent  60\% of the total variance, whereas row and column shocks represent 20\% of the variance each:
	\begin{equation}\label{eq:DGPsim}
Y_{1,(j_1,j_2)}=\frac{1}{\sqrt{5}}\left(U_{j_1,0}+U_{0,j_2}+\sqrt{3}U_{j_1,j_2}\right), \quad \left(U_{j_1,0},U_{0,j_2},U_{j_1,j_2}\right)\sim \mathcal{N}(0,I_3).
\end{equation}
The parameter of interest is $\theta_0=\E(Y_{1,\un})$ and we consider $C_1=C_2$ taking values in $\{5,10,30,50,100\}$. %For each case, we simulate 1000 samples according to \eqref{eq:DGPsim}.
\item ``two-way, w/o adjust'': this scenario is as the baseline, except that we compute variance estimators without including the finite-sample corrections described below. The purpose is to investigate the effects of this correction in practice.%do not include the in the the computation of differs from the baseline scenario simply in the computation by omission of the factor $\frac{C_i}{C_i-1}$ in the definition of $\widehat{\Sigma}_i$ and omission of the factor $\frac{C_1C_2}{C_1C_2-1}$ in the definition of $\widehat{\Sigma}_12$.
\item ``two-way, binary'': this scenario is as the baseline, except that $\theta_0=\E(\mathds{1}\left\{Y_{1,\un}>0\right\})$. The goal is to investigate whether accuracy of inference in our baseline scenario is driven by the fact that $\widehat{\theta}$ itself is normally distributed.
\item ``two-way, probit'': in this scenario, we consider a simple probit model, with random cell sizes. Namely, we suppose that the $(N_{\j})_{\j\geq \un}$ are independent, with $N_{\j}\sim 1+\mathcal{P}(5)$. Our outcome variable is then defined by
$$\tilde{Y}_{\ell, (j_1, j_2)}=\mathds{1}\left\{\beta_0+\theta_0 X_{\ell,(j_1, j_2)}+\frac{1}{\sqrt{6}}\left(U_{(j_1,0)}+U_{(0,j_2)}+U_{(j_1,j_2)}\right) + \frac{U_{\ell,\j}}{\sqrt{2}}>0\right\},$$
where the $(X_{\ell,\j})_{\ell\geq 1,\j\geq\un}$, $(U_{\j})_{\j\in \mathbb{N}^2}$ and $(U_{\ell,\j})_{\ell\geq 1,\j\geq\un}$ are mutually independent and standard normal variables (as above, the $(U_{\j})_{\j\in \mathbb{N}^2}$ and $(U_{\ell,\j})_{\ell\geq 1,\j\geq\un}$ are assumed unobserved). $\theta_0$ is again our parameter of interest and $(\beta_0,\theta_0)=(0,1)$. In these simulations $(\widehat{\beta},\, \widehat{\theta})$ is the pseudo-maximum likelihood estimator of $(\beta_0,\, \theta_0)$, i.e. the usual probit estimator obtained on the pooled sample. The first aim of this scenario is to study the sensitivity of inference to the non-linearity of the estimator. The second aim is to investigate the sensitivity of  inference to randomness in cell sizes $N_{\j}$.
\item ``three-way, Gaussian'': this scenario differs from the baseline in that we consider three-way clustering. As in the baseline, $N_{\j}=1$, $Y_{\j}$ is Gaussian and 60\% of the variance is due to cell shocks. 6,67\% of the variance is due to shocks specific to dimension 1, 2 or 3 of the clustering. The remaining variance is due to shocks common to dimensions 1 and 2, 2 and 3, and 1 and 3. Specifically,
$$Y_{1,\j}=\frac{1}{\sqrt{15}}\left(U_{(j_1,0,0)}+U_{(0,j_2,0)}+U_{(0,0,j_3)} + U_{(j_1,j_2,0)}+U_{(j_1,0,j_3)} + U_{(0,j_2,j_3)} + 3U_{\j}\right),$$
where the $(U_{\j})_{\j \in \mathbb{N}^3}$ are independent standard normal variables. We consider $C_1=C_2=C_3$ taking values in $\{3,5,10,30,50\}$. These values were chosen so as to correspond roughly to the same number of cells as in the corresponding cases under our baseline scenario.
\end{enumerate}

For each scenario, we compute four confidence intervals. The first three are based on the asymptotic normality of $\widehat{\theta}$ and the consistent estimators $\widehat{V}_1$, $\widehat{V}_2$ and $\widehat{V}_{\text{cgm}}$ of the asymptotic variance. The fourth is based on the pigeonhole bootstrap. As explained above, the variance estimator $\widehat{V}_1$ is very easy to compute with popular econometric softwares such as Stata or R, as it satisfies:
$$\widehat{V}_1=\underline{C}\sum_{i=1}^k\widehat{\Sigma}_i,$$
where $\widehat{\Sigma}_{i}$ is the clustered estimated variance with respect to the $i$-th dimension of clustering.\footnote{The term $\underline{C}$ accounts for the fact that $\widehat{V}_1$ estimates the asymptotic variance of $\widehat{\theta}$ rather than its variance.} In other words, one has only to compute $k$ variances under one-way clustering and add them to get a consistent estimate of variance under multiway clustering.  $\widehat{V}_{\text{cgm}}$ takes a similar form except that one has to consider additional terms, since it is based on the inclusion-exclusion principle (see \eqref{eq:CGM} above). For instance, with $k=2$,  $\widehat{V}_{\text{cgm}}= \widehat{V}_1 - \underline{C}\widehat{\Sigma}_{12}$, where $\widehat{\Sigma}_{12}$ corresponds to the variance under one-way clustering with clusters defined by the intersection of dimensions 1 and 2 (namely, cells with $k=2$). Finally, $\widehat{V}_2$ can also be written as $\widehat{V}_2=\underline{C}\sum_{i=1}^k\widetilde{\Sigma}_i$, but $\widetilde{\Sigma}_i$ does not correspond to the usual estimator of variance under one-way clustering along dimension $i$.

\medskip
The small-sample correction $C_i/(C_i-1)$ is often used by default for the computation of the clustered variance $\widehat{\Sigma}_i$, and we follow this practice hereafter (also for $\widehat{\Sigma}_{12}$ and $\widetilde{\Sigma}_i$), except in Scenario 2. Hence, in the baseline scenario, we have
\begin{align*}
\widehat{\Sigma}_1 & =\frac{C_1}{C_1-1} \frac{1}{\Pi_C^2}\sum_{j_1=1}^{C_1} \left(\sum_{j_2=1}^{C_2}\sum_{\ell=1}^{N_{(j_1,j_2)}}(Y_{\ell,(j_1,j_2)}-\widehat{\theta})\right)^2 \\
%\widehat{\Sigma}_2 & =\frac{C_2}{C_2-1} \frac{1}{\left(\sum_{\un \leq \j \leq \C}N_{\j}\right)^2}\sum_{j_2=1}^{C_2} \left(\sum_{j_1=1}^{C_1}\sum_{\ell=1}^{N_{(j_1,j_2)}}(Y_{\ell,(j_1,j_2)}-\widehat{\theta})\right)^2,\\
\widetilde{\Sigma}_1& =\frac{C_1}{C_1-1} \frac{1}{C_1^2C_2(C_2-1)}\sum_{j_1=1}^{C_1} \left(\sum_{\substack{1\leq j_2,j_2'\leq C_2\\
		j_2'\neq j_2}}(Y_{1,(j_1,j_2)}-\widehat{\theta})(Y_{1,(j_1,j_2')}-\widehat{\theta})\right),\\
%\widetilde{\Sigma}_2& =\frac{C_2}{C_2-1}\frac{1}{\left(\sum_{\un \leq \j \leq \C}N_{\j}\right)^2}\sum_{j_2=1}^{C_2} \left(\sum_{\substack{1\leq j_1,j_1'\leq C_1\\
%		j_1'\neq j_1}}(Y_{1,(j_1,j_2)}-\widehat{\theta})(Y_{1,(j_1',j_2)}-\widehat{\theta})\right),\\
\widehat{\Sigma}_{12}&=\frac{C_1C_2}{C_1C_2-1} \frac{1}{\Pi_C^2}\sum_{\un \leq \j\leq \C} \left(\sum_{\ell=1}^{N_{\j}}(Y_{\ell,\j}-\widehat{\theta})\right)^2.
\end{align*}
$\widehat{\Sigma}_2$ and $\widetilde{\Sigma}_2$ satisfy the same formulas, up to inverting the roles of $C_1$ and $C_2$, and $j_1$ and $j_2$ in the summations. The formulas are identical for the third scenario. In the second scenario, the formulas remain also the same, except that we remove the correction terms $C_i/(C_i-1)$ and $C_1C_2/(C_1C_2-1)$.   Finally, the corresponding formulas for the fourth and fifth scenarios are detailed in Appendix \ref{sec:additional_details_on_the_simulations}. Note that when $\widehat{V}_2$ or $\widehat{V}_{\text{cgm}}$ are negative, we simply set the confidence intervals to the point estimates.

\medskip
We also compute Efron's percentile bootstrap confidence interval based on the pigeonhole bootstrap presented in Section \ref{sec:boot}:
$$\text{IC}_{\text{boot}}=\left[q^{*}_{0.025};q^{*}_{0.975}\right],\text{ with } q^*_{\alpha} \text{ the quantile or order } \alpha \text{ of } \theta^{*}|(N_{\j},(Y_{\ell,\j})_{\ell\leq N_{\j}})_{\un \leq \j\leq \C}.$$
This confidence interval is valid since the asymptotic distribution of $\widehat{\theta}$ is symmetric. To simulate the distribution of $\theta^{*}|(N_{\j},(Y_{\ell,\j})_{\ell\leq N_{\j}})_{\un \leq \j\leq \C}$, we use 1,000 bootstrap replications for each initial sample we draw.

\medskip
The results are displayed in Table \ref{tab:MC_results}. In the first scenario, the actual coverage  when using our preferred estimator of variance and the pigeonhole bootstrap  is always very close to the nominal one, even for $\underline{C}$ as small as 5. It is often considered that between 30 and 50 clusters are necessary with one-way clustering to get reliable confidence intervals \citep{bertrand2004, cameron2015}. Here, we find that even when 40\% of the variance of the cells is related to cluster shocks, 25 cells resulting from a $5\times5$ design are sufficient to get reliable inference, at least with fixed cell sizes and when the estimator $\widehat{\theta}$ is Gaussian.

\medskip

For the same design and still $\underline{C}=5$, inference based on the estimator of \cite{cameron2011} leads to an actual coverage of around $88\%$, for a nominal coverage of $95\%$. The confidence intervals based on $\widehat{V}_2$ perform poorly with small samples. In a $5\times 5$ design, the actual coverage is only around 62\%. This is partly but not entirely due to the fact that for 16\% of the simulations, we get a negative estimator of the variance, implying that we do not cover $\theta_0$. On the other hand, and in line with the theory, we do observe that the coverage rate of IC$_2$ converges to 95\% as $\underline{C}$ grows.

\medskip
The results corresponding to the second scenario show that excluding the small-sample correction deteriorates significantly the coverage rates for $\underline{C} = 5$ and also, when considering IC$_2$ and IC$_{\text{cgm}}$, for $\underline{C} = 10$. IC$_1$ is less sensitive to the adjustment than IC$_2$ and IC$_{\text{cgm}}$. The correction does not have a notable influence when $C_1,C_2\geq 30$ for any of the confidence intervals. But overall, our simulations suggest that this small-sample correction is desirable.

\medskip
Results with binary outcomes are qualitatively similar to our baseline simulations: the coverage rates of $\text{IC}_1$ and $\text{IC}_{\text{boot}}$ are closer to the nominal rate than those of $\text{IC}_2$ and $\text{IC}_{\text{cgm}}$. But quantitatively, the coverage rate of $\text{IC}_2$ and $\text{IC}_{\text{cgm}}$
are even further away from the nominal rate, falling respectively under 57\% and 84\% in the $5\times 5$ design. On the other hand, the coverage rate of  $\text{IC}_1$ and $\text{IC}_{\text{boot}}$ remain close to the nominal rate (93,5\% and 95,2\% in the $5 \times 5$ design). Contrary to the baseline case, we observe that  $\text{IC}_{\text{boot}}$ and $\text{IC}_1$ (for $\underline{C}\geq 10$) are slightly conservative here.

\medskip
The results of the probit model give rise to similar conclusions. $\text{IC}_2$ performs even worse with $\underline{C}=5$, with more than 75\% of the variance estimates being negative, but somewhat better than previously with $\underline{C}\geq 10$. The coverage rates of $\text{IC}_{\text{cgm}}$ are very close to those observed in the third scenario. Finally, $\text{IC}_1$ and $\text{IC}_{\text{boot}}$ are again closer to the nominal coverage rates, even if they tend to be slightly more conservative than previously.

\medskip
Finally, our results with three-way clustering show again the very good performance of IC$_1$ and IC$_{\text{boot}}$ for $\underline{C}$ as small as 3. They also emphasize that even with a normal sample average, IC$_2$ and IC$_{\text{cgm}}$ can still severely undercover with a small number of clusters. In particular, neglecting asymptotically negligible terms as done in $\widehat{V}_2$ leads to 85\% of negative estimates with $\underline{C}=3$.

\medskip
Overall, our results suggest that contrary to IC$_2$ and, to a lesser extent, IC$_{\text{cgm}}$, IC$_1$ and IC$_{\text{boot}}$ may be generally reliable, even with few clusters. As explained above, another advantage of IC$_1$ is that it is even simpler to compute than IC$_{\text{cgm}}$. IC$_{\text{boot}}$ may also be useful in particular for estimators whose asymptotic variance takes a complicated form. %, as is the case when such estimators are obtained through multiple steps.   

\begin{table}[H]
\caption{Coverage rates on the five scenarios (nominal coverage rate: 0.95)}
	\label{tab:MC_results}

\begin{center}	
\begin{tabular}{c c c c ccccc}
	\hline \hline
	
 && & &\multicolumn{5}{c}{$\underline{C}$}\\
% \cline{5-10}
Scenario& & Interval &&5&10&30&50&100	\\
\hline%\cline{1-1}\cline{3-3}\cline{5-9}
%[1.00e+00 6.53e-01 8.70e-01 9.26e-01 9.43e-01 9.49e-01 1.62e-01 0.00e+00 0.00e+00 0.00e+00 0.00e+00]
%[1.00e+00 8.75e-01 9.16e-01 9.36e-01 9.52e-01 9.52e-01 5.00e-03 0.00e+00 0.00e+00 0.00e+00 0.00e+00]

	 & &  $\text{IC}_{\text{boot}}$ & &0.929& 0.94&  0.948& 0.952& 0.951 \\
	two-way, Gaussian &&   $\text{IC}_1$ & &0.935 &0.939 &0.949& 0.957& 0.955 \\
	&&   $\text{IC}_2$ & &$\underset{\text{\tiny{[16.2\%]}}}{0.653^*}$& 0.87&  0.926& 0.943& 0.949\\
	 &&   $\text{IC}_{\text{cgm}}$ & &$\underset{\text{\tiny{[0.5\%]}}}{0.875^*}$ &0.916 &0.936& 0.952& 0.952\\[3mm]
	\cline{3-9}
	
%&&&&&&&&\\
%[2.00e+00 6.26e-01 8.53e-01 9.22e-01 9.42e-01 9.49e-01 1.62e-01 0.00e+00 0.00e+00 0.00e+00 0.00e+00]
%[2.00e+00 8.16e-01 8.97e-01 9.30e-01 9.45e-01 9.49e-01 1.40e-02 0.00e+00 0.00e+00 0.00e+00 0.00e+00]

%	&&    $\text{IC}_{\text{boot}}$ & &0.929& 0.94&  0.948& 0.952& 0.951\\
	&&    $\text{IC}_1$ & &0.904& 0.933& 0.945& 0.955& 0.952 \\
	two-way, w/o adjust.&&    $\text{IC}_2$ & &$\underset{\text{\tiny{[16.2\%]}}}{0.626^*}$ &0.853& 0.922& 0.942& 0.949\\
	&&    $\text{IC}_{\text{cgm}}$ & &$\underset{\text{\tiny{[1.4\%]}}}{0.816^*}$ &0.897& 0.93& 0.945& 0.949 \\[3mm]
	\cline{3-9}
%&&&&&&&&\\
%[3.00e+00 5.61e-01 8.40e-01 9.25e-01 9.26e-01 9.46e-01 3.18e-01 8.00e-03 0.00e+00 0.00e+00 0.00e+00]
%[3.00e+00 8.37e-01 9.21e-01 9.40e-01 9.44e-01 9.48e-01 8.00e-03 0.00e+00 0.00e+00 0.00e+00 0.00e+00]

	&&    $\text{IC}_{\text{boot}}$ & &0.952& 0.97&  0.955& 0.951& 0.953\\
	two-way, binary &&    $\text{IC}_1$ & &0.937 &0.959& 0.957 &0.955& 0.952\\
    &&$\text{IC}_2$&&$\underset{\text{\tiny{[31.8\%]}}}{0.561^*}$& $\underset{\text{\tiny{[0.8\%]}}}{0.84^*}$ & 0.925 &0.926 &0.946\\
	&&    $\text{IC}_{\text{cgm}}$ & &$\underset{\text{\tiny{[0.8\%]}}}{0.837^*}$ &0.921& 0.94&  0.944& 0.948\\[3mm]
	\cline{3-9}	
%&&&&&&&&\\
%[4.00e+00 1.65e-01 2.95e-01 6.97e-01 8.29e-01 8.98e-01 7.72e-01 5.69e-01 6.80e-02 1.00e-03 0.00e+00]
%[4.00e+00 7.55e-01 8.72e-01 9.25e-01 9.35e-01 9.36e-01 9.00e-02 1.50e-02 0.00e+00 0.00e+00 0.00e+00]

	&&    $\text{IC}_{\text{boot}}$ & &0.938& 0.977& 0.982& 0.976& 0.964\\
	two-way, probit &&    $\text{IC}_1$ & &0.97&  0.977& 0.978& 0.977& 0.959\\
	 &&$\text{IC}_2$&&$\underset{\text{\tiny{[77.2\%]}}}{0.165^*}$ &$\underset{\text{\tiny{[56.9\%]}}}{0.295^*}$& $\underset{\text{\tiny{[6.8\%]}}}{0.697^*}$& $\underset{\text{\tiny{[0.1\%]}}}{0.829^*}$& 0.898\\
	&&    $\text{IC}_{\text{cgm}}$ & &$\underset{\text{\tiny{[9.0\%]}}}{0.755^*}$&$\underset{\text{\tiny{[1.5\%]}}}{0.872^*}$ &0.925 &0.935& 0.936\\[3mm]
	\cline{3-9}
%&&&&&&&&\\
& &&&\multicolumn{5}{c}{$\underline{C}$}\\
%\cline{5-10}
& & & &3&5&10&15&20	\\
\cline{5-9}
%[5.00e+00 9.58e-01 9.66e-01 9.60e-01 9.56e-01 9.57e-01 0.00e+00 0.00e+00
%0.00e+00 0.00e+00 0.00e+00]
%[5.00e+00 9.42e-01 9.56e-01 9.57e-01 9.52e-01 9.58e-01 0.00e+00 0.00e+00
%0.00e+00 0.00e+00 0.00e+00]
%[5.00e+00 9.60e-02 4.84e-01 8.60e-01 9.14e-01 9.25e-01 8.54e-01 2.55e-01
%0.00e+00 0.00e+00 0.00e+00]
%[5.00e+00 7.69e-01 8.59e-01 9.19e-01 9.34e-01 9.37e-01 5.60e-02 5.00e-03
%0.00e+00 0.00e+00 0.00e+00]]

	&&    $\text{IC}_{\text{boot}}$ & &0.958 &0.966 &0.960 &0.956 &0.957\\
	three-way, Gaussian &&   $\text{IC}_1$ & &0.942 &0.956 &0.957 &0.952 &0.958\\
	 &&   $\text{IC}_2$ & &$\underset{\text{\tiny{[85.4\%]}}}{0.096^*}$ &$\underset{\text{\tiny{[25.5\%]}}}{0.484^*}$ &0.86 & 0.914 & 0.925\\
	&&   $\text{IC}_{\text{cgm}}$ & &$\underset{\text{\tiny{[5.6\%]}}}{0.769^*}$ &$\underset{\text{\tiny{[0.5\%]}}}{0.859^*}$ &0.919 &0.934 &0.937\\
\hline \hline
\end{tabular}
\end{center}
\begin{tablenotes}
	Coverage rate estimated on 1000 simulations. The bootstrap confidence intervals are based on 1000 bootstrap samples. ${}^*$ indicates that some estimated variance were negative, in which case the share of negative variance is reported in brackets below. When an estimated variance is negative, we set the corresponding confidence interval to the point estimate.
\end{tablenotes}
\end{table}

\section{Conclusion}
\label{sec:conclu}
In this paper, we have shown two weak convergence results under multiway clustering. The first implies not only simple central limit theorems, but also asymptotic normality of various nonlinear estimators. The second implies the general validity of the pigeonhole bootstrap under multiway clustering. We also establish the consistency of three variance estimators. Inference based on either the pigeonhole bootstrap or asymptotic normality and our preferred variance estimator works very well in simulations, with coverage rates close to their nominal values for no more than five clusters in each dimension with two-way clustering, or even three with three-way clustering. 
%%%%%%%%%%%%%%%%%%%%%%%%
\newpage
\bibliography{biblio}

\newpage
\appendix
%%%%%%%%%%%%%%%%%%%%%%%%%%

\section{Weak convergence under another restriction on $\mathcal{F}$} % (fold)
\label{app:smoothness_class}

In this appendix, we show that similar results as those obtained so far can be obtained under another restriction on the class $\mathcal{F}$. For this purpose, let us consider the norm $||\cdot||_{\infty,\beta}$ defined by 
$$||f||_{\infty,\beta}=\sup_{y\in \mathcal{Y}}\left|f(y)(1+|y|^2)^{\beta/2}\right|$$ 
for $\beta\in\mathbb{R}$. When $\beta=0$, $\left|\left|\cdot\right|\right|_{\infty,\beta}$ corresponds to the standard supremum norm. When $\beta<0$, the norm $\left|\left|\cdot\right|\right|_{\infty,\beta}$ is convenient for classes of smooth but unbounded functions that diverge in the tails at an appropriate rate. When $\beta>0$, $\left|\left|\cdot\right|\right|_{\infty,\beta}$ is useful when the class $\mathcal{F}$ consists of uniformly bounded functions decaying sufficiently fast in the tails \citep[see e.g.][for more details]{freyberger2015compactness}.

\medskip
We then consider the following restriction on $\mathcal{F}$. 

\setcounter{hyp}{2}
\begin{hyp}\hspace{-0.2cm}\textbf{'}\label{as:smooth}
$\mathcal{F}$ admits an envelope $F$ with $\abs{\abs{F}}_{\infty,\beta}<+\infty$ and
\begin{align*}
\mathbb{E}\left[\left(\sum_{\ell=1}^{N_{\un}}(1+|Y_{\ell,\un}|^2)^{-\beta/2}\right)^2\right]& <+\infty,\\
	\int_0^{+\infty}\sqrt{\log N(\epsilon||F||_{\infty,\beta},\mathcal{F},||.||_{\infty,\beta})}d\epsilon & <+\infty.
\end{align*}
\end{hyp}

Compared to Assumption \ref{as:vc},  Assumption~\ref{as:smooth}' is useful to establish
asymptotic normality in models involving infinite-dimensional but smooth parameters. In the i.i.d. setup, \cite{nickl2007} show that under Assumption~\ref{as:smooth}' and a moment condition, the $L_2(P)$ bracketing integral of many well-known classes of smooth functions is finite, which is a key ingredient in proving asymptotic normality results for estimators of smooth functional parameters.\footnote{Here $P$ refers to the probability measure of $(N_{\un}, (Y_{\ell,\un})_{\ell\geq 1})$. We refer to, e.g., \cite{vanderVaartWellner1996} for a definition of bracketing integrals.} The two main types of smoothness classes of functions used in practice are Sobolev classes and H\"older classes. The former have been used for instance by \cite{newey2003} in the nonparametric instrumental variable model and \cite{gallant1987} in a semi-nonparametric maximum likelihood estimation framework. The H\"older classes have been used for instance by \cite{chen2015} in the context of nonparametric quantile instrumental variable models. For more details on the use of (weighted-)nonparametric smoothness classes in econometrics, we refer to \cite{chen2007} and \cite{freyberger2015compactness}.

\medskip
We obtain the same result as Theorems \ref{thm:unifTCL} and \ref{thm:boot_unif} when replacing Assumption \ref{as:vc} by Assumption \ref{as:smooth}'.

\begin{thm} \label{thm:unifTCL_bis}
Suppose that Assumptions~\ref{as:dgp}-\ref{as:measurability} and \ref{as:smooth}' hold. Then the process $\mathbb{G}_C$ converges weakly to a centered Gaussian process $\mathbb{G}$ on $\mathcal{F}$ as $\underline{C}$ tends to infinity. Moreover, the covariance kernel $K$ of $\mathbb{G}$ satisfies:
$$K(f_1,f_2) = \sum_{i=1}^{k}\lambda_i \Cov\left(\sum_{\ell=1}^{N_{\un}}f_1(Y_{\ell,\un}), \sum_{\ell=1}^{N_{\deux}}f_2(Y_{\ell,\deux})\right).$$
\end{thm}

\begin{thm}
Suppose that Assumptions \ref{as:dgp}-\ref{as:measurability} and \ref{as:smooth}' hold. Then $\mathbb{G}^{\ast}_{C}$ converges weakly to $\mathbb{G}$ in probability, namely \eqref{eq:conv_boot_proc} holds.
\label{thm:boot_unif_bis}
\end{thm}

Asymptotic normality of GMM estimators also holds if we replace Assumption \ref{as:GMM}.5 by a condition involving Assumption \ref{as:smooth}' instead of Assumption \ref{as:vc}.

\begin{hyp}\hspace{-0.2cm}\textbf{'} Assumptions \ref{as:GMM}.1-\ref{as:GMM}.4 and \ref{as:GMM}.6 hold and for all $s=1,...,L$ the class $\mathcal{F}_s=\{y\mapsto m_s(y,\theta):\theta \in \Theta\}$ fulfills Assumptions \ref{as:measurability}-\ref{as:smooth}'.
%	\end{enumerate}
	\label{as:GMMsmooth}
\end{hyp}

\begin{thm}
	Suppose that Assumption \ref{as:dgp} and \ref{as:GMMsmooth}' hold.  Then $\widehat{\theta}$ is well-defined with probability approaching one and
	%$$\sqrt{\underline{C}}\left(\widehat{\theta}-\theta_0\right)=-(J' \Xi J)^{-1}J'\Xi \mathbb{H}_C(\theta_0)+o_p(1),$$
	$$\sqrt{\underline{C}}\left(\widehat{\theta}-\theta_0\right)\convD\mathcal{N}\left(0,V_0\right),$$
	where $V_0 = (J'\Xi J)^{-1} J' \Xi  H \Xi J (J'\Xi J)^{-1}$ and
	$$H = \sum_{i=1}^k \lambda_i  \E\left[\left(\sum_{\ell=1}^{N_{\un}}m(Y_{\ell,\un},\theta_0)\right)\left(\sum_{\ell=1}^{N_{\deux}}m(Y_{\ell,\deux},\theta_0)\right)'\right].$$
	\label{thm:AN_GMMsmooth}
\end{thm}

Consistent inference for the GMM is also ensured if in Assumption \ref{as:varGMM}.2, we basically replace Assumption \ref{as:vc} by Assumption \ref{as:smooth}'.

\begin{hyp}\hspace{-0.2cm}\textbf{'}
	Assumptions \ref{as:varGMM}.1 and \ref{as:varGMM}.3-\ref{as:varGMM}.4 hold and for all $(r,s)\in\{1,...,p\}\times\{1,...,L\}$, the class $\mathcal{F}_{r,s}=\{y\mapsto d_{r,s}(y,\theta):\theta \in \Theta\}$ fulfills Assumption \ref{as:measurability} and admits an envelope $F_{r,s}$ such that 		
 $\abs{\abs{F}}_{\infty,\beta}<+\infty$ and $N(\epsilon||F||_{\infty,\beta},\mathcal{F},||.||_{\infty,\beta})<+\infty$ for every $\epsilon>0$.
%	\end{enumerate}
	\label{as:varGMMsmooth}
\end{hyp}

\begin{thm}
	\label{thm:var_gmmsmooth}
	Assume that Assumptions \ref{as:dgp} and \ref{as:GMMsmooth}' hold and $H$ is positive definite. Then:
	\begin{enumerate}
		\item If Assumption \ref{as:varGMMsmooth}' holds as well, $\widehat{V}\convP V_0$ and confidence regions and tests on $\theta_0$ based on asymptotic normality and $\widehat{V}$ are asymptotically valid.
		\item Confidence regions and tests on $\theta_0$ based on the pigeonhole bootstrap are asymptotically valid.
	\end{enumerate}
\end{thm}
\setcounter{hyp}{6}

Last, we can notice that the asymptotic normality of the GMM holds if we mix Assumptions \ref{as:GMM} and \ref{as:GMMsmooth}'. Namely, we could consider instead that there exist subsets $S$ and $S'$ of $\{1,...,L\}$ such that $S\cup S'=\{1,...,L\}$ and the classes $\mathcal{F}_s$ fulfill Assumption \ref{as:vc} for $s\in S$ and Assumption \ref{as:smooth} for $s\in S'$. This is because the asymptotic normality essentially follows from the fact that each moment function belongs to a class satisfying a uniform CLT, like Theorems \ref{thm:unifTCL} or \ref{thm:unifTCL_bis}. Similarly, validity of the inference still holds if we mix Assumptions \ref{as:varGMMsmooth} and \ref{as:varGMMsmooth}'.

% section weak_convergence_under_another_restriction_on_mathcal_f (end)

\newpage
\section{Additional details on the simulations} % (fold)
\label{sec:additional_details_on_the_simulations}

\subsection{Probit in a two-way design}

We adapt the formulas for $\widehat{\Sigma}_i$, $\widetilde{\Sigma}_i$ and $\widehat{\Sigma}_{12}$ using Equations (19-21) and (19-23) in \cite{greene2000} for the score and the Hessian matrix of the log-likelihood. Specifically, let
\begin{align*}
\lambda_{\ell,\j}(\widehat{\beta}) & =(2Y_{\ell,\j}-1)\frac{\phi\left((2Y_{\ell,\j}-1)(\widehat{\beta}_0+\widehat{\beta}_1 X_{\ell,\j})\right)}{\Phi\left((2Y_{\ell,\j}-1)(\widehat{\beta}_0+\widehat{\beta}_1 X_{\ell,\j})\right)}, \quad s_{\ell,\j}(\widehat{\beta})=\lambda_{\ell,\j}(\widehat{\beta})\left(\begin{array}{c}1\\X_{\ell,\j}\end{array}\right) \\
J & =-\frac{1}{\Pi_C}\sum_{\un \leq \j \leq \C}\sum_{\ell=1}^{N_{\j}}\lambda_{\ell,\j}(\widehat{\beta})\left(\widehat{\beta}_0+\widehat{\beta}_1X_{\ell,\j}+\lambda_{\ell,\j}(\widehat{\beta})\right) \left(\begin{array}{cc}1&X_{\ell,\j}\\X_{\ell,\j}&X_{\ell,\j}^2\end{array}\right).
\end{align*}
Then $\widehat{\Sigma}_i$, $\widetilde{\Sigma}_i$ and $\widehat{\Sigma}_{12}$ satisfy
\begin{align*}
\widehat{\Sigma}_1&=\frac{C_1}{C_1-1}J^{-1}\frac{1}{\Pi_C^2}\left[\sum_{j_1=1}^{C_1}\left(\sum_{j_2=1}^{C_2} \sum_{\ell=1}^{N_{(j_1,j_2)}}s_{\ell,(j_1,j_2)}(\widehat{\beta})\right)\left(\sum_{j_2=1}^{C_2} \sum_{\ell=1}^{N_{(j_1,j_2)}}s'_{\ell,(j_1,j_2)}(\widehat{\beta})\right)\right]J^{-1},\\
\widehat{\Sigma}_{12}&=\frac{C_1C_2}{C_1C_2-1}J^{-1}\frac{1}{\Pi_C^2}\sum_{\un \leq \j \leq \C}\left[ \sum_{\ell=1}^{N_{\j}}s_{\ell,\j}(\widehat{\beta}) \sum_{\ell=1}^{N_{\j}}s'_{\ell,\j}(\widehat{\beta})\right]J^{-1},\\
\widetilde{\Sigma}_1&=\frac{C_1}{C_1-1}J^{-1}\frac{1}{C_1^2C_2(C_2-1)}\left[\sum_{j_1=1}^{C_1}\sum_{\substack{1\leq j_2,j_2'\leq C_2\\
		j_2'\neq j_2}} \left(\sum_{\ell=1}^{N_{(j_1,j_2)}}s_{\ell,(j_1,j_2)}(\widehat{\beta}) \sum_{\ell=1}^{N_{(j_1,j_2')}}s'_{\ell,(j_1,j_2')}(\widehat{\beta})\right)\right]J^{-1}.
	\end{align*}

\subsection{Average in the three-way design}

In this design, $\widehat{V}_1$ (resp. $\widehat{V}_2$) remains as in our baseline scenario up to the additional term $\widehat{\Sigma}_3$ (resp. $\widetilde{\Sigma}_3$), which is similar to $\widehat{\Sigma}_1$ (resp. $\widetilde{\Sigma}_1$). The expression of $\widehat{V}_{cgm}$ is slightly more complex. By the inclusion-exclusion principle,
$$\widehat{V}_{cgm}=\widehat{V}_1-\underline{C}\widehat{\Sigma}_{12}-\underline{C}\widehat{\Sigma}_{23}-\underline{C}\widehat{\Sigma}_{13}+\underline{C}\widehat{\Sigma}_{123},$$
with:
$$\widehat{\Sigma}_{12}=\frac{C_1C_2}{C_1C_2-1} \frac{1}{\Pi_C^2}\sum_{j_1=1}^{C_1}\sum_{j_2=1}^{C_2} \left(\sum_{j_3=1}^{C_3}\sum_{\ell=1}^{N_{\j}}(Y_{\ell,\j}-\widehat{\theta})\right)^2,$$
$$\widehat{\Sigma}_{123}=\frac{C_1C_2C_3}{C_1C_2C_3-1} \frac{1}{\Pi_C^2}\sum_{\un \leq \j\leq \C} \left(\sum_{\ell=1}^{N_{\j}}(Y_{\ell,\j}-\widehat{\theta})\right)^2.$$ 

% section additional_details_on_the_simulations (end)

\newpage
\section{Proofs of the main results} % (fold)
\label{sec:proofs_of_the_main_results}

% section proofs_of_the_main_results (end)

\subsection{Preliminaries}
\label{sec:notation}

\subsubsection{Notation} % (fold)
\label{sub:notation}

We first introduce or recall the notation used throughout the proofs.

\medskip
\textbf{Algebra in $\mathbb{N}^k$}

\begin{longtable}{lp{13cm}}
	 $\j, \j', \e, \C$... & elements of $\mathbb{N}^k$, with respective component $(j_1,...,j_k)$, $(j'_1,...,j'_k)$, $(e_1,...,e_k)$, $(C_1,...,C_k)$. Hereafter, $\e$ is always in $\{0,1\}^k$ \\
	$\zero,\un,\deu$ &  respectively $(0,...,0)$, $(1,...,1)$, $(2,...,2)$
	\\ $\j\leq \j'$ & for all $i=1,...,k$, $j_i\leq j'_i$
	\\ $\j< \j'$ & $\j\leq \j'$ and $\j \neq \j'$
	\\ $\odot$ &the Hadamard product, i.e. $\j \odot \j'=(j_1 j'_1,...,j_k j'_k)$
	\\ $\vee$ and $\wedge$ & the componentwise maximum and minimum, respectively. 
	\\ $\mathcal{E}_{i}$ & $\{\e \in\{0;1\}^k: \sum_{i'=1}^k e_{i'}=i\}$ for $i=1,...,k$
	\\ $\mathcal{I}_i$ &   $\{\j \odot \e: \un \leq \j, \e \in \mathcal{E}_i\}$ for $i=1,...,k$	
	\\ $\mathcal{I}_i(\C)$ &  $\{\j \odot \e: \un \leq \j\leq \C, \, \e \in \mathcal{E}_i\}$ for $i=1,...,k$
	\\ $\e \preceq \e'$ & either $\sum_{i=1}^k \mathds{1}\{e_i=0\} > \sum_{i=1}^k \mathds{1}\{e'_i=0\} $, or $\sum_{i=1}^k \mathds{1}\{e_i=0\} = \sum_{i=1}^k \mathds{1}\{e'_i=0\} $ and $\sum_{i=1}^k e_i \times 10^i\leq \sum_{i=1}^k e'_i \times 10^i$.
	\\ $\e \prec \e'$ & $\e \preceq \e'$ and $\e \neq \e'$.\\
	$\mathcal{A}_{\e}$& 	$\{(\j,\j'):\un \leq \j,\j'\leq \C: \forall i=1,...,k, ~e_i=1 \Leftrightarrow j_i=j'_i\}$ \\
	$\mathcal{B}_{\e}$&	$\{(\j,\j'):\un \leq \j,\j'\leq \C: \forall i=1,...,k, ~e_i=1 \Rightarrow j_i=j'_i\}$ \\
	$\mathcal{A}_{i}$ and $\mathcal{B}_{i}$& $\mathcal{A}_{\e}$ and $\mathcal{B}_{\e}$ for $\e=(0,...,0,1,0,...,0)$ with 1 located at component $i$.
\end{longtable}

\medskip
\textbf{Classes of functions}

\begin{longtable}{lp{13cm}}
$\mathcal{F}^2$& $\{f^2: f\in \mathcal{F}\}$\\
$\mathcal{F}\times \mathcal{F}$ or $(\mathcal{F})^2$&  $\{(f_1, f_2): f_1\in \mathcal{F}, f_2\in \mathcal{F}\}$\\
%$(\mathcal{F})^m$& the m-ary Cartesian power of $\mathcal{F}$ for any $m\geq 2$\\
$\mathcal{F}_{\delta} $& $\left\{h=f_1-f_2:(f_1,f_2)\in\mathcal{F}\times\mathcal{F},\E\left(\left( \sum_{\ell=1}^{N_{\un}} h(Y_{\ell,\un})\right)^2\right)\leq\delta^2\right\}.$\\
$\mathcal{F}_{\infty} $& $\left\{h=f_1-f_2:(f_1,f_2)\in\mathcal{F}\times\mathcal{F}\right\}.$ \\
$\mathcal{F}_{\infty}^2$& $\left\{(f_1-f_2)^2: (f_1,f_2)\in \mathcal{F}\times \mathcal{F}\right\}$\\
$\widetilde{f}$& the function $\left(n,\left(y_l\right)_{\ell\geq 1}\right)\mapsto\sum_{\ell=1}^nf(y_l)$\\
$\widetilde{\mathcal{F}}$&  $\left\{(n,(y_{\ell})_{\ell \in \mathbb{N}})\in \mathbb{N}\times \mathcal{Y}^{\mathbb{N}}\mapsto \sum_{\ell=1}^{n} f(y_{\ell}):f\in\mathcal{F}\right\}$. \\
Id & The identity function. \\
\end{longtable}

\medskip
\textbf{Additional random variables and probability measures}
\medskip

Note that we sometimes need to evaluate random variables at some specific value of the probability space. We denote by $\omega$ elements of this probability space $\Omega$.

\begin{longtable}{lp{13cm}}
 $\vec{Y}_{\j}$& %the sequence of the $Y$ within the cell $\j$: 
 $\left(Y_{\ell,\j}\right)_{\ell\geq 1}$\\
 $\bm{Z}$ & $(N_{\j},\vec{Y}_{\j})_{\j\geq \un}$ \\
 $A_F$& $N_{\un}\sum_{\ell=1}^{N_{\un}}F\left(Y_{\ell,\un}\right)^2$\\ 
 $A_{\beta}$&$\sum_{\ell=1}^{N_{\un}}(1+|Y_{\ell,\un}|^2)^{-\beta/2}$\\
$A_r$&$\frac{1}{\Pi_C}\sum_{\un\leq\j\leq\C}\left|\sum_{\ell=1}^{N_{\j}}(1+|Y_{\ell,\j}|^2)^{-\beta/2}\right|^r$ for $r\in \mathbb{N}$\\ 
$\overline{N}_r$&$\frac{1}{\Pi_C}\sum_{\un\leq \j \leq \C}N_{\j}^r$ for $r\in \mathbb{N}$\\
 $\sigma^2_C$&$\sup_{f\in \mathcal{F}_{\delta}} \frac{1}{\Pi_C}\sum_{\un\leq\j\leq\C}\left(\sum_{\ell=1}^{N_{\j}}f(Y_{\ell,\j})\right)^2$ \\
$(\epsilon_{\c})_{\c\geq \zero}$ & a Rademacher process on $\mathbb{N}^k$, i.e. a set of independent random variables such that $\mathbb{P}(\epsilon_{\c}=1)=\mathbb{P}(\epsilon_{\c}=-1)=1/2$ for any $\c\in \mathbb{N}^k$ (such variables are called Rademacher variables) \\
$P$ & The probability distribution of $(N_{\un},\vec{Y}_{\un})$ \\
$\mu_C$ & $\frac{1}{\Pi_C}\sum_{\un\leq\j\leq\C}\delta_{N_{\j},\vec{Y}_{\j}}$, with $\delta_a$ the Dirac measure at $a$ \\
$\mathbb{Q}_C^r$&$\frac{1}{\overline{N}_r\Pi_C}\sum_{\un\leq\j\leq\C}N_{\j}^{r-1}\sum_{\ell=1}^{N_{\j}}\delta_{Y_{\ell,\j}}$, with $r\in \mathbb{N}$.
\end{longtable}
note that if $N_{\j}=0$ for every $\j\leq \C$, the random variables $A_F$, $A_{\beta}$, $\overline{N}_r$, $A_r$,  and $\sigma^2_C$ are equal to 0, and the random measures $\mu_C$ and $\mathbb{Q}_C^r$ are the null measures.

\medskip
\textbf{Linear algebra and norms}
\medskip
 
\begin{longtable}{lp{13cm}}
$B^{1/2}$ & the square root of $B$, for any symmetric, positive matrix $B$ \\
$\rho(B)$ & the largest modulus of the eigenvalues of a symmetric matrix $B$ \\
$|b|$ & the euclidean norm of a vector $b$ \\
 $||g||_{\mu, r}$& $\left(\int |g|^r d\mu\right)^{1/r}$ for $\mu$ a measure and $r\geq 1$ \\
  $\left|\left|g\right|\right|_{\e, r}^r$&%the random variable 
  $\frac{1}{\prod_{s:\e_s=1}C_s}\sum_{ \c: \e\leq\c\leq\C\odot \e}\left|\frac{1}{\prod_{s:\e_s=0}C_s}\sum_{\c':(\un-\e)\leq\c'\leq\C\odot(\un-\e)}g(N_{\c+\c'},\vec{Y}_{\c+\c'})\right|^r$ \\
\end{longtable}

Our notation implies for instance that $||\widetilde{f}||^r_{P, r} = \E\left[\left|\sum_{\ell=1}^{N_{\un}} f(Y_{\ell, \un})\right|^r\right]$ and
$$	||\widetilde{f}||^r_{\mu_C, r}= \left[\frac{1}{\Pi_C}\sum_{\un\leq\j\leq\C}\left|\sum_{\ell=1}^{N_{\j}} f(Y_{\ell, \j}) \right|^r\right].$$

\medskip
\textbf{Covering numbers}

\begin{longtable}{lp{13cm}}
$N(\varepsilon,\mathcal{F},\abs{\abs{.}})$& the covering numbers, i.e the minimal number of closed balls for the semi-norm $\abs{\abs{.}}$, of radius $\varepsilon$ and centers in $\mathcal{F}$ that is necessary to cover $\mathcal{F}$. We follow here the convention adopted by \cite{GineNickl2015} or \cite{Kato2016}, which has the advantage of automatically dealing with some degenerate cases.\footnote{In particular, if the semi-norm  $\abs{\abs{.}}$ is null on $\mathcal{F}$ then $N(\epsilon,\mathcal{F},\abs{\abs{.}})=1$ for any $\varepsilon\geq 0$.} \\
$J_{p,\mathcal{F}}(u)$&$\int_0^u\sup_Q\left(\log N\left(\varepsilon\left|\left|F\right|\right|_{Q,p},\mathcal{F},\left|\left|.\right|\right|_{Q,p}\right)\right)^{1/2}d\varepsilon$ where the supremum is taken on the set of measures on $\mathcal{Y}$ with finite support (including the null measure).\\
 $J_{\infty,\beta,\mathcal{F}}(u)$&$\int_0^u\left(\log N(\varepsilon\abs{\abs{F}}_{\infty,\beta},\mathcal{F},\abs{\abs{.}}_{\infty,\beta})\right)^{1/2}d\varepsilon$.
\end{longtable}

\medskip
\textbf{The bootstrap}
  
Generally speaking, we simply add a star to indicate the bootstrap counterpart of a random variable. We also define the following elements.

\begin{longtable}{lp{13cm}}
$\stackrel{d^*}{\longrightarrow}$ & convergence in distribution conditional on $(N_{\j},\vec{Y}_{\j})_{\j\geq \un}$.\\
	$N^{*}_{\j},\vec{Y}^*_{\j}$&the bootstraped cell $\j$, corresponding to the intersection of the $j_1$-th draw in dimension $i=1$, the $j_2$-th draw in dimension $i=2$ etc. \\
	$N^{\e *}_{\j},\vec{Y}^{\e *}_{\j}$& for $\zero \leq \e \leq \un$, the cell whose component $j_i$ corresponds to the component of the $j_i$-th draw in dimension $i$ if $e_i=1$, the component $j_i$ if $e_i=0$. In particular: 	$(N^{\un *}_{\j},\vec{Y}^{\un *}_{\j})=(N^{*}_{\j},\vec{Y}^{*}_{\j})$ and $(N^{\zero *}_{\j},\vec{Y}^{\zero *}_{\j})=(N_{\j},\vec{Y}_{\j})$\\
	 $A^*_F$& $N^*_{\un}\sum_{\ell=1}^{N^*_{\un}}F\left(Y_{\ell,\un}^*\right)^2$\\ $A^*_{\beta}$&$\sum_{\ell=1}^{N^*_{\un}}(1+|Y^*_{\ell,\un}|^2)^{-\beta/2}$
\end{longtable}

% subsection notation (end)

\subsubsection{Measurability issues} % (fold)
\label{sub:measurability_issues}

If $\mathcal{F}$ is pointwise measurable, there exists a countable subclass $\mathcal{H}$ of $\mathcal{F}$ such that for any $f\in \mathcal{F}$, there exist $(f_k)_{k\in\N} \in\mathcal{H}^{\N}$ converging pointwise to $f$. For any $g\in \widetilde{\mathcal{F}}$, there exists $f\in \mathcal{F}$ such that $g(n,\vec{y})=\sum_{\ell=1}^{+\infty}f(y_{\ell})\mathds{1}_{\{\ell\leq n\}}$, and next there exists $(f_k)_{k\in\N} \in\mathcal{H}^{\N}$ such that $\sum_{\ell=1}^{+\infty}f_k(y_{\ell})\mathds{1}_{\{\ell\leq n\}}$ converges to $\sum_{\ell=1}^{+\infty}f(y_{\ell})\mathds{1}_{\{\ell\leq n\}}$, for any $(n,\vec{y})$. Because $$\widetilde{\mathcal{H}}=\left\{h(n,\vec{y})=\sum_{\ell=1}^{+\infty}h(y_{\ell})\mathds{1}_{\{\ell\leq n\}}:h\in\mathcal{H}\right\}$$ 
is a countable subclass of $\widetilde{\mathcal{F}}$, we deduce that $\widetilde{\mathcal{F}}$ is pointwise measurable. Lemma 8.10 in \cite{Kosorok2006} also ensures that $\widetilde{\mathcal{F}}_{\infty}=[\widetilde{\mathcal{F}}]_{\infty}$, $\widetilde{\mathcal{F}^2}$ and $\widetilde{\mathcal{F}_{\infty}}^2$ are pointwise measurable. If $F$ is an envelope of $\mathcal{F}$ such that $\E\left[\left(\sum_{\ell=1}^{N_{\un}}F(Y_{\ell,\un})\right)^2\right]<+\infty$, pointwise measurability of $\widetilde{\mathcal{F}}$ also extends to $\widetilde{\mathcal{F}}_{\delta}=[\widetilde{\mathcal{F}}]_{\delta}$ for any $\delta>0$ \citep[see for instance Proposition 8.11 in][]{Kosorok2006}. These properties ensure that we can ignore measurability issue and consider usual expectations and probabilities in the following proofs, instead of outer expectations or outer probabilities.

% subsection measurability_issues (end)

\subsubsection{Representation lemma}

%\subsection{Representation Lemma}
\label{sec:representation_lem}

\begin{lem}
	Let Assumptions \ref{as:dgp}.1 and \ref{as:dgp}.2 hold. Then there exists a measurable function $\tau$ such that
    \begin{align*}
    	&\left\{N_{\j},\vec{Y}_{\j}\right\}_{\j\geq\un}\overset{a.s.}{=}\left\{\tau\left(\left(U_{\j\odot\e}\right)_{\zero\prec\e\preceq \un}\right)\right\}_{\j\geq\un},
    \end{align*}
    where $\left(U_{\c}\right)_{\c > \zero}$ is a family of mutually independent random variables, which are uniform on $[0,1]$.
    \label{lem:representation}
\end{lem}

The result follows directly from the equivalence between Conditions $(ii)$ and $(iii)$ in Lemma 7.35 of \cite{kallenberg05}.

%\subsection{Proof of Theorem \ref{thm:TCL}}\label{ssec:proof_TCL}

\subsection{Proof of Theorems~\ref{thm:unifTCL} and \ref{thm:unifTCL_bis}}
\label{ssec:proof_unifTCL}

The proof consists in three important steps. We first prove the asymptotic normality of $(\mathbb{G}_Cf_1,...,\mathbb{G}_Cf_m)$ for any $m\geq 1$ and $(f_1,...,f_m)\in \mathcal{F}\times ... \times \mathcal{F} $. Second, we prove the asymptotic equicontinuity of $\mathbb{G}_C$, as a process indexed by functions in $\widetilde{\mathcal{F}}$. Third, we check the total boundedness of $\widetilde{\mathcal{F}}$ for the norm $\abs{\abs{.}}_{P,2}$ (recall that $\abs{\abs{\widetilde{f}}}^2_{P,2}=\E\left[\left(\sum_{\ell=1}^{N_{\un}}f(Y_{\ell,\un})\right)^2\right]$). By Theorems 1.5.4 and 1.5.7 in \cite{vanderVaartWellner1996}, this shows the weak convergence of $\mathbb{G}_C$ towards a centered Gaussian process. %The first and second steps are themselves divided into several substeps.
 %The form of the kernel $K$ is established in Step \ref{ssec:cramer_wold_multiway} below. 

\subsubsection{Asymptotic normality of $(\mathbb{G}_Cf_1,...,\mathbb{G}_Cf_m)$}

By the Cram\'er-Wold device, it suffices to prove the asymptotic normality for any single function $f$ of the form $f=\sum_{s=1}^{m}t_s f_s$, with $(t_s)_{s=1,...,m}\in \mathbb{R}^{m}$. Note for such a $f$, we have $\E\left(\left|\sum_{\ell=1}^{N_{\un}}f(Y_{\ell,\un})\right|^2\right)\leq \sum_{s=1}^{m}|t_s|^2\E\left(\left|\sum_{\ell=1}^{N_{\un}}F(Y_{\ell,\un})\right|^2\right)\leq  \sum_{s=1}^{m}|t_s|^2 \E\left(N_{\un}\sum_{\ell=1}^{N_{\un}}F^2(Y_{\ell,\un})\right)$, which is finite by Assumption \ref{as:vc}. We prove the result in two steps. First, we show that the H\'ajek projection $H_1 (f)$ of $\mathbb{G}_C f$ on a  suitable subspace is asymptotically normal, we compute its variance and we show that $\V(\mathbb{G}_C f)= \V(H_1 f)(1+o(1))$. Second, we show asymptotic normality of $\mathbb{G}_Cf$, using its asymptotic equivalence with $H_1 f$.

 \medskip
 \textbf{a. H\'ajek Projection and comparison of variances}
 
 \medskip
Let $H_1 f$ denote the H\'ajek projection  of $\mathbb{G}_C f$ on the linear subspace of random variables $\sum_{\c \in \mathcal{I}_1(\C)} g_{\c}(U_{\c})$, for $g_{\c}\in L^2([0,1])$ and $(U_{\c})_{\c > \zero}$ the random variables defined in the representation Lemma \ref{lem:representation}. We have $H_1 f=\sum_{\c\in \mathcal{I}_1(\C)}\E\left(\mathbb{G}_C f | U_{\c}\right)$. Moreover, Lemma \ref{lem:hajek} applied to $r=\underline{r}=1$ implies
\begin{align}
H_1 f & \convD \mathcal{N}\left(0, \sum_{i=1}^{k} \lambda_i \Cov\left(\sum_{\ell=1}^{N_{\un}}f(Y_{\ell,\un}),\sum_{\ell=1}^{N_{\deux}}f(Y_{\ell,\deux})\right)\right) \nonumber \\	
\V(H_1 f)&=\sum_{i=1}^{k}\frac{\underline{C}}{C_i} \Cov\left(\sum_{\ell=1}^{N_{\un}}f(Y_{\ell,\un}), \sum_{\ell=1}^{N_{\deux}}f(Y_{\ell,\deux})\right). \label{eq:variance_H1}
\end{align}

Now, let us expand $\mathbb{V}\left(\mathbb{G}_C f\right)$ using the fact that cells without common component are independent:
	\begin{align*}
 \mathbb{V}\left(\mathbb{G}_C f\right)
=&\frac{\underline{C}}{\left(\Pi_C\right)^2}\sum_{\un\leq\j, \j'\leq\C}\Cov\left(\sum_{\ell=1}^{N_{\j}}f(Y_{\ell,\j}),\sum_{\ell=1}^{N_{\j'}}f(Y_{\ell,\j'})\right) \\
=& \frac{\underline{C}}{\left(\Pi_C\right)^2}\sum_{\e\in\mathcal{E}_1}\sum_{(\j,\j')\in \mathcal{A}_{\e}}\Cov\left(\sum_{\ell=1}^{N_{\j}}f(Y_{\ell,\j}),\sum_{\ell=1}^{N_{\j'}}f(Y_{\ell,\j'})\right) \\
&+ \frac{\underline{C}}{\left(\Pi_C\right)^2}\sum_{r=2}^k \sum_{\e\in\mathcal{E}_r}\sum_{(\j,\j')\in \mathcal{A}_{\e}}\Cov\left(\sum_{\ell=1}^{N_{\j}}f(Y_{\ell,\j}),\sum_{\ell=1}^{N_{\j'}}f(Y_{\ell,\j'})\right).
\end{align*}
Let us call $R$ the second term of the last right-hand side. By the Cauchy-Schwarz inequality and exchangeability of the cells,
\begin{align*}
|R|&\leq \frac{\underline{C}}{\left(\Pi_C\right)^2} \sum_{\e \in \cup_{l=2}^k \mathcal{E}_l} |\mathcal{A}_{\e}| \V\left(\sum_{\ell=1}^{N_{\un}}f(Y_{\ell,\un})\right).
\end{align*}

For all $r \geq 1$ and $\e \in \mathcal{E}_r$, we have 
\begin{equation}
|\mathcal{A}_{\e}|=\Pi_C \times \prod_{s:e_s=0}(C_s-1).
	\label{eq:enum_Ae}
\end{equation}
Thus, $R=O(\underline{C}^{-1})$. Moreover, let $\e \in \mathcal{E}_1$ and $i$ denotes its component such that $e_i=1$. By the exchangeability assumption and \eqref{eq:enum_Ae}, 
\begin{align*}
\frac{\underline{C}}{\left(\Pi_C\right)^2}\sum_{(\j,\j')\in \mathcal{A}_{\e}}\Cov\left(\sum_{\ell=1}^{N_{\j}}f(Y_{\ell,\j}),\sum_{\ell=1}^{N_{\j'}}f(Y_{\ell,\j'})\right)
&=\frac{\underline{C}}{\left(\Pi_C\right)^2}|\mathcal{A}_{\e}|\Cov\left(\sum_{\ell=1}^{N_{\un}}f(Y_{\ell,\un}),\sum_{\ell=1}^{N_{\deu-\e}}f(Y_{\ell,\deu-\e})\right)\\
%&=\lambda_i\Cov\left(\sum_{\ell=1}^{N_{\un}}f(Y_{\ell,\un}),\sum_{\ell=1}^{N_{\deux}}f(Y_{\ell,\deux})\right)+o(1)\\
&=\frac{\underline{C}}{C_i}\Cov\left(\sum_{\ell=1}^{N_{\un}}f(Y_{\ell,\un}),\sum_{\ell=1}^{N_{\deux}}f(Y_{\ell,\deux})\right)+O(\underline{C}^{-1}).
\end{align*}
Together with  $R=O(\underline{C}^{-1})$, this implies $\mathbb{V}\left(\mathbb{G}_C f\right)=\V\left(H_1 f\right)+o(1)$.
	
\medskip	
\textbf{b. Asymptotic normality of $\mathbb{G}_{C}f$}
	
\medskip	
By \eqref{eq:variance_H1}
$$\lim_{\underline{C}\rightarrow \infty}\V(H_1 f)=\sum_{i=1}^{k} \lambda_i \Cov\left(\sum_{\ell=1}^{N_{\un}}f(Y_{\ell,\un}),\sum_{\ell=1}^{N_{\deux}}f(Y_{\ell,\deux})\right)<\infty.$$ 
If $\sum_{i=1}^{k} \lambda_i \Cov\left(\sum_{\ell=1}^{N_{\un}}f(Y_{\ell,\un}),\sum_{\ell=1}^{N_{\deux}}f(Y_{\ell,\deux})\right)=0$ then $\lim_{\underline{C}\rightarrow \infty}\V(\mathbb{G}_Cf)=0$. Because $\mathbb{G}_Cf$ is centered, this means that $\mathbb{G}_Cf$ converges in $L^2$, and thus in distribution, to $0\overset{d}{=}\mathcal{N}(0,0)$. Otherwise, by Step b, we have $\mathbb{V}\left(\mathbb{G}_{C} f\right)/\V(H_1 f)\rightarrow 1$. Then, by Theorem 11.2 of \cite{vanderVaart2000},
$$\frac{\mathbb{G}_C f}{\V(\mathbb{G}_C f)^{1/2}}-\frac{H_1f }{\V(H_1f)^{1/2}}=o_p(1).$$
By Slutsky's Lemma and the asymptotic normality of $H_1f/\V(H_1f)^{1/2}$, $$\frac{\mathbb{G}_{C}f}{\V(\mathbb{G}_{C}f)^{1/2}}\convD\mathcal{N}\left(0,1\right).$$ 
Finally, by the previous step again, 
$$\V(\mathbb{G}_{C}f)=\V(H_1f)+o(1)=\sum_{i=1}^{k} \lambda_i \Cov\left(\sum_{\ell=1}^{N_{\un}}f(Y_{\ell,\un}),\sum_{\ell=1}^{N_{\deux}}f(Y_{\ell,\deux})\right)+o(1).$$ 
Hence, by Slutsky's Lemma,
	$$\mathbb{G}_{C}f\convD\mathcal{N}\left(0,\sum_{i=1}^{k} \lambda_i \Cov\left(\sum_{\ell=1}^{N_{\un}}f(Y_{\ell,\un}),\sum_{\ell=1}^{N_{\deux}}f(Y_{\ell,\deux})\right)\right).$$

\subsubsection{Asymptotic equicontinuity}\label{sssec:equicontinuity}

The asymptotic equicontinuity of $\mathbb{G}_C$ can be stated as
\begin{align}\label{eq:multi_unif_res_1}
&\lim_{\delta\to 0}\limsup_{\underline{C}\to +\infty}\mathbb{P}\left(\sup_{f\in \mathcal{F}_{\delta}}\left|\mathbb{G}_Cf\right|>\epsilon\right)=0.
\end{align}

By Markov's inequality, for any $\epsilon>0$, $$\mathbb{P}\left(\sup_{\mathcal{F}_{\delta}}\left|\mathbb{G}_Cf\right|>\epsilon\right)\leq\frac{1}{\epsilon}\mathbb{E}\left[\sup_{\mathcal{F}_{\delta}}\left|\mathbb{G}_Cf\right|\right].$$ 
Thus, it is sufficient to show \begin{align}&\lim_{\delta\to 0}\limsup_{\underline{C}\to +\infty}\mathbb{E}\left[\sup_{\mathcal{F}_{\delta}}\left|\mathbb{G}_Cf\right|\right].\label{eq:asym_equic}\end{align}
 We first establish an inequality on this expectation, which involves 
 $$\sigma^2_C=\sup_{\mathcal{F}_{\delta}} \frac{1}{\Pi_C}\sum_{\un\leq\j\leq\C}\left(\sum_{\ell=1}^{N_{\un}}f(Y_{\ell,\un})\right)^2.$$ 
 Next, we show that $\lim_{\delta\to 0}\limsup_{\underline{C}\to+\infty}\mathbb{E}\left[\sigma_{C}^2\right]=0$. We conclude in a third step.

\medskip
\textbf{1. Upper bound on the supremum of the empirical process}

\medskip
By Lemma \ref{lem:representation}, we have $\left(N_{\j},\vec{Y}_{\j}\right)_{\j\geq\un}\overset{a.s.}{=} \left(\tau\left(\left(U_{\j\odot\e}\right)_{\zero\prec \e \preceq \un}\right)\right)_{\j\geq\un}$ with $(U_{\c})_{\c>\zero}$ independent and uniform variables. Then, by Lemma \ref{lem:sym} applied to $Z_{\j}=\left(N_{\j},\vec{Y}_{\j}\right)$, $\mathcal{G}=\widetilde{\mathcal{F}_{\delta}}=
\{g\left(N_{\j},\vec{Y}_{\j}\right)=\sum_{\ell=1}^{N_{\j}}f(Y_{\ell,\j}): f\in \mathcal{F}_{\delta}\}$ and $\Phi=\Id$, we have
$$\mathbb{E}\left[\sup_{\mathcal{F}_{\delta}}\left|\mathbb{G}_Cf\right|\right]\leq 2\sqrt{\overline{C}} \sum_{\zero\prec \e \preceq \un}\mathbb{E}\left[\sup_{\mathcal{F}_{\delta}}\left|\frac{1}{\Pi_C}\sum_{\un\leq\j\leq\C}\epsilon_{\j\odot\e}\sum_{\ell=1}^{N_{\j}}f\left(Y_{\ell,\j}\right)\right|\right],$$
where $(\epsilon_{\c})_{\c\geq \zero}$ is a Rademacher process, independent of $(Z_{\j})_{\j\geq \un}$. By Lemma~\ref{lemma:donsker_entropy_bound}, we can control the right-hand side under Assumptions~\ref{as:dgp}-\ref{as:measurability} and \ref{as:vc} or \ref{as:smooth}'. 
Specifically, let  $A_F=N_{\un}\sum_{\ell=1}^{N_{\un}}F\left(Y_{\ell,\un}\right)^2$, $A_{\beta}=\sum_{\ell=1}^{N_{\un}}(1+|Y_{\ell,\un}|^2)^{-\beta/2}$,  and 
$$\sigma^2_C=\sup_{\mathcal{F}_{\delta}} \frac{1}{\Pi_C}\sum_{\un\leq\j\leq\C}\left(\sum_{\ell=1}^{N_{\un}}f(Y_{\ell,\un})\right)^2.$$ 
Under Assumptions~\ref{as:dgp}-\ref{as:measurability} and \ref{as:vc} , there exists $K(F,P)$, depending only of the envelope $F$ of $\mathcal{F}$ and the distribution $P$ of $(N_{\un},\vec{Y}_{\un})$, such that $$\mathbb{E}\left[\sup_{\mathcal{F}_{\delta}}\left|\mathbb{G}_Cf\right|\right]\leq  K(F,P)\left\{\sqrt{\mathbb{E}\left[\sigma_{C}^2\right]}+ J_{2,\mathcal{F}}\left(\frac{1}{4}\sqrt{\frac{\mathbb{E}\left[\sigma_{C}^2\right]}{\E\left[A_F\right]}}\right)\right\}.$$
Similarly, under Assumptions~\ref{as:dgp}-\ref{as:measurability} and \ref{as:smooth}', 
$$\mathbb{E}\left[\sup_{\mathcal{F}_{\delta}}\left|\mathbb{G}_Cf\right|\right]\leq  K(F,P) \left\{\sqrt{\mathbb{E}\left[\sigma_{C}^2\right]}+J_{\infty,\beta,\mathcal{F}}\left(\frac{1}{4}\sqrt{\frac{\mathbb{E}\left[\sigma_{C}^2\right]}{\left|\left|F\right|\right|^2_{\infty,\beta}\mathbb{E}\left[A^2_{\beta}\right]}}\right)\right\}.$$

\medskip
\textbf{2. $\lim_{\delta\to 0}\limsup_{\underline{C}\to+\infty}\mathbb{E}\left[\sigma_{C}^2\right]=0$.} 

\medskip
We have
\begin{align*}
\sigma^2_C&=\sup_{\mathcal{F}_{\delta}} \frac{1}{\Pi_C}\sum_{\un\leq\j\leq\C}\left(\sum_{\ell=1}^{N_{\un}}f(Y_{\ell,\un})\right)^2\\
&\leq \sup_{\mathcal{F}_{\delta}}\left\{\left|\text{ }\frac{1}{\Pi_C}\sum_{\un\leq\j\leq\C}\left(\sum_{\ell=1}^{N_{\j}}f(Y_{\ell,\j})\right)^2-\E\left[\left(\sum_{\ell=1}^{N_{\un}}f(Y_{\ell,\un})\right)^2\right]\text{ }\right|+\E\left[\left(\sum_{\ell=1}^{N_{\un}}f(Y_{\ell,\un})\right)^2\right]\right\}\\
&\leq \sup_{\mathcal{F}_{\delta}}\left\{\left|\text{ }\frac{1}{\Pi_C}\sum_{\un\leq\j\leq\C}\left(\sum_{\ell=1}^{N_{\j}}f(Y_{\ell,\j})\right)^2-\E\left[\left(\sum_{\ell=1}^{N_{\un}}f(Y_{\ell,\un})\right)^2\right]\text{ }\right|\right\} \\ & +\sup_{\mathcal{F}_{\delta}}\left\{\E\left[\left(\sum_{\ell=1}^{N_{\un}}f(Y_{\ell,\un})\right)^2\right]\right\}\\
&\leq \sup_{\mathcal{F}_{\infty}}\left|\text{ }\frac{1}{\Pi_C}\sum_{\un\leq\j\leq\C}\left(\sum_{\ell=1}^{N_{\j}}f(Y_{\ell,\j})\right)^2-\E\left[\left(\sum_{\ell=1}^{N_{\un}}f(Y_{\ell,\un})\right)^2\right]\text{ }\right|+\delta^2.
\end{align*}
Thus, it suffices to show that 
\begin{equation}
\sup_{\mathcal{F}_{\infty}}\left|\text{ }\frac{1}{\Pi_C}\sum_{\un\leq\j\leq\C}\left(\sum_{\ell=1}^{N_{\j}}f(Y_{\ell,\j})\right)^2-\E\left[\left(\sum_{\ell=1}^{N_{\un}}f(Y_{\ell,\un})\right)^2\right]\text{ }\right|\xrightarrow{L_1}0.	
	\label{eq:ineg_for_sigma2}
\end{equation}
Lemma \ref{lem:sym} applied to $Z_{\j}=\left(N_{\j},\vec{Y}_{\j}\right)$, $\mathcal{G}=\widetilde{\mathcal{F}_{\infty}}^2 
=\left\{g\left(N_{\j},\vec{Y}_{\j}\right)=\left(\sum_{\ell=1}^{N_{\j}}f(Y_{\ell,\j})\right)^2: f\in \mathcal{F}_{\infty}\right\}$ and $\Phi=\Id$ implies
\begin{align}\label{eq:multi_up_bound_3}
    &\mathbb{E}\left[\sup_{\mathcal{F}_{\infty}}\left|\text{ }\frac{1}{\Pi_C}\sum_{\un\leq\j\leq\C}\left(\sum_{\ell=1}^{N_{\j}}f(Y_{\ell,\j})\right)^2-\E\left[\left(\sum_{\ell=1}^{N_{\un}}f(Y_{\ell,\un})\right)^2\right] \right|\right] \nonumber\\
    \leq & 2\sum_{\zero\prec \e \preceq \un}\mathbb{E}\left[\sup_{\mathcal{F}_{\infty}}\left|\frac{1}{\Pi_C}\sum_{\un\leq\j\leq\C}\epsilon_{\j\odot\e}\left(\sum_{\ell=1}^{N_{\j}}f\left(Y_{\ell,\j}\right)\right)^2\right|\right].
\end{align}

Let $\widetilde{F}\left(N_{\j},\vec{Y}_{\j}\right)=\sum_{\ell=1}^{N_{\j}}F(Y_{\ell,\j})$. For every $f\in \mathcal{F}_{\infty}$ we have $|f|\leq 2 F$, $\E(\widetilde{F}^2)\leq \E(A_F^2)<\infty$ under Assumption \ref{as:vc} and $\E(\widetilde{F}^2)\leq||F||^2_{\infty,\beta}\E(A_{\beta})<\infty$ under Assumption \ref{as:smooth}'. Next, we split the expectations in the upper bound into two, depending on whether $4\widetilde{F}^2\leq M$ or not, for some arbitrary $M$. For every $\e$ such that $\zero\prec\e \preceq \un$,  by the triangle inequality,
\begin{align*}   \mathbb{E}\left[\sup_{\mathcal{F}_{\infty}}\left|\frac{1}{\Pi_C}\sum_{\un\leq\j\leq\C}\epsilon_{\j\odot\e}\left(\sum_{\ell=1}^{N_{\j}}f\left(Y_{\ell,\j}\right)\right)^2\mathds{1}\left\{4\widetilde{F}^2> M\right\}\right|\right]
&\leq  4\mathbb{E}\left[\widetilde{F}\left(N_{\un},\vec{Y}_{\un}\right)^2\mathds{1}\left\{4\widetilde{F}^2> M\right\}\right].
\end{align*}
Therefore, 
\begin{align*}  &  2\sum_{\zero\prec \e \preceq \un}\mathbb{E}\left[\sup_{\mathcal{F}_{\infty}}\left|\frac{1}{\Pi_C}\sum_{\un\leq\j\leq\C}\epsilon_{\j\odot\e}\left(\sum_{\ell=1}^{N_{\j}}f\left(Y_{\ell,\j}\right)\right)^2\mathds{1}\left\{4\widetilde{F}^2> M\right\}\right|\right] \\
\leq  & 8  (2^{k}-1)\mathbb{E}\left[\widetilde{F}\left(N_{\un},\vec{Y}_{\un}\right)^2\mathds{1}\left\{4\widetilde{F}^2> M\right\}\right],
\end{align*}
which vanishes when $M\to +\infty$ by the dominated convergence theorem. 

\medskip
By Lemma~\ref{lemma:glivenko_squares}, there exists a non-increasing function $u$ from $]0,+\infty[$ to $[0,+\infty[$ and a non-random $K'(\mathcal{F},P)$ such that for every $M>0$ and $\eta>0$,  
\begin{align*}
	& 2\sum_{\zero\prec \e \preceq \un}\mathbb{E}\left[\sup_{\mathcal{F}_{\infty}}\left|\frac{1}{\Pi_C}\sum_{\un\leq\j\leq\C}\epsilon_{\j\odot\e}\left(\sum_{\ell=1}^{N_{\j}}f\left(Y_{\ell,\j}\right)\right)^2\right|\mathds{1}\{4\widetilde{F}^2\leq M\}\right]=K'(\mathcal{F},P)\left(\frac{\sqrt{M} u(\eta)}{\sqrt{\underline{C}}}+\eta\right).
\end{align*}

It follows that for every $M>0$ and $\eta>0$,
\begin{align*}
&\mathbb{E}\left[\sup_{\mathcal{F}_{\infty}}\left|\text{ }\frac{1}{\Pi_C}\sum_{\un\leq\j\leq\C}\left(\sum_{\ell=1}^{N_{\j}}f(Y_{\ell,\j})\right)^2-\E\left[\left(\sum_{\ell=1}^{N_{\un}}f(Y_{\ell,\un})\right)^2\right]\text{ }\right|\right] \\
=& K''(\mathcal{F},P)\left(\mathbb{E}\left[\widetilde{F}\left(N_{\un},\vec{Y}_{\un}\right)^2 \mathds{1}\left\{4\widetilde{F}^2> M\right\}\right]+\frac{M u(\eta)}{\sqrt{\underline{C}}}+\eta\right),
\end{align*}
for some non-random $K''(\mathcal{F},P)$. Now, fix $M$ and $\eta$ such that $\mathbb{E}\left[\widetilde{F}\left(N_{\un},\vec{Y}_{\un}\right)^2\mathds{1}\left\{4\widetilde{F}^2> M\right\}\right]+\eta$ is arbitrary small. This implies that for $\underline{C}$ sufficiently large, $\mathbb{E}\left[\widetilde{F}\left(N_{\un},\vec{Y}_{\un}\right)^2\mathds{1}\left\{4\widetilde{F}^2> M\right\}\right]+\frac{M u(\eta)}{\sqrt{\underline{C}}}+\eta$ is arbitrary small. As a result, \eqref{eq:ineg_for_sigma2} holds, and the result follows.

\medskip
\textbf{3. Conclusion on asymptotic equicontinuity}

\medskip
Under Assumptions~\ref{as:dgp}-\ref{as:vc}, we have shown in the previous step that $\mathbb{E}\left[\sigma_{C}^2\right]=\delta^2+o(1)$  and
\begin{equation}
	\mathbb{E}\left[\sup_{\mathcal{F}_{\delta}}\left|\mathbb{G}_Cf\right|\right]\leq K(F,P)\left(\sqrt{\E(\sigma_C^2)}+ J_{2,\mathcal{F}}\left(\frac{1}{4}\sqrt{\frac{\mathbb{E}\left[\sigma_{C}^2\right]}{\E(A_F)}}\right)\right).
	\label{eq:for_as_equi}
\end{equation}
Then, by continuity of $J_{2,\mathcal{F}}$ at 0, 
$\lim_{\delta\to 0}\limsup_{\underline{C}\to +\infty}\mathbb{E} \left[\sup_{\mathcal{F}_{\delta}}\left|\mathbb{G}_Cf\right|\right]=0$. This proves asymptotic equicontinuity. 

\medskip
Inequality \eqref{eq:for_as_equi} also holds when Assumption \ref{as:smooth}' holds instead of \ref{as:vc}, by just replacing  $J_{2,\mathcal{F}}$ by  $J_{\infty,\beta,\mathcal{F}}$ and $\E(A_F)$ by $||F||^2_{\infty,\beta}\mathbb{E}\left[A^2_{\beta}\right]$. The result follows similarly.

\subsubsection{Total boundedness}\label{sec:totbound}

Let $\left|\left|f\right|\right|_{P,2}=\mathbb{E}\left[\left(\sum_{\ell=1}^{N_{\un}} f(Y_{\ell,\un})\right)^2\right]^{1/2}$. We first check total boundedness under Assumptions~\ref{as:dgp}-\ref{as:vc}. %the proof of the total boundedness of $\mathcal{F}$ under the $\left|\left|\cdot\right|\right|_{P,2}$ semimetric is similar to that in \cite{vanderVaartWellner1996}, adapted to our framework. We first 
	We have to show that $N\left(\varepsilon,\mathcal{F},\left|\left|\cdot\right|\right|_{P,2}\right)<\infty$ for any 
$\eps>0$. In the previous step, we have shown that $$\sup_{\mathcal{F}_{\infty}}\left|\text{ }\frac{1}{\Pi_C}\sum_{\un\leq\j\leq\C}\left(\sum_{\ell=1}^{N_{\j}}f(Y_{\ell,\j})\right)^2-\E\left[\left(\sum_{\ell=1}^{N_{\un}}f(Y_{\ell,\un})\right)^2\right]\text{ }\right|\xrightarrow{L_1}0.$$

As a result, for any $\eps>0$, the previous supremum is bounded by $\varepsilon^2$ with probability approaching one. In other words, with probability approaching one,  for every $(f,g)\in\left(\mathcal{F}\right)^2$,
 \begin{equation*}
     0\leq\left|\left|\widetilde{f}-\widetilde{g}\right|\right|_{P,2}^2\leq \left|\left|\widetilde{f}-\widetilde{g}\right|\right|_{\mu_C,2}^2+\varepsilon^2.
 \end{equation*}
 This implies that  
 \begin{equation}
 N\left(\varepsilon,\widetilde{\mathcal{F}},\left|\left|\cdot\right|\right|_{P,2}\right)\leq N\left(\frac{\varepsilon}{\sqrt{2}},\widetilde{\mathcal{F}},\left|\left|\cdot\right|\right|_{\mu_C,2}\right)+o_p(1).	
	\label{eq:borne_N}
 \end{equation}
If $\overline{N}_2=0$ then $\mu_C$ is the null measure on $\widetilde{\mathcal{F}}$, so that 
\begin{equation}
N\left(\varepsilon,\widetilde{\mathcal{F}},\left|\left|\cdot\right|\right|_{P,2}\right)= N\left(\frac{\varepsilon}{\sqrt{2}},\widetilde{\mathcal{F}},\left|\left|\cdot\right|\right|_{\mu_C,2} \right)=1.	
	\label{eq:N_trivial_case}
\end{equation}
If $\overline{N}_2>0$, Lemmas \ref{lemma:bound_empirical_entropy_dgg} $i)$ and \ref{lem:csteF}  and the uniform entropy condition imply 
\begin{align} N\left(\frac{\varepsilon}{\sqrt{2}},\widetilde{\mathcal{F}},\left|\left|\cdot\right|\right|_{\mu_C,2}\right) &\leq N\left(\frac{\varepsilon}{\sqrt{2\overline{N}_2}},\mathcal{F}, \left|\left|\cdot\right|\right|_{\mathbb{Q}_C^2,2}\right) \nonumber\\
%& = N\left(\frac{\varepsilon  }{\sqrt{2\overline{N}_2} \left|\left|F\right|\right|_{\mathbb{Q}_C^2,2}} \left|\left|F\right|\right|_{\mathbb{Q}_C^2,2},\mathcal{F},\left|\left|\cdot\right|\right|_{\mathbb{Q}_C^2,2}\right)\\
& \leq \sup_{Q}N\left(\frac{\varepsilon  }{\sqrt{2\overline{N}_2} \left|\left|F\right|\right|_{\mathbb{Q}_C^2,2}} \abs{\abs{F}}_{Q,2},\mathcal{F},\left|\left|\cdot\right|\right|_{Q,2}\right) \nonumber\\
& < +\infty. 	\label{eq:N_nontrivial_case}
\end{align}
where the supremum is taken over all finitely supported measures. Together, \eqref{eq:N_trivial_case} and \eqref{eq:N_nontrivial_case} imply that $N\left(\frac{\varepsilon}{\sqrt{2}},\widetilde{\mathcal{F}},\left|\left|\cdot\right|\right|_{\mu_C,2}\right)=O_p(1)$. In turn, it follows from \eqref{eq:borne_N} that %
%with $\overline{N}_2=\frac{1}{\Pi_C}\sum_{\un\leq\j\leq\C}N_{\j}^2$ and $\left|\left|f\right|\right|_{\mathbb{Q}_C^2,2}^2=\frac{1}{\overline{N}_2\Pi_C}\sum_{\un\leq\j\leq\C}N_{\j}\sum_{\ell=1}^{N_{\j}}f(Y_{\ell,\j})^2$.
%for any $\eta>0$, $$N\left(\eta\abs{\abs{F}}_{\mathbb{Q}_C^2,2},\mathcal{F},\left|\left|\cdot\right|\right|_{\mathbb{Q}_C^2,2}\right)\leq $$
%This is in particular true for $\eta=\frac{\varepsilon  }{\sqrt{2\overline{N}_2} \left|\left|F\right|\right|_{\mathbb{Q}_C^2,2}}$, 
 for any $\varepsilon>0$, $N\left(\varepsilon,\widetilde{\mathcal{F}},\left|\left|\cdot\right|\right|_{P,2}\right)<+\infty$, meaning that 
$(\widetilde{\mathcal{F}},||.||_{P,2})$ is totally bounded under Assumption \ref{as:vc}.

\medskip
We now check total boundedness of $(\widetilde{\mathcal{F}},||.||_{P,2})$ under Assumptions~\ref{as:dgp}, \ref{as:measurability} and \ref{as:smooth}'. First, that for every $(f,g)\in\left(\mathcal{F}\right)^2$
\begin{align*}
	\left|\left|f\right|\right|_{P,2}\leq\left|\left|f\right|\right|_{\infty,\beta}\sqrt{\mathbb{E}\left[A_{\beta}^2\right]}.
\end{align*}
Then, by Lemma \ref{lemma:ch_norme_class}-i) and ii),
\begin{align*}
N(\varepsilon,\widetilde{\mathcal{F}},||.||_{P,2})&\leq 
N\left(\varepsilon,\mathcal{F},\sqrt{\mathbb{E}\left[A_{\beta}^2\right]}||.||_{\infty,\beta}\right)\\
	&\leq N\left(\frac{\varepsilon}{\sqrt{\mathbb{E}\left[A_{\beta}^2\right]}||F||_{\infty,\beta}}||F||_{\infty,\beta},\mathcal{F},||.||_{\infty,\beta}\right) \\
	& < +\infty,
\end{align*}
where the last inequality follows by Assumption \ref{as:smooth}'. The result follows.  %$(\widetilde{\mathcal{F}},||.||_{P,2})$ is totally bounded under Assumption \ref{as:smooth}'.

\subsection{Proof of Theorems \ref{thm:boot_unif} and \ref{thm:boot_unif_bis}}

The proof is divided in several steps that mirror the steps of the proof of Theorem \ref{thm:unifTCL}. We first prove the consistency of  $(\mathbb{G}^{\ast}_{C}f_1,...,\mathbb{G}^{\ast}_{C}f_m)$ for any $m\geq 1$ and $(f_1,...,f_m)\in (\mathcal{F})^m$. In a second step, we prove the asymptotic equicontinuity of the boostrap process. Note that the total boundedness is a property of $\mathcal{F}$ that has already been established in the proof of Theorem \ref{thm:unifTCL}. To simplify notation, we let $\bm{Z}=\left(N_{\j},\vec{Y}_{\j}\right)_{\j\geq\un}$ hereafter.

\subsubsection{Conditional convergence of $(\mathbb{G}^{\ast}_{C}f_1,...,\mathbb{G}^{\ast}_{C}f_m)$}
\label{sec:boot_marg}
The Cram\'er-Wold device ensures that we only have to prove the asymptotic normality for a single function $f$ such that $\E\left(\left(\sum_{\ell=1}^{N_{\un}}f(Y_{\ell,\un})\right)^{2}\right)<\infty$. The proof of the finite-dimensional convergence of the bootstrap process follows the same steps as for the initial process: characterization of the H\'ajek projection, comparison of variances and conclusion.

\medskip
\textbf{a. H\'ajek Projection}

\medskip
For $\c \in \mathcal{I}_r$, let $W^{\C}_{\c}=\prod_{i:c_i=1} W^i_{c_i}$ (when $\c\geq \un$, we have $W^{\C}_{\c}=\prod_{i=1}^k W^i_{c_i}$). Because the $(W^i_j)_{j\geq 1}$ are mutually independent across $i$ and independent of $\bm{Z}$, the H\'ajek projection of $\mathbb{G}_{C}^{\ast}(f)$ on $\sum_{\e\in \mathcal{E}_1} g_{\e}\left(\left(W^{\C}_{\c}\right)_{\c: \c\wedge \un=\e}, \bm{Z}\right)$ with $g_{\e}$ square integrable functions, is
\begin{align*}
\sum_{\e\in \mathcal{E}_1}^{k}\E\left(\mathbb{G}_{C}^{\ast}(f)\bigg| \left(W^{\C}_{\c}\right)_{\c:\c\wedge \un=\e}, \bm{Z}\right)&
=\sum_{\e\in \mathcal{E}_1}\frac{\sqrt{\underline{C}}}{\prod_{i:e_i=1}C_i}\sum_{\e\leq \c\leq \C\odot \e}\left(W^{\C}_{\c}-1\right)a^{\C}_{\e}(\c)\\
&=\sum_{\e\in \mathcal{E}_1}\frac{\sqrt{\underline{C}}}{\prod_{i:e_i=1}\sqrt{C_i}}\mathbb{H}_{\e}(f),
\end{align*}
with $a^{\C}_{\e}(\c)=\frac{1}{\prod_{s: e_s=0}C_s}\sum_{\un-\e\leq \c' \leq \C \odot (\un-\e)} \sum_{\ell=1}^{N_{\c+\c'}}f(Y_{\ell,\c+\c'})-\frac{1}{\Pi_C}\sum_{\un\leq \j\leq \C}\sum_{\ell\geq 1}^{N_{\j}}f(Y_{\ell,\j}),$
and $\mathbb{H}_{\e}(f)=\frac{1}{\prod_{i:e_i=1}\sqrt{C_i}}\sum_{\e\leq \c\leq \C\odot \e}\left(W^{\C}_{\c}-1\right)a^{\C}_{\e}(\c)$. Because the $(W^i_j)_{j\geq 1}$ are mutually independent across $i$ and independent of $\bm{Z}$, we have
\begin{align*}
&\V\left( \sum_{\e\in \mathcal{E}_1}\frac{\sqrt{\underline{C}}}{\prod_{i:e_i=1}\sqrt{C_i}}\mathbb{H}_{\e}(f)\bigg| \bm{Z} \right)\\&=\sum_{\e \in \mathcal{E}_1} \frac{\underline{C}}{\prod_{i:e_i=1}C_i^2}\sum_{\e\leq \c,\c'\leq \C \odot \e}\E\left(\left(W^{\C}_{\c}-1\right)\left(W^{\C}_{\c'}-1\right)\right)a_{\e}^{\C}(\c)a_{\e}^{\C}(\c').
\end{align*}

Noting that $\E(W^i_j)=1$, $\E(W^i_j W^i_{j'})=\mathds{1}_{j=j'}+1-C_i^{-1}$, and for any $\e\in \mathcal{E}_1$ and $\c,\c'$ such that $\e\leq \c,\c'\leq \C\odot \e$, we have
$$\E\left(\left(W^{\C}_{\c}-1\right)\left(W^{\C}_{\c'}-1\right)\right)=\E\left(W^{\C}_{\c}W^{\C}_{\c'}\right)-1=\mathds{1}_{\{\c=\c'\}}-\frac{1}{\prod_{i:e_i=1}C_i},$$
we have
\begin{align*}
&\V\left( \sum_{\e\in \mathcal{E}_1}\frac{\sqrt{\underline{C}}}{\prod_{i:e_i=1}\sqrt{C_i}}\mathbb{H}_{\e}(f)\bigg| \bm{Z} \right)\\&=\sum_{\e \in \mathcal{E}_1}\frac{\underline{C}}{\prod_{i:e_i=1}C_i}\left(\frac{1}{\prod_{i:e_i=1}C_i}\sum_{\e\leq \c\leq \C\odot \e}^{C_i}\left(a_{\e}^{\C}(\c)\right)^2-\left[\frac{1}{\prod_{i:e_i=1}C_i}\sum_{\e\leq \c\leq \C\odot \e}a_{\e}^{\C}(\c)\right]^2\right).\end{align*}
Because $\frac{1}{\prod_{i:e_i=1}C_i}\sum_{\e\leq \c \leq \C}a_{\e}^{\C}(\c)=0$, we have
\begin{align*}
\V\left( \sum_{\e\in \mathcal{E}_1}\frac{\sqrt{\underline{C}}}{\prod_{i:e_i=1}\sqrt{C_i}}\mathbb{H}_{\e}(f)\bigg| \bm{Z} \right)&=\sum_{\e \in \mathcal{E}_1}\frac{\underline{C}}{\prod_{i:e_i=1}C_i^2}\sum_{\e\leq \c\leq \C\odot \e}\left(a_{\e}^{\C}(\c)\right)^2.
\end{align*}

Lemma \ref{lem:boot} ensures that
$$\V\left( \sum_{\e\in \mathcal{E}_1}\frac{\sqrt{\underline{C}}}{\prod_{i:e_i=1}\sqrt{C_i}}\mathbb{H}_{\e}(f)\bigg| \bm{Z} \right)\convAS \sum_{i=1}^k \lambda_i \Cov\left(\sum_{\ell=1}^{N_{\un}}f(Y_{\ell,\un}),\sum_{\ell=1}^{N_{\deux}}f(Y_{\ell,\deux})\right)$$

We have to prove the asymptotic normality conditional on $\bm{Z}$. 
To do so, we apply the Lindeberg-Feller CLT. Let $(N^{\e \ast}_{\j},(Y^{\e \ast}_{\ell,\j})_{\ell \geq 1})_{\C\geq \j\geq 1}$ denote the bootstrap sample obtained by the selection of the cells when sampling clusters of component $i$ corresponding to the non-zero component of $\e$, $\mathbb{H}_{\e}(f)$ is also equal to
\begin{align*}\mathbb{H}_{\e}(f)&=\sum_{\e\leq \c\leq \C\odot\e}\frac{1}{\prod_{i: e_i=1}\sqrt{C_i}}\left(a^{\ast \C}_{\e}(\c)-a^{\C}_{\e}(\c)\right)\\
&=\sum_{\e\leq \c\leq \C\odot\e}\frac{1}{\prod_{i: e_i=1}\sqrt{C_i}}\left[a^{\ast \C}_{\e}(\c)-\E\left(a^{\ast \C}_{\e}(\c)\mid \bm{Z}\right)\right],
\end{align*}
with $a^{\ast \C}_{\e}(\c)=\frac{1}{\prod_{s: e_s=0}C_s}\sum_{\un-\e\leq \c' \leq \C \odot (\un-\e)} \sum_{\ell=1}^{N^{\e \ast}_{\c+\c'}}f(Y^{\e \ast}_{\ell,\c+\c'})-\frac{1}{\Pi_C}\sum_{\un\leq \j\leq \C}\sum_{\ell\geq 1}^{N_{\j}}f(Y_{\ell,\j})$.
Because the bootstrap sampling in each component are done with replacement and equal probability, for any function $g$ and any $\e\in \mathcal{E}_1$, we have $$\E\left[g(a^{\ast \C}_{\e}(\c))\bigg| \bm{Z}\right]=\frac{1}{\prod_{i:e_i=1}C_i}\sum_{\e\leq \j \leq \C\odot \e}g(a^{\C}_{\e}(\j)).$$
%\begin{align*}
%\V\left(\mathbb{H}_{\e}(f)\bigg| \bm{Z} \right)&=.
%\end{align*}
It follows that
\begin{align*}
\sum_{\e\leq \c\leq \C\odot\e}\V\left(\frac{a^{\ast \C}_{\e}(\c)}{\prod_{i: e_i=1}\sqrt{C_i}}\bigg| \bm{Z}\right)&=\frac{1}{\prod_{i: e_i=1}C_i}\sum_{\e\leq \j\leq \C\odot\e}\left(a^{\C}_{\e}(\j)\right)^2 -\left(\frac{1}{\prod_{i: e_i=1}C_i}\sum_{\e\leq \j\leq \C\odot\e}a^{\C}_{\e}(\j)\right)^2\\
&=\frac{1}{\prod_{i: e_i=1}C_i}\sum_{\e\leq \j\leq \C\odot\e}\left(a^{\C}_{\e}(\j)\right)^2
\end{align*}
and for any $\varepsilon>0$
\begin{align*}
&\frac{1}{\prod_{i: e_i=1}C_i}\sum_{\e\leq \c\leq \C\odot\e}\E\left(\left(a^{\ast \C}_{\e}(\c)\right)^2\mathds{1}\left\{\left|a^{\ast \C}_{\e}(\c)\right|>\left(\Pi_{i: e_i=1}C_i\right)^{1/2}\varepsilon\right\}\bigg| \bm{Z}\right)\\
=& \frac{1}{\prod_{i: e_i=1}C_i}\sum_{\e\leq \c\leq \C\odot\e}\frac{1}{\prod_{i: e_i=1}C_i}\sum_{\e\leq \j\leq \C\odot\e}\left(a^{\C}_{\e}(\j)\right)^2\mathds{1}\left\{\left|a^{ \C}_{\e}(\j)\right|>\left(\Pi_{i: e_i=1}C_i\right)^{1/2}\varepsilon\right\}\\
= & \frac{1}{\prod_{i: e_i=1}C_i}\sum_{\e\leq \j\leq \C\odot\e}\left(a^{\C}_{\e}(\j)\right)^2\mathds{1}\left\{\left|a^{ \C}_{\e}(\j)\right|>\left(\Pi_{i: e_i=1}C_i\right)^{1/2}\varepsilon\right\}.
\end{align*}
Lemma \ref{lem:boot} ensures that
 $$\frac{1}{\prod_{i: e_i=1}C_i}\sum_{\e\leq \j\leq \C\odot\e}\left(a^{\C}_{\e}(\j)\right)^2 \convAS \Cov\left(\sum_{\ell=1}^{N_{\un}}f(Y_{\ell,\un}),\sum_{\ell=1}^{N_{\deu-\e}}f(Y_{\ell,\deu-\e})\right)<\infty,$$
$$\frac{1}{\prod_{i: e_i=1}C_i}\sum_{\e\leq \j\leq \C\odot\e}\left(a^{\C}_{\e}(\j)\right)^2\mathds{1}\left\{\left|a^{ \C}_{\e}(\j)\right|>\left(\Pi_{i: e_i=1}C_i\right)^{1/2}\varepsilon\right\}\convAS 0.$$

Then, by the Lindeberg-Feller Theorem \citep[see, e.g.][Section 2.8]{vanderVaart2000}, 
$$\mathbb{H}_{\e}(f) \stackrel{d^*}{\longrightarrow} \mathcal{N}\left(0,\Cov\left(\sum_{\ell=1}^{N_{\un}}f(Y_{\ell,\un}),\sum_{\ell=1}^{N_{\deu -\e}}f(Y_{\ell,\deu -\e})\right)\right).$$ 

Because the $\mathbb{H}_{\e}(f)$ are mutually independent conditional on $\bm{Z}$, we have the joint asymptotic normality.  Next, by Slutsky's Lemma,
$$\sum_{\e\in \mathcal{E}_1}\frac{\sqrt{\underline{C}}}{\prod_{i:e_i=1}\sqrt{C_i}}\mathbb{H}_{\e}(f) \stackrel{d^*}{\longrightarrow} \mathcal{N}\left(0,\sum_{i=1}^{k} \lambda_i \Cov\left(\sum_{\ell=1}^{N_{\un}}f(Y_{\ell,\un}),\sum_{\ell=1}^{N_{\deux}}f(Y_{\ell,\deux})\right)\right).$$

\medskip
\textbf{b. Comparison of variances}

\medskip
%For $\e$ such that $\zero\leq \e\leq \un$, let $\mathcal{A}_{\e}=\left\{(\j,\j')\in (\mathbb{N}^{\ast})^k: \un \leq \j,\j'\leq\C, \forall i=1,...,k~~ e_i=\mathds{1}_{\{j_i=j'_i\}} \right\}$.
Let $V=\V\left(\mathbb{G}_C^{\ast}(f)\bigg|\bm{Z} \right)$ and
$$\Delta_{\j} = \sum_{\ell=1}^{N_{\j}}f(Y_{\ell,{\j}})-\frac{1}{\Pi_C}\sum_{\un\leq \j\leq \C} \sum_{\ell=1}^{N_{\j}}f(Y_{\ell,{\j}}).$$
We have
\begin{align*}
V&=\frac{\underline{C}}{(\Pi_C)^2}\sum_{\un\leq \j,\j'\leq \C}\left[\E(W_{\j}W_{\j'}^{\C})-1\right]\Delta_{\j}
\Delta_{\j'}\\
&=\frac{\underline{C}}{(\Pi_C)^2}\sum_{\un \leq \j,\j'\leq \C}\left[\prod_{i=1}^{k}\left(\mathds{1}_{j_i=j'_i}+1-\frac{1}{C_i}\right)-1\right]\Delta_{\j}\Delta_{\j'}\\
&=\frac{\underline{C}}{(\Pi_C)^2}\sum_{\zero\leq\e\leq \un}\sum_{(\j,\j')\in \mathcal{A}_{\e}}\left[\prod_{i:e_i=1}\left(2-\frac{1}{C_i}\right)\prod_{i:e_i=0}\left(1-\frac{1}{C_i}\right)-1\right] \Delta_{\j}\Delta_{\j'}.
\end{align*}

Let us focus on the term corresponding to $\e=\zero$. Because $\sum_{\un\leq \j,\j'\leq \C}\Delta_{\j}\Delta_{\j'}=0$, we have %We can assume without loss of generality that $\E\left(\sum_{\ell=1}^{N_{\un}}f(Y_{\ell,\un})\right)=0$ (otherwise rescale $f$ as $f-\frac{1}{\E(N_{\un})}\E\left(\sum_{\ell=1}^{N_{\un}}f(Y_{\ell,\un})\right)$).
$$\sum_{(\j,\j')\in \mathcal{A}_{\zero}}\Delta_{\j} \Delta_{\j'}\\
=-\sum_{\zero<\e\leq \un}\sum_{(\j,\j')\in \mathcal{A}_{\e}}\Delta_{\j}\Delta_{\j'}.$$
Then
$$V=\sum_{\zero<\e\leq \un}\left[\prod_{i:e_i=0}\left(1-\frac{1}{C_i}\right)\left(\prod_{i:e_i=1}\left(2-\frac{1}{C_i}\right)-\prod_{i:e_i=1}\left(1-\frac{1}{C_i}\right)\right)\right]\frac{\underline{C}}{(\Pi_C)^2}\sum_{(\j,\j')\in \mathcal{A}_{\e}}\Delta_{\j}\Delta_{\j'}.$$
We have $\mathcal{B}_{\e}\cap \mathcal{B}_{\e'}= \mathcal{B}_{\e \vee \e'}$ and $\mathcal{B}_{\e}=\mathcal{A}_{\e}\cup \left(\cup_{\e<\e'\leq \un}\mathcal{B}_{\e'}\right)$. Because $\mathcal{A}_{\e}\cap \mathcal{B}_{\e'}=\emptyset$ for any $\e'>\e$, we have $\mathds{1}_{\mathcal{A}_{\e}}=\mathds{1}_{\mathcal{B}_{\e}}-\mathds{1}_{\cup_{\e<\e'\leq \un}\mathcal{B}_{\e'}}$. Moreover, by the inclusion-exclusion principle,
$$\mathds{1}_{\mathcal{A}_{\e}}=\mathds{1}_{\mathcal{B}_{\e}}-\sum_{\e<\e' \leq \un}(-1)^{\sum_{i=1}^k e'_i}\mathds{1}_{\mathcal{B}_{\e'}}.$$
Hence,
$$\frac{\underline{C}}{(\Pi_C)^2}\sum_{(\j,\j')\in \mathcal{A}_{\e}}\Delta_{\j}\Delta_{\j'}
=\frac{\underline{C}}{(\Pi_C)^2}\sum_{(\j,\j')\in \mathcal{B}_{\e}}\Delta_{\j}\Delta_{\j}-\sum_{\e<\e'\leq \un} (-1)^{\sum_{i=1}^k e'_i}\frac{\underline{C}}{(\Pi_C)^2}\sum_{(\j,\j')\in \mathcal{B}_{\e'}}\Delta_{\j}\Delta_{\j'}.$$

Let $r=1,...,k-1$ and $\e\in \mathcal{E}_r$. By Lemma \ref{lem:boot}, $$\frac{\underline{C}}{(\Pi_C)^2}\sum_{(\j,\j')\in \mathcal{B}_{\e}}\Delta_{\j}\Delta_{\j'}
=\frac{\underline{C}}{\prod_{i:e_i=1}C_i^2}\sum_{\e \leq \c \leq \C\odot \e}a^{\C}_{\e}(\c)^2=O_{as}(\underline{C}^{1-r}),$$
where $O_{as}(\underline{C}^{1-r})$ denotes a sequence of random variable that is uniformly bounded by $\underline{C}^{1-r}$ on a set of probability one. Moreover, by Lemma \ref{lem:boot} again,  the almost-sure limit when $r=1$ is $$\lambda_{i}\Cov\left(\sum_{\ell=1}^{N_{\un}}f(Y_{\ell,\un}),\sum_{\ell=1}^{N_{\deux}}f(Y_{\ell,\deux})\right),$$ 
with $i$ the non-zero component of $\e$. By Lemma 7.35 in \cite{kallenberg05}, we also have 
\begin{align*}
\frac{1}{\Pi_C}\sum_{\un \leq \j \leq \C}\sum_{\ell=1}^{N_{\j}}f(Y_{\ell,\j}) & \convAS \E\left(\sum_{\ell=1}^{N_{\un}}f(Y_{\ell,\un})\right), \\
\frac{1}{\Pi_C}\sum_{\un \leq \j \leq \C}\left(\sum_{\ell=1}^{N_{\j}}f(Y_{\ell,\j})\right)^2 & \convAS \E\left(\left(\sum_{\ell=1}^{N_{\un}}f(Y_{\ell,\un})\right)^2\right).
\end{align*}
Combining all these results, we obtain
$$\frac{\underline{C}}{(\Pi_C)^2}\sum_{(\j,\j')\in \mathcal{B}_{\un}}\Delta_{\j}\Delta_{\j'}
	=  \frac{\underline{C}}{(\Pi_C)^2}\sum_{\un \leq \j \leq \C}\Delta_{\j}^2=O_{as}(\underline{C}^{1-k}).$$
Finally, for any $\e\in \mathcal{E}_r$, we have
$$\lim_{\underline{C}\rightarrow \infty}\left[\prod_{i:e_i=0}\left(1-\frac{1}{C_i}\right)\left(\prod_{i:e_i=1}\left(1-\frac{1}{C_i}\right)-\prod_{i:e_i=1}\left(1-\frac{1}{C_i}\right)\right)\right]=2^r-1+O_{as}(\underline{C}^{-1}).$$ 
This implies that
\begin{align*}
V&=\sum_{\zero<\e\leq \un}\left[\prod_{i:e_i=0}\left(1-\frac{1}{C_i}\right)\left(\prod_{i:e_i=1}\left(1-\frac{1}{C_i}\right)-\prod_{i:e_i=1}\left(1-\frac{1}{C_i}\right)\right)\right]\frac{\underline{C}}{(\Pi_C)^2}\sum_{(\j,\j')\in \mathcal{A}_{\e}}\Delta_{\j}\Delta_{\j'}\\
&=\sum_{\e\in \mathcal{E}_1}\frac{\underline{C}}{(\Pi_C)^2}\sum_{(\j,\j')\in \mathcal{A}_{\e}}\Delta_{\j}\Delta_{\j'}+O_{as}(\underline{C}^{-1})\\
&=\V\left( \sum_{\e\in \mathcal{E}_1}\frac{\sqrt{\underline{C}}}{\prod_{i:e_i=1}\sqrt{C_i}}\mathbb{H}_{\e}(f)\bigg| \bm{Z} \right)+O_{as}(\underline{C}^{-1}).
\end{align*}

\medskip
\textbf{c. Asymptotic normality of $\mathbb{G}^{\ast}_{C}f$.}

\medskip
In Step a, we have proved that
$$\V\left( \sum_{\e\in \mathcal{E}_1}\frac{\sqrt{\underline{C}}}{\prod_{i:e_i=1}\sqrt{C_i}}\mathbb{H}_{\e}(f)\bigg| \bm{Z} \right)\convAS \sum_{i=1}^k\lambda_i \Cov\left(\sum_{i=1}^{N_{\un}}f(Y_{\ell,\un}),\sum_{i=1}^{N_{\deux}}f(Y_{\ell,\deux})\right)<\infty.$$

In Step b, we have shown that 
$$\V\left(\mathbb{G}_C^{\ast}(f)\bigg| \bm{Z} \right)-\V\left( \sum_{\e\in \mathcal{E}_1}\frac{\sqrt{\underline{C}}}{\prod_{i:e_i=1}\sqrt{C_i}}\mathbb{H}_{\e}(f)\bigg| \bm{Z} \right)\convAS 0$$

If $\sum_{i=1}^k\lambda_i \Cov\left(\sum_{i=1}^{N_{\un}}f(Y_{\ell,\un}),\sum_{i=1}^{N_{\deux}}f(Y_{\ell,\deux})\right)=0$, then, since $\E\left(\mathbb{G}_C^{\ast}(f)\bigg|\bm{Z}\right)=0$, $\mathbb{G}^{\ast}_{C}f$ converges in $L^2$ conditional on the data to 0. Therefore, $\mathbb{G}^{\ast}_{C}f\convDboot\mathcal{N}(0,0)$.

\medskip
Now, if $\sum_{i=1}^k\lambda_i \Cov\left(\sum_{i=1}^{N_{\un}}f(Y_{\ell,\un}),\sum_{i=1}^{N_{\deux}}f(Y_{\ell,\deux})\right)>0$, because $\sum_{\e\in \mathcal{E}_1}\frac{\sqrt{\underline{C}}}{\prod_{i:e_i=1}\sqrt{C_i}}\mathbb{H}_{\e}(f)$ is asymptotically normal, with the asymptotic variance being the almost-sure limit of 
$$\V\left( \sum_{\e\in \mathcal{E}_1}\frac{\sqrt{\underline{C}}}{\prod_{i:e_i=1}\sqrt{C_i}}\mathbb{H}_{\e}(f)\bigg| \bm{Z} \right),$$ 
Theorem 11.2 in \cite{vanderVaart2000} combined with Slustky's Lemma implies that
$$\mathbb{G}^{\ast}_{C}f\convDboot\mathcal{N}\left(0,\sum_{i=1}^k\lambda_i \Cov\left(\sum_{i=1}^{N_{\un}}f(Y_{\ell,\un}),\sum_{i=1}^{N_{\deux}}f(Y_{\ell,\deux})\right)\right).$$

\subsubsection{Asymptotic equicontinuity}\label{sssec:boot_equicontinuity}

In this section, we show an analog of Formula (\ref{eq:asym_equic}): $$\lim_{\delta\rightarrow 0}\limsup_{\underline{C}\to +\infty}\mathbb{E}\left[\sup_{\mathcal{F}_{\delta}}\left|\mathbb{G}_C^*f\right|\mid \bm{Z}\right]\convP 0.$$
We follow the strategy used in Section \ref{sssec:equicontinuity}: we first bound $\mathbb{E}\left[\sup_{\mathcal{F}_{\delta}}\left|\mathbb{G}_C^*f\right|\mid \bm{Z}\right]$ with an expression involving $\mathbb{E}\left[\sigma_C^{*2}\big| \bm{Z}\right]$ and next we show that $\mathbb{E}\left[\sigma_C^{*2}\big| \bm{Z}\right]\convP 0$ when $\underline{C}\rightarrow \infty$ followed by $\delta\rightarrow 0$.

\medskip
\textbf{1. Upper bound in the supremum of the bootstrap process}

\medskip
Conditional on $\bm{Z}$, $\left(N_{\j}^*,\vec{Y}_{\j}^*\right)_{\j\geq\un}$ is a separately exchangeable and dissociated random array  \citep[for a definition of dissociation, see e.g.][p. 339]{kallenberg05}. As a result, Lemma \ref{lem:representation} applies and we have
$\left(N_{\j}^*,\vec{Y}_{\j}^*\right)_{\j\geq\un}\overset{a.s.}{=}\left(\tau\left(\left(U_{\j\odot\e}\right)_{\zero\prec \e \preceq \un}\right)\right)_{\j\geq\un}$, where $(U_{\c})_{\c>\zero}$ is a family of i.i.d., uniform random variables and  $\tau$ depends on $\bm{Z}$. 

\medskip
In turn, by Lemma \ref{lem:sym} applied to $Z_{\j}=\left(N_{\j}^*,\vec{Y}_{\j}^*\right)$, $\mathcal{G}=\widetilde{\mathcal{F}_{\delta}}$
and $\Phi=\Id$, we have
	\begin{align*}\mathbb{E}\left[\sup_{\mathcal{F}_{\delta}}\left|\mathbb{G}_C^*f\right|\bigg| \bm{Z}\right] %=&\mathbb{E}\left[\sup_{\mathcal{F}_{\delta}}\left|\mathbb{G}_C^*f| \bm{Z}\right]\right|\bigg| \bm{Z}\right] \\
\leq&  2 \sqrt{\underline{C}} \sum_{\zero\prec \e \preceq \un}\mathbb{E}\left[\sup_{\mathcal{F}_{\delta}} \left|\frac{1}{\Pi_C}\sum_{\un\leq\j\leq\C}\epsilon_{\j\odot\e}\sum_{\ell=1}^{N_{\j}^*}f\left(Y_{\ell,\j}^*\right)\right|\bigg| \bm{Z}\right],  \end{align*}
where $(\epsilon_{\c})_{\c\geq \zero}$ is a Rademacher process, independent of $\left(N_{\j}^*,\vec{Y}_{\j}^*\right)_{\j\geq\un}$ conditional on $\bm{Z}$. 

Then, by applying Lemma \ref{lemma:donsker_entropy_bound} conditional on $\bm{Z}$, we get, under Assumptions~\ref{as:dgp}-\ref{as:vc},
\begin{align}\label{eq:donker_boot_01}
    &\mathbb{E}\left[\sup_{\mathcal{F}_{\delta}}\left|\mathbb{G}_C^*f\right|\bigg| \bm{Z}\right]\nonumber \\
    \leq& 2\times 2^{k-1}\left\{4\sqrt{2\mathbb{E}\left[\sigma_C^{*2}\big| \bm{Z}\right]\log 2}
+ 32\sqrt{2\mathbb{E}\left[A_F^*\big| \bm{Z}\right]} J_{2,\mathcal{F}}\left(\frac{1}{4}\sqrt{\frac{\mathbb{E}\left[\sigma_C^{*2}\big| \bm{Z}\right]}{\E\left[A_F^*\big| \bm{Z}\right]}}\right)\right\},
\end{align}
where $A_F^*=N_{\un}^*\sum_{\ell=1}^{N_{\un}^*}F\left(Y_{\ell,\un}^*\right)^2$ and $\sigma^{*2}_C=\sup_{\mathcal{F}_{\delta}} \frac{1}{\Pi_C}\sum_{\un\leq\j\leq\C}\left(\sum_{\ell=1}^{N_{\j}^*}f(Y_{\ell,\j}^*)\right)^2$. Similarly, letting $A_{\beta}^*=\sum_{\ell=1}^{N_{\un}^*}(1+|Y_{1,\un}|^{*2})^{-\beta/2}$, we have, under Assumptions~\ref{as:dgp}, \ref{as:measurability} and \ref{as:smooth}',
\begin{align}
    \mathbb{E}\left[\sup_{\mathcal{F}_{\delta}}\left|\mathbb{G}_C^*f\right|\big| \bm{Z}\right]  \leq 2\times 2^{k-1} & \left\{ 4\sqrt{2\mathbb{E}\left[\sigma_C^{*2}\big| \bm{Z}\right]\log 2} +32\abs{\abs{F}}_{\infty,\beta}\sqrt{2\mathbb{E}\left[A_{\beta}^{*2}\big| \bm{Z}\right]} \right. \nonumber\\ 
	& \; \times \left. J_{\infty,\beta,\mathcal{F}}\left(\frac{1}{4}\sqrt{\frac{\mathbb{E}\left[\sigma_C^{*2}\big| \bm{Z}\right]}{\left|\left|F\right|\right|_{\infty,\beta}\mathbb{E}\left[A_{\beta}^{*2}\big| \bm{Z}\right]}}\right)\right\}. \label{eq:donker_boot_02}
\end{align}

Moreover, by definition of the bootstrap scheme and Lemma 7.35 in \cite{kallenberg05},
\begin{align*}
\E\left[A_F^*\big| \bm{Z}\right]=\frac{1}{\Pi_C}\sum_{\un\leq\j\leq\C}N_{\j}\sum_{\ell=1}^{N_{\j}}F(Y_{\ell,\j})^2 & \convAS \mathbb{E}\left[N_{\un}\sum_{\ell=1}^{N_{\un}}F(Y_{\ell,\un})^2\right]>0, \\
\E\left[A_{\beta}^{*2}\big| \bm{Z}\right]=\frac{1}{\Pi_C}\sum_{\un\leq\j\leq\C}\left(\sum_{\ell=1}^{N_{\j}}(1+ |Y_{\ell,\j}|^2)^{-\beta/2}\right)^2 & \convAS \mathbb{E}\left[\left(\sum_{\ell=1}^{N_{\un}}(1+ |Y_{\ell,\un}|^2)^{-\beta/2}\right)^2\right]>0.
\end{align*}

\textbf{2. Control of $\mathbb{E}\left[\sigma_C^{*2}\big| \bm{Z}\right]$.}

Observe that
\begin{align}\label{eq:donsker_boot_1}
&\mathbb{E}\left[\sigma_C^{*2}\big| \bm{Z}\right]\nonumber \\
\leq &  \mathbb{E}\left[\sup_{\mathcal{F}_{\delta}}\left|\frac{1}{\Pi_C}\sum_{\un\leq\j\leq\C}\left(\sum_{\ell=1}^{N_{\j}^*}f(Y_{\ell,\j}^*)\right)^2-\frac{1}{\Pi_C}\sum_{\un\leq\j\leq\C}\left(\sum_{\ell=1}^{N_{\j}}f(Y_{\ell,\j})\right)^2\right|\bigg| \bm{Z}\right]\nonumber \\
&+\sup_{\mathcal{F}_{\delta}}\left|\text{ }\frac{1}{\Pi_C}\sum_{\un\leq\j\leq\C}\left(\sum_{\ell=1}^{N_{\j}}f(Y_{\ell,\j})\right)^2-\E\left[\left(\sum_{\ell=1}^{N_{\un}}f(Y_{\ell,\un})\right)^2\right]\text{ }\right|+\sup_{\mathcal{F}_{\delta}}\E\left[\left(\sum_{\ell=1}^{N_{\un}}f(Y_{\ell,\un})\right)^2\right]\nonumber \\
\leq & \mathbb{E}\left[\sup_{\mathcal{F}_{\infty}}\left|\frac{1}{\Pi_C}\sum_{\un\leq\j\leq\C}\left(\sum_{\ell=1}^{N_{\j}^*}f(Y_{\ell,\j}^*)\right)^2-\frac{1}{\Pi_C}\sum_{\un\leq\j\leq\C}\left(\sum_{\ell=1}^{N_{\j}}f(Y_{\ell,\j})\right)^2\right|\bigg| \bm{Z}\right]\nonumber \\
&+\sup_{\mathcal{F}_{\infty}}\left|\frac{1}{\Pi_C}\sum_{\un\leq\j\leq\C}\left(\sum_{\ell=1}^{N_{\j}}f(Y_{\ell,\j})\right)^2-\E\left[\left(\sum_{\ell=1}^{N_{\un}}f(Y_{\ell,\un})\right)^2\right]\right|+\delta^2.  
\end{align}

In Section \ref{sssec:equicontinuity}, we showed that
\begin{align}\label{eq:donsker_boot_2}
	&\sup_{\mathcal{F}_{\infty}}\left|\frac{1}{\Pi_C}\sum_{\un\leq\j\leq\C}\left(\sum_{\ell=1}^{N_{\j}}f(Y_{\ell,\j})\right)^2-\E\left[\left(\sum_{\ell=1}^{N_{\un}}f(Y_{\ell,\un})\right)^2\right]\right|\convP 0.
\end{align}

As a result, we only need to control
\begin{align*}
	&\mathbb{E}\left[\sup_{\mathcal{F}_{\infty}}\left|\frac{1}{\Pi_C}\sum_{\un\leq\j\leq\C}\left(\sum_{\ell=1}^{N_{\j}^*}f(Y_{\ell,\j}^*)\right)^2-\frac{1}{\Pi_C}\sum_{\un\leq\j\leq\C}\left(\sum_{\ell=1}^{N_{\j}}f(Y_{\ell,\j})\right)^2\right|\bigg| \bm{Z}\right].
\end{align*}

By Lemma \ref{lem:sym} applied to $Z_{\j}=\left(N_{\j}^*,\vec{Y}_{\j}^*\right)$, $\mathcal{G}=\widetilde{\mathcal{F}_{\infty}}^2$
and $\Phi=\Id$, we obtain
\begin{align*} &\mathbb{E}\left[\sup_{\mathcal{F}_{\infty}}\left|\frac{1}{\Pi_C}\sum_{\un\leq\j\leq\C}\left(\sum_{\ell=1}^{N_{\j}^*}f(Y_{\ell,\j}^*)\right)^2-\frac{1}{\Pi_C}\sum_{\un\leq\j\leq\C}\left(\sum_{\ell=1}^{N_{\j}}f(Y_{\ell,\j})\right)^2\right|\bigg| \bm{Z}\right] \\
    \leq & 2\sum_{\zero\prec \e \preceq \un}\mathbb{E}\left[\sup_{\mathcal{F}_{\infty}}\left|\frac{1}{\Pi_C}\sum_{\un\leq\j\leq\C}\epsilon_{\j\odot\e}\left(\sum_{\ell=1}^{N_{\j}^*}f\left(Y_{\ell,\j}^*\right)\right)^2\right|\bigg| \bm{Z}\right].
\end{align*}

Let $\widetilde{F}\left(N_{\j}^*,\vec{Y}_{\j}^*\right)=\sum_{\ell=1}^{N_{\j}^*}F(Y_{\ell,\j}^*)$. Note that for every $f\in \mathcal{F}_{\infty}$ we have $|f|\leq 2 F$. We split the expectations in the upper bound into two, depending on whether $4\widetilde{F}^2\leq M$ or not, for some arbitrary $M>0$. First, for every $\e$ such that $\zero\prec\e \preceq \un$,
\begin{align*}   &\mathbb{E}\left[\sup_{\mathcal{F}_{\infty}}\left|\frac{1}{\Pi_C}\sum_{\un\leq\j\leq\C}\epsilon_{\j\odot\e}\left(\sum_{\ell=1}^{N_{\j}^*}f\left(Y_{\ell,\j}^*\right)\right)^2\mathds{1}\left\{4\widetilde{F}\left(N_{\j}^*,\vec{Y}_{\j}^*\right)^2> M\right\}\right|\bigg| \bm{Z}\right] \\
\leq & 4\mathbb{E}\left[\widetilde{F}\left(N_{\un}^*,\vec{Y}_{\un}^*\right)^2 \mathds{1}\left\{4\widetilde{F}\left(N_{\un}^*,\vec{Y}_{\un}^*\right)^2> M\right\}\bigg| \bm{Z}\right] \\
= & \frac{1}{\Pi_C}\sum_{\un\leq\j\leq\C}\widetilde{F}\left(N_{\j},\vec{Y}_{\j}\right)^2 \mathds{1}\left\{4\widetilde{F}(N_{\j},\vec{Y}_{\j})^2> M\right\}.
\end{align*}
Therefore, 
\begin{align}\label{eq:donsker_boot_3}   
&2\sum_{\zero\prec \e \preceq \un}\mathbb{E}\left[\sup_{\mathcal{F}_{\infty}}\left|\frac{1}{\Pi_C}\sum_{\un\leq\j\leq\C}\epsilon_{\j\odot\e}\left(\sum_{\ell=1}^{N_{\j}^*}f\left(Y_{\ell,\j}^*\right)\right)^2\mathds{1}\left\{4\widetilde{F}\left(N_{\j}^*,\vec{Y}_{\j}^*\right)^2> M\right\}\right|\bigg| \bm{Z}\right]\nonumber \\
\leq &  8(2^{k}-1)\frac{1}{\Pi_C}\sum_{\un\leq\j\leq\C}\widetilde{F}\left(N_{\j},\vec{Y}_{\j}\right)^2\mathds{1}\left\{4\widetilde{F}(N_{\j},\vec{Y}_{\j})^2> M\right\},
\end{align}
which converges almost surely to $8(2^{k}-1) \E\left(\widetilde{F}\left(N_{\un},\vec{Y}_{\un}\right)^2\mathds{1}\left\{4\widetilde{F}(N_{\un},\vec{Y}_{\un})^2> M\right\}\right)$ by Lemma 7.35 in \cite{kallenberg05}. %and the Dominated Convergence Theorem (under Assumption~\ref{as:vc} or \ref{as:smooth}').

\medskip
Under Assumptions~\ref{as:dgp}, \ref{as:measurability} and \ref{as:vc}, Lemma~\ref{lemma:glivenko_squares} ensures the existence of $K(\mathcal{F})$ a non-negative number depending on the class $\mathcal{F}$ only such that for every $M>0$ and every $\eta>0$,  
\begin{align}\label{eq:donsker_boot_4} 
	&2\sum_{\zero\prec \e \preceq \un}\mathbb{E}\left[\sup_{\mathcal{F}_{\infty}}\left|\frac{1}{\Pi_C}\sum_{\un\leq\j\leq\C}\epsilon_{\j\odot\e}\left(\sum_{\ell=1}^{N_{\j}^*}f\left(Y_{\ell,\j}^*\right)\right)^2\mathds{1}\left\{4\widetilde{F}\left(N_{\j}^*,\vec{Y}_{\j}^*\right)^2\leq M\right\}\right|\bigg| \bm{Z}\right]\nonumber \\
    \leq &  K(\mathcal{F})\left(\frac{\sqrt{M}}{\sqrt{\underline{C}}}\left(1+\frac{1}{\eta}\right)+\eta\mathbb{E}\left[A_F^*\big| \bm{Z}\right]\right)\nonumber \\
    \leq &  K(\mathcal{F})\left(\frac{\sqrt{M}}{\sqrt{\underline{C}}}\left(1+\frac{1}{\eta}\right)+\eta\frac{1}{\Pi_C}\sum_{\un\leq\j\leq\C}N_{\j}\sum_{\ell=1}^{N_{\j}}F(Y_{\ell,\j})^2\right)\nonumber\\
    \leq & K(\mathcal{F})\left(\frac{\sqrt{M}}{\sqrt{\underline{C}}}\left(1+\frac{1}{\eta}\right)+\eta \left(\mathbb{E}\left[A_{F}\right]+o_{a.s}(1)\right)\right),
\end{align}
where the last inequality follows from Lemma 7.35 in \cite{kallenberg05}.
Fix $M$ and $\eta$ such that $8(2^{k}-1) \E\left(\widetilde{F}\left(N_{\un},\vec{Y}_{\un}\right)^2\mathds{1}\left\{4\widetilde{F}(N_{\un},\vec{Y}_{\un})^2> M\right\}\right)+K(\mathcal{F})\eta \mathbb{E}\left[N_{\un}\sum_{\ell=1}^{N_{\un}}F(Y_{\ell,\un})^2\right]$ is arbitrarily small and consider that $\underline{C}\rightarrow +\infty$ to conclude that
\begin{align}\label{eq:donsker_boot_6}
&\mathbb{E}\left[\sup_{\mathcal{F}_{\infty}}\left|\frac{1}{\Pi_C}\sum_{\un\leq\j\leq\C}\left(\sum_{\ell=1}^{N_{\j}^*}f(Y_{\ell,\j}^*)\right)^2-\frac{1}{\Pi_C}\sum_{\un\leq\j\leq\C}\left(\sum_{\ell=1}^{N_{\j}}f(Y_{\ell,\j})\right)^2\right|\bigg| \bm{Z}\right]\convAS 0,
\end{align}
and next, $\mathbb{E}\left[\sigma_C^{*2}\big| \bm{Z}\right]=\delta^2+o_p(1)$.

\medskip
Under Assumptions~\ref{as:dgp}, \ref{as:measurability} and \ref{as:smooth}', we follow the same reasoning, with Inequality (\ref{eq:donsker_boot_4}) replaced by 
\begin{align}\label{eq:donsker_boot_5} 
	&2\sum_{\zero\prec \e \preceq \un}\mathbb{E}\left[\sup_{\mathcal{F}_{\infty}}\left|\frac{1}{\Pi_C}\sum_{\un\leq\j\leq\C}\epsilon_{\j\odot\e}\left(\sum_{\ell=1}^{N_{\j}^*}f\left(Y_{\ell,\j}^*\right)\right)^2\mathds{1}\left\{4\widetilde{F}\left(N_{\j}^*,\vec{Y}_{\j}^*\right)^2\leq M\right\}\right|\bigg| \bm{Z}\right]\nonumber \\
    \leq & K(\mathcal{F})\left(\frac{\sqrt{M}}{\sqrt{\underline{C}}}\left(1+\frac{1}{\eta}\right)+\eta\left(\E\left(A_{\beta}^2\right)+o_{a.s.}(1)\right)\right).
\end{align}

\textbf{3. Conclusion on asymptotic equicontinuity}

Under Assumptions~\ref{as:dgp}, \ref{as:measurability} and \ref{as:vc} (respectively \ref{as:smooth}'), because $\mathbb{E}\left[\sigma_{C}^{*2}\bigg| \bm{Z}\right]=\delta^2+o_p(1)$, $\mathbb{E}[A_F^*\big| \bm{Z}]=\mathbb{E}\left[A_F\right]+o_{a.s}(1)$ (respectively $\mathbb{E}[A_{\beta}^{*2}\big| \bm{Z}]=\mathbb{E}\left[A_{\beta}^{2}\right]+o_{a.s}(1)$), and $J_{2,\mathcal{F}}$ (respectively $J_{\infty,\beta,\mathcal{F}}$) is continuous at 0, we obtain, by the continuous mapping theorem in probability and \eqref{eq:donker_boot_01} (respectively \eqref{eq:donker_boot_02}), 
$$\lim_{\delta \rightarrow 0} \limsup_{\underline{C}\to +\infty}\mathbb{E}\left[\sup_{\mathcal{F}_{\delta}}\left|\mathbb{G}_C^*f\right|\bigg| \bm{Z}\right]\convP 0.$$

\subsection{Proof of Proposition \ref{prop:as_var}}

To prove convergence of any symmetric matrix $\widehat{V}$ towards $V$, it suffices to show the convergence towards zero of $t'(\widehat{V}-V)t$. The latter corresponds to $\widehat{V}_t-V_t$ with $V_t$ (resp. $\widehat{V}_t$) the asymptotic variance of the average of the $t'Y_{\ell,\j}$ (resp. the estimator of $V_t$). Hence, we can suppose without loss of generality that $Y_{\ell,\j}\in \R$.

\medskip
Now, let $\bm{b}_i=(0,...,0,1,0,...,0)$ with $1$ at the $i$-th coordinate. 
Note that $|\mathcal{B}_i|=C_i \prod_{s\neq i} C_s^2$ and
\begin{align*}\widehat{V}_1&=\sum_{i=1}^k\frac{\underline{C}}{C_i}\frac{1}{|\mathcal{B}_i|}\left[\left(\sum_{(\j,\j')\in \mathcal{B}_i} S_{\j}S_{\j'}\right)-2\widehat{\theta}\left(\prod_{s\neq i}C_s\right) \left(\sum_{\un\leq \j\leq \C} S_{\j}\right)+|\mathcal{B}_i|\widehat{\theta}^2\right]\\
&=\sum_{i=1}^k\frac{\underline{C}}{C_i}\left(\frac{1}{|\mathcal{B}_i|}\sum_{(\j,\j')\in \mathcal{B}_i} S_{\j}S_{\j'}-\widehat{\theta}^2\right).
\end{align*}
The set $\{S_{\j}S_{\j'}: (\j,\j')\in \mathcal{B}_i\}$ is equal to
$\{S_{\c+\c'}S_{\c+\c''}: \bm{b}_i\leq \c\leq C_i\bm{b}_i ; \un-\bm{b}_i\leq \c',\c''\leq \C \odot(\un-\bm{b}_i)\}$, so this is a 2$k$-1 dimensional array indexed by the non-zero component of $\c$, $\c'$ and $\c''$. This array is jointly exchangeable and dissociated \citep[for a definition of jointly exchangeable arrays, see, e.g.,][p.300]{kallenberg05}. 
Lemma 7.35 in \cite{kallenberg05} ensures that this array is ergodic so $\frac{1}{|\mathcal{B}_i|}\sum_{(\j,\j')\in \mathcal{B}_i} S_{\j}S_{\j'}$ converges in $L^1$ and almost surely to a constant. Moreover, by the first part of Lemma \ref{lemma:glivenko_prod_fin} applied to $\mathcal{F}=\mathcal{G}=\{\Id\}$, 
 $\frac{1}{|\mathcal{B}_i|}\sum_{(\j,\j')\in \mathcal{B}_i} S_{\j}S_{\j'}-\frac{1}{|\mathcal{A}_i|}\sum_{(\j,\j')\in \mathcal{A}_i} S_{\j}S_{\j'}$ converges in $L^1$ to 0. Assumption \ref{as:dgp} and the representation Lemma \ref{lem:representation} ensure that $\E\left(\frac{1}{|\mathcal{A}_i|}\sum_{(\j,\j')\in \mathcal{A}_i} S_{\j}S_{\j'}\right)=\E\left(S_{\un}S_{\deux}\right)$. As a result, $\frac{1}{|\mathcal{B}_i|}\sum_{(\j,\j')\in \mathcal{B}_i} S_{\j}S_{\j'}=\E\left(S_{\un}S_{\deux}\right)+o_p(1)$. The asymptotic normality of $\widehat{\theta}$ implies that $\widehat{\theta}=\theta_0+o_p(1)$. Thus, by the continuous mapping theorem
 \begin{align*}\widehat{V}_1&=\sum_{i=1}^{k}\lambda_i \left(\E\left(S_{\un}S_{\deux}\right)-\theta_0^2\right)+o_p(1)\\
 &=\sum_{i=1}^{k}\lambda_i \Cov\left(S_{\un},S_{\deux}\right)+o_p(1).
\end{align*}

Next, consider $\widehat{V}_2$. We have
\begin{align*}
	\widehat{V}_2=& \sum_{i=1}^k\frac{\underline{C}}{C_i}\frac{1}{|\mathcal{A}_i|}\sum_{(\j,\j')\in \mathcal{A}_i} (S_{\j}-\widehat{\theta})(S_{\j'}-\widehat{\theta}) \\
	=& \sum_{i=1}^k\frac{\underline{C}}{C_i}\left(\frac{1}{|\mathcal{A}_i|}\sum_{(\j,\j')\in \mathcal{A}_i} S_{\j}S_{\j'}-\widehat{\theta}^2\right). 
\end{align*}
Then, by the triangle inequality and the first part of Lemma \ref{lemma:glivenko_prod_fin} (with $\mathcal{F}=\mathcal{G}=\{\Id\}$): $$\E\left(\left|\widehat{V}_1-\widehat{V}_2\right|\right)=\E\left(\left|\sum_{i=1}^k\frac{\underline{C}}{C_i}\frac{1}{|\mathcal{B}_i|}\sum_{(\j,\j')\in \mathcal{B}_i} S_{\j}S_{\j'}-\sum_{i=1}^k\frac{\underline{C}}{C_i}\frac{1}{|\mathcal{A}_i|}\sum_{(\j,\j')\in \mathcal{A}_i} S_{\j}S_{\j'}\right|\right)=o(1).$$
Finally, to show the consistency of $\widehat{V}_{\text{cgm}}$, note that
\begin{align*}
\left|\widehat{V}_{\text{cgm}}-\widehat{V}_{1}\right|&\leq \frac{\underline{C}}{(\Pi_C)^2}\sum_{\e \in \cup_{r=2}^k\mathcal{E}_r}\left|\sum_{(\j,\j')\in \mathcal{B}_{\e}}(S_{\j}-\widehat{\theta})(S_{\j'}-\widehat{\theta})\right|\\
&=  \frac{\underline{C}}{(\Pi_C)^2}\sum_{\e \in \cup_{r=2}^k\mathcal{E}_r}\left|\sum_{(\j,\j')\in \mathcal{B}_{\e}}S_{\j}S_{\j'}-\widehat{\theta}^2\underline{C}\sum_{\e \in \cup_{r=2}^k\mathcal{E}_r}\frac{1}{\prod_{s:e_s=1}C_s}\right|\\
&\leq  \frac{\underline{C}}{(\Pi_C)^2}\sum_{\e \in \cup_{r=2}^k\mathcal{E}_r}\left|\sum_{(\j,\j')\in \mathcal{B}_{\e}}S_{\j}S_{\j'}\right|+\widehat{\theta}^2\underline{C}^{-1}(2^k-k-1).
\end{align*}
Because $\widehat{\theta}=O_p(1)$, the second term on the right-hand side tends to 0 in probability. Moreover, 
by the triangle inequality and the second part of Lemma \ref{lemma:glivenko_prod_fin}, $$\mathbb{E}\left(\frac{\underline{C}}{(\Pi_C)^2}\sum_{\e \in \cup_{r=2}^k\mathcal{E}_r}\left|\sum_{(\j,\j')\in \mathcal{B}_{\e}}S_{\j}S_{\j'}\right|\right)=O(\underline{C}^{-1}).$$
Hence, $\left|\widehat{V}_{\text{cgm}}-\widehat{V}_{1}\right|=o_p(1)$, and $\widehat{V}_{\text{cgm}}$ is consistent. 

\medskip
Finally, to prove the second result, remark that if $V$ is positive definite, $\widehat{V}_{k}$ is also positive definite with probability approaching one and by the continuous mapping theorem,
$$ \underline{C}(\theta-\widehat{\theta})' \widehat{V}_k^{-1} (\theta-\widehat{\theta})\convD \chi^2_L.$$
The result follows.

\subsection{Proof of Proposition \ref{prop:boot_mean}}

Let $W\sim \mathcal{N}(0,V)$. By Theorem \ref{thm:boot_unif} applied to the class $\mathcal{F}=\{\Id\}$, we have
$$\sqrt{\underline{C}} \left(\widehat{\theta}^*-\widehat{\theta}\right) \stackrel{d^*}{\longrightarrow} W.$$
Then, by the continuous mapping theorem, $\underline{C} \left|\widehat{\theta}^*-\widehat{\theta}\right|^2 \stackrel{d^*}{\longrightarrow} |W|^2$. Because $V$ is symmetric positive, $|W|^2$ admits a positive density everywhere on $R^{*+}$. Therefore, we have $q_{1-\alpha}^*\convP q_{1-\alpha}$, where $q_{1-\alpha}$ is the quantile of order $1-\alpha$ of $|W|^2$. The result follows.

\subsection{Proof of Propositions \ref{prop:linear} and \ref{prop:linear2}}

Because $\E\left((\sum_{\ell=1}^{N_{\un}}|Y_{\ell,\un}|^2)^2\right)<\infty$, we have $\E\left((\sum_{\ell=1}^{N_{\un}}X_{r,\ell,\un}^2)^2\right)<\infty$ for $X_{r,\ell,\un}$ any component of the $X_{\ell,\un}$. Because $u_{\ell,\j}$ is a linear combination of the components of $Y_{\ell,\j}$, we also have $\E\left((\sum_{\ell=1}^{N_{\un}}u_{\ell,\un}^2)^2\right)<\infty$, and next by the Cauchy-Schwarz inequality, we also have 
$\E\left((\sum_{\ell=1}^{N_{\un}}|X_{r,\ell,\un}u_{\ell,\un}|)^2\right)<\infty$ and $\E\left((\sum_{\ell=1}^{N_{\un}}|X_{r,\ell,\un}X_{r',\ell,\un}|)^2\right)<\infty$ for any $r,r'$. Last, we also have 
$\E\left((\sum_{\ell=1}^{N_{\un}}|X_{\ell,\un}| |u_{\ell,\un}|)^2\right)<\infty$ and $\E\left((\sum_{\ell=1}^{N_{\un}}|X_{\ell,\un}|^2)^2\right)<\infty$.

For any $r,r'$, Theorem \ref{thm:unifTCL} applied to the class $\mathcal{F}=\{f(Y_{\ell,\j})=X_{r,\ell,\un}X_{r',\ell,\un}\}$ for all $(r,r')$ ensures that $$\frac{1}{\Pi_C}\sum_{\un \leq \j \leq \C}\sum_{\ell=1}^{N_{\j}}X_{\ell,\j}X'_{\ell,\j}=\E\left(\sum_{\ell=1}^{N_{\j}}X_{\ell,\j}X_{\ell,\j}\right)+o_p(1).$$
Similarly, Theorem \ref{thm:boot_unif} ensures that 
$$\widehat{J}^* = \frac{1}{\Pi_C}\sum_{\un \leq \j \leq \C}\sum_{\ell=1}^{N_{\j}}X^*_{\ell,\j}X^*_{\ell,\j}{}'=\E\left(\sum_{\ell=1}^{N_{\j}}X_{\ell,\j}X_{\ell,\j}'\right) +o^*_p(1),$$ 
where $A=o^*_p(1)$ means that conditional on the data $\bm{Z}$, $A$ converges weakly to 0 in probability when $\underline{C}\rightarrow \infty$. Then, by the continuous mapping theorem, $\widehat{J}^{-1}=J^{-1}+o_p(1)$ and $\widehat{J}^{*-1}=J^{-1}+o^*_p(1)$.

\medskip
Now, consider a vector $\mu$ with the same dimension as $\theta_0$ and $f_{\mu}(Y_{\ell,\j})= \mu' X_{\ell,\j} u_{\ell,j}$. Theorems \ref{thm:unifTCL} applied to $\mathcal{F}=\{f_{\mu}\}$ implies 
$$\frac{\sqrt{\underline{C}}}{\Pi_C} \sum_{\un \leq \j \leq \C} \sum_{\ell=1}^{N_{\j}}f_{\mu}(\widetilde{Y}_{\ell,\j})\convD \mathcal{N}\left(0, \sum_{i=1}^k \lambda_i\Cov\left(\sum_{\ell=1}^{N_{\un}}f_{\mu}(Y_{\ell,\un}),\sum_{\ell=1}^{N_{\deux}}f_{\mu}(Y_{\ell,\deux})\right)\right).$$
Because $\sum_{i=1}^k \lambda_i\Cov\left(\sum_{\ell=1}^{N_{\un}}f_{\mu}(Y_{\ell,\un}), \sum_{\ell=1}^{N_{\deux}}f_{\mu}(Y_{\ell,\deux})\right)=\mu'H\mu$, by the Cram\'er-Wold device,
$$\frac{\sqrt{\underline{C}}}{\Pi_C} \sum_{\un \leq \j \leq \C} \sum_{\ell=1}^{N_{\j}}X_{\ell,\j} u_{\ell,\j} \convD \mathcal{N}(0,H).$$
Similarly, by Theorem \ref{thm:boot_unif}, $\frac{\sqrt{\underline{C}}}{\Pi_C} \sum_{\un \leq \j \leq \C} \sum_{\ell=1}^{N_{\j}}f_{\mu}(\widetilde{Y}^*_{\ell,\j})$ converges weakly conditional on $\bm{Z}$ to the same limit. 

\medskip
Next, by Slutsky's lemma,
$$\sqrt{\underline{C}}(\widehat{\theta}-\theta_0) = \left[\frac{1}{\Pi_C}\sum_{\un \leq \j \leq \C}\sum_{\ell=1}^{N_{\j}}X_{\ell,\j}X_{\ell,\j}'\right]^{-1}\left[\frac{\sqrt{\underline{C}}}{\Pi_C} \sum_{\un \leq \j \leq \C} \sum_{\ell=1}^{N_{\j}} X_{\ell,\j} u_{\ell,\j}\right] \convD \mathcal{N}(0,V).$$
Similarly, by Theorem \ref{thm:boot_unif}, $\sqrt{\underline{C}}(\widehat{\theta}^{*}-\theta_0)$ converges weakly to $\mathcal{N}(0,V)$ conditional on $\bm{Z}$.

\medskip
It remains to show that $\widehat{V}$ is consistent or, equivalently since $\widehat{J}^{-1}=J^{-1}+o_P(1)$, that $\widehat{H}$ is consistent. Because $\widehat{u}_{\ell,\j}={u}_{\ell,\j}+X_{\ell,\j}'(\theta_0-\widehat{\theta})$, we  have $\widehat{H}=T_1+T_2+T_3+T_4$, with
{\small
\begin{align*}
T_1&=\sum_{i=1}^k\frac{\underline{C}}{C_i} \frac{1}{C_i}\sum_{j'_i=1}^{C_i} \left(\frac{1}{\prod_{s\neq i}C_s}\sum_{\j:j_i=j'_i} \sum_{\ell=1}^{N_{\j}}X_{\ell,\j}u_{\ell,\j}\right)
\left(\frac{1}{\prod_{s\neq i}C_s}\sum_{\j:j_i=j'_i} \sum_{\ell=1}^{N_{\j}} u_{\ell,\j}X_{\ell,\j}'\right),\\
T_2 & =\sum_{i=1}^k\frac{\underline{C}}{C_i} \frac{1}{C_i}\sum_{j'_i=1}^{C_i} \left(\frac{1}{\prod_{s\neq i}C_s}\sum_{\j:j_i=j'_i} \sum_{\ell=1}^{N_{\j}}X_{\ell,\j}X_{\ell,\j}'\right)
(\theta_0-\widehat{\theta}) (\theta_0-\widehat{\theta})'\left(\frac{1}{\prod_{s\neq i}C_s}\sum_{\j:j_i=j'_i} \sum_{\ell=1}^{N_{\j}} X_{\ell,\j}X_{\ell,\j}'\right),\\
T_3 & =\sum_{i=1}^k\frac{\underline{C}}{C_i} \frac{1}{C_i}\sum_{j'_i=1}^{C_i} \left(\frac{1}{\prod_{s\neq i}C_s}\sum_{\j:j_i=j'_i} \sum_{\ell=1}^{N_{\j}}X_{\ell,\j}X_{\ell,\j}'\right)
(\theta_0-\widehat{\theta})\left(\frac{1}{\prod_{s\neq i}C_s}\sum_{\j:j_i=j'_i} \sum_{\ell=1}^{N_{\j}} u_{\ell,\j} X_{\ell,\j}'\right),\\
T_4 & =\sum_{i=1}^k\frac{\underline{C}}{C_i} \frac{1}{C_i}\sum_{j'_i=1}^{C_i} \left(\frac{1}{\prod_{s\neq i}C_s}\sum_{\j:j_i=j'_i} \sum_{\ell=1}^{N_{\j}}X_{\ell,\j}u_{\ell,\j}\right)
(\theta_0-\widehat{\theta})'\left(\frac{1}{\prod_{s\neq i}C_s}\sum_{\j:j_i=j'_i} \sum_{\ell=1}^{N_{\j}} X_{\ell,\j}X_{\ell,\j}'\right).
\end{align*}}
Proposition \ref{prop:as_var} ensures that $T_1\convP H$. Next, we show that the three remaining terms converge to zero. Let $T_2=\sum_{i=1}^k (\underline{C}/C_i) T_{2i}$. By the triangle inequality, sub-multiplicativity of Frobenius norm and Cauchy-Schwarz inequality,
\begin{align*}
%&\left|\frac{1}{C_i}\sum_{j'_i=1}^{C_i} \left(\frac{1}{\prod_{s\neq i}C_s}\sum_{\j:j_i=j'_i} \sum_{\ell=1}^{N_{\j}}X_{\ell,\j}X_{\ell,\j}'\right)
%(\theta_0-\widehat{\theta}) (\theta_0-\widehat{\theta})'\left(\frac{1}{\prod_{s\neq i}C_s}\sum_{\j:j_i=j'_i} \sum_{\ell=1}^{N_{\j}} X_{\ell,\j}X_{\ell,\j}'\right)\right|\\
T_{2i} &\leq \frac{1}{C_i}\sum_{j'_i=1}^{C_i} \left|\left(\frac{1}{\prod_{s\neq i}C_s}\sum_{\j:j_i=j'_i} \sum_{\ell=1}^{N_{\j}}X_{\ell,\j}X_{\ell,\j}'\right)
(\theta_0-\widehat{\theta}) (\theta_0-\widehat{\theta})'\left(\frac{1}{\prod_{s\neq i}C_s}\sum_{\j:j_i=j'_i} \sum_{\ell=1}^{N_{\j}} X_{\ell,\j}X_{\ell,\j}'\right)\right|\\
&\leq \frac{1}{C_i}\sum_{j'_i=1}^{C_i} \left|\left(\frac{1}{\prod_{s\neq i}C_s}\sum_{\j:j_i=j'_i} \sum_{\ell=1}^{N_{\j}}X_{\ell,\j}X_{\ell,\j}'\right)\right|^2
|\theta_0-\widehat{\theta}|^2\\
&\leq \frac{1}{C_i}\sum_{j'_i=1}^{C_i} \left(\frac{1}{\prod_{s\neq i}C_s}\sum_{\j:j_i=j'_i} \sum_{\ell=1}^{N_{\j}}|X_{\ell,\j}'|^2\right)^2
|\theta_0-\widehat{\theta}|^2\\
&\leq \frac{1}{C_i}\sum_{j'_i=1}^{C_i} \frac{1}{\prod_{s\neq i}C_s} \sum_{\j:j_i=j'_i} \left( \sum_{\ell=1}^{N_{\j}}|X_{\ell,\j}'|^2\right)^2
|\theta_0-\widehat{\theta}|^2\\
&=\frac{1}{\Pi_C}\sum_{\un \leq \j \leq \C}\left( \sum_{\ell=1}^{N_{\j}}|X_{\ell,\j}'|^2\right)^2 |\theta_0-\widehat{\theta}|^2
\end{align*}
Because $\E\left(( \sum_{\ell=1}^{N_{\un}}|X_{\ell,\un}|^2)^2\right)<\infty$, Lemma 7.35 of \cite{kallenberg05} implies
$$\frac{1}{\Pi_C}\sum_{\un \leq \j \leq \C}\left( \sum_{\ell=1}^{N_{\j}}|X_{\ell,\j}'|^2\right)^2=\E\left(\left( \sum_{\ell=1}^{N_{\un}}|X_{\ell,\un}|^2\right)^2\right)+o_p(1)=O_p(1).$$ 
Moreover, $|\widehat{\theta}-\theta_0|=o_p(1)$ and for any $i$, $\underline{C}/C_i=\lambda_i+o(1)$. As a result, $T_2=o_p(1)$. Next, $T_3$ and $T_4$ are bounded by
$$\sum_{i=1}^k\frac{\underline{C}}{C_i}\left[\frac{1}{\Pi_C}\sum_{\un \leq \j \leq \C} \left(\sum_{\ell=1}^{N_{\j}}|X_{\ell,\j}|^2\right)^2\right]^{1/2}\left[\frac{1}{\Pi_C}\sum_{\un \leq \j \leq \C} \left(\sum_{\ell=1}^{N_{\j}}|X_{\ell,\j}| |u_{\ell,\j}|\right)^2\right]^{1/2}|\widehat{\theta}-\theta_0|,$$
and again Lemma 7.35 in Kallenberg implies $$\left[\frac{1}{\Pi_C}\sum_{\un \leq \j \leq \C} \left(\sum_{\ell=1}^{N_{\j}}|X_{\ell,\j}|^2\right)^{2}\right]^{1/2}\left[\frac{1}{\Pi_C}\sum_{\un \leq \j \leq \C}\left( \sum_{\ell=1}^{N_{\j}}|X_{\ell,\j}| |u_{\ell,\j}|\right)^2\right]^{1/2}=O_p(1).$$ Thus, $T_3=o_p(1)$ and $T_4=o_p(1)$, implying that $\widehat{H}=H+o_p(1)$. The result follows.

{\subsection{Proof that moments of quantile IV satisfy Assumption \ref{as:GMM}.3-\ref{as:GMM}.5} % (fold)
\label{sub:quantile_IV}

We check Assumptions \ref{as:GMM}.3-\ref{as:GMM}.5 assuming that the $(Y_{\ell, \un})_{\ell\geq 1}$ are identically distributed and under the following conditions: 

\begin{hyp}
	\label{as:quantile_IV}
	~
\begin{enumerate}
	\item $\theta_0$ belongs to the interior of $\Theta$, a compact subset of $\R^p$.
	\item The support of $X$ is a compact subset of $\R^p$.
	\item $\E[N_{\un}^2 (1+|Z_{1,\un}|^2)]<+\infty$.
	\item The conditional cdf $F_{W_{1,\un}|X_{1,\un},Z_{1,\un}}(\cdot|X_{1,\un},Z_{1,\un})$ is continuous everywhere, $X_{1,\un},Z_{1,\un}$-almost surely.
	\item There exists $r>0$ such that for almost all $(x,z)$, $F_{W_{1,\un}|X_{1,\un},Z_{1,\un}}(\cdot|x,z)$ is differentiable on $\{x'\theta_0 + t, t\in [-r,r]\}$ and
	\begin{equation}
	\sup_{(x,z,t)\in \Supp(X_{1,\un},Z_{1,\un})\times[-r,r]} f_{W_{1,\un}|X_{1,\un},Z_{1,\un}}(x'\theta_0+t|x,z) < +\infty,
	\label{eq:bounded_density}	
	\end{equation}
where $\Supp(X_{1,\un},Z_{1,\un})$ denotes the support of $(X_{1,\un},Z_{1,\un})$. 	
%	\item $\E[N_{\un} |X_{1,\un} Z'_{1,\un}|]<+\infty$;
	\item The rank of $\E\left[N_{\un} X_{1,\un} Z'_{1,\un}f_{W_{1,\un}|X_{1,\un},Z_{1,\un}}(X'_{\ell, \j}\theta_0|X_{1,\un},Z_{1,\un})\right]$ is equal to $p$.
\end{enumerate}
\end{hyp}

With a slight abuse of notation, we denote by $(a,b]$ either the interval $(a,b]$ if $a<b$, or $(b,a]$ if $b<a$. First, we have
\begin{align*}
&  \E\left[\left|\sum_{\ell=1}^{N_{\un}}m(Y_{\ell,\un},\theta')-\sum_{\ell=1}^{N_{\un}}m(Y_{\ell,\un},\theta)\right|^2 \right] \\
\leq &  \E\left[N_{\un}\sum_{\ell=1}^{N_{\un}}|Z_{\ell,\un}|^2  \mathds{1}\{W_{\ell, \j} \in (X'_{\ell, \j}\theta, X'_{\ell, \j}\theta']\} \right] \\
\leq  & \E\left[N_{\un}^2 |Z_{\ell,\un}|^2  \left|F_{W_{1,\un}|X_{1,\un},Z_{1,\un}}(X'_{\ell, \j}\theta)- F_{W_{1,\un}|X_{1,\un},Z_{1,\un}}(X'_{\ell, \j}\theta')\right| \right],
\end{align*}
where the first inequality follows by the Cauchy-Schwarz inequality and the second uses the fact that the $(Y_{\ell, \un})_{\ell\geq 1}$ are identically distributed and the law of iterated expectation. Then because $\E[N_{\un}^2 |Z_{1,\un}|^2]<+\infty$ and $F_{W_{1,\un}|X_{1,\un},Z_{1,\un}}(\cdot|X_{1,\un},Z_{1,\un})$ is continuous everywhere,
Assumption \ref{as:GMM}.3 follows by the dominated convergence theorem. 

\medskip
Turning to \ref{as:GMM}.4, by the same arguments as those used to obtain the second inequality above,
$$\E\left(\sum_{\ell=1}^{N_{\un}}m(Y_{\ell,\un},\theta)\right)=\E\left[N_{\un} Z_{1,\un}\left(\tau - F_{W_{1,\un}|X_{1,\un},Z_{1,\un}}(X'_{\ell, \j}\theta|X_{1,\un},Z_{1,\un})\right)\right].$$
By Assumptions \ref{as:quantile_IV}.2-\ref{as:quantile_IV}.3 and the Cauchy-Schwarz inequality, $\E[N_{\un} |X_{1,\un} Z'_{1,\un}|]<+\infty$. Moreover, still by Assumption \ref{as:quantile_IV}.2 and the Cauchy-Schwarz inequality, there exists a neighborhood $\mathcal{V}$ of $\theta_0$ such that for any $\theta\in \mathcal{V}$, $|x'\theta - x'\theta_0|\leq r$ for all $x$ in the support of $X_{1,\un}$ with $r$ defined in Assumption \ref{as:quantile_IV}.5. Then, by Assumption \ref{as:quantile_IV}.5, we can apply the dominated convergence theorem to $\theta \mapsto \E\left(\sum_{\ell=1}^{N_{\un}}m(Y_{\ell,\un},\theta)\right)$ defined on $\mathcal{V}$. This implies that  $\theta\mapsto \E\left(\sum_{\ell=1}^{N_{\un}}m(Y_{\ell,\un},\theta)\right)$ is differentiable at $\theta_0$, with a Jacobian matrix $J$ satisfying 
$$J=\E\left[N_{\un} X_{1,\un} Z'_{1,\un}f_{W_{1,\un}|X_{1,\un},Z_{1,\un}}(X'_{\ell, \j}\theta_0|X_{1,\un},Z_{1,\un})\right].$$

\medskip
Finally, let us check \ref{as:GMM}.5. We have to prove that Assumptions \ref{as:measurability}-\ref{as:vc} hold for the class
$$\mathcal{F}_s= \left\{(w,x,z_s) \mapsto z_s\left(\tau - \mathds{1}\{w - x'\theta\leq 0\}\right), \theta \in \Theta \right\}, \; s\in\{1,...,L\}.$$
Reasoning as in the proof of Lemma 8.12 in \cite{Kosorok2006}, the class $\left\{\mathds{1}\{w - x'\theta\leq 0\}, \theta \in \Theta \right\}$ is pointwise measurable. By Lemma 8.10 in \cite{Kosorok2006}, $\mathcal{F}_s$ is pointwise measurable as well. 

\medskip
We now check Assumption \ref{as:vc} for $\mathcal{F}_s$. We have $E[N_{\un}^2]<+\infty$ and by Assumption \ref{as:quantile_IV}.3, the envelope function $F_s(w,x,z)=2|z_s|$ satisfies $\E\left[N_{\un}\left(\sum_{\ell=1}^{N_{\un}} F(Y_{\ell,\un}) \right)^2\right]<+\infty$. Turning to the entropy condition, by Theorem 9.3 in \cite{Kosorok2006}, it suffices to prove that $\mathcal{F}_s$ is a VC class \citep[for a definition of VC classes, see e.g.][Section 9.1.1]{Kosorok2006}. The class  $\mathcal{G}_s=\{(w,x,z)\mapsto \mathds{1}\{w-x'\theta\leq 0\}, \theta\in \Theta\}$ is a subset of 
$$\left\{(w,x,z)\mapsto \mathds{1}\{w\eta + x'\theta + z'\gamma\leq \delta\}, (\eta, \theta, \gamma, \delta)\in \R\times \R^p \times \R^L \times \R\right\},$$ 
which is VC by Lemma 9.8 and 9.12 in \cite{Kosorok2006}. Hence, $\mathcal{G}_s$ is VC as well. By Lemma 9.9-(v) and (vi), $\mathcal{F}_s$ is also VC. The result follows.

% subsection proof_that_moments_of_quantile_iv_satisfy_assumption_ref_as_gmm (end)

\subsection{Proof of Theorem \ref{thm:AN_GMM} }

The proof is a combination of those of Theorem 1 in \cite{hahn1996}, Theorem 3.3 in \cite{PakesPollard89} and Theorem 5.21 in \cite{vanderVaart2000}.

Hereafter, for any $\theta\in \Theta$, $\mathbb{M}_{C}(\theta)$ denotes the multidimensional random process from $\Theta$ to $\mathbb{R}^{L}$:
$$\mathbb{M}_{C}(\theta)=\mathbb{G}_Cm(.,\theta)=\frac{\underline{C}^{1/2}}{\Pi_C}\sum_{\un\leq \j \leq \C}\sum_{\ell=1}^{N_{\j}}\left[m(Y_{\ell,\j},\theta)-\E\left(\sum_{\ell=1}^{N_{\un}}m(Y_{\ell,\un},\theta)\right)\right].$$
%Note that Theorem \ref{thm:unifTCL} implies that $\mathbb{M}_{C}(\theta)=O_p(1)$ for any $\theta$ and even $\sup_{\theta \in \Theta}\left|\mathbb{M}_{C}(\theta)\right|=O_p(1)$.
Because $\Xi$ is a symmetric positive definite matrix, $\widehat{\Xi}$ is also symmetric positive definite with probability tending to 1.%, and $\mathds{1}_{\{\det(\widehat{\Xi})>0\}}=1+o_p(1)$ and $\mathds{1}_{\{\det(\widehat{\Xi})\leq 0\}}=o_p(1)$. 
Thus, its square root $\widehat{\Xi}^{1/2}$ is well-defined with probability tending to 1. Then, for any $\theta\in \Theta$, let
\begin{align*}
M(\theta) &= \left|\Xi^{1/2}\E\left(\sum_{\ell=1}^{N_{\un}}m(Y_{\ell,\un},\theta)\right)\right|, \\ 
M_C(\theta) & =\left|\widehat{\Xi}^{1/2}\left(\frac{1}{\Pi_C}\sum_{\un\leq \j \leq \C}\sum_{\ell=1}^{N_{\j}}m(Y_{\ell,\j},\theta)\right)\right|, \\ 
\overline{M_C}(\theta) & =\left|\widehat{\Xi}^{1/2}\E\left(\sum_{\ell=1}^{N_{\un}}m(Y_{\ell,\un},\theta)\right)\right|.	
\end{align*}
When $\det(\widehat{\Xi})\leq 0$, $M_C(\theta)$ and $\overline{M_C}(\theta)$ are defined arbitrarily.

\medskip
\textbf{1. Consistency}

\medskip
For any symmetric matrix $B$, let $\rho(B)$ denote the largest modulus of its eigenvalues. By the triangle inequality,
\begin{align*}
&\left|M_{C}(\theta)-M(\theta)\right| \mathds{1}_{\{\det(\widehat{\Xi})> 0\}}\\
%&\leq \left|M_{C}(\theta)-M(\theta)\right|\mathds{1}_{\{\det(\widehat{\Xi})> 0\}}+ \left|M_{C}(\theta)-M(\theta)\right| \mathds{1}_{\{\det(\widehat{\Xi})\leq 0\}}\\
\leq & \left(\left|M_{C}(\theta)-\overline{M_C}(\theta)\right|+\left|\overline{M_C}(\theta)-M(\theta)\right|\right)\mathds{1}_{\{\det(\widehat{\Xi})> 0\}}\\
\leq & \left(\rho(\widehat{\Xi}^{1/2})\underline{C}^{-1/2}\left|\mathbb{M}_{C}(\theta) \right|+\rho(\widehat{\Xi}^{1/2}-\Xi^{1/2})\left|\E\left(\sum_{\ell=1}^{N_{\un}}m(Y_{\ell,\un},\theta)\right) \right|\right)\times \mathds{1}_{\{\det(\widehat{\Xi})> 0\}}
\end{align*}

Assumption \ref{as:GMM}.5 ensures that Theorem \ref{thm:unifTCL} applies to every $m_s$ for $s=1,...,L$, which yields
$$ \underline{C}^{-1/2}\sup_{\theta\in \Theta}\left|\mathbb{M}_{C}(\theta) \right|=O_p(\underline{C}^{-1/2})=o_p(1).$$

By Assumption \ref{as:GMM}.6 and the continuity of $S\mapsto \rho(S)$ and $S\mapsto S^{1/2}$, we have $\rho(\widehat{\Xi}^{1/2}-\Xi^{1/2})=o_p(1)$ and $\rho(\widehat{\Xi}^{1/2})=O_p(1)$. Assumptions \ref{as:GMM}.1 and \ref{as:GMM}.3 together also imply that
$\sup_{\theta \in \Theta}\left|\E\left(\sum_{\ell=1}^{N_{\un}}m(Y_{\ell,\un},\theta)\right) \right|<\infty$. As a result, 
\begin{align}\label{eq:proofGMM1}
\sup_{\theta \in \Theta}|M_C(\theta)-M(\theta)| \mathds{1}_{\{\det(\widehat{\Xi})> 0\}}=o_p(1).
\end{align}

Because $M$ is continuous on $\Theta$ compact and reaches its minimum only at $\theta_0$, we have, for any $\varepsilon>0$:
$$\inf_{\theta\in \Theta: |\theta-\theta_0|>\varepsilon}M(\theta)>M(\theta_0).$$

This means that for any $\varepsilon>0$ there exists $\eta>0$ such that $|\theta-\theta_0|>\varepsilon$ implies $M(\theta)>M(\theta_0)+\eta$. Then, because $\widehat{\theta}=\arg \min_{\theta}M_C(\theta)$, 
\begin{align*}
\{|\widehat{\theta}-\theta_0|>\varepsilon\}\cap\{\det(\widehat{\Xi})> 0\}&\subset\{M(\widehat{\theta})>M(\theta_0)+\eta\}\cap\{\det(\widehat{\Xi})> 0\}\\
&\subset\{M(\widehat{\theta})-M_C(\widehat{\theta})+M_C(\widehat{\theta})-M(\theta_0)>\eta\}\cap\{\det(\widehat{\Xi})> 0\}\\
&\subset\{M(\widehat{\theta})-M_C(\widehat{\theta})+M_C(\theta_0)-M(\theta_0)>\eta\}\cap\{\det(\widehat{\Xi})> 0\}\\
&\subset\{2\sup_{\theta \in \Theta}|M_C(\theta)-M(\theta)|\mathds{1}_{\{\det(\widehat{\Xi})> 0\}}>\eta\}.
\end{align*}
As a result, 
$$\mathbb{P}(|\widehat{\theta}-\theta_0|>\varepsilon)\leq \mathbb{P}(2\sup_{\theta \in \Theta}|M_C(\theta)-M(\theta)| \mathds{1}_{\{\det(\widehat{\Xi})> 0\}}>\eta)+\mathbb{P}(\det(\widehat{\Xi})\leq 0).$$
Consistency of $\widehat{\theta}$ follows by Equation (\ref{eq:proofGMM1}) and continuity of $S\mapsto \det(S)$.

\medskip
\textbf{2. Asymptotic normality}

\medskip
Theorem \ref{thm:unifTCL} and the Cram\'er-Wold device ensure that $\mathbb{M}_{C}(\theta)$ converges weakly to a centered $L$-multidimensional gaussian process with covariance kernel:
$$\sum_{i=}^k\lambda_i\Cov\left(\sum_{\ell=1}^{N_{\un}}m(Y_{\ell,\un},\theta_1), \sum_{\ell=1}^{N_{\deux}}m(Y_{\ell,\deux},\theta_2)\right).$$

%Let $Q$ the probability measure defined by $dQ(n,y_1,...,y_n)=dP_{Y_{1,\un},...,Y_{n,\un}|N_{\un}=n}(y_1,...,y_n) dP_{N_{\un}}(n)$.

Consider $\widetilde{\theta}=\theta_0+o_p(1)$. By Assumptions \ref{as:GMM}.1, \ref{as:GMM}.3 and the continuous mapping theorem, for any $s=1,...,L$, 
$$\E\left[ \left|\sum_{\ell=1}^{N_{\un}}m_s(Y_{\ell,\un},\theta)-m_s(Y_{\ell,\un},\theta_0)\right|^{2}\right]_{\theta=\widetilde{\theta}}=o_p(1).$$ 
Next, by Lemma 19.24 in \cite{vanderVaart2000}, %for any $s=1,...,L$:
%$$\mathbb{G}_{C}\left(m_s(.,\widehat{\theta})\right)-\mathbb{G}_{C}\left(m_s(.,\theta_0)\right)=o_p(1).$$
%By the Slutsky lemma, this means that:
\begin{align}\label{eq:proofGMM2}\mathbb{M}_{C}(\theta_0)&=\mathbb{M}_{C}(\widetilde{\theta})+o_p(1). 
\end{align}
By Assumption \ref{as:GMM}.4,
$$\E\left( \sum_{\ell=1}^{N_{\un}} m(Y_{\ell,\un},\theta)\right)_{\theta=\widetilde{\theta}}=J (\widetilde{\theta}-\theta_0)+o_p(|\widetilde{\theta}-\theta_0|),$$
and next
\begin{align}\mathbb{M}_{C}(\theta_0)&=\frac{\underline{C}^{1/2}}{\Pi_C}\sum_{\un\leq \j \leq \C}\sum_{\ell=1}^{N_{\j}}m(Y_{\ell,\j},\widetilde{\theta})-\underline{C}^{1/2}J (\widetilde{\theta}-\theta_0)+o_p(\underline{C}^{1/2}|\widetilde{\theta}-\theta_0|)+o_p(1).\label{eq:gmm_proof1} \end{align}

If $\det(\widehat{\Xi})>0$, let $L_C(\theta)=\Xi^{1/2}J(\theta-\theta_0)+\underline{C}^{-1/2}\widehat{\Xi}^{1/2}\mathbb{M}_{C}(\theta_0)$ (otherwise define $L_C(\theta)$ arbitrarily). By the triangle inequality,
\begin{align*}&\left|L_{C}(\widetilde{\theta})-\widehat{\Xi}^{1/2}\frac{1}{\Pi_C}\sum_{\un\leq \j \leq \C} \sum_{\ell=1}^{N_{\j}}m(Y_{\ell,\j},\widetilde{\theta})\right| \mathds{1}_{\{\det(\widehat{\Xi})> 0\}} \\
\leq & \left(\underline{C}^{-1/2} \rho(\widehat{\Xi}^{1/2})  \left|\mathbb{M}_{C}(\widetilde{\theta})-\mathbb{M}_{C}(\theta_0)\right|+
\underline{C}^{-1/2}\rho(\widehat{\Xi}^{1/2}-{\Xi}^{1/2}) \left|\mathbb{M}_{C}(\theta_0)\right|\right)\mathds{1}_{\{\det(\widehat{\Xi})> 0\}}+o_p(\underline{C}^{-1/2})
\end{align*}

Equation (\ref{eq:proofGMM2}) ensures that $\left|\mathbb{M}_{C}(\widetilde{\theta})-\mathbb{M}_{C}(\theta_0)\right|=o_p(1)$. We also have $\rho(\widehat{\Xi}^{1/2})=O_p(1)$, $\rho(\widehat{\Xi}^{1/2}-{\Xi}^{1/2}) \left|\mathbb{M}_{C}(\theta_0)\right|=o_p(1) O_p(1)=o_p(1)$ and $\mathds{1}_{\{\det(\widehat{\Xi})> 0\}}=1+o_p(1)$. Therefore, 
\begin{align}\left|L_{C}(\widetilde{\theta})-\widehat{\Xi}^{1/2}\frac{1}{\Pi_C}\sum_{\un\leq \j \leq \C} \sum_{\ell=1}^{N_{\j}}m(Y_{\ell,\j},\widetilde{\theta})\right|&=o_p(\underline{C}^{-1/2}).\label{eq:gmm_proof2} \end{align}

We have
\begin{align*}\left|\frac{\underline{C}^{1/2}}{\Pi_C}\sum_{\un\leq \j \leq \C}\sum_{\ell=1}^{N_{\j}}m(Y_{\ell,\j},\widehat{\theta})\right|\mathds{1}_{\{\det(\widehat{\Xi})> 0\}}&\leq \rho(\widehat{\Xi}^{-1/2}) \underline{C}^{1/2}M_C(\widehat{\theta})\mathds{1}_{\{\det(\widehat{\Xi})> 0\}}\\
&\leq \rho(\widehat{\Xi}^{-1/2}) \underline{C}^{1/2} M_C(\theta_0)\mathds{1}_{\{\det(\widehat{\Xi})> 0\}}\\
&\leq \rho(\widehat{\Xi}^{-1/2}) \rho(\Xi^{1/2})\left|\mathbb{M}_{C}(\theta_0)\right|\mathds{1}_{\{\det(\widehat{\Xi})> 0\}}=O_p(1).\end{align*}
Because $\mathds{1}_{\{\det(\widehat{\Xi})> 0\}}=1+o_p(1)$, we deduce that $\left|\frac{\underline{C}^{1/2}}{\Pi_C}\sum_{\un\leq \j \leq \C}\sum_{\ell=1}^{N_{\j}}m(Y_{\ell,\j},\widehat{\theta})\right|=O_p(1)$.
Now, if we consider $\widetilde{\theta}=\widehat{\theta}$ in Equation (\ref{eq:gmm_proof1}), because $\mathbb{M}_{C}(\theta_0)=O_p(1)$, we have
$$O_p(1)=O_p(1)-\underline{C}^{1/2}J (\widehat{\theta}-\theta_0)+o_p(\underline{C}^{1/2}|\widetilde{\theta}-\theta_0|)+o_p(1)$$
Since $J$ is full column rank, we deduce $\widehat{\theta}-\theta_0=O_p(\underline{C}^{-1/2})$.

\medskip
Equation (\ref{eq:gmm_proof2}) holds for $\widetilde{\theta}=\widehat{\theta}$ and $\widetilde{\theta}=\theta^{\ast}=\theta_0-\underline{C}^{-1/2}(J'\Xi J)^{-1}J' \Xi^{1/2}\widehat{\Xi}^{1/2}\mathbb{M}_{C}(\theta_0)$. Because $\widehat{\theta}=\argmin_{\theta}M_{C}(\theta)$ and $\theta^{\ast}=\arg \min_{\theta}\left| L_C(\theta) \right|$, we have
$$\left| L_C(\widehat{\theta})\right|-o_p(\underline{C}^{1/2})\leq M_C(\widehat{\theta})\leq M_{C}(\theta^{\ast})\leq \left|L_{C}(\theta^{\ast})\right|+o_p(\underline{C}^{-1/2})\leq \left|L_{C}(\widehat{\theta})\right|+o_p(\underline{C}^{-1/2}).$$
It follows that
$$\left| L_C(\widehat{\theta})\right|=\left|L_{C}(\theta^{\ast})\right|+o_p(\underline{C}^{-1/2}).$$
Then, because $L_{C}(\theta^{\ast})=\underline{C}^{-1/2}\left(I-\Xi^{1/2}J(J' \Xi J)^{-1}J'\Xi^{1/2}\right)\widehat{\Xi}^{1/2}\mathbb{M}_{C}(\theta_0)=O_p(\underline{C}^{-1/2})$,
$$\left| L_C(\widehat{\theta})\right|^2=\left|L_{C}(\theta^{\ast})\right|^2+o_p(\underline{C}^{-1}),$$
Moreover, $J' \Xi L_{C}(\theta^{\ast})=0$ implies that
$$\left|L_C(\widehat{\theta})\right|^2=\left|L_C(\theta^{\ast})\right|^2+\left|\Xi^{1/2}J(\widehat{\theta}-\theta^{\ast})\right|^2.$$
Combining the two previous equations, we get $\left|\Xi^{1/2}J(\widehat{\theta}-\theta^{\ast})\right|=o_p(\underline{C}^{-1/2})$. Next, because $\Xi^{1/2}$ is non-singular and $J$ is full column rank, we have $\widehat{\theta}-\theta^{\ast}=o_p(\underline{C}^{-1/2})$. It follows that:
\begin{align*}
\underline{C}^{1/2}(\widehat{\theta}-\theta_0)&=\underline{C}^{1/2}(\theta^{\ast}-\theta_0)+o_p(1)\\
&=-(J'\Xi J)^{-1}J' \Xi^{1/2}\widehat{\Xi}^{1/2}\mathbb{M}_{C}(\theta_0)+o_p(1)\\
&=-(J'\Xi J)^{-1}J' \Xi\mathbb{M}_{C}(\theta_0)+o_p(1)
\end{align*}
We already know that $\mathbb{M}_{C}(\theta_0) \convD \mathcal{N}(0,H)$. The result follows by Slutsky's Lemma and the continuous mapping theorem.

\subsection{Proof of Theorem \ref{thm:var_gmm}}

\subsubsection{Inference based on asymptotic normality}

Slutsky's Lemma and the continuous mapping theorem ensure that $\widehat{V}\convP V_0$ if $\widehat{J}\convP J$ and $\widehat{H}\convP H$.

We first consider $\widehat{J}$. We have $\widehat{J}=\widehat{J}(\widehat{\theta})$, with $\widehat{J}(\theta)=\frac{1}{\Pi_C}\sum_{\un\leq \j \leq \C} \sum_{\ell=1}^{N_{\j}}d(Y_{\ell,\j},\theta)$. Moreover,
\begin{align*}
	&\left|\widehat{J}-J\right|\leq\left|\widehat{J}-\mathbb{E}[\widehat{J}(\theta)]_{\theta=\widehat{\theta}}\right|+\left|\mathbb{E}[\widehat{J}(\theta)]_{\theta=\widehat{\theta}}-J\right|.
\end{align*}

Because $\widehat{\theta}=\theta_0+o_p(1)$ and Assumption \ref{as:varGMM}.3 holds, the continuous mapping theorem  ensures that $\left|\mathbb{E}[\widehat{J}(\theta)]_{\theta=\widehat{\theta}}-J\right|=o_p(1)$. We also have:
\begin{align*}
	&\mathbb{E}\left[\left|\widehat{J}-\mathbb{E}[\widehat{J}(\theta)]_{\theta=\widehat{\theta}}\right|\right] \\
\leq &\sqrt{Lp} \mathbb{E}\left[\sum_{s=1}^{L}\sum_{r=1}^p\left|\frac{1}{\Pi_C}\sum_{\un\leq\j\leq\C}\sum_{\ell=1}^{N_{\j}}d_{r,s}(Y_{\ell,\j},\widehat{\theta})-\mathbb{E}\left[\sum_{\ell=1}^{N_{\un}}d_{r,s}(Y_{\ell,\un},\theta)\right]_{\theta=\widehat{\theta}}\right|\right] \\
    \leq &(Lp)^{\frac{3}{2}}\max_{1\leq r\leq p,1\leq s\leq L}\mathbbm{E}\left[\left|\frac{1}{\Pi_C}\sum_{\un\leq\j\leq\C}\sum_{\ell=1}^{N_{\j}}d_{r,s}(Y_{\ell,\j},\widehat{\theta})-\mathbb{E}\left[\sum_{\ell=1}^{N_{\un}}d_{r,s}(Y_{\ell,\un},\theta)\right]_{\theta=\widehat{\theta}}\right|\right] \\
    \leq &(Lp)^{\frac{3}{2}}\max_{1\leq r\leq p,1\leq s\leq L}\mathbbm{E}\left[\sup_{\Theta}\left|\frac{1}{\Pi_C}\sum_{\un\leq\j\leq\C}\sum_{\ell=1}^{N_{\j}}d_{r,s}(Y_{\ell,\j},\theta)-\mathbb{E}\left[\sum_{\ell=1}^{N_{\un}}d_{r,s}(Y_{\ell,\un},\theta)\right]\right|\right].
\end{align*}

The assumptions on the classes $\mathcal{G}_{r,s}=\{y\mapsto d_{r,s}(y,\theta):\theta \in \Theta\}$ are sufficient to apply Lemma~\ref{lemma:glivenko}, implying that 
\begin{align*}
    &\max_{r,s} \mathbbm{E}\left[\sup_{\Theta}\left|\frac{1}{\Pi_C}\sum_{\un\leq\j\leq\C}\sum_{\ell=1}^{N_{\j}}d_{r,s}(Y_{\ell,\j},\theta)-\mathbb{E}\left[\sum_{\ell=1}^{N_{\un}}d_{r,s}(Y_{\ell,\un},\theta)\right]\right|\right]=o(1).
\end{align*}
Hence, $\widehat{J}$ is consistent. 

\medskip
Let us turn to $\widehat{H}$. We define
\begin{align*}
	\widehat{H}_i(\theta) & = \frac{1}{C_i\prod_{s\neq i}C_s^2}\sum_{(\j,\j')\in\mathcal{B}_i}\sum_{\ell=1}^{N_{\j}}m\left(Y_{\ell,\j},\theta\right)\sum_{\ell=1}^{N_{\j'}}m'\left(Y_{\ell,\j'},\theta\right), \\
	H_i(\theta) & = \E\left[\left(\sum_{\ell=1}^{N_{\un}}m(Y_{\ell,\un},\theta)\right)\left(\sum_{\ell=1}^{N_{\deux}} m(Y_{\ell,\deux},\theta)\right)'\right].
\end{align*}
We have $H=\sum_i \lambda_i H_i(\theta_0)$ and $\widehat{H}=\sum_i (\underline{C}/C_i) \widehat{H}_i(\widehat{\theta})$. As $\underline{C}/C_i=\lambda_i+o_p(1)$, we only need to show that  for every $i=1,...,k$, $\widehat{H}_i(\widehat{\theta})=H_i(\theta_0)+o_p(1)$. By Lemma \ref{lemma:glivenko_prod_fin2},  $\E(\sup_{\theta\in \Theta}|\widehat{H}_i(\theta)-H_i(\theta)|)=o(1)$. Thus, by Markov inequality, $\widehat{H}_i(\widehat{\theta})=H_i(\widehat{\theta})+o_p(1)$. Then, by Assumption \ref{as:varGMM}.3 and the continuous mapping theorem, $H_i(\widehat{\theta})=H_i(\theta_0)+o_p(1)$. The result follows. % We will in fact show a stronger convergence: $\E\left(\left|\widehat{H}_i-H_i\right|\right)=o(1)$.

\subsubsection{Inference based on the bootstrap}

We remark that in the proof of Theorem 1 in \cite{hahn1996}, the assumption of i.i.d. data is only used to ensure the weak convergence of the bootstrap empirical process conditional on the data. Since we show weak convergence of $\mathbb{G}_C^{\ast}$ in Section \ref{sec:boot}, we can directly follow the proof of Theorem 1 in \cite{hahn1996} to conclude that conditional on $\bm{Z}$ and with probability approaching one,
\begin{align*}
	&\sqrt{\underline{C}}\left(\widehat{\theta}^*-\widehat{\theta}\right)\convD\mathcal{N}(0,V_0).
\end{align*}
The result follows as in the proof of Proposition \ref{prop:boot_mean}.

\subsection{Proof of Proposition \ref{prop:functional_dm}} % (fold)
\label{sub:proof_of_proposition_ref_prop}

\textbf{1. Weak convergence of $\widehat{F_Y}$.} Let $\overline{\R}^k=\R^k \cup \{+\infty,...,+\infty\}$. By, e.g., Examples 2.6.1 and 2.10.7 in \cite{vanderVaartWellner1996}, the class $\mathcal{F}=\{t\mapsto \mathds{1}\{t\leq y\}, y\in \overline{\R}^k\}$, where the inequality sign is understood componentwise, satisfies the entropy condition in Assumption \ref{as:vc}. The other condition of Assumption \ref{as:vc} holds by taking the envelope function $F(t)=1$, using $\E(N_{1,\un}^2)<+\infty$.  Hence, by Theorem \ref{thm:unifTCL}, $\mathbb{G}_C$ defined on $\mathcal{F}$ converges to a gaussian process, with kernel
\begin{equation}
\widetilde{K}(y_1,y_2) = \sum_{i=1}^{k}\lambda_i \Cov\left(\sum_{\ell=1}^{N_{\un}}\mathds{1}\{Y_{\ell,\un}\leq y_1\}, \sum_{\ell=1}^{N_{\deux}}\mathds{1}\{Y_{\ell,\deux}\leq y_2\}\right).
\label{eq:Ktilde}
\end{equation}
Then, for any function $G$ from $\overline{\R}^k$ to $\R$, let us define $f(G)$ by $f(G)(y)=G(y)/G(+\infty,...,+\infty)$. This function is Hadamard differentiable at any $G$ such that $G(+\infty,...,+\infty)\neq 0$, with 
\begin{equation}
df_G(h)= \frac{1}{G(+\infty,...,+\infty)} \left[h - f(G) h(+\infty,...,+\infty)\right].
\label{eq:df_G}
\end{equation}
Remark that $\widehat{F_Y}=f(\mathbb{G}_C)$ and $\mathbb{G}_C(+\infty,...,+\infty)\neq 0$ with probability 1 by Assumption \ref{as:dgp}. Then, by the functional delta method \citep[see, e.g.,][Theorem 20.8]{vanderVaart2000}, $\sqrt{\underline{C}}(\widehat{F_Y}-F_Y)$ converges to a gaussian process. Equations \eqref{eq:Ktilde} and \eqref{eq:df_G} implies that its kernel $K$ satisfies \eqref{eq:kernel_had_diff}.

\medskip
\textbf{2. Asymptotic normality of $\widehat{\theta}$.} This follows directly by Point 1 and the functional delta method. 

\medskip
\textbf{3. Consistency of the pigeonhole bootstrap.} By the functional delta method for the bootstrap \citep[see, e.g.,][Theorem 3.9.11]{vanderVaartWellner1996}, conditional on the data, $\sqrt{\underline{C}} \left(\widehat{\theta}^*-\widehat{\theta}\right)$ converges to the same limit as $\sqrt{\underline{C}} \left(\widehat{\theta}-\theta_0\right)$. The result follows as in Proposition \ref{prop:boot_mean}. 
% subsection proof_of_proposition_ref_prop (end) 

\subsection{Proof that $\mathbb{G}$ is continuous at $\theta_0$ in the quantile example} % (fold)
\label{app:example_quantile}

We first prove that $\rho(t)=\V\left[\mathbb{G}(t)-\mathbb{G}(\theta_0)\right]^{1/2}$ is continuous at $\theta_0$. Given the expression of $K$, it suffices to prove that for all $i\in\{1,...,k\}$, $K_i(t)= \Cov(S_{\un}(t)-S_{\un}(\theta_0),S_{\deux}(t)-S_{\deux}(\theta_0))$ is continuous at $\theta_0$, with
$S_{\j}(y)=\sum_{\ell=1}^{N_{\j}}(\mathds{1}\{Y_{\ell,\j}\leq y\} -F(y))$. Let us consider $t>\theta_0$ (the proof is similar otherwise). We have
\begin{align}
	|K_i(t)| & \leq \V(S_{\un}(t)-S_{\un}(\theta_0)) \nonumber  \\
	& \leq \E\left[N_{\un}|S_{\un}(t)-S_{\un}(\theta_0)|\right] \nonumber \\
	& \leq \E\left[N_{\un}\left(\sum_{\ell=1}^{N_{\un}}\left(\mathds{1}\{Y_{\ell,\un}\leq t\} - \mathds{1}\{Y_{\ell,\un}\leq \theta_0\}\right)\right)\right] + \E(N^2_{\un})\left(F(t)-F(\theta_0)\right). \label{eq:ineq_quant1}
\end{align}
The first inequality follows by the Cauchy-Schwarz inequality and Assumption \ref{as:dgp}. The second follows from $\E(S_{\un}(t)-S_{\un}(\theta_0))=0$ and 
$$|S_{\un}(t)-S_{\un}(\theta_0)| \leq \sum_{\ell=1}^{N_{\un}}\left|\mathds{1}\{Y_{\ell,\un}\in (\theta_0, t]\}  - (F(t)-F(\theta_0)\right| \leq N_{\un}.$$ 
Finally, the third inequality is based on the triangle inequality. Moreover, letting $\alpha=1/(1+\zeta)$, we have, by H\"older's inequality,
\begin{align}
& \E\left[N_{\un}\left(\sum_{\ell=1}^{N_{\un}}\left(\mathds{1}\{Y_{\ell,\un}\leq t\} - \mathds{1}\{Y_{\ell,\un}\leq \theta_0\}\right)\right)\right] \nonumber \\
=& \E\left[N_{\un}^{1+\alpha} \left(\frac{1}{N_{\un}^{\alpha}}\sum_{\ell=1}^{N_{\un}}\left(\mathds{1}\{Y_{\ell,\un}\leq t\} - \mathds{1}\{Y_{\ell,\un}\leq \theta_0\}\right)\right)\right] \nonumber \\
\leq & \E\left[N_{\un}^{(1+\alpha)(1+\zeta)}\right]^{1/(1+\zeta)}  \E\left[\left(\frac{1}{N_{\un}^\alpha}\sum_{\ell=1}^{N_{\un}}\left(\mathds{1}\{Y_{\ell,\un}\leq t\} - \mathds{1}\{Y_{\ell,\un}\leq \theta_0\}\right)\right)^{1+1/\zeta}\right]^{\zeta/(1+\zeta)}. \label{eq:ineq_quant2}
\end{align}
Now, remark that $\E\left[N_{\un}^{(1+\alpha)(1+\zeta)}\right]=\E\left[N_{\un}^{2+\zeta}\right]<+\infty$ and
$$\left(\frac{1}{N_{\un}^\alpha}\sum_{\ell=1}^{N_{\un}}\left(\mathds{1}\{Y_{\ell,\un}\leq t\} - \mathds{1}\{Y_{\ell,\un}\leq \theta_0\}\right)\right)^{1/\zeta}\leq N_{\un}^{\frac{1-\alpha}{\zeta}} = N_{\un}^{\alpha}.$$
Hence,
\begin{align}
& \E\left[\left(\frac{1}{N_{\un}^\alpha}\sum_{\ell=1}^{N_{\un}}\left(\mathds{1}\{Y_{\ell,\un}\leq t\} - \mathds{1}\{Y_{\ell,\un}\leq \theta_0\}\right)\right)^{1+1/\zeta}\right] \nonumber \\
\leq & \E\left[\left(\sum_{\ell=1}^{N_{\un}}\left(\mathds{1}\{Y_{\ell,\un}\leq t\} - \mathds{1}\{Y_{\ell,\un}\leq \theta_0\}\right)\right)\right] \nonumber \\
%\leq & \E\left[\sum_{\ell=1}^{N_{\un}}\left(\mathds{1}\{\min(t,\theta_0)<Y_{\ell,\un}\leq \max(t,\theta_0)\} \right)\right] \nonumber \\
%= & \left(F(\max(t,\theta_0))-F(\min(t,\theta_0))\right) \nonumber\\
=& \E(N_{\un}) \left(F(t)-F(\theta_0)\right)\label{eq:ineq_quant3}
\end{align}
Combining \eqref{eq:ineq_quant1}-\eqref{eq:ineq_quant3} and continuity of $F$ at $\theta_0$ shows that $K_i$, and therefore $\rho$, is continuous at $\theta_0$. 

\medskip
Now fix $\eps>0$. By Lemma 18.15 of \cite{vanderVaart2000}, for almost all $\omega\in\Omega$, $\mathbb{G}(\omega)$ is uniformly continuous on any compact set $T$ equipped with the metric $\widetilde{\rho}(s,t)=\V\left[\mathbb{G}(t)-\mathbb{G}(\theta_0)\right]^{1/2}$. Then, picking $T$ such that $\theta_0$ belongs to its an interior point, there exists $\delta>0$ such that $\rho(t)=\widetilde{\rho}(t,\theta_0)<\delta$ implies $|\mathbb{G}(\omega)(t) - G(\omega)(\theta_0)|<\eps$. Moreover, by what precedes, there exists $\delta'>0$ such that $|t-\theta_0|<\delta'$ implies that $\rho(t)<\delta$. Hence, for this $\delta'$, $|\mathbb{G}(\omega)(t) - G(\omega)(\theta_0)|<\eps$. The result follows.

% subsection proof_that_mathbb_g_is_continuous_at_theta_0_in_the_quantile_example (end)

\newpage
\section{Technical lemmas}

\subsection{Lemma on Assumption \ref{as:vc}}

\begin{lem}\label{lem:csteF} Let $\mathcal{F}$ be an infinite class of functions from $\mathcal{Y}$ to $\mathbb{R}$.
	\begin{itemize}
	\item[i)] If Assumption \ref{as:measurability} holds for $\mathcal{F}$, it also holds for $\mathcal{F}\cup\{1\}$. 
	\item[ii)] If Assumption \ref{as:vc} holds for $\mathcal{F}$, it also holds for $\mathcal{F}\cup\{1\}$. 
	\end{itemize}
	Consequently, under Asumptions \ref{as:measurability} and \ref{as:vc}, if the class $\mathcal{F}$ is infinite, we can assume without loss of generality that $F$ is bounded below by 1.
	\end{lem}
\subsubsection{Proof of Lemma \ref{lem:csteF}}
Because the constant function $1$ is measurable Assumption \ref{as:measurability} holds for $\mathcal{F}'=\mathcal{F}\cup \{1\}$ if it holds for $\mathcal{F}$.

Assume that Assumption \ref{as:vc} holds for $\mathcal{F}$.
Let $F=\sup_{f\in \mathcal{F}}|f|$ and $F'=\max(F,1)$ then we have:$$ \int_0^{+\infty}\sup_Q\sqrt{\log N\left(\varepsilon||F'||_{Q,2},\mathcal{F}',||.||_{Q,2}\right)}d\varepsilon=\int_0^{2}\sup_Q\sqrt{\log N\left(\varepsilon||F'||_{Q,2},\mathcal{F}',||.||_{Q,2}\right)}d\varepsilon.$$ We can always complete any covering of $\mathcal{F}$ with a ball centered on $1$ and next $N\left(\eta,\mathcal{F}',||.||_{Q,2}\right)\leq 1+N\left(\eta,\mathcal{F},||.||_{Q,2}\right)$ for any $\eta>0$. It follows that \begin{align*}N\left(\varepsilon||F'||_{Q,2},\mathcal{F}',||.||_{Q,2}\right)&\leq 1+N\left(\varepsilon||F'||_{Q,2},\mathcal{F},||.||_{Q,2}\right)\\&\leq 2N\left(\varepsilon||F'||_{Q,2},\mathcal{F},||.||_{Q,2}\right)\text{ because }N\left(\varepsilon||F'||_{Q,2},\mathcal{F},||.||_{Q,2}\right)\geq 1 \text{ for }\varepsilon>0\\&\leq 2N\left(\varepsilon||F||_{Q,2},\mathcal{F},||.||_{Q,2}\right)\text{ because }||F||_{Q,2}\leq ||F'||_{Q,2}.
\end{align*}
Then we have \begin{align*}
 \int_0^{+\infty}\sup_Q\sqrt{\log N\left(\varepsilon||F||_{Q,2},\mathcal{F},||.||_{Q,2}\right)}d\varepsilon&= \int_0^{2}\sup_Q\sqrt{\log N\left(\varepsilon||F'||_{Q,2},\mathcal{F}',||.||_{Q,2}\right)}d\varepsilon\\
 &\leq \int_0^{2}\sup_Q\sqrt{\log 2 N\left(\varepsilon||F||_{Q,2},\mathcal{F},||.||_{Q,2}\right)}d\varepsilon\\
  &\leq 2\sqrt{\log(2)}+\int_0^{2}\sup_Q\sqrt{\log N\left(\varepsilon||F||_{Q,2},\mathcal{F},||.||_{Q,2}\right)}d\varepsilon\\
 &\leq 2\sqrt{\log(2)}+\int_0^{+\infty}\sup_Q\sqrt{\log N\left(\varepsilon||F||_{Q,2},\mathcal{F},||.||_{Q,2}\right)}d\varepsilon\\
 & <\infty.
 \end{align*}
  And next, if the integral condition holds for $\mathcal{F}$, this is also the case for $\mathcal{F}'$. The moment condition holds for $F'$ if and only if it holds for $F$ and $\E(N_{\un}^2)<\infty$.

\subsection{Lemma on H\'ajek projections}

In the following lemma, we consider the H\'ajek projection of statistics of random variables sampled according to the representation lemma (Lemma \ref{lem:representation}). For any $r=1,...,k$, we let $\mathcal{E}_r=\{\e\in \{0;1\}^k: \sum_{i=1}^k \e_i=r\}$ and $\mathcal{I}_{r}(\C)=\{\c=\j \odot \e: \e\in\mathcal{E}_r, \un \leq \j \leq \C \}$, with $\odot$ the Hadamard product on $\mathbb{R}^k$.

%A REPRENDRE AVEC $\prec$ ET $\preceq$...
\begin{lem}\label{lem:hajek}$\ \ $\\
Let $\left(N_{\j},\vec{Y}_{\j}\right)_{\j\geq\un}$ denote a family of random variables such that $$\left\{N_{\j},\vec{Y}_{\j}\right\}_{\j\geq\un}=\left\{\tau\left(\left(U_{\j\odot\e}\right)_{\zero\prec \e \preceq \un}\right)\right\}_{\j\geq\un},$$
for some measurable function $\tau$ and $\left(U_{\c}\right)_{\c \geq \zero}$ a family of mutually independent 
uniform random variables on $[0,1]$. Let $f$ be such that $\E\left[\left(\sum_{\ell=1}^{N_{\un}}f(Y_{\ell,\un})\right)^2\right]<\infty$, and assume that $\underline{C}\rightarrow \infty$ and for every $\e\in \mathcal{E}_r$, $\frac{\underline{C}^r}{\prod_{i: \e_i=1}C_i}\rightarrow \lambda_{\e}\geq 0$. Then $H_r f$, the H\'ajek projection of $\mathbb{G}_{C}f$ on the set of statistics of the form $\sum_{\c\in \mathcal{I}_r(\C)}g_{\c}(U_{\c})$ (with $g_{\c}(U_{\c})$ square integrable), satisfies
\begin{align*}
&H_r f=\sum_{\c \in \mathcal{I}_r(\C)}\frac{\sqrt{\underline{C}}}{\prod_{i: \c_i\neq 0}C_i} \left(\E\left(\sum_{\ell=1}^{N_{\c \vee \un}}f(Y_{\ell,\c \vee \un})\bigg|U_{\c}\right)-\E\left(\widetilde{f}(N_{\un},\vec{Y}_{\un})\right)\right),\\
	&\underline{C}^{(r-1)/2} H_r f  \convD\mathcal{N}\left(0,\sum_{\e\in \mathcal{E}_r} \lambda_{\e} \V\left(\E\left(\sum_{\ell=1}^{N_{\un}}f(Y_{\ell,\un})\bigg|U_{ \e}\right)\right)\right), \\
&\V\left(\underline{C}^{(r-1)/2}H_r f\right)  = \sum_{\e \in \mathcal{E}_r}\frac{\underline{C}^r}{\prod_{i: \e_i=1}C_i}\V\left(\E\left(\sum_{\ell=1}^{N_{\un}}f(Y_{\ell,\un})\bigg|U_{ \e}\right)\right),
\end{align*}
If $\left\{N_{\j},\vec{Y}_{\j}\right\}_{\j\geq\un}=\left\{\tau\left(\left(U_{\j\odot\e}\right)_{ \e\in \cup_{r=\underline{r}}^k \mathcal{E}_r}\right)\right\}_{\j\geq\un}$ for $\underline{r}\geq 1$, we also get for every $\e\in \mathcal{E}_{\underline{r}}$
$$\V\left(\E\left(\sum_{\ell=1}^{N_{\un}}f(Y_{\ell,\un})\bigg|U_{ \e}\right)\right)=\Cov\left(\sum_{\ell=1}^{N_{\un}}f(Y_{\ell,\un}), \sum_{\ell=1}^{N_{\deu-\e}}f(Y_{\ell,\deu-\e})\right).$$
\end{lem}

\subsubsection{Proof of Lemma \ref{lem:hajek}}

The H\'ajek projection $H_r f$ is characterized by
$$\E\left( \left(\mathbb{G}_C f-H_r f\right) \times \sum_{\c\in \mathcal{I}_r(\C)}g_{\c}(U_{\c})\right)=0 \text{ for any }(g_{\c})_{\c \in \mathcal{I}_r(\C)}\in \left(L^{2}([0;1])\right)^{|\mathcal{I}_r(\C)|}.$$
As a result, we have
$$\E\left(\mathbb{G}_C f|U_{\c}\right)=\E(H_r f|U_{\c}), \text{ for any }\c \in \mathcal{I}_r(\C).$$
Because the range of $H_r$ is a closed subspace of the space of square integrable random variables, $H_r f$ is equal to its H\'ajek projection:
$$H_r f=\sum_{\c \in \mathcal{I}_r(\C)}\E(H_r f|U_{\c}).$$
Next
$$H_r f=\sum_{\c \in \mathcal{I}_r(\C)}\E\left(\mathbb{G}_C f|U_{\c}\right).$$
Note that for any $\c \in \mathcal{I}_r(\C)$, $\c\wedge \un$ is the unique element $\e\in\mathcal{E}_r$ such that $\c=\j\odot \e$ for some $\j$ (note that $\j$ is not unique).
Moreover, for any $\c \in \mathcal{I}_r(\C)$ independence between the $U$'s ensures that $\sum_{\ell=1}^{N_{\j}}f(Y_{\ell,\j})\indep U_{\c}$ if $\j\odot \e \neq \c $. This implies
\begin{eqnarray*}
	\E\left(\mathbb{G}_C f|U_{\c}\right)&=&\frac{\sqrt{\underline{C}}}{\Pi_C}\sum_{\un \leq \j \leq \C}\E\left(\sum_{\ell=1}^{N_{\j}}f(Y_{\ell,\j})-\E\left(\widetilde{f}(N_{\un},\vec{Y}_{\un})\right)|U_{\c}\right)\\
&=&\frac{\sqrt{\underline{C}}}{\Pi_C}\sum_{\un \leq \j \leq \C}\mathds{1}\{\j\odot \e=\c\}\E\left(\sum_{\ell=1}^{N_{\j}}f(Y_{\ell,\j})-\E\left(\widetilde{f}(N_{\un},\vec{Y}_{\un})\right)|U_{\c}\right).
\end{eqnarray*}
The representation of $\left\{N_{\j},\vec{Y}_{\j}\right\}_{\j\geq\un}$ in terms of the $U$'s implies that 
$$\E\left(\sum_{\ell=1}^{N_{\j}}f(Y_{\ell,\j})-\E\left(\widetilde{f}(N_{\un},\vec{Y}_{\un})\right)|U_{\c}\right)=\E\left(\sum_{\ell=1}^{N_{\c \vee \un}}f(Y_{\ell,\c \vee \un})-\E\left(\widetilde{f}(N_{\un},\vec{Y}_{\un})\right)|U_{\c}\right)	
$$
for any $\j$ such that $\j\odot \e=\c$. Moreover,
\begin{eqnarray*}
	\E\left(\mathbb{G}_C f|U_{\c}\right)&=&\frac{\sqrt{\underline{C}}}{\Pi_C}\sum_{\un \leq \j \leq \C}\mathds{1}\{\j\odot \e=\c\}\E\left(\sum_{\ell=1}^{N_{\c \vee \un}}f(Y_{\ell,\c \vee \un})-\E\left(\widetilde{f}(N_{\un},\vec{Y}_{\un})\right)|U_{\c}\right)\\
&=&\frac{\sqrt{\underline{C}}\prod_{i: \c_i= 0}C_i}{\Pi_C}\E\left(\sum_{\ell=1}^{N_{\c \vee \un}}f(Y_{\ell,\c \vee \un})-\E\left(\widetilde{f}(N_{\un},\vec{Y}_{\un})\right)|U_{\c}\right)\\
&=&\frac{\sqrt{\underline{C}}}{\prod_{i: \c_i\neq 0}C_i} \left(\E\left(\sum_{\ell=1}^{N_{\c \vee \un}}f(Y_{\ell,\c \vee \un})|U_{\c}\right)-\E\left(\widetilde{f}(N_{\un},\vec{Y}_{\un})\right)\right).
\end{eqnarray*}
It follows that
\begin{align*}
&H_r f=\sum_{\c \in \mathcal{I}_r(\C)}\frac{\sqrt{\underline{C}}}{\prod_{i: \c_i\neq 0}C_i} \left(\E\left(\sum_{\ell=1}^{N_{\c \vee \un}}f(Y_{\ell,\c \vee \un})\bigg|U_{\c}\right)-\E\left(\widetilde{f}(N_{\un},\vec{Y}_{\un})\right)\right).
\end{align*}

Note that for any $\c \in \mathcal{I}_r(\C)$ and any $i=1,...,k$, we have $\c_i\neq0$ if and only if $(\c\wedge \un)_i =1$.  By rearrangin the terms in $H_r f$, we get
$$H_r f=\sum_{\e\in \mathcal{E}_r} \frac{\sqrt{\underline{C}}}{\prod_{i: \e_i=1}C_i} \sum_{ \c \in \mathcal{I}_r(\C): \c\wedge \un=\e} \left(\E\left(\sum_{\ell=1}^{N_{\c \vee \un}}f(Y_{\ell,\c \vee \un})\bigg|U_{\c}\right)-\E\left(\widetilde{f}(N_{\un},\vec{Y}_{\un})\right)\right).$$

By independence of the $U$'s, the random variables $\left\{\E\left(\sum_{\ell=1}^{N_{\c \vee \un}}f(Y_{\ell,\c \vee \un})|U_{\c}\right)-\E\left(\widetilde{f}(N_{\un},\vec{Y}_{\un})\right)\right\}$ are i.i.d. across $\c \in \mathcal{I}_r(\C)$ such that $\c \wedge \un=\e$. They are also centered with common variance 
\begin{align}\label{eq:lem_hajek_1}
\V\left(\E\left(\sum_{\ell=1}^{N_{\c \vee \un}}f(Y_{\ell,\c \vee \un})\bigg|U_{\c}\right)\right)&=\V\left(\E\left(\sum_{\ell=1}^{N_{\un}}f(Y_{\ell,\un})\bigg|U_{\c \wedge \un}\right)\right).
\end{align}
To prove this last equality, note that by the representation Lemma \ref{lem:representation}, there exists $h$ such that
\begin{align*}
\E\left(\sum_{\ell=1}^{N_{\c \vee \un}}f(Y_{\ell,\c \vee \un})|U_{\c}\right)&=\E\left(h\left((U_{(\c \vee \un)\odot \e'})_{\e'\in \cup_{r'=1}^k \mathcal{E}_{r'}}\right)|U_{\c}\right).
\end{align*}
Moreover, using the fact that the $U$'s are i.i.d, we get, for every $\c'\in\mathcal{I}_r(\C)$,
\begin{align*}
	\E\left(h\left((U_{(\c \vee \un)\odot \e'})_{\e'\in \cup_{r'=1}^k \mathcal{E}_{r'}}\right)|U_{\c}\right)&\overset{d}{=}\E\left(h\left((U_{(\c' \vee \un)\odot \e'})_{\e'\in \cup_{r'=1}^k \mathcal{E}_{r'}}\right)|U_{\c'}\right).
\end{align*}
Then, picking $\c'=\c\wedge\un$ and remarking that $(\c\wedge \un)\vee \un=\un$, we obtain
\begin{align*}
\E\left(\sum_{\ell=1}^{N_{\c \vee \un}}f(Y_{\ell,\c \vee \un})|U_{\c}\right)&=\E\left(\sum_{\ell=1}^{N_{\un}}f(Y_{\ell,\un})|U_{\c\wedge \un}\right),
\end{align*}
which is sufficient to prove that the equality in (\ref{eq:lem_hajek_1}) is true.

\medskip
Since for a given $\e \in \mathcal{E}_r$ we have $\#\{\c \in \mathcal{I}_r(\C): \c\wedge \un=\e\}=\prod_{i: \e_i=1}C_i$, the CLT ensures that
\begin{align*}
& \frac{1}{\left(\prod_{i: \e_i=1}C_i\right)^{1/2}} \sum_{ \c \in \mathcal{I}_r(\C): \c\wedge \un=\e} \left(\E\left(\sum_{\ell=1}^{N_{\c \vee \un}}f(Y_{\ell,\c \vee \un})\bigg|U_{\c}\right)-\E\left(\widetilde{f}(N_{\un},\vec{Y}_{\un})\right)\right) \\
\convD & \mathcal{N}\left(0,\V\left(\E\left(\sum_{\ell=1}^{N_{\un}}f(Y_{\ell,\un})\bigg|U_{\e}\right)\right)\right).	
\end{align*}
Moreover, since the families $(U_{\c})_{\c \in \mathcal{I}_r(\C):\c \wedge \un=\e}$ are mutually independent across $\e\in \mathcal{E}_r$, we have
\begin{align*} \underline{C}^{(r-1)/2} H_r f = & \sum_{e\in \mathcal{E}_r} \left(\frac{\underline{C}^r}{\prod_{i: \e_i=1}C_i}\right)^{1/2}\frac{1}{\left(\prod_{i: \e_i=1}C_i\right)^{1/2}} \sum_{ \c \in \mathcal{I}_r(\C): \c\wedge \un=\e} \left(\E\left(\sum_{\ell=1}^{N_{\c \vee \un}}f(Y_{\ell,\c \vee \un})\bigg|U_{\c}\right)-\E\left(\widetilde{f}(N_{\un},\vec{Y}_{\un})\right)\right)\\
	 \convD & \mathcal{N}\left(0,\sum_{\e\in \mathcal{E}_r} \lambda_{\e} \V\left(\E\left( \sum_{\ell=1}^{N_{\un}}f(Y_{\ell,\un})\bigg|U_{ \e}\right)\right)\right).
\end{align*}
Moreover, 
$$\V\left(\underline{C}^{(r-1)/2}H_r f\right)=\sum_{\e \in \mathcal{E}_r}\frac{\underline{C}^r}{\prod_{i: \e_i=1}C_i}\V\left(\E\left(\sum_{\ell=1}^{N_{\un}}f(Y_{\ell,\un})\bigg|U_{\e}\right)\right).$$

To get the last result of the lemma, we have to show that for $\e \in \mathcal{E}_{\underline{r}}$
$$\V\left(\E\left(\sum_{\ell=1}^{N_{\un}}f(Y_{\ell,\un})\bigg|U_{ \e}\right)\right)=\Cov\left(\sum_{\ell=1}^{N_{\un}}f(Y_{\ell,\un}), \sum_{\ell=1}^{N_{\deu-\e}}f(Y_{\ell,\deu-\e})\right).$$

As $\left\{N_{\j},(Y_{\ell,\j})_{\ell\geq 1}\right\}_{\j \geq \un}=\left\{\tau\left((U_{\j \odot \e})_{\e\in \cup_{r=\underline{r}}^{k}\mathcal{E}_r}\right)\right\}_{\j \geq \un}$ with i.i.d. $U$'s, we have
$\E\left(\sum_{\ell=1}^{N_{\un}}f(Y_{\ell,\un})\bigg|U_{\e}\right)=\E\left(\sum_{\ell=1}^{N_{\j}}f(Y_{\ell,\j})\bigg|U_{\e}\right)$ for any $\j$ such that $\j \odot \e=\un \odot \e=\e$. Because $(\deu-\e)\odot \e=\e$, we have $\V\left(\E\left(\sum_{\ell=1}^{N_{\un}}f(Y_{\ell,\un})\bigg|U_{\e}\right)\right)=\Cov\left(\E\left(\sum_{\ell=1}^{N_{\un}}f(Y_{\ell,\un})\bigg| U_{\e}\right), \E\left(\sum_{\ell=1}^{N_{\deu-\e}}f(Y_{\ell,\deu-\e})\bigg| U_{\e}\right)\right)$.

For any $\e \in \mathcal{E}_{\underline{r}}$, we have $\deu-\e \neq \un$, so that independence of the $U$'s ensures 
$$(U_{\un \odot \e'})_{\e'\in \cup_{r=\underline{r}}^{k}\mathcal{E}_r\setminus{\e}}\indep (U_{(\deu-\e) \odot \e'})_{\e'\in \cup_{r=\underline{r}}^{k}\mathcal{E}_r\setminus{\e}}|U_{\e}$$ and next $\sum_{\ell=1}^{N_{\un}}f(Y_{\ell,\un}) \indep \sum_{\ell=1}^{N_{\deu-\e}}f(Y_{\ell,\deu-\e}) | U_{\e}$. 

Hence, for $\e \in \mathcal{E}_{\underline{r}}$
$$\E\left(\Cov\left(\sum_{\ell=1}^{N_{\un}}f_1(Y_{\ell,\un}),\sum_{\ell=1}^{N_{\deu-\e}}f_2(Y_{\ell,\deu-\e})\bigg| U_{\e}\right)\right)=0.$$
By the law of total covariance, we ultimately have
$$\V\left(\E\left(\sum_{\ell=1}^{N_{\un}}f(Y_{\ell,\un})\bigg|U_{ \e}\right)\right)=\Cov\left(\sum_{\ell=1}^{N_{\un}}f(Y_{\ell,\un}), \sum_{\ell=1}^{N_{\deu-\e}}f(Y_{\ell,\deu-\e})\right).$$

\subsection{Symmetrization}
In this section, we extend the standard symmetrization lemma for empirical processes based on independent observations \citep[see for instance Lemma 2.3.1 in][]{vanderVaartWellner1996} to empirical processes built from separately exchangeable arrays of observations.

\begin{lem}\label{lem:sym}$\ \ $\\
	Let $\left(Z_{\j}\right)_{\j\geq\un}$ a family of random variables indexed by $\j\in (\mathbb{N}^{\ast})^k$ and with values in a Polish space, such that $$\left(Z_{\j}\right)_{\j\geq\un}\overset{a.s.}{=}\left(\tau\left(\left(U_{\j\odot\e}\right)_{\zero\prec \e \preceq \un}\right)\right)_{\j\geq\un}$$ for $(U_{\c})_{\c>\zero}$ a family of mutually independent, uniform random variables on $[0,1]$ and some measurable function $\tau$.
	%$\left(W_{\j}\right)_{\j\geq\un}$ is separately echangeable:
	%$$\{W_{\j}\}_{\j \geq \un}\overset{d}{=}\{W_{\pi(\j)}\}_{\j \geq \un},$$
%with $\pi(\j)=\pi(j_1,...,j_k)=(\pi_1(j_1),...,\pi_k(j_k))$ for any $(\pi_1,...,\pi_k)$ $k$-tuple of permutations of $\mathbb{N}^{(1)}$.
Let $\mathcal{G}$ a pointwise measurable class of integrable functions of $Z_{\un}$, and $\Phi$ a non-decreasing convex function from $\mathbb{R}^{+}$ to $\mathbb{R}$. We have
\begin{align*}&\E \left[ \Phi \left(\sup_{g \in \mathcal{G}}\left|\frac{1}{\Pi_C} \sum_{\un \leq \j \leq \C} g\left(Z_{\j}\right)-\E\left[g\left(Z_{\un}\right)\right]\right|\right)\right]\\
&~~~~\leq \frac{1}{2^k-1}\sum_{\zero\prec \e\preceq \un}\E \left[ \Phi \left(2(2^k-1)\sup_{g \in \mathcal{G}}\left|\frac{1}{\Pi_C} \sum_{\un \leq \j \leq \C}\epsilon_{\j \odot \e} g\left(Z_{\j}\right)\right|\right)\right],\end{align*}
with $(\epsilon_{\c})_{\c\geq \zero}$ a Rademacher process, independent of $\left(Z_{\j}\right)_{\j\geq\un}$.
\end{lem}

\subsubsection{Proof of Lemma \ref{lem:sym}}

 Let $(U^{(1)}_{\c})_{\c>\zero}$ an independent family of uniform-$(0,1)$, also independent of $(U_{\c})_{\c>\zero}$, and let $\left(Z_{\j}^{(1)}\right)_{\j\geq\un}=\left(\tau\left(\left(U_{\j\odot\e'}^{(1)}\right)_{\zero \prec \e'\preceq \un}\right)\right)_{\j\geq\un}$. 

We have $\E\left[g\left(Z_{\un}\right)\right]=\frac{1}{\Pi_C} \sum_{\un \leq \j \leq \C} \E\left[g\left(Z^{(1)}_{\j}\right)\right]=\frac{1}{\Pi_C} \sum_{\un \leq \j \leq \C} \E\left[g\left(Z^{(1)}_{\j}\right)\mid\left(Z_{\j'}\right)_{\j'\geq\un}\right]$. This plus the Jensen inequality applied repeatedly with the convex functions $|.|$, $\sup_{g \in \mathcal{G}}$ and $\Phi$ ensures
\begin{align*}		&\E \left[ \Phi \left(\sup_{g \in \mathcal{G}}\left|\frac{1}{\Pi_C} \sum_{\un \leq \j \leq \C} g\left(Z_{\j}\right)-\E\left[g\left(Z_{\un}\right)\right]\right|\right)\right]\leq
\E \left[ \Phi \left(\sup_{g \in \mathcal{G}}\left|\frac{1}{\Pi_C} \sum_{\un \leq \j \leq \C} g\left(Z_{\j}\right)-g\left(Z^{(1)}_{\j}\right)\right|\right)\right].
\end{align*}

For $\e$ such that $\zero\prec \e\preceq \un$, let $\left(Z_{\j}(\e)\right)_{\j\geq\un}=\left(\tau\left(\left(U_{\j\odot\e'}\right)_{0\prec \e'\preceq \e },\left(U_{\j\odot\e'}^{(1)}\right)_{\e \prec \e'\preceq \un}\right)\right)_{\j\geq\un}$ and $\left(Z_{\j}^{(1)}(\e)\right)_{\j\geq\un}=\left(\tau\left(\left(U_{\j\odot\e'}\right)_{0\prec \e'\prec \e },\left(U_{\j\odot\e'}^{(1)}\right)_{\e \preceq \e'\preceq \un}\right)\right)_{\j\geq\un}$. 

We have $\left(Z_{\j}(\un)\right)_{\j\geq\un}=\left(Z_{\j}\right)_{\j\geq\un}$. Moreover, if $s(\e)$ is the successor of $\e$ for the total order $\prec$, $\left(Z_{\j}^{(1)}(s(\zero))\right)_{\j\geq\un}=\left(Z_{\j}^{(1)}\right)_{\j\geq\un}$ and for $\zero \prec \e \prec \un$ $\left(Z_{\j}^{(1)}(s(\e))\right)_{\j\geq\un}=\left(Z_{\j}(\e)\right)_{\j\geq\un}$.

It follows that
\begin{align*}
&\mathbb{E}\left[\Phi\left(\sup_{g\in \mathcal{G}}\left|\frac{1}{\Pi_C}\sum_{\un\leq\j\leq\C}\left(g(Z_{\j})-g(Z^{(1)}_{\j})\right)\right|\right)\right] \\
&=\mathbb{E}\left[\Phi\left(\sup_{g \in \mathcal{G}}\left|\sum_{\zero \prec \e \preceq \un}\frac{1}{\Pi_C}\sum_{\un\leq\j\leq\C}\left(g(Z_{\j}(\e))-g(Z^{(1)}_{\j}(\e))\right)\right|\right)\right] \\
&\leq \mathbb{E}\left[\Phi\left(\sum_{\zero \prec \e \preceq \un} \sup_{g \in \mathcal{G}}\left|\frac{1}{\Pi_C}\sum_{\un\leq\j\leq\C}\left(g(Z_{\j}(\e))-g(Z^{(1)}_{\j}(\e))\right)\right|\right)\right] \\
&\leq\frac{1}{2^k-1}\sum_{\zero\prec \e \preceq \un}\mathbb{E}\left[\Phi\left((2^k-1)\sup_{g \in \mathcal{G}}\left|\frac{1}{\Pi_C}\sum_{\un\leq\j\leq\C}\left(g(Z_{\j}(\e))-g(Z^{(1)}_{\j}(\e))\right)\right|\right)\right].
\end{align*}

For any $\e$ such that $\zero\prec \e \preceq \un$, we have $\j=\j\odot\e+\j\odot(\un-\e)$, and observe that for any function $q$ on $(\mathbb{N}^{*})^k$
$$\sum_{\un\leq \j \leq \C}q(\j)=\sum_{\e\leq \c \leq \C\odot \e}\sum_{\un-\e\leq \c' \leq \C\odot (\un-\e)}q(\c+\c').$$

Then for any $\e$ such that $\zero\prec \e \preceq \un$
\begin{align*}
&\sum_{\un\leq\j\leq\C}\left(g(Z_{\j}(\e))-g(Z^{(1)}_{\j}(\e))\right) =\sum_{\e\leq\c\leq\C\odot\e}~~~~\sum_{(\un-\e)\leq\c'\leq\C\odot(\un-\e)}\left(g(Z_{\c+\c'}(\e))-g(Z^{(1)}_{\c+\c'}(\e))\right).
\end{align*}
 Because $(\c+\c')\odot \e=\c\odot\e=\c$ for any $(\c,\c')$ such that $\e\leq\c\leq\C\odot\e$ and $(\un-\e)\leq\c'\leq\C\odot(\un-\e)$, we have
\begin{align*}
&g(Z_{\c+\c'}(\e))=g\circ \tau\left(\left(U_{(\c+\c')\odot\e'}\right)_{\e'\prec \e}, U_{\c},\left(U_{(\c+\c')\odot\e'}^{(1)}\right)_{\e\prec \e'}\right) \\
&\text{and }g(Z^{(1)}_{\c+\c'}(\e))=g\circ \tau\left(\left(U_{(\c+\c')\odot\e'}\right)_{\e'\prec \e}, U^{(1)}_{\c},\left(U_{(\c+\c')\odot\e'}^{(1)}\right)_{\e\prec \e'}\right).
\end{align*}

Let $R_{\e}=\left(\left(U_{\j\odot\e'}\right)_{\e'\prec \e},\left(U_{\j\odot\e'}^{(1)}\right)_{\e\prec \e'}\right)_{\j\geq \un}$.
For any $\c$ such that $\e\leq \c \leq \C\odot \e$
$$\E\left(\sum_{(\un-\e)\leq\c'\leq\C\odot(\un-\e)}g(Z_{\c+\c'}(\e))-g(Z^{(1)}_{\c+\c'}(\e))\bigg| R_{\e}\right)=0,$$
and for any $\c^1,\c^2$ such that $\e\leq \c^1,\c^2 \leq \C\odot \e$
$$\left(\sum_{(\un-\e)\leq\c'\leq\C\odot(\un-\e)}g(Z_{\c^1+\c'}(\e))-g(Z^{(1)}_{\c^1+\c'}(\e))\right)\indep\left( \sum_{(\un-\e)\leq\c'\leq\C\odot(\un-\e)}g(Z_{\c^2+\c'}(\e))-g(Z^{(1)}_{\c^2+\c'}(\e))\right)\bigg| R_{\e}.$$

For any $\e$, $\#\{\c: \e\leq\c\leq\C\odot\e\}=\prod_{i:e_i=1}C_i$. By symmetry, for any $\epsilon_{\c}\in \{-1;1\}^{\prod_{i:e_i=1}C_i}$
\begin{align*}&\left(\sum_{(\un-\e)\leq\c'\leq\C\odot(\un-\e)}g(Z_{\c+\c'}(\e))-g(Z^{(1)}_{\c+\c'}(\e))\right)_{\e \leq \c \leq \C\odot \e}\bigg| R_{\e} 
\\&\overset{d}{=}\left(\epsilon_{\c} \sum_{(\un-\e)\leq\c'\leq\C\odot(\un-\e)}g(Z_{\c+\c'}(\e))-g(Z^{(1)}_{\c+\c'}(\e))\right)_{\e \leq \c \leq \C\odot \e}\bigg| R_{\e}\end{align*}

Then introducing $(\epsilon_{\j})_{\zero\leq \j}$ a  Rademacher process, independent of the $U$'s and the $U^{(1)}$'s, we get
\begin{align*}
&\mathbb{E}\left[\Phi\left((2^k-1)\sup_{g \in \mathcal{G}}\left|\frac{1}{\Pi_C}\sum_{\un\leq\j\leq\C}\left(g(Z_{\j}(\e))-g(Z^{(1)}_{\j}(\e))\right)\right|\right)\right]\\
&=\mathbb{E}\left[\Phi\left((2^k-1)\sup_{g \in \mathcal{G}}\left|\frac{1}{\Pi_C}\sum_{\e\leq\c\leq\C\odot \e}\sum_{\un-\e\leq\c'\leq\C \odot(\un-\e)}\left(g(Z_{\c+\c'}(\e))-g(Z^{(1)}_{\c+\c'}(\e))\right)\right|\right)\right]\\
&=\mathbb{E}\left[\E\left[\Phi\left((2^k-1)\sup_{g \in \mathcal{G}}\left|\frac{1}{\Pi_C}\sum_{\e\leq\c\leq\C\odot \e}\sum_{\un-\e\leq\c'\leq\C \odot(\un-\e)}\left(g(Z_{\c+\c'}(\e))-g(Z^{(1)}_{\c+\c'}(\e))\right)\right|\right)\bigg| (\epsilon_{\j})_{\un \leq \j \leq \C}, R_{\e}\right]\right]\\
&=\mathbb{E}\left[\mathbb{E}\left[\Phi\left((2^k-1)\sup_{g \in \mathcal{G}}\left|\frac{1}{\Pi_C}\sum_{\e\leq\c\leq\C\odot \e}\epsilon_{\c}\sum_{\un-\e\leq\c'\leq\C \odot(\un-\e)}\left(g(Z_{\c+\c'}(\e))-g(Z^{(1)}_{\c+\c'}(\e))\right)\right|\right)\bigg| (\epsilon_{\j})_{\un \leq \j \leq \C}, R_{\e}\right]\right]\\
&= \mathbb{E}\left[\Phi\left((2^k-1)\sup_{g \in \mathcal{G}}\left|\frac{1}{\Pi_C}\sum_{\un\leq\j\leq\C}\epsilon_{\j \odot \e}\left(g(Z_{\j}(\e))-g(Z^{(1)}_{\j}(\e))\right)\right|\right)\right].
\end{align*}
The triangle inequality and the convexity of $\Phi$ ensures
\begin{align*}
& \Phi\left((2^k-1)\sup_{g \in \mathcal{G}}\left|\frac{1}{\Pi_C}\sum_{\un\leq\j\leq\C}\epsilon_{\j \odot \e}\left(g(Z_{\j}(\e))-g(Z^{(1)}_{\j}(\e))\right)\right|\right)\\
&\leq \Phi\left((2^k-1)\sup_{g \in \mathcal{G}}\left|\frac{1}{\Pi_C}\sum_{\un\leq\j\leq\C}\epsilon_{\j \odot \e}\left(g(Z_{\j}(\e))\right)\right|+(2^k-1)\sup_{g \in \mathcal{G}}\left|\frac{1}{\Pi_C}\sum_{\un\leq\j\leq\C}\epsilon_{\j \odot \e}\left(g(Z^{(1)}_{\j}(\e))\right)\right|\right)\\
&\leq \frac{1}{2}\Phi\left(2(2^k-1)\sup_{g \in \mathcal{G}}\left|\frac{1}{\Pi_C}\sum_{\un\leq\j\leq\C}\epsilon_{\j \odot \e}\left(g(Z_{\j}(\e))\right)\right|\right)+\frac{1}{2}\Phi\left(2(2^k-1)\sup_{g \in \mathcal{G}}\left|\frac{1}{\Pi_C}\sum_{\un\leq\j\leq\C}\epsilon_{\j \odot \e}\left(g(Z^{(1)}_{\j}(\e))\right)\right|\right).
\end{align*}
Because $\sum_{\un\leq\j\leq\C}\epsilon_{\j \odot \e}\left(g(Z_{\j}(\e))\right)\overset{d}{=}\sum_{\un\leq\j\leq\C}\epsilon_{\j \odot \e}\left(g(Z^{(1)}_{\j}(\e))\right)\overset{d}{=}\sum_{\un\leq\j\leq\C}\epsilon_{\j \odot \e}\left(g(Z_{\j})\right)$, we conclude
\begin{align*}
& \E\left[\Phi\left((2^k-1)\sup_{g \in \mathcal{G}}\left|\frac{1}{\Pi_C}\sum_{\un\leq\j\leq\C}\left(g(Z_{\j}(\e))-g(Z^{(1)}_{\j}(\e))\right)\right|\right)\right]\\
&\leq \E\left[\Phi\left(2(2^k-1)\sup_{g \in \mathcal{G}}\left|\frac{1}{\Pi_C}\sum_{\un\leq\j\leq\C}\epsilon_{\j \odot \e}g(Z_{\j}(\e))\right|\right)\right]\\
&= \E\left[\Phi\left(2(2^k-1)\sup_{g \in \mathcal{G}}\left|\frac{1}{\Pi_C}\sum_{\un\leq\j\leq\C}\epsilon_{\j \odot \e}g(Z_{\j})\right|\right)\right].
\end{align*}

\subsection{Lemmas for uniform CLT and inference based on asymptotic normality}

%We state the lemmas we use in our proof, before showing them in turn. 

\begin{lem}\label{lemma:lemma_222_vdvw}
	Let $\{a_j\}_{j=1}^m$ be a sequence of $n$-dimensional Euclidean vectors and $\{\epsilon_i\}_{i=1}^n$ independent Rademacher random variables. For every $m\geq 1$,
    \begin{align*}
    	&\mathbb{E}\left[\max_{j\in\{1,...,m\}}\left|\sum_{i=1}^n\epsilon_i a_{ji}\right|\right]\leq \sqrt{2\log 2m}\max_{j\in\{1,...,m\}}\abs{a_j}.
    \end{align*}
\end{lem}

\begin{lem}\label{lemma:ch_norme_class} Let $\varepsilon>0$,
	\begin{enumerate}	
		\item[(i)] if $||\cdot||$ is a pseudo-norm on $\mathcal{G}$ and $\lambda>0$, then
		$$N(\varepsilon,\mathcal{G},\lambda||.||)= N(\varepsilon/\lambda,\mathcal{G},||.||).$$

	\item[(ii)] if $||.||_a$ and $||.||_b$ are two pseudo-norms on a class $\mathcal{G}$ of functions such that $||g||_a\leq ||g||_b$ for any $g\in \mathcal{G}$, then
	$$N(\varepsilon,\mathcal{G},||.||_a)\leq N(\varepsilon,\mathcal{G},||.||_b).$$

\item[(iii)] if $||\cdot||$ is a pseudo-norm on $\mathcal{G}'\supset \mathcal{G}$, then
	$$N(\varepsilon,\mathcal{G},||.||)\leq N(\varepsilon/2,\mathcal{G}',||.||).$$
	
\item[(iv)] if $||\cdot||$ is a pseudo-norm on $\mathcal{G}$ and $\mathcal{G}_{\infty}$, then
$$N(\varepsilon,\mathcal{G}_{\infty},||.||)\leq N^2(\varepsilon/2,\mathcal{G},||.||).$$
\end{enumerate}
\end{lem}

\begin{lem}\label{lemma:bound_empirical_entropy_dgg}
	For any $\varepsilon>0$, $\delta\in]0,+\infty]$ and $r\geq 1$ 
    \begin{enumerate}
    	\item[(i)] if $\overline{N}_r>0$, then $N\left(\varepsilon,\widetilde{\mathcal{F}},\left|\left|\cdot\right|\right|_{\mathbb{\mu}_C,r}\right)\leq N\left(\frac{\varepsilon}{\overline{N}_r^{\frac{1}{r}}},\mathcal{F},\left|\left|\cdot\right|\right|_{\mathbb{Q}_C^r,r}\right)$
    	
    	\item[(ii)] if $\overline{N}_r>0$, then $N\left(\varepsilon,\widetilde{\mathcal{F}_{\delta}},\left|\left|\cdot\right|\right|_{\mu_C,r}\right)\leq N^2\left(\frac{\varepsilon}{4\overline{N}_r^{\frac{1}{r}}},\mathcal{F},\left|\left|\cdot\right|\right|_{\mathbb{Q}_C^r,r}\right)$
        \item[(iii)] if $A_r>0$, then $N\left(\varepsilon,\widetilde{\mathcal{F}},\left|\left|\cdot\right|\right|_{\mathbb{\mu}_C,r}\right)\leq N\left(\frac{\varepsilon}{A_r^{\frac{1}{r}}},\mathcal{F},\left|\left|\cdot\right|\right|_{\infty,\beta}\right)$.
    
    \item[(iv)] if $A_r>0$, then $N\left(\varepsilon,\widetilde{\mathcal{F}_{\delta}},\left|\left|\cdot\right|\right|_{\mu_C,r}\right)\leq N^2\left(\frac{\varepsilon}{4A_r^{\frac{1}{r}}},\mathcal{F},\left|\left|\cdot\right|\right|_{\infty,\beta}\right).$

    \end{enumerate}
\end{lem}

\begin{lem}\label{lemma:bound_empirical_entropy_kato}
	For every $\varepsilon>0$
 \begin{align*}   &N\left(2\varepsilon\abs{\abs{\widetilde{F}^2}}_{\mu_C,1},\widetilde{\mathcal{F}_{\infty}}^2,\left|\left|\cdot\right|\right|_{\mu_C,1}\right)\leq N^2\left(\varepsilon\abs{\abs{\widetilde{F}} }_{\mathbb{\mu}_C,2},\widetilde{\mathcal{F}},\left|\left|\cdot\right|\right|_{\mu_C,2}\right).
\end{align*}

\end{lem}

\begin{lem}\label{lemma:donsker_entropy_bound}

Let $\e\in\cup_{i=1}^k\mathcal{E}_i$. and $(\epsilon_{\c})_{\c\geq \zero}$ a Rademacher process independent from $(N_{\j}, \vec{Y}_{\j})_{\j \geq \un}$. Under Assumptions~\ref{as:dgp}, \ref{as:measurability} and \ref{as:vc}
\begin{align*}
&\mathbb{E}\left[\sup_{\mathcal{F}_{\delta}}\left|\frac{1}{\Pi_C}\sum_{\un\leq\j\leq\C}\epsilon_{\j\odot\e}\sum_{\ell=1}^{N_{\j}}f(Y_{\ell,\j})\right|\right]\leq 4\sqrt{\frac{2\mathbb{E}\left[\sigma_C^2\right]\log 2}{\prod_{s:\e_s=1}C_s}}+32\sqrt{\frac{\mathbb{E}\left[A_F\right]}{\prod_{s:\e_s=1}C_s}}J_{2,\mathcal{F}}\left(\frac{1}{4}\sqrt{\frac{\mathbb{E}\left[\sigma_C^2\right]}{\E\left[A_F\right]}}\right),
\end{align*}
where $A_F:=N_{\un}\sum_{\ell=1}^{N_{\un}}F\left(Y_{\ell,\un}\right)^2$. 

Similarly, under Assumptions~\ref{as:dgp}, \ref{as:measurability} and \ref{as:smooth}'
\begin{align*}
&\mathbb{E}\left[\sup_{\mathcal{F}_{\delta}}\left|\frac{1}{\Pi_C}\sum_{\un\leq\j\leq\C}\epsilon_{\j\odot\e}\sum_{\ell=1}^{N_{\j}}f(Y_{\ell,\j})\right|\right] \\
\leq & 4\sqrt{\frac{2\mathbb{E}\left[\sigma_C^2\right]\log 2}{\prod_{s:\e_s=1}C_s}}+32\abs{\abs{F}}_{\infty,\beta}\times\sqrt{\frac{\mathbb{E}\left[A_{\beta}^2\right]}{\prod_{s:\e_s=1}C_s}}J_{\infty,\beta,\mathcal{F}}\left(\frac{1}{4\left|\left|F\right|\right|_{\infty,\beta}}\sqrt{\frac{\mathbb{E}\left[\sigma_C^2\right]}{\mathbb{E}\left[A_{\beta}^2\right]}}\right),
\end{align*}
where $A_{\beta}:=\sum_{\ell=1}^{N_{\un}}(1+|Y_{1,\un}|^2)^{-\beta/2}$.
\end{lem}

\begin{lem}\label{lemma:glivenko_squares}

	Let $M>0$, $\eta>0$ and $\e\in\cup_{i=1}^k\mathcal{E}_i$ and $(\epsilon_{\c})_{\c\geq \zero}$ a Rademacher process independent from $(N_{\j}, \vec{Y}_{\j})_{\j \geq \un}$. Under Assumptions~\ref{as:dgp}, \ref{as:measurability} and \ref{as:vc}
\begin{align*}
	&\mathbb{E}\left(\sup_{\mathcal{F}_{\infty}}\left|\text{ }\frac{1}{\Pi_C}\sum_{\un\leq\j\leq\C}\epsilon_{\j\odot\e}\left(\sum_{\ell=1}^{N_{\j}}f(Y_{\ell,\j})\right)^2\mathds{1}\left\{\sum_{\ell=1}^{N_{\j}}F(Y_{\ell,\j})\leq M\right\}\text{ }\right|\right) \\
    \leq & \frac{M}{\sqrt{\prod_{s:\e_s=1}C_s}}\times\left\{2\sqrt{2\log 2}+\frac{4}{\eta}J_{2,\mathcal{F}}(\infty)\right\}+2\eta \E\left(A_F\right).
\end{align*}
Similarly, under Assumptions~\ref{as:dgp}, \ref{as:measurability} and \ref{as:smooth}'
\begin{align*}
	&\mathbb{E}\left(\sup_{\mathcal{F}_{\infty}}\left|\text{ }\frac{1}{\Pi_C}\sum_{\un\leq\j\leq\C}\epsilon_{\j\odot\e}\left(\sum_{\ell=1}^{N_{\j}}f(Y_{\ell,\j})\right)^2\mathds{1}\left\{\sum_{\ell=1}^{N_{\j}}F(Y_{\ell,\j})\leq M\right\}\text{ }\right|\right) \\
    \leq & \frac{M}{\sqrt{\prod_{s:\e_s=1}C_s}}\times\left\{2\sqrt{2\log 2}+\frac{4}{\eta} J_{\infty,\beta,\mathcal{F}}(\infty)\right\}+4\eta\left|\left|F\right|\right|_{\infty,\beta}^2\mathbb{E}\left[A_{\beta}^2\right].
\end{align*}
\end{lem}

\begin{lem}\label{lemma:glivenko_prod_fin}
	Let $\mathcal{F}$ and $\mathcal{G}$ be two pointwise measurable classes of functions with respective envelope $F$ and $G$.\\ Under Assumption~\ref{as:dgp}, if $\E\left[\left(\sum_{\ell=1}^{N_{\un}}F(Y_{\ell,\un})\right)^2\right]\vee\E\left[\left(\sum_{\ell=1}^{N_{\un}}G(Y_{\ell,\un})\right)^2\right]<\infty$, then for any $i\in\{1,...,k\}$
	\begin{align*}%\label{eq:glivenko_prod_1}
	&\mathbb{E}\left[\sup_{\mathcal{F}\times\mathcal{G}}\left|\frac{1}{|\mathcal{B}_i|}\sum_{(\j,\j')\in\mathcal{B}_i}\sum_{\ell=1}^{N_{\j}}f(Y_{\ell,\j})\sum_{\ell=1}^{N_{\j'}}g(Y_{\ell,\j'})-\frac{1}{|\mathcal{A}_i|}\sum_{(\j,\j')\in\mathcal{A}_i}\sum_{\ell=1}^{N_{\j}}f(Y_{\ell,\j})\sum_{\ell=1}^{N_{\j'}}g(Y_{\ell,\j'})\right|\right]=o(1)\\
	\end{align*}
and for any $\e\in \mathcal{E}_r$ with $r\geq 1$
\begin{align*}
\left(\Pi_C\right)^{-2}\E\left(\sup_{\mathcal{F}\times \mathcal{G}}\left|\sum_{(\j,\j')\in \mathcal{B}_{\e}}\sum_{\ell=1}^{N_{\j}}f(Y_{\ell,\j})\sum_{\ell'=1}^{N_{\j'}}g(Y_{\ell',\j'})\right|\right)=O(\underline{C}^{-r}).
\end{align*}	

\end{lem}

\begin{lem}\label{lemma:glivenko_prod_fin2}
	Let $\mathcal{F}$ and $\mathcal{G}$ be two classes of functions that satisfy Assumptions~\ref{as:dgp}, \ref{as:measurability} and either Assumptions~\ref{as:vc} or \ref{as:smooth}'. 
	Then for every $i\in\{1,...,k\}$
\begin{align*}
&\lim_{\underline{C}\to +\infty}\mathbb{E}\left[\sup_{\mathcal{F}\times\mathcal{G}}\left|\frac{1}{|\mathcal{B}_i|}\sum_{(\j,\j')\in\mathcal{B}_i}\sum_{\ell=1}^{N_{\j}}f(Y_{\ell,\j})\sum_{\ell=1}^{N_{\j'}}g(Y_{\ell,\j'})-\mathbb{E}\left[\sum_{\ell=1}^{N_{\un}}f(Y_{\ell,\un})\sum_{\ell=1}^{N_{\deux}}g(Y_{\ell,\deux})\right]\right|\right]=0,
\end{align*}
and
\begin{align*}
&\lim_{\underline{C}\to +\infty}\mathbb{E}\left[\sup_{\mathcal{F}\times\mathcal{G}}\left|\frac{1}{|\mathcal{A}_i|}\sum_{(\j,\j')\in\mathcal{A}_i}\sum_{\ell=1}^{N_{\j}}f(Y_{\ell,\j})\sum_{\ell=1}^{N_{\j'}}g(Y_{\ell,\j'})-\mathbb{E}\left[\sum_{\ell=1}^{N_{\un}}f(Y_{\ell,\un})\sum_{\ell=1}^{N_{\deux}}g(Y_{\ell,\deux})\right]\right|\right]=0.
\end{align*}
\end{lem}

\begin{lem}\label{lemma:glivenko}
Let $\mathcal{F}$ a class of functions that fulfills Assumption \ref{as:measurability} and such that\\
$i)$ $\mathcal{F}$ admits an envelop function $F$ with $\E\left[\sum_{\ell=1}^{N_{\un}}F(Y_{\ell,\un})\right]<\infty$, $\sup_{Q}\log N(\eta ||.||_{Q,1},\mathcal{F}, ||.||_{Q,1})<\infty$ for any $\eta>0$, \\
or\\
$ii)$ $\mathcal{F}$ admits an envelop function $F$ and there exists for some $\beta$ such that $\abs{\abs{F}}_{\infty,\beta}<+\infty$,
$\mathbb{E}\left[\left|\sum_{\ell=1}^{N_{\un}}(1+|Y_{\ell,\un}|^2)^{-\beta/2}\right|\right] <+\infty$ and $N(\eta||F||_{\infty,\beta},\mathcal{F},||.||_{\infty,\beta})<\infty$ for any $\eta>0$.\\
Then under Assumption \ref{as:dgp}, we have
$$\E\left[\sup_{\mathcal{F}}\left|\frac{1}{\Pi_C}\sum_{\un\leq \j \leq \C}\sum_{\ell=1}^{N_{\j}}f(Y_{\ell,\j})-\E\left[\sum_{\ell=1}^{N_{\un}}f(Y_{\ell,\un})\right]\right|\right]=o(1).$$ 
\end{lem}

\subsubsection{Proof of Lemma~\ref{lemma:lemma_222_vdvw}}
A random variable $V$ is sub-Gaussian with parameter $\sigma>0$ if for any $\lambda \in \mathbb{R}$, $\E\left(e^{\lambda V}\right)\leq e^{\lambda^2 \sigma^2/2}$. A Rademacher variable is sub-Gaussian of parameter 1. By independence of the $\epsilon_i$, it follows that $\sum_{i=1}^n \epsilon_i a_{ji}$ is sub-Gaussian of parameter $|a_j|$ for any $j$. Lemma 2.3.4 in \cite{GineNickl2015} ensures the result. 

\subsubsection{Proof of Lemma~\ref{lemma:ch_norme_class}}
(i): A ball of radius $\varepsilon/\lambda$ for $||.||$ is also a ball of radius $\varepsilon$ for $\lambda||.||$.\\
(ii): Take a collection of closed balls $(B^b_{n})_{n=1,...,N}$ for $||.||_b$, with radius $\varepsilon$ and centers in $\mathcal{G}$, that covers $\mathcal{G}$. The balls $B^a_{n}, n=1,...,N$ with the same centers and same radius for the norm $||.||_a$ are such that $B^b_n\subset B^a_n$. We conclude that 	$N(\varepsilon,\mathcal{G},||.||_a)\leq N(\varepsilon,\mathcal{G},||.||_b)$.\\
(iii): Take a collection of closed balls $(B'_{n})_{n=1,...,N}$ for $||.||_b$, with radius $\varepsilon/2$ and centers in $\mathcal{G}'$, that covers $\mathcal{G}'$. %They also cover $\mathcal{G}$ but we are not ensured that their centers are also in $\mathcal{G}$. 
For all balls $B'_{n}$ with non-empty intersection with $\mathcal{G}$, select an element of $\mathcal{G}\cap B'_{n}$ as a center of $B_{n}$, a new ball of radius $\varepsilon$. Then $B'_{n}\subset B_n$, and since $\mathcal{G}\subset \mathcal{G}'$, the collection of such balls $B_n$ covers $\mathcal{G}$ and balls $B_n$ have centers in $\mathcal{G}$.\\
(iv): Take a finite collection of closed balls $B_n$, $n=1,...,N$ for $||.||$, with radius $\varepsilon/2$ and centers $c_n$ in $\mathcal{G}$ that covers $\mathcal{G}$. Then the collection of 
$\Delta_{n,n'}=\{f-g, (f,g)\in B_n\times B_{n'}\}$, for $n,n'=1,...,N$, covers $\mathcal{G}_{\infty}$. Moreover any $\Delta_{n,n'}$ is included in a ball of $\mathcal{G}_{\infty}$ centered at $c_n-c_{n'}\in \mathcal{G}_{\infty}$ of radius $\varepsilon$.

\subsubsection{Proof of Lemma~\ref{lemma:bound_empirical_entropy_dgg}}
(i): For any $f\in \mathcal{F}$
\begin{align*}%\label{eq:emp_entropy_eq_1}
\left|\left|\widetilde{f}\right|\right|_{\mu_C,r}^r&=\frac{1}{\Pi_C}\sum_{\un \leq \j \leq C} \left|\sum_{\ell=1}^{N_{\j}}f(Y_{\ell,\j})\right|^r\leq\frac{1}{\Pi_C}\sum_{\un \leq \j \leq C}N_{\j}^{r-1} \sum_{\ell=1}^{N_{\j}}\left|f(Y_{\ell,\j})\right|^r\leq \overline{N}_r \left|\left| f\right|\right|_{\mathbb{Q}^r_C,r}^r.
\end{align*}
It follows that $||\widetilde{.} ||_{\mu_C,r}\leq \overline{N}_r^{1/r} ||.||_{\mathbb{Q}^r_C,r}$. Lemma \ref{lemma:ch_norme_class} $i)$ and $ii)$ ensures that
$$N\left(\varepsilon,\mathcal{F},||\widetilde{.} ||_{\mu_C,r}\right)\leq N\left(\frac{\varepsilon}{\overline{N}_r^{\frac{1}{r}}},\mathcal{F},\left|\left|\cdot\right|\right|_{\mathbb{Q}_C^r,r}\right).$$
Because $N(\varepsilon,\mathcal{F},||\widetilde{.}||_{\mu_C,r})=N(\varepsilon,\widetilde{\mathcal{F}},|| . ||_{\mu_C,r})$, we have $$N\left(\varepsilon,\widetilde{\mathcal{F}},\left|\left|\cdot\right|\right|_{\mathbb{\mu}_C,r}\right)\leq N\left(\frac{\varepsilon}{\overline{N}_r^{\frac{1}{r}}},\mathcal{F},\left|\left|\cdot\right|\right|_{\mathbb{Q}_C^r,r}\right).$$
(ii): Because $\widetilde{\mathcal{F}_{\delta}}=[\widetilde{\mathcal{F}}]_{\delta}\subset [\widetilde{\mathcal{F}}]_{\infty}=\widetilde{\mathcal{F}_{\infty}}$, Lemma.\ref{lemma:ch_norme_class} $iii)$, $iv)$ and Lemma \ref{lemma:bound_empirical_entropy_dgg} $i)$ ensure that
\begin{align*}N\left(\varepsilon,\widetilde{\mathcal{F}_{\delta}},||.||_{\mu_C,r}\right)&\leq N\left(\frac{\varepsilon}{2},\widetilde{\mathcal{F}}_{\infty},||.||_{\mu_C,r}\right)\\
&\leq  N^2\left(\frac{\varepsilon}{4},\widetilde{\mathcal{F}},||.||_{\mu_C,r}\right)\\&\leq N^2\left(\frac{\varepsilon}{4\overline{N}_r^{\frac{1}{r}}},\mathcal{F},\left|\left|\cdot\right|\right|_{\mathbb{Q}_C^r,r}\right).\end{align*}
The proofs of $iii)$ and $iv)$ follow the same line as the proof of $i)$ and $ii)$ after having noted that $||\widetilde{.}||_{\mu_C,r}\leq A_r^{1/r}||.||_{\infty,\beta}$.

\subsubsection{Proof of Lemma~\ref{lemma:bound_empirical_entropy_kato}}

For all $f\in \mathcal{F}$, we have $\left|\left|\widetilde{f}\right|\right|^2_{\mu_C,2}=\left|\left|\widetilde{f}^2\right|\right|_{\mu_C,1}$ and $\left|\left|\widetilde{f}\right|\right|_{\mu_C,2}/\abs{\abs{\widetilde{F}}}_{\mu_C,2}\leq 1$, which implies $\left|\left|\widetilde{f}^2\right|\right|_{\mu_C,1} / \abs{\abs{\widetilde{F}^2}}_{\mu_C,1} = \left|\left|\widetilde{f}\right|\right|^2_{\mu_C,2}/\abs{\abs{\widetilde{F}}}^2_{\mu_C,2} \leq \left|\left|\widetilde{f}\right|\right|_{\mu_C,2}/\abs{\abs{\widetilde{F}}}_{\mu_C,2}$. Then applying successively Lemma \ref{lemma:ch_norme_class} iv), i), ii) and i) again, we get
    \begin{align*}   	N\left(2\varepsilon\abs{\abs{\widetilde{F}^2}}_{\mu_C,1},\widetilde{\mathcal{F}_{\infty}}^2,\left|\left|\cdot\right|\right|_{\mu_C,1}\right)&=N\left(2\varepsilon\abs{\abs{\widetilde{F}^2}}_{\mu_C,1},\mathcal{F}_{\infty},\left|\left|  \widetilde{\cdot}^2\right|\right|_{\mu_C,1}\right)\\
    &\leq N^2\left(\varepsilon\abs{\abs{\widetilde{F}^2}}_{\mu_C,1},\mathcal{F},\left|\left|\widetilde{\cdot}^2\right|\right|_{\mu_C,1}\right)\\
    &=N^2\left(\varepsilon,\mathcal{F},\frac{\left|\left|\widetilde{\cdot}^2 \right|\right|_{\mu_C,1}}{\abs{\abs{\widetilde{F}^2}}_{\mu_C,1}}\right)\\
    &\leq N^2\left(\varepsilon,\mathcal{F},\frac{\left|\left|\widetilde{\cdot}\right|\right|_{\mu_C,2}}{\abs{\abs{\widetilde{F}}}_{\mu_C,2}}\right)\\
    &= N^2\left(\varepsilon\abs{\abs{\widetilde{F}}}_{\mathbb{\mu}_C,2},\widetilde{\mathcal{F}},\left|\left|\cdot\right|\right|_{\mu_C,2}\right).
\end{align*}

\subsubsection{Proof of Lemma~\ref{lemma:donsker_entropy_bound}}

A weighted sum of Rademacher variables is sub-Gaussian with respect to the Euclidean norm $|.|$ of the vectors of weights. So, as a process indexed by the vector of weights. this is a sub-Gaussian process for the Euclidean norm of the weights. Then, conditional on the original data, we can apply Theorem 2.3.6 in \cite{GineNickl2015}. It follows that we have for any $\e\in\cup_{i=1}^k\mathcal{E}_i$  
	\begin{align*}
	&\mathbb{E}\left[\sup_{\mathcal{F}_{\delta}}\left|\frac{1}{\prod_{s:\e_s=1}C_s}\sum_{\e\leq\c\leq\C\odot\e}\epsilon_{\c}\frac{1}{\prod_{s:\e_s=0}C_s}\sum_{(\un-\e)\leq\c'\leq\C\odot(\un-\e)}\sum_{\ell=1}^{N_{\c+\c'}}f(Y_{\ell,\c+\c'})\right|\text{ }\mid \bm{Z}\right] \\
	&\leq \frac{4\sqrt{2}}{\sqrt{\prod_{s:\e_s=1}C_s}}\int_0^{\sigma_C^{\e}}\sqrt{\log 2N\left(\varepsilon,\widetilde{\mathcal{F}_{\delta}},\left|\left|\cdot\right|\right|_{\e,2}\right)}d\varepsilon
	\end{align*}
The Jensen inequality ensures $||.||_{\e,2}\leq ||.||_{\mu_C,2}$ and $\sigma_C^{\e}\leq \sigma_C$. Linearity of the integration and Lemma \ref{lemma:ch_norme_class} $ii)$ ensure
	\begin{align}\label{eq:donsker_thm_eq_1}
&\mathbb{E}\left[\sup_{\mathcal{F}_{\delta}}\left|\frac{1}{\prod_{s:\e_s=1}C_s}\sum_{\e\leq\c\leq\C\odot\e}\epsilon_{\c}\frac{1}{\prod_{s:\e_s=0}C_s}\sum_{(\un-\e)\leq\c'\leq\C\odot(\un-\e)}\sum_{\ell=1}^{N_{\c+\c'}}f(Y_{\ell,\c+\c'})\right|\text{ }\mid \bm{Z}\right]\nonumber \\
&\leq \frac{4\sqrt{2}}{\sqrt{\prod_{s:\e_s=1}C_s}}\int_0^{\sigma_C}\sqrt{\log 2N\left(\varepsilon,\widetilde{\mathcal{F}_{\delta}},\left|\left|\cdot\right|\right|_{\mu_C,2}\right)}d\varepsilon \nonumber\\
    &= \frac{4\sqrt{2 \log(2)} \sigma_{C}}{\sqrt{\prod_{s:\e_s=1}C_s}}+\frac{4\sqrt{2}}{\sqrt{\prod_{s:\e_s=1}C_s}}\int_0^{\sigma_C}\sqrt{\log N\left(\varepsilon,\widetilde{\mathcal{F}_{\delta}},\left|\left|\cdot\right|\right|_{\mu_C,2}\right)}d\varepsilon.
\end{align}

$\bullet$ We first consider that $\mathcal{F}$ fulfills Assumption \ref{as:vc}.
If $\overline{N}_{2}>0$, Lemma \ref{lemma:bound_empirical_entropy_dgg} $ii)$ ensures

\begin{align*}
\int_0^{\sigma_C}\sqrt{\log N\left(\varepsilon,\widetilde{\mathcal{F}_{\delta}},\left|\left|\cdot\right|\right|_{\mu_C,2}\right)}d\varepsilon&\leq\sqrt{2}\int_0^{\sigma_C}\sqrt{\log N\left(\frac{\varepsilon}{4\sqrt{\overline{N}_2}},\mathcal{F},\left|\left|\cdot\right|\right|_{\mathbb{Q}_C^2,2}\right)}d\varepsilon 		\end{align*}
Lemma \ref{lem:csteF} implies that we can assume $\left|\left|F\right|\right|_{\mathbb{Q}_C^2,2}\geq 1$ without loss of generality. Next we apply the change of variable $\varepsilon'=\frac{\varepsilon}{4\sqrt{\overline{N}_2}\left|\left|F\right|\right|_{\mathbb{Q}^2_C,2}}$ to get
\begin{align*}
\int_0^{\sigma_C}\sqrt{\log N\left(\varepsilon,\widetilde{\mathcal{F}_{\delta}},\left|\left|\cdot\right|\right|_{\mu_C,2}\right)}d\varepsilon&\leq 4\sqrt{2}\sqrt{\overline{N}_2}\left|\left|F\right|\right|_{\mathbb{Q}_C^2,2} J_{2,\mathcal{F}}\left(\frac{\sigma_C}{4\sqrt{\overline{N}_2}\left|\left|F\right|\right|_{\mathbb{Q}_C^2,2}}\right).
\end{align*}
Because $u\in ]0,+\infty[\mapsto J_{2,\mathcal{F}}(u/4)$ is an increasing and concave function,
$(x,y)\in ]0,+\infty[^2\mapsto \sqrt{y}J_{2,\mathcal{F}}\left(\frac{\sqrt{x}}{4\sqrt{y}}\right)$ is concave and it follows from the Jensen inequality that
\begin{align*}
\E\left(\sqrt{\overline{N}_2}\left|\left|F\right|\right|_{\mathbb{Q}_C^2,2} J_{2,\mathcal{F}}\left(\frac{\sigma_C}{4\sqrt{\overline{N}_2}\left|\left|F\right|\right|_{\mathbb{Q}_C^2,2}}\right)\bigg| \overline{N}_2>0\right)&\leq \sqrt{\E\left(\overline{N}_2\left|\left|F\right|\right|^2_{\mathbb{Q}_C^2,2}|\overline{N}_2>0\right)}\\
&~~~~~~~J_{2,\mathcal{F}}\left(\frac{1}{4}\sqrt{\frac{\E(\sigma^2_C|\overline{N}_2>0)}{\E(\overline{N}_2\left|\left|F\right|\right|^2_{\mathbb{Q}_C^2,2}|\overline{N}_2>0)}}\right).
\end{align*}

Since all the random variables in the expectations of the previous inequality are null when $\overline{N}_2=0$, we have
\begin{align*}
\E\left(\sqrt{\overline{N}_2}\left|\left|F\right|\right|_{\mathbb{Q}_C^2,2} J_{2,\mathcal{F}}\left(\frac{\sigma_C}{4\sqrt{\overline{N}_2}\left|\left|F\right|\right|_{\mathbb{Q}_C^2,2}}\right)\right)&\leq\sqrt{\mathbb{P}(\overline{N}_2>0)} \sqrt{\E\left(\overline{N}_2\left|\left|F\right|\right|^2_{\mathbb{Q}_C^2,2}\right)}\\
&~~~~~~~J_{2,\mathcal{F}}\left(\frac{1}{4}\sqrt{\frac{\E(\sigma^2_C)}{\E(\overline{N}_2\left|\left|F\right|\right|^2_{\mathbb{Q}_C^2,2})}}\right)\\
&\leq \sqrt{\E\left(A_F\right)}J_{2,\mathcal{F}}\left(\frac{1}{4}\sqrt{\frac{\E(\sigma^2_C)}{\E(A_F)}}\right),
\end{align*}
because $\E(\overline{N}_2\left|\left|F\right|\right|^2_{\mathbb{Q}_C^2,2})=\E(A_F)$.
This implies that
\begin{align*}
&\mathbb{E}\left[\sup_{\mathcal{F}_{\delta}}\left|\frac{1}{\Pi_C}\sum_{\un\leq\j\leq\C}\epsilon_{\j\odot\e}\sum_{\ell=1}^{N_{\j}}f(Y_{\ell,\j})\right|\right]\leq 4\sqrt{\frac{2\mathbb{E}\left[\sigma_C^2\right]\log 2}{\prod_{s:\e_s=1}C_s}}+32\sqrt{\frac{\mathbb{E}\left[A_F\right]}{\prod_{s:\e_s=1}C_s}}J_{2,\mathcal{F}}\left(\frac{1}{4}\sqrt{\frac{\mathbb{E}\left[\sigma_C^2\right]}{\E\left[A_F\right]}}\right).
\end{align*}

$\bullet$ We now consider that $\mathcal{F}$ fulfills Assumption \ref{as:smooth}'.
Let $A_2=\frac{1}{\Pi_C}\sum_{\un\leq\j\leq\C}\left(\sum_{\ell=1}^{N_{\j}}(1+|Y_{\ell,\j}|^2)^{-\beta/2}\right)^2$. Under Assumption \ref{as:smooth}', we deduce from Lemma~\ref{lemma:bound_empirical_entropy_dgg} $iv)$ that
\begin{align*}
	\int_0^{\sigma_C}\sqrt{\log N\left(\varepsilon,\widetilde{\mathcal{F}_{\delta}},\left|\left|\cdot\right|\right|_{\mu_C,2}\right)}d\varepsilon     &\leq\sqrt{2}\int_0^{\sigma_C}\sqrt{\log N\left(\frac{\varepsilon}{4\sqrt{A_2}},\mathcal{F},\abs{\abs{\cdot}}_{\infty,\beta}\right)}d\varepsilon \\
    &\leq 4 \sqrt{2}\abs{\abs{F}}_{\infty,\beta}\sqrt{A_2}J_{\infty,\beta,\mathcal{F}}\left(\frac{1}{4 \abs{\abs{F}}_{\infty,\beta}}\sqrt{\frac{\sigma_C^2}{A_2}}\right).	
\end{align*}

Integration of Inequality (\ref{eq:donsker_thm_eq_1}) over $\left(N_{\j},\vec{Y}_{\j}\right)_{\j\geq \un}$, combined with Jensen inequality ensures
\begin{align*}
&\mathbb{E}\left[\sup_{\mathcal{F}_{\delta}}\left|\frac{1}{\Pi_C}\sum_{\un\leq\j\leq\C}\epsilon_{\j\odot\e}\sum_{\ell=1}^{N_{\j}}f(Y_{\ell,\j})\right|\right] \\
&\leq 4\sqrt{\frac{2\mathbb{E}\left[\sigma_C^2\right]\log 2}{\prod_{s:\e_s=1}C_s}}+32\abs{\abs{F}}_{\infty,\beta}\times\sqrt{\frac{\mathbb{E}\left[A_{\beta}^2\right]}{\prod_{s:\e_s=1}C_s}}J_{\infty,\beta,\mathcal{F}}\left(\frac{1}{4\left|\left|F\right|\right|_{\infty,\beta}}\sqrt{\frac{\mathbb{E}\left[\sigma_C^2\right]}{\mathbb{E}\left[A_{\beta}^2\right]}}\right).
\end{align*}

\subsubsection{Proof of Lemma~\ref{lemma:glivenko_squares}}

Let $\mathds{1}_M(\j):=\mathds{1}\left\{\sum_{\ell=1}^{N_{\j}}F(Y_{\ell,\j})\leq M\right\}$ and $\left(\widetilde{\mathcal{F}_{\infty}}^2\right)\mathds{1}_M:=\left\{\widetilde{f}^2\mathds{1}\left\{
\widetilde{F}\leq M\right\}:f\in\mathcal{F}_{\infty}\right\}$.

Let $m=N\left(\eta_1,\left(\widetilde{\mathcal{F}_{\infty}}^2\right)\mathds{1}_M,\left|\left|\cdot\right|\right|_{\e,1}\right)$ balls that cover the class $\left(\widetilde{\mathcal{F}_{\infty}}^2\right)\mathds{1}_M$. In each ball $B_{\stackrel{\circ}{m}}$ ($\stackrel{\circ}{m}=1,...,m$) of the covering, we can select its center $f^*_{\stackrel{\circ}{m}}\in \left(\widetilde{\mathcal{F}_{\infty}}^2\right)\mathds{1}_M$. The triangle inequality implies
\begin{align*}
&\mathbb{E}\left[\sup_{\mathcal{F}_{\infty}}\left|\frac{1}{\Pi_C}\sum_{\e\leq\c\leq\C\odot\e}\epsilon_{\c}\sum_{\un-\e\leq \c'\leq \C\odot(\un-\e)}\left(\sum_{\ell=1}^{N_{\c+\c'}}f(Y_{\ell,\c+\c'})\right)^2\mathds{1}_M(\c+\c')\right|\text{ }\bigg| \bm{Z}\right] \\
&\leq \mathbb{E}\left[\sup_{\stackrel{\circ}{m}=1,...,m}\left|\frac{1}{\Pi_C}\sum_{\e\leq\c\leq\C\odot\e}\epsilon_{\c}\sum_{\un-\e\leq \c'\leq \C\odot(\un-\e)}f^{\ast}_{\stackrel{\circ}{m}}(N_{\c+\c'},\vec{Y}_{\c+\c'})\right|\text{ }\bigg| \bm{Z}\right]+\eta_1.
\end{align*}

As the Euclidean norm of $\left(\frac{1}{\prod_{s:\e_s=0}C_s}\sum_{\un-\e\leq \c'\leq \C\odot(\un-\e)}f^{\ast}_{\stackrel{\circ}{m}}(N_{\c+\c'},\vec{Y}_{\c+\c'})\right)_{\e\preceq \c \preceq \C \odot \e}$ is bounded by $2\sqrt{\prod_{s:\e_s=1}C_s} \times M$ for any $m$, Lemma~\ref{lemma:lemma_222_vdvw} ensures for any $\eta_1>0$
\begin{align*}
&\mathbb{E}\left[\sup_{\mathcal{F}_{\infty}}\left|\frac{1}{\Pi_C}\sum_{\un\odot\e\leq\c\leq\C\odot\e}\epsilon_{\c}\sum_{\un-\e\leq \c'\leq \C\odot(\un-\e)}\left(\sum_{\ell=1}^{N_{\c+\c'}}f(Y_{\ell,\c+\c'})\right)^2\mathds{1}_M(\c+\c')\right|\text{ }\bigg| \bm{Z}\right] \\
&\leq 2\sqrt{2\log 2N\left(\eta_1,\left(\widetilde{\mathcal{F}_{\infty}}^2\right)\mathds{1}_M,\left|\left|\cdot\right|\right|_{\e,1}\right)}\frac{M}{\sqrt{\prod_{s:\e_s=1}C_s}}+\eta_1 \end{align*}
Because $||.||_{\e,1}\leq ||.||_{\mu_C,1}$, Lemma \ref{lemma:ch_norme_class} $ii)$ implies
\begin{align}
&\mathbb{E}\left[\sup_{\mathcal{F}_{\infty}}\left|\frac{1}{\Pi_C}\sum_{\un\leq\j\leq\C}\epsilon_{\j\odot \e}\left(\sum_{\ell=1}^{N_{\j}}f(Y_{\ell,\j})\right)^2\mathds{1}_M(\j)\right|\text{ }\bigg| \bm{Z}\right] \nonumber \\
&=\mathbb{E}\left[\sup_{\mathcal{F}_{\infty}}\left|\frac{1}{\Pi_C}\sum_{\un\odot\e\leq\c\leq\C\odot\e}\epsilon_{\c}\sum_{\un-\e\leq \c'\leq \C\odot(\un-\e)}\left(\sum_{\ell=1}^{N_{\c+\c'}}f(Y_{\ell,\c+\c'})\right)^2\mathds{1}_M(\c+\c')\right|\text{ }\bigg| \bm{Z}\right] \nonumber\\
&\leq 2\sqrt{2\log 2N\left(\eta_1,\widetilde{\mathcal{F}_{\infty}}^2,\left|\left|\cdot\right|\right|_{\mu_C,1}\right)}\frac{M}{\sqrt{\prod_{s:\e_s=1}C_s}}+\eta_1.
\label{eq:ineg_lem}
\end{align}

Note that if $N_{\j}=0$ for all $\j$, then the random measure $\mu_C$ is null and the previous inequality also hold for $\eta_1=0$.

We fix $\eta>0$.

$\bullet$ Let us first focus on Assumption \ref{as:vc}. 

We apply the Inequality (\ref{eq:ineg_lem}) to the non-negative random variable $$\eta_1=2\eta \abs{\abs{F}}_{\mathbb{Q}_C^2,2}\sqrt{\overline{N}_2}\abs{\abs{\widetilde{F}}}_{\mu_C,2}=2\eta \abs{\abs{F}}_{\mathbb{Q}_C^2,2}\sqrt{\overline{N}_2}\frac{\abs{\abs{\widetilde{F}^2}}_{\mu_C,1}}{\abs{\abs{\widetilde{F}}}_{\mu_C,2}}.$$
Note that Lemma \ref{lem:csteF} ensures that $\eta_1=0$ if and only if $\overline{N}_{2}=0$. When $\overline{N}_{2}>0$, we use Lemma \ref{lemma:bound_empirical_entropy_kato} to deduce that
\begin{align*}
&\mathbb{E}\left[\sup_{\mathcal{F}_{\infty}}\left|\frac{1}{\Pi_C}\sum_{\un\leq\j\leq\C}\epsilon_{\j\odot \e}\left(\sum_{\ell=1}^{N_{\j}}f(Y_{\ell,\j})\right)^2\mathds{1}_M(\j)\right|\text{ }\bigg| \bm{Z}\right] \\
&\leq 2\sqrt{2\log 2N\left(2\eta \abs{\abs{F}}_{\mathbb{Q}_C^2,2}\sqrt{\overline{N}_2}\frac{\abs{\abs{\widetilde{F}^2}}_{\mu_C,1}}{\abs{\abs{\widetilde{F}}}_{\mu_C,2}},\widetilde{\mathcal{F}_{\infty}}^2,\left|\left|\cdot\right|\right|_{\mu_C,1}\right)}\frac{M}{\sqrt{\prod_{s:\e_s=1}C_s}}\\&+2\eta \abs{\abs{F}}_{\mathbb{Q}_C^2,2}\sqrt{\overline{N}_2}\abs{\abs{\widetilde{F}}}_{\mu_C,2} \\
&\leq \frac{M}{\sqrt{\prod_{s:\e_s=1}C_s}}\times\left\{2\sqrt{2\log 2}+4\sqrt{\log N\left(\eta \abs{\abs{F}}_{\mathbb{Q}_C^2,2}\sqrt{\overline{N}_2},\widetilde{\mathcal{F}},\left|\left|\cdot\right|\right|_{\mu_C,2}\right)}\right\}\\&+2\eta \abs{\abs{F}}_{\mathbb{Q}_C^2,2}\sqrt{\overline{N}_2}\abs{\abs{\widetilde{F}}}_{\mu_C,2}. 
\end{align*}
Using Lemma \ref{lemma:bound_empirical_entropy_dgg} $i)$ for the first term and the Cauchy-Schwarz inequality for the second term, we deduce
\begin{align*}
&\mathbb{E}\left[\sup_{\mathcal{F}_{\infty}}\left|\frac{1}{\Pi_C}\sum_{\un\leq\j\leq\C}\epsilon_{\j\odot \e}\left(\sum_{\ell=1}^{N_{\j}}f(Y_{\ell,\j})\right)^2\mathds{1}_M(\j)\right|\text{ }\bigg| \bm{Z}\right] \\
&\leq \frac{M}{\sqrt{\prod_{s:\e_s=1}C_s}}\times\left\{2\sqrt{2\log 2}+4\sqrt{\log N\left(\eta \abs{\abs{F}}_{\mathbb{Q}_C^2,2}, \mathcal{F},\abs{\abs{\cdot}}_{\mathbb{Q}_C^2,2}\right)}\right\}\\&+2\eta \abs{\abs{F}}^2_{\mathbb{Q}_C^2,2}\overline{N}_2.
\end{align*}

Since $\eta\mapsto\sqrt{\log N\left(\eta \abs{\abs{F}}_{\mathbb{Q}_C^2,2},\widetilde{\mathcal{F}},\abs{\abs{\cdot}}_{\mathbb{Q}_C^2,2}\right)}$ is decreasing, we have
\begin{align*}\sqrt{\log N\left(\eta \abs{\abs{F}}_{\mathbb{Q}_C^2,2},\widetilde{\mathcal{F}},\abs{\abs{\cdot}}_{\mathbb{Q}_C^2,2}\right)}&\leq \frac{1}{\eta}\int_0^{\eta}\sqrt{\log N\left(u \abs{\abs{F}}_{\mathbb{Q}_C^2,2},\widetilde{\mathcal{F}},\abs{\abs{\cdot}}_{\mathbb{Q}_C^2,2}\right)}du\\
&\leq \frac{1}{\eta}J_{2,\mathcal{F}}(\eta)\\
&\leq \frac{1}{\eta}J_{2,\mathcal{F}}(\infty).\end{align*}
Note that $J_{2,\mathcal{F}}(\infty)<\infty$ by Assumption \ref{as:vc}. When $\overline{N}_2=0$, then $N_{\j}=0$ for any $\j$ and next the measure $\mathbb{Q}_C^2$ is null (by convention) and $\sqrt{\log N\left(\eta \abs{\abs{F}}_{\mathbb{Q}_C^2,2},\widetilde{\mathcal{F}},\abs{\abs{\cdot}}_{\mathbb{Q}_C^2,2}\right)}=0\leq \frac{1}{\eta}J_{2,\mathcal{F}}(\infty)$.
Last, by integration with respect to $\bm{Z}$, we get
\begin{align*}
&\mathbb{E}\left[\sup_{\mathcal{F}_{\infty}}\left|\frac{1}{\Pi_C}\sum_{\un\leq\j\leq\C}\epsilon_{\j\odot \e}\left(\sum_{\ell=1}^{N_{\j}}f(Y_{\ell,\j})\right)^2\mathds{1}_M(\j)\right|\text{ }\right] \\
&\leq  \frac{M}{\sqrt{\prod_{s:\e_s=1}C_s}}\times\left\{2\sqrt{2\log 2}+\frac{4}{\eta}J_{2,F}(\infty)\right\}+2\eta \E\left(\abs{\abs{F}}^2_{\mathbb{Q}_C^2,2}\overline{N}_2\right)\\
&=\frac{M}{\sqrt{\prod_{s:\e_s=1}C_s}}\times\left\{2\sqrt{2\log 2}+\frac{4}{\eta}J_{2,F}(\infty)\right\}+2\eta \E\left(N_{\un} \sum_{\ell=1}^{N_{\un}}F^2(Y_{\ell,\un})\right)
\end{align*}

This concludes the proof of the first part of the Lemma.

$\bullet$ Under Assumption~\ref{as:smooth}', for any pair $(f,g)\in\left(\mathcal{F}\right)^2$, we have
\begin{align}\label{eq:square_class_eq_1}
\left|\left|(\widetilde{f}-\widetilde{g})^2\right|\right|_{\mu_C,1}
&\leq A_2\times \left|\left|f-g\right|\right|_{\infty,\beta}^2 \nonumber\\
&\leq 2\left|\left|F\right|\right|_{\infty,\beta}A_2\times \left|\left|f-g\right|\right|_{\infty,\beta},
\end{align}
where $A_2=\frac{1}{\Pi_C}\sum_{\un\leq\j\leq\C}\left(\sum_{\ell=1}^{N_{\j}}(1+|Y_{\ell,\j}|^2)^{-\beta/2}\right)^2$.

From Equation (\ref{eq:square_class_eq_1}), Lemma \ref{lemma:ch_norme_class} $ii)$, $i)$ and $iv)$ we deduce that for any $\eta_1>0$
\begin{align*}
N\left(\eta_1,\widetilde{\mathcal{F}_{\infty}}^2,\left|\left|\cdot\right|\right|_{\mu_C,1}\right)&\leq N\left(\eta_1,\mathcal{F}_{\infty},2\left|\left|F\right|\right|_{\infty,\beta}A_2 \left|\left|\cdot\right|\right|_{\infty,\beta}\right)\\
&\leq N\left(\frac{\eta_1}{2\left|\left|F\right|\right|_{\infty,\beta}A_2},\mathcal{F}_{\infty}, \left|\left|\cdot\right|\right|_{\infty,\beta}\right)\\ &\leq N^2\left(\frac{\eta_1}{4\left|\left|F\right|\right|_{\infty,\beta}A_2},\mathcal{F},\left|\left|\cdot\right|\right|_{\infty,\beta}\right).
\end{align*}

Now consider $\eta_1=4\eta\left|\left|F\right|\right|_{\infty,\beta}^2A_2$ in Equation (\ref{eq:ineg_lem}), to deduce that
\begin{align*}
&\mathbb{E}\left[\sup_{\mathcal{F}_{\infty}}\left|\frac{1}{\Pi_C}\sum_{\un\leq\j\leq\C}\epsilon_{\j\odot \e}\left(\sum_{\ell=1}^{N_{\j}}f(Y_{\ell,\j})\right)^2\mathds{1}_M(\j)\right|\text{ }\bigg| \bm{Z}\right]\\
&\leq 2\sqrt{2\log 2N\left(4\eta\left|\left|F\right|\right|_{\infty,\beta}^2A_2,\widetilde{\mathcal{F}_{\infty}}^2,\left|\left|\cdot\right|\right|_{\mu_C,1}\right)}\frac{M}{\sqrt{\prod_{s:\e_s=1}C_s}}+4\eta\left|\left|F\right|\right|_{\infty,\beta}^2A_2 \\
&\leq \frac{M}{\sqrt{\prod_{s:\e_s=1}C_s}}\times\left\{2\sqrt{2\log 2}+4\sqrt{\log N\left(\eta\left|\left|F\right|\right|_{\infty,\beta},\mathcal{F},\left|\left|\cdot\right|\right|_{\infty,\beta}\right)}\right\}+4\eta\left|\left|F\right|\right|_{\infty,\beta}^2A_2.
\end{align*}

Because $\eta \mapsto \sqrt{\log N\left(\eta\left|\left|F\right|\right|_{\infty,\beta},\mathcal{F},\left|\left|\cdot\right|\right|_{\infty,\beta}\right)}$ is decreasing, we have
$$\sqrt{\log N\left(\eta\left|\left|F\right|\right|_{\infty,\beta},\mathcal{F},\left|\left|\cdot\right|\right|_{\infty,\beta}\right)} \leq \frac{1}{\eta}J_{\infty,\beta,\mathcal{F}}(\eta)\leq \frac{1}{\eta}J_{\infty,\beta,\mathcal{F}}(\infty)$$

Integration over $\left(N_{\j},\vec{Y}_{\j}\right)_{\j\geq \un}$ implies the result. 

\subsubsection{Proof of Lemma~\ref{lemma:glivenko_prod_fin}}

Let $F$ and $G$ the respective envelopes of $\mathcal{F}$ and $\mathcal{G}$. 

Recall that $$\mathcal{A}_i:=\left\{(\j,\j'):\un\leq\j\leq\C,\un\leq\j'\leq\C,j_i=j_i',j_s\neq j_s'\forall s\neq i\right\}$$ and $$\mathcal{B}_i:=\left\{(\j,\j'):\un\leq\j\leq\C,\un\leq\j'\leq\C,j_i=j_i'\right\}.$$

Because $\mathcal{A}_i \subset \mathcal{B}_i$, we have
\begin{align*}%\label{eq:glivenko_prod_1}
&\mathbb{E}\left[\sup_{\mathcal{F}\times\mathcal{G}}\left|\frac{1}{|\mathcal{B}_i|}\sum_{(\j,\j')\in\mathcal{B}_i}\sum_{\ell=1}^{N_{\j}}f(Y_{\ell,\j})\sum_{\ell=1}^{N_{\j'}}g(Y_{\ell,\j'})-\frac{1}{|\mathcal{A}_i|}\sum_{(\j,\j')\in\mathcal{A}_i}\sum_{\ell=1}^{N_{\j}}f(Y_{\ell,\j})\sum_{\ell=1}^{N_{\j'}}g(Y_{\ell,\j'})\right|\right]\\
&\leq \left|\frac{1}{|\mathcal{B}_i|}-\frac{1}{|\mathcal{A}_i|}\right|\mathbb{E}\left[\sup_{\mathcal{F}\times\mathcal{G}}\left|\sum_{(\j,\j')\in\mathcal{A}_i}\sum_{\ell=1}^{N_{\j}}f(Y_{\ell,\j})\sum_{\ell=1}^{N_{\j'}}g(Y_{\ell,\j'})\right|\right]\\
&+\frac{1}{|\mathcal{B}_i|}\mathbb{E}\left[\sup_{\mathcal{F}\times\mathcal{G}}\left|\sum_{(\j,\j')\in\mathcal{B}_i\setminus \mathcal{A}_i}\sum_{\ell=1}^{N_{\j}}f(Y_{\ell,\j})\sum_{\ell=1}^{N_{\j'}}g(Y_{\ell,\j'})\right|\right]\\
&\leq\left|\frac{1}{|\mathcal{B}_i|}-\frac{1}{|\mathcal{A}_i|}\right|\times|\mathcal{A}_i|\times \sqrt{\mathbb{E}\left[\left(\sum_{\ell=1}^{N_{\un}}F(Y_{\ell,\un})\right)^2\right]}\sqrt{\mathbb{E}\left[\left(\sum_{\ell=1}^{N_{\un}}G(Y_{\ell,\un})\right)^2\right]}\\
&+\frac{1}{|\mathcal{B}_i|}\times|\mathcal{B}_i\setminus \mathcal{A}_i|\times \sqrt{\mathbb{E}\left[\left(\sum_{\ell=1}^{N_{\un}}F(Y_{\ell,\un})\right)^2\right]}\sqrt{\mathbb{E}\left[\left(\sum_{\ell=1}^{N_{\un}}G(Y_{\ell,\un})\right)^2\right]}.
\end{align*}
We have $|\mathcal{A}_i|=C_i\prod_{s\neq i}C_s(C_s-1)$, $|\mathcal{B}_i|=C_i\prod_{s\neq i}C_s^2$ and $|\mathcal{B}_i\setminus \mathcal{A}_i|=|\mathcal{B}_i|-|\mathcal{A}_i|$. This implies that $\lim_{\underline{C}\rightarrow \infty}\frac{|\mathcal{A}_i|}{|\mathcal{B}_i|}=1$. Next

\begin{align*}%\label{eq:glivenko_prod_1}
&\mathbb{E}\left[\sup_{\mathcal{F}\times\mathcal{G}}\left|\frac{1}{|\mathcal{B}_i|}\sum_{(\j,\j')\in\mathcal{B}_i}\sum_{\ell=1}^{N_{\j}}f(Y_{\ell,\j})\sum_{\ell=1}^{N_{\j'}}g(Y_{\ell,\j'})-\frac{1}{|\mathcal{A}_i|}\sum_{(\j,\j')\in\mathcal{A}_i}\sum_{\ell=1}^{N_{\j}}f(Y_{\ell,\j})\sum_{\ell=1}^{N_{\j'}}g(Y_{\ell,\j'})\right|\right]=o(1).
\end{align*}

For the second part of the Lemma, note that the envelope condition and the Cauchy-Schwarz inequality ensure
\begin{align*}
&\E\left(\sup_{\mathcal{F}\times \mathcal{G}}\left|\sum_{(\j,\j')\in \mathcal{B}_{\e}}\sum_{\ell=1}^{N_{\j}}f(Y_{\ell,\j})\sum_{\ell=1}^{N_{\j'}}g(Y_{\ell,\j'})\right|\right)\\&\leq \E\left(\sum_{(\j,\j')\in \mathcal{B}_{\e}}\sum_{\ell=1}^{N_{\j}}F(Y_{\ell,\j})\sum_{\ell=1}^{N_{\j'}}G(Y_{\ell,\j'})\right)\\
&\leq \E\left(\sum_{\e \leq \c\leq \C \odot \e}\sum_{\un-\e \leq \c'\leq \C \odot(\un- \e)}\sum_{\ell=1}^{N_{\c+\c'}}F(Y_{\ell,\c+\c'})\sum_{\un-\e \leq \c''\leq \C \odot(\un- \e)}\sum_{\ell=1}^{N_{\c+\c''}}G(Y_{\ell,\c+\c''})\right)\\
&\leq \E\left(\sum_{\e \leq \c\leq \C \odot \e}\left(\sum_{\un-\e \leq \c'\leq \C \odot(\un- \e)}\sum_{\ell=1}^{N_{\c+\c'}}F(Y_{\ell,\c+\c'})\right)^2\right)^{1/2}\\
&~~~\times \E\left(\sum_{\e \leq \c\leq \C \odot \e}\left(\sum_{\un-\e \leq \c''\leq \C \odot(\un- \e)}\sum_{\ell=1}^{N_{\c+\c''}}G(Y_{\ell,\c+\c''})\right)^2\right)^{1/2}\\
&\leq \left(\prod_{s:e_i=0}C_s\right) \times \E\left(\sum_{\un \leq \j\leq \C}\left(\sum_{\ell=1}^{N_{\j}}F(Y_{\ell,\j})\right)^2\right)^{1/2}\times \E\left(\sum_{\e \leq \j\leq \C}\left(\sum_{\ell=1}^{N_{\j}}G(Y_{\ell,\j})\right)^2\right)^{1/2}\\
&\leq \left(\prod_{s:e_i=0}C_s\right) \times \Pi_C \times \E\left(\left(\sum_{\ell=1}^{N_{\un}}F(Y_{\ell,\un})\right)^2\right)^{1/2}\times \E\left(\left(\sum_{\ell=1}^{N_{\un}}G(Y_{\ell,\un})\right)^2\right)^{1/2}
\end{align*}

\subsubsection{Proof of Lemma~\ref{lemma:glivenko_prod_fin2}}
We first prove the result for $\mathcal{A}_i$ with $i=1,...,k$. The result for $\mathcal{B}_i$ follows from Lemma \ref{lemma:glivenko_prod_fin}.

The representation Lemma \ref{lem:representation} ensures that
$\bm{Z}\overset{a.s}{=}\left(\tau\left(\left(U_{\j\odot\e}\right)_{\zero<\e\leq \un}\right)\right)_{\j\geq\un}$, for some mutually independent uniform random variables $(U_{\c})_{\c>\zero} $. If $(U^{(1)}_{\c})_{\c>\zero}$ is an independent copy of $(U_{\c})_{\c>\zero} $, then $\left(N_{\j}^{(1)},\vec{Y}_{\j}^{(1)}\right)_{\j\geq\un}\overset{a.s}{=}\left(\tau\left(\left(U_{\j\odot\e}^{(1)}\right)_{\zero<\e\leq \un}\right)\right)_{\j\geq\un}$ is an independent copy of $\bm{Z}$.
Because the array is separately exchangeable, we have $$\mathbb{E}\left[\sum_{\ell=1}^{N_{\un}}f(Y_{\ell,\un})\sum_{\ell=1}^{N_{\deux}}g(Y_{\ell,\deux})\right]=\mathbb{E}\left[\sum_{\ell=1}^{N_{\j}}f(Y_{\ell,\j})\sum_{\ell=1}^{N_{\j'}}g(Y_{\ell,\j'})\right]$$ for any $(\j,\j')\in \mathcal{A}_i $ and next
\begin{align*}\mathbb{E}\left[\sum_{\ell=1}^{N_{\un}}f(Y_{\ell,\un})\sum_{\ell=1}^{N_{\deux}}g(Y_{\ell,\deux})\right]=\mathbb{E}\left[\frac{1}{|\mathcal{A}_i|}\sum_{(\j,\j')\in\mathcal{A}_i}\sum_{\ell=1}^{N_{\j}}f(Y_{\ell,\j})\sum_{\ell=1}^{N_{\j'}}g(Y_{\ell,\j'})\right]\\
=\E\left[\frac{1}{|\mathcal{A}_i|}\sum_{(\j,\j')\in\mathcal{A}_i}\sum_{\ell=1}^{N^{(1)}_{\j}}f^{(1)}(Y_{\ell,\j})\sum_{\ell=1}^{N^{(1)}_{\j'}}g(Y^{(1)}_{\ell,\j'}) \bigg| \bm{Z}\right].\end{align*}

Monotonicity of the expectation and the law of iterated expectations ensure

\begin{align}
	&\mathbb{E}\left[\sup_{\mathcal{F}\times\mathcal{G}}\left|\frac{1}{|\mathcal{A}_i|}\sum_{(\j,\j')\in\mathcal{A}_i}\sum_{\ell=1}^{N_{\j}}f(Y_{\ell,\j})\sum_{\ell=1}^{N_{\j'}}g(Y_{\ell,\j'})-\mathbb{E}\left[\sum_{\ell=1}^{N_{\un}}f(Y_{\ell,\j})\sum_{\ell=1}^{N_{\deux}}g(Y_{\ell,\j'})\right]\right|\right] \nonumber \\
&\leq \mathbb{E}\left[\sup_{\mathcal{F}\times\mathcal{G}}\left|\frac{1}{|\mathcal{A}_i|}\sum_{(\j,\j')\in\mathcal{A}_i}\left(\sum_{\ell=1}^{N_{\j}}f(Y_{\ell,\j})\sum_{\ell=1}^{N_{\j'}}g(Y_{\ell,\j'})-\sum_{\ell=1}^{N_{\j}^{(1)}}f(Y_{\ell,\j}^{(1)})\sum_{\ell=1}^{N_{\j'}^{(1)}}g(Y_{\ell,\j'}^{(1)})\right)\right|\right].\label{eq:glivenko_prod_4}
\end{align}

Moreover, Lemma \ref{lemma:glivenko_prod_fin} combined with the triangle inequality ensures that
\begin{align}
&\mathbb{E}\left[\sup_{\mathcal{F}\times\mathcal{G}}\left|\frac{1}{|\mathcal{A}_i|}\sum_{(\j,\j')\in\mathcal{A}_i}\sum_{\ell=1}^{N_{\j}}f(Y_{\ell,\j})\sum_{\ell=1}^{N_{\j'}}g(Y_{\ell,\j'})-\mathbb{E}\left[\sum_{\ell=1}^{N_{\un}}f(Y_{\ell,\j})\sum_{\ell=1}^{N_{\deux}}g(Y_{\ell,\j'})\right]\right|\right] \nonumber \\
&\leq \mathbb{E}\left[\sup_{\mathcal{F}\times\mathcal{G}}\left|\frac{1}{|\mathcal{B}_i|}\sum_{(\j,\j')\in\mathcal{B}_i}\left(\sum_{\ell=1}^{N_{\j}}f(Y_{\ell,\j})\sum_{\ell=1}^{N_{\j'}}g(Y_{\ell,\j'})-\sum_{\ell=1}^{N_{\j}^{(1)}}f(Y_{\ell,\j}^{(1)})\sum_{\ell=1}^{N_{\j'}^{(1)}}g(Y_{\ell,\j'}^{(1)})\right)\right|\right]+o(1).\label{eq:glivenko_prod_41}
\end{align}

Let $S^f_{\j}=\sum_{\ell=1}^{N_{\j}}f(Y_{\ell,\j})$, $\widetilde{S}^{f}_{\j}(\zero)=\sum_{\ell=1}^{N_{\j}^{(1)}}f(Y_{\ell,\j}^{(1)})$. For any $\e$ such that $\zero \leq \e\leq \un$ let $S^f_{\j}(\e)=\sum_{\ell=1}^{N_{\j}(\e)}f(Y_{\ell,\j}(\e))$, $\widetilde{S}^f_{\j}(\e)=\sum_{\ell=1}^{\widetilde{N}_{\j}(\e)}f(\widetilde{Y}_{\ell,\j}(\e))$ where 
\begin{align*}&\left(N_{\j}(\e),\vec{Y}_{\j}(\e)\right)_{\j\geq\un}\overset{a.s}{=}\left(\tau\left(\left(U_{\j\odot\e'}\right)_{\zero\prec\e'\preceq\e},\left(U_{\j\odot\e'}^{(1)}\right)_{\e\prec\e'\preceq\un}\right)\right)_{\j\geq\un}\\
\text{and }&\left(\widetilde{N}_{\j}(\e),\vec{\widetilde{Y}}_{\j}(\e)\right)_{\j\geq\un}\overset{a.s}{=}\left(\tau\left(\left(U_{\j\odot\e'}\right)_{\zero\prec\e'\prec\e},\left(U_{\j\odot\e'}^{(1)}\right)_{\e\preceq\e'\preceq\un}\right)\right)_{\j\geq\un}.
\end{align*} 
Observe that $S^f_{\j}(\un)S^g_{\j}(\un)=S^f_{\j} S^g_{\j}$ and $\widetilde{S}^{f}_{\j}(\zero)\widetilde{S}^{g}_{\j}(\zero)=\widetilde{S}^{f}_{\j}\widetilde{S}^{g}_{\j}$.
The triangle inequality ensures
\begin{align}
&\mathbb{E}\left[\sup_{\mathcal{F}\times\mathcal{G}}\left|\frac{1}{|\mathcal{B}_i|}\sum_{(\j,\j')\in\mathcal{B}_i}\left(\sum_{\ell=1}^{N_{\j}}f(Y_{\ell,\j})\sum_{\ell=1}^{N_{\j'}}g(Y_{\ell,\j'})-\sum_{\ell=1}^{N_{\j}^{(1)}}f(Y_{\ell,\j}^{(1)})\sum_{\ell=1}^{N_{\j'}^{(1)}}g(Y_{\ell,\j'}^{(1)})\right)\right|\right] \nonumber\\
&\leq \sum_{\bm{b}_i\leq \e\leq \un}\mathbb{E}\left[\sup_{\mathcal{F}\times\mathcal{G}}\left|\frac{1}{C_i}\sum_{\bm{b}_i\leq \c\leq C_i \bm{b}_i}\frac{1}{\prod_{s\neq i}C^2_s}\sum_{\un-\bm{b}_i\leq \c',\c''\leq \C \odot (\un-\bm{b}_i)}S^f_{\c+\c'}(\e)S^g_{\c+\c''}(\e)-\widetilde{S}^f_{\c+\c'}(\e)\widetilde{S}^g_{\c+\c''}(\e)\right|\right] \nonumber\\
&+\mathbb{E}\left[\sup_{\mathcal{F}\times\mathcal{G}}\left|\frac{1}{C_i}\sum_{\bm{b}_i\leq \c\leq C_i \bm{b}_i}\frac{1}{\prod_{s\neq i}C^2_s}\sum_{\un-\bm{b}_i\leq \c',\c''\leq \C \odot (\un-\bm{b}_i)}S^f_{\c+\c'}(\zero)S^g_{\c+\c''}(\zero)-\widetilde{S}^f_{\c+\c'}(\zero)\widetilde{S}^g_{\c+\c''}(\zero)\right|\right].\nonumber 
\end{align}
Let $R_{\e}=\left(\left(U_{\j\odot\e'}\right)_{\zero\prec\e'\prec\e},\left(U_{\j\odot\e'}^{(1)}\right)_{\e\prec\e'\preceq\un}\right)_{\j\geq\un}$. For any $(\e,\c)$ such that $\zero \leq \e \leq \un$ and $\bm{b}_i\leq \c\leq C_i \bm{b}_i$, the terms
$\left(\sum_{\un-\bm{b}_i\leq \c',\c''\leq \C \odot (\un-\bm{b}_i)}S^f_{\c+\c'}(\zero)S^g_{\c+\c''}(\zero)-\widetilde{S}^f_{\c+\c'}(\e)\widetilde{S}^g_{\c+\c''}(\e)\right)$ are independent across $\c$ conditionally on $R_{\e}$ and have a symmetric conditional distribution. It follows that for any Rademacher process $\epsilon_{\c,\e}$ indexed by $(\c,\e)$ and independent form $(U_{\j}, U^{(1)}_{\j})_{\j\geq \zero}$, we have
\begin{align}\label{eq:glivenko_prod_411}
&\mathbb{E}\left[\sup_{\mathcal{F}\times\mathcal{G}}\left|\frac{1}{|\mathcal{B}_i|}\sum_{(\j,\j')\in\mathcal{B}_i}\left(\sum_{\ell=1}^{N_{\j}}f(Y_{\ell,\j})\sum_{\ell=1}^{N_{\j'}}g(Y_{\ell,\j'})-\sum_{\ell=1}^{N_{\j}^{(1)}}f(Y_{\ell,\j}^{(1)})\sum_{\ell=1}^{N_{\j'}^{(1)}}g(Y_{\ell,\j'}^{(1)})\right)\right|\right] \nonumber\\
&\leq \sum_{\bm{b}_i\leq \e\leq \un}\mathbb{E}\left[\sup_{\mathcal{F}\times\mathcal{G}}\left|\frac{1}{C_i}\sum_{\bm{b}_i\leq \c\leq C_i \bm{b}_i}\frac{\epsilon_{\c,\e}}{\prod_{s\neq i}C^2_s}\sum_{\un-\bm{b}_i\leq \c',\c''\leq \C \odot (\un-\bm{b}_i)}S^f_{\c+\c'}(\e)S^g_{\c+\c''}(\e)-\widetilde{S}^f_{\c+\c'}(\e)\widetilde{S}^g_{\c+\c''}(\e)\right|\right] \nonumber\\
&+\mathbb{E}\left[\sup_{\mathcal{F}\times\mathcal{G}}\left|\frac{1}{C_i}\sum_{\bm{b}_i\leq \c\leq C_i \bm{b}_i}\frac{\epsilon_{\c,\zero}}{\prod_{s\neq i}C^2_s}\sum_{\un-\bm{b}_i\leq \c',\c''\leq \C \odot (\un-\bm{b}_i)}S^f_{\c+\c'}(\zero)S^g_{\c+\c''}(\zero)-\widetilde{S}^f_{\c+\c'}(\zero)\widetilde{S}^g_{\c+\c''}(\zero)\right|\right]\nonumber\\
&\leq 2\sum_{\bm{b}_i\leq \e\leq \un}\E\left[\sup_{\mathcal{F}\times\mathcal{G}}\left|\frac{1}{C_i}\sum_{\bm{b}_i\leq \c\leq C_i \bm{b}_i}\epsilon_{\c,\e} \frac{1}{\prod_{s\neq i}C^2_s}\sum_{\un-\bm{b}_i\leq \c',\c''\leq \C \odot (\un-\bm{b}_i)}S^f_{\c+\c'}S^g_{\c+\c''}\right|\right]\nonumber\\
&+2\E\left[\sup_{\mathcal{F}\times\mathcal{G}}\left|\frac{1}{C_i}\sum_{\bm{b}_i\leq \c\leq C_i \bm{b}_i}\epsilon_{\c,\zero} \frac{1}{\prod_{s\neq i}C^2_s}\sum_{\un-\bm{b}_i\leq \c',\c''\leq \C \odot (\un-\bm{b}_i)}S^f_{\c+\c'}S^g_{\c+\c''}\right|\right].
\end{align}
The last inequality follows from the triangle inequality, independence between $\epsilon_{.,.}$ and the $S^.(.)$ and the fact that $\left(S^f_{\j}(\e)S^g_{\j}(\e)\right)_{\j\geq \un}\overset{d}{=}\left(\widetilde{S}^f_{\j}(\e)\widetilde{S}^g_{\j}(\e)\right)_{\j\geq \un}\overset{d}{=}\left(S^f_{\j}S^g_{\j}\right)_{\j\geq \un}$. \\
Let $\mathds{1}_{M}(\c):=\mathds{1}\left\{ \frac{1}{\prod_{s\neq i}C_s}\sum_{\un-\bm{b}_i\leq \c'\leq \C \odot (\un-\bm{b}_i)}S^F_{\c+\c'}\vee \frac{1}{\prod_{s\neq i}C_s}\sum_{\un-\bm{b}_i\leq \c''\leq \C \odot (\un-\bm{b}_i)}S^G_{\c+\c''}\leq M\right\}$. For any $\e$ we have
\begin{align}\label{eq:glivenko_prod_42}
&\E\left[\sup_{\mathcal{F}\times\mathcal{G}}\left|\frac{1}{C_i}\sum_{\bm{b}_i\leq \c\leq C_i \bm{b}_i}\epsilon_{\c,\e} \frac{1}{\prod_{s\neq i}C^2_s}\sum_{\un-\bm{b}_i\leq \c',\c''\leq \C \odot (\un-\bm{b}_i)}S^f_{\c+\c'}S^g_{\c+\c''}\right|\right] \nonumber\\
&\leq \E\left[\frac{1}{C_i}\sum_{\bm{b}_i\leq \c\leq C_i \bm{b}_i} \frac{1}{\prod_{s\neq i}C^2_s}\sum_{\un-\bm{b}_i\leq \c',\c''\leq \C \odot (\un-\bm{b}_i)}S^F_{\c+\c'}S^G_{\c+\c''} (1-\mathds{1}_{M}(\c))\right] \nonumber\\
&+\E\left[\sup_{\mathcal{F}\times\mathcal{G}}\left|\frac{1}{C_i}\sum_{\bm{b}_i\leq \c\leq C_i \bm{b}_i}\epsilon_{\c,\e} \frac{1}{\prod_{s\neq i}C^2_s}\sum_{\un-\bm{b}_i\leq \c',\c''\leq \C \odot (\un-\bm{b}_i)}S^f_{\c+\c'}S^g_{\c+\c''} \mathds{1}_{M}(\c)\right|\right].\end{align} 
The first term tends to 0 as $M$ increases. Lemma \ref{lemma:lemma_222_vdvw} and the inequality $\sqrt{x+y}\leq \sqrt{x}+\sqrt{y}$ ensure that for any $M>0$ and $\eta>0$
\begin{align}\label{eq:glivenko_prod_43}&\E\left[\sup_{\mathcal{F}\times\mathcal{G}}\left|\frac{1}{C_i}\sum_{\bm{b}_i\leq \c\leq C_i \bm{b}_i}\epsilon_{\c,\e} \frac{1}{\prod_{s\neq i}C^2_s}\sum_{\un-\bm{b}_i\leq \c',\c''\leq \C \odot (\un-\bm{b}_i)}S^f_{\c+\c'}S^g_{\c+\c''} \mathds{1}_{M}(\c)\right|\bigg| (N_{\j}, \vec{Y}_{\j})_{\j \geq \un}\right] \nonumber \\
&\leq \eta+ M^2 \sqrt{\frac{2 \ln(2)+2\ln N(\eta, \mathcal{H}, ||.||_{q})}{C_i}}\nonumber \\
&\leq \eta+ \frac{M^2 }{\sqrt{C_i}}\left(\sqrt{2 \ln(2)}+\sqrt{2\ln N(\eta, \mathcal{H}, ||.||_{q})}\right),
\end{align}
with $\mathcal{H}$ the class of functions on the $(k-1)^2$-dimensional subarrays $A_{\c}$ indexed by $\c\in \mathbb{N}^+ \bm{b}_i$ defined by $$\mathcal{H}=\left\{r(A_{\c})=\frac{1}{\prod_{s\neq i}C^2_s}\sum_{\un-\bm{b}_i\leq \c',\c''\leq \C \odot (\un-\bm{b}_i)}S^f_{\c+\c'}S^g_{\c+\c''} \mathds{1}_{M}(\c), (f,g)\in \mathcal{F}\times \mathcal{G}\right\},$$ for  $$A_{\c}=\left\{(\vec{N}_{\c+\c'},\vec{Y}_{\c+\c'}, \vec{N}_{\c+\c''},\vec{Y}_{\c+\c''}), 1-\bm{b}_i\leq\c',\c''\leq \C \odot (1-\bm{b}_i)\right\},$$ and $||r||_q=\frac{1}{C_i}\sum_{\bm{b}_i\leq \c\leq C_i \bm{b}_i}|r(A_{\c})|$.
The Cauchy-Schwarz inequality applied repeatedly ensures
\begin{align*}||r||_q&\leq \frac{1}{C_i}\sum_{\bm{b}_i\leq \c\leq C_i \bm{b}_i} \frac{1}{\prod_{s\neq i} C_s^2}\left|\sum_{\un-\bm{b}_i\leq \c'\leq \C \odot (\un-\bm{b}_i)}S^f_{\c+\c'}\right| \times \left|\sum_{\un-\bm{b}_i\leq \c''\leq \C \odot (\un-\bm{b}_i)}S^g_{\c+\c''}\right|\\
&\leq \left(\frac{1}{C_i}\sum_{\bm{b}_i\leq \c\leq C_i \bm{b}_i} \frac{1}{\prod_{s\neq i} C_s^2}\left|\sum_{\un-\bm{b}_i\leq \c'\leq \C \odot (\un-\bm{b}_i)}S^f_{\c+\c'}\right|^2\right)^{1/2}\\
& \times \left(\frac{1}{C_i}\sum_{\bm{b}_i\leq \c\leq C_i \bm{b}_i} \frac{1}{\prod_{s\neq i} C_s^2}\left|\sum_{\un-\bm{b}_i\leq \c'\leq \C \odot (\un-\bm{b}_i)}S^g_{\c+\c'}\right|^2\right)^{1/2}\\
&\leq \left(\frac{1}{C_i}\sum_{\bm{b}_i\leq \c\leq C_i \bm{b}_i} \frac{1}{\prod_{s\neq i} C_s}\sum_{\un-\bm{b}_i\leq \c'\leq \C \odot (\un-\bm{b}_i)}\left(S^f_{\c+\c'}\right)^2\right)^{1/2}\\
& \times \left(\frac{1}{C_i}\sum_{\bm{b}_i\leq \c\leq C_i \bm{b}_i} \frac{1}{\prod_{s\neq i} C_s}\sum_{\un-\bm{b}_i\leq \c'\leq \C \odot (\un-\bm{b}_i)}\left(S^g_{\c+\c'}\right)^2\right)^{1/2}\\
&=||\widetilde{f}||_{\mu_{\C},2}\times ||\widetilde{g}||_{\mu_{\C},2}.\end{align*}
It follows that for any $\eta>0$, we have \begin{align}\label{eq:glivenko_prod_44}N(\eta, \mathcal{H}, ||.||_{q})\leq N(\eta^{1/2}, \widetilde{\mathcal{F}}, ||.||_{\mu_{\C},2}) \times N(\eta^{1/2}, \widetilde{\mathcal{G}}, ||.||_{\mu_{\C},2}).\end{align}
We are now lead back to a case study, depending if $\mathcal{F}$ and $\mathcal{G}$ fulfill Assumptions \ref{as:vc} or \ref{as:smooth}'.
We will only treat the case where $\mathcal{F}$ fulfills Assumption \ref{as:vc} and $\mathcal{G}$ fulfills Assumption \ref{as:smooth}', other cases can be treated similarly up to a simple adaptation.
If $\overline{N}_2=0$ (respectively $A_2=0$) then  $N(\eta^{1/2}, \widetilde{\mathcal{F}}, ||.||_{\mu_{\C},2})=1$ (respectively $N(\eta^{1/2}, \widetilde{\mathcal{G}}, ||.||_{\mu_{\C},2})=1$). Otherwise, Lemma \ref{lemma:bound_empirical_entropy_dgg} i) combined with Assumption \ref{as:vc} and Lemma \ref{lemma:bound_empirical_entropy_dgg} ii) combined with Assumption \ref{as:smooth} ensure respectively
\begin{align}\label{eq:glivenko_prod_45}\sqrt{\log N\left(\eta^{1/2}, \widetilde{\mathcal{F}}, ||.||_{\mu_{\C},2}\right)}&\leq \sqrt{\log N\left(\frac{\eta^{1/2}}{\sqrt{\overline{N}_2}},\mathcal{F},||.||_{\mathbb{Q}_{C}^2,2}\right)}\nonumber \\&\leq \frac{\overline{N}_2^{1/2}}{\eta^{1/2}}\int_0^{\frac{\eta^{1/2}}{\overline{N}_2^{1/2}}}\sqrt{\log N(u,\mathcal{F},||.||_{\mathbb{Q}_{C}^2,2})}du \nonumber \\&\leq \frac{\overline{N}_2^{1/2}||F||_{\mathbb{Q}_{C}^2,2}}{\eta^{1/2}}J_{2,\mathcal{F}}(\infty),\end{align}
\begin{align}\label{eq:glivenko_prod_46}\sqrt{\log N\left(\eta^{1/2}, \widetilde{\mathcal{G}}, ||.||_{\mu_{\C},2} \right)}&\leq \sqrt{\log N\left(\frac{\eta^{1/2}}{A_2^{1/2}}, \mathcal{G}, ||.||_{\mu_{\C},2} \right)}\nonumber \\
&\leq \frac{A_2^{1/2}||G||_{\infty,\beta}}{\eta^{1/2}}J_{\infty,\beta}(\infty).\end{align}

The combination of Inequalities (\ref{eq:glivenko_prod_42}), (\ref{eq:glivenko_prod_43}) , (\ref{eq:glivenko_prod_44}), (\ref{eq:glivenko_prod_45}) (\ref{eq:glivenko_prod_46}) and the inequality $\sqrt{x+y}\leq \sqrt{x}+\sqrt{y}$ ensure that for any $\e$ we have
\begin{align*}
&\E\left[\sup_{\mathcal{F}\times\mathcal{G}}\left|\frac{1}{C_i}\sum_{\bm{b}_i\leq \c\leq C_i \bm{b}_i}\epsilon_{\c,\e} \frac{1}{\prod_{s\neq i}C^2_s}\sum_{\un-\bm{b}_i\leq \c',\c''\leq \C \odot (\un-\bm{b}_i)}S^f_{\c+\c'}S^g_{\c+\c''}\right|\right] \nonumber\\
&\leq \E\left[\frac{1}{C_i}\sum_{\bm{b}_i\leq \c\leq C_i \bm{b}_i} \frac{1}{\prod_{s\neq i}C^2_s}\sum_{\un-\bm{b}_i\leq \c',\c''\leq \C \odot (\un-\bm{b}_i)}S^F_{\c+\c'}S^G_{\c+\c''} (1-\mathds{1}_{M}(\c))\right]+\eta \nonumber\\
&+\frac{M^2}{\sqrt{C_i}}\sqrt{2}\left(\sqrt{\log(2)}+\frac{\E\left(\overline{N}_2^{1/2}||F||_{\mathbb{Q}_{C}^2,2}\right)}{\eta^{1/2}}J_{2,\mathcal{F}}(\infty)+\frac{\E(A_2^{1/2})||G||_{\infty,\beta}}{\eta^{1/2}}J_{\infty,\beta}(\infty)\right).\end{align*} 
Fix $M$ sufficiently large and $\eta$ sufficiently small to ensure that $$\E\left[\frac{1}{C_i}\sum_{\bm{b}_i\leq \c\leq C_i \bm{b}_i} \frac{1}{\prod_{s\neq i}C^2_s}\sum_{\un-\bm{b}_i\leq \c',\c''\leq \C \odot (\un-\bm{b}_i)}S^F_{\c+\c'}S^G_{\c+\c''} (1-\mathds{1}_{M}(\c))\right]+\eta$$ is arbitrarily small. Jensen's inequality ensures that $\E\left(\overline{N}^{1/2}_2||F||_{\mathbb{Q}_C^2,2}\right)\leq \E\left(\overline{N}_2||F||^2_{\mathbb{Q}_C^2,2}\right)^{1/2}=\E\left(N_{\un}\sum_{\ell=1}^{N_{\un}}F(Y_{\ell,\un})^2\right)^{1/2}$ and $\E\left(A_2^{1/2}\right)\leq \E\left(A_2\right)^{1/2}=\E\left[\left(\sum_{\ell=1}^{N_{\un}}\left(1+ |Y_{\ell,\un}|^2\right)^{-\beta/2}\right)^2\right]^{1/2}$  which are finite by assumption. Then we deduce that when $\underline{C}\rightarrow \infty$, we have for any $\e$
\begin{align*}
\E\left[\sup_{\mathcal{F}\times\mathcal{G}}\left|\frac{1}{C_i}\sum_{\bm{b}_i\leq \c\leq C_i \bm{b}_i}\epsilon_{\c,\e} \frac{1}{\prod_{s\neq i}C^2_s}\sum_{\un-\bm{b}_i\leq \c',\c''\leq \C \odot (\un-\bm{b}_i)}S^f_{\c+\c'}S^g_{\c+\c''}\right|\right] =o(1),
\end{align*}
it follows from (\ref{eq:glivenko_prod_41}) and (\ref{eq:glivenko_prod_411}) that
\begin{align*}
&\mathbb{E}\left[\sup_{\mathcal{F}\times\mathcal{G}}\left|\frac{1}{|\mathcal{A}_i|}\sum_{(\j,\j')\in\mathcal{A}_i}\sum_{\ell=1}^{N_{\j}}f(Y_{\ell,\j})\sum_{\ell=1}^{N_{\j'}}g(Y_{\ell,\j'})-\mathbb{E}\left[\sum_{\ell=1}^{N_{\un}}f(Y_{\ell,\j})\sum_{\ell=1}^{N_{\deux}}g(Y_{\ell,\j'})\right]\right|\right] =o(1).\end{align*}
Lemma \ref{lemma:glivenko_prod_fin} combined with triangle inequality ensures that
\begin{align*}
&\mathbb{E}\left[\sup_{\mathcal{F}\times\mathcal{G}}\left|\frac{1}{|\mathcal{B}_i|}\sum_{(\j,\j')\in\mathcal{B}_i}\sum_{\ell=1}^{N_{\j}}f(Y_{\ell,\j})\sum_{\ell=1}^{N_{\j'}}g(Y_{\ell,\j'})-\mathbb{E}\left[\sum_{\ell=1}^{N_{\un}}f(Y_{\ell,\j})\sum_{\ell=1}^{N_{\deux}}g(Y_{\ell,\j'})\right]\right|\right] =o(1).\end{align*}

\subsection{Proof of Lemma~\ref{lemma:glivenko}}
Lemmas \ref{lem:representation}, \ref{lem:sym} together ensure that
\begin{align*}
&\E\left[\sup_{\mathcal{F}}\left|\frac{1}{\Pi_C}\sum_{\un\leq \j \leq \C}\sum_{\ell=1}^{N_{\j}}f(Y_{\ell,\j})-\E\left[\sum_{\ell=1}^{N_{\un}}f(Y_{\ell,\un})\right]\right|\right]\\
&\leq 2\sum_{\zero\prec \e\preceq \un}\E \left[  \sup_{\mathcal{F}}\left|\frac{1}{\Pi_C} \sum_{\un \leq \j \leq \C}\epsilon_{\j \odot \e} \sum_{\ell=1}^{N_{j}}f\left(Y_{\ell,\j}\right)\right|\right],\end{align*}
with $(\epsilon_{\c})_{\c\geq \zero}$ a Rademacher process, independent of $\bm{Z}$.

Because $\overline{N}_1=0$ implies that $\sup_{\mathcal{F}}\left|\frac{1}{\Pi_C} \sum_{\un \leq \j \leq \C}\epsilon_{\j \odot \e} \sum_{\ell=1}^{N_{j}}f\left(Y_{\ell,\j}\right)\right|=0$, we also have
\begin{align*}
&\E\left[\sup_{\mathcal{F}}\left|\frac{1}{\Pi_C}\sum_{\un\leq \j \leq \C}\sum_{\ell=1}^{N_{\j}}f(Y_{\ell,\j})-\E\left[\sum_{\ell=1}^{N_{\un}}f(Y_{\ell,\un})\right]\right|\right]\\
&\leq 2\sum_{\zero\prec \e\preceq \un}\E \left[  \sup_{\mathcal{F}}\left|\frac{1}{\Pi_C} \sum_{\un \leq \j \leq \C}\epsilon_{\j \odot \e} \sum_{\ell=1}^{N_{j}}f\left(Y_{\ell,\j}\right)\mathds{1}_{\{\overline{N}_1>0\}}\right|\right].\end{align*}

The triangle inequality, the Lemma \ref{lemma:lemma_222_vdvw} and inequality $\sqrt{x+y}\leq \sqrt{x}+\sqrt{y}$ together ensure that for any $M>0$ and any random $\eta_1>0$
\begin{align*}
&\E\left[\sup_{\mathcal{F}}\left|\frac{1}{\Pi_C}\sum_{\un\leq \j \leq \C}\sum_{\ell=1}^{N_{\j}}f(Y_{\ell,\j})-\E\left[\sum_{\ell=1}^{N_{\un}}f(Y_{\ell,\un})\right]\right|\right]\\
& \leq 2 (2^k-1)\E\left[\frac{1}{\Pi_C}\sum_{1\leq \j \leq \C}\sum_{\ell=1}^{N_{\j}}F(Y_{\ell,\j})(1-\mathds{1}_M(\j))\right]\\
&+2(2^k-1)\E\left[\left(\eta_1+\frac{M \sqrt{2}}{\sqrt{\Pi_C}}\left(\sqrt{\log 2}+\sqrt{\log N(\eta_1, \widetilde{\mathcal{F}}, ||.||_{\mu_C,1})}\right)\mathds{1}_{\{\overline{N}_1>0\}}\right)\right],
\end{align*}
with $\mathds{1}_M(\j)=\mathds{1}\{\sum_{\ell=1}^{N_{\j}}F(Y_{\ell,\j})\leq M\}$.

If the Condition $i)$ holds, 
consider $\eta_1=\eta \overline{N}_1 ||F||_{Q_{C}^1,1}=\eta \frac{1}{\Pi_C}\sum_{\un \leq \j \leq \C}^{N_{\j}}F(Y_{\ell,\j})$ and use Lemma \ref{lemma:bound_empirical_entropy_dgg} i) to deduce that
$\sqrt{\log N(\eta_1, \widetilde{\mathcal{F}}, ||.||_{\mu_C,1})}\mathds{1}_{\{\overline{N}_1>0\}}\leq \sqrt{\sup_Q \log N(\eta||F||_{Q}, \mathcal{F}, ||.||_{Q})} <\infty$. Moreover we have $\E(\eta_1)=\eta \E\left(\sum_{\ell=1}^{N_{\un}}F(Y_{\ell,\un})\right)<\infty$.

If the Condition $ii)$ holds, note that $\mathds{1}_{\{\overline{N}_1>0\}}=\mathds{1}_{\{A_1>0\}}$. Consider $\eta_1=\eta A_1 ||F||_{\infty,\beta}$ and use Lemma \ref{lemma:bound_empirical_entropy_dgg} iii) to deduce that
$\sqrt{\log N(\eta_1, \widetilde{\mathcal{F}}, ||.||_{\mu_C,1})}\mathds{1}_{\{A_1>0\}}\leq \sqrt{\log N(\eta||F||_{\infty,\beta}, \mathcal{F}, ||.||_{\infty,\beta})} <\infty$.
Moreover we have $\E(\eta_1)=\eta ||F||_{\infty,\beta}\E\left(A_1\right)<\infty$.

We also have $\E\left[\frac{1}{\Pi_C}\sum_{1\leq \j \leq \C}\sum_{\ell=1}^{N_{\j}}F(Y_{\ell,\j})(1-\mathds{1}_M(\j))\right]=\E\left[\sum_{\ell=1}^{N_{\un}}F(Y_{\ell,\un})(1-\mathds{1}_M(\un))\right]$ which converges to 0 when $M$ tends to $\infty$ by the dominated convergence theorem. 

So, fixing M sufficiently large and $\eta$ sufficiently small first ensures that\\ $\E\left[\sup_{\mathcal{F}}\left|\frac{1}{\Pi_C}\sum_{\un\leq \j \leq \C}\sum_{\ell=1}^{N_{\j}}f(Y_{\ell,\j})-\E\left[\sum_{\ell=1}^{N_{\un}}f(Y_{\ell,\un})\right]\right|\right]$ is arbitrary small for a sufficiently large $\underline{C}$. This means
$$\lim_{\underline{C}\rightarrow \infty}\E\left[\sup_{\mathcal{F}}\left|\frac{1}{\Pi_C}\sum_{\un\leq \j \leq \C}\sum_{\ell=1}^{N_{\j}}f(Y_{\ell,\j})-\E\left[\sum_{\ell=1}^{N_{\un}}f(Y_{\ell,\un})\right]\right|\right]=0.$$

\subsection{Lemma for the bootstrap}

\begin{lem}
	Let $\left(N_{\j},\vec{Y}_{\j}\right)_{\j\geq\un}$ a family of random variables such that $$\left\{N_{\j},\vec{Y}_{\j}\right\}_{\j\geq\un}=\left\{\tau\left(\left(U_{\j\odot\e}\right)_{\zero\prec \e \preceq \un}\right)\right\}_{\j\geq\un},$$
	for some measurable function $\tau$ and $\left(U_{\c}\right)_{\c \geq \zero}$ a family of mutually independent 
	uniform random variables on $(0,1)$. Let $f$ such that $\E\left[\left(\sum_{\ell=1}^{N_{\un}}f(Y_{\ell,\un})\right)^2\right]<\infty$.
	\begin{comment}
	Let $\left\{N_{\j},\left(Y_{\ell,\j}\right)_{\ell \geq 1}\right\}_{\j\geq\un}$ such that there exists a measurable function $\tau$ such that
	\begin{align*}
	&\left\{N_{\j},\left(Y_{\ell,\j}\right)_{\ell \geq 1}\right\}_{\j\geq\un}\overset{a.s.}{=}\left\{\tau\left(\left(U_{\j\odot\e}\right)_{\zero\prec\e\preceq \un}\right)\right\}_{\j\geq\un},
	\end{align*}
	where $\left(U_{\j}\right)_{\j > \zero}$ is a family of mutually independent 
	Uniform-$(0,1)$ random variables. Let $f$ such that $\E\left[\left(\sum_{\ell=1}^{N_{\un}}f(Y_{\ell,\un})\right)^2\right]<\infty$.
	\end{comment}
	For every $(\e,\c)$ that satisfy $\zero <\e < \un$ and $\c\wedge 1=\e$, let 
	$$a^{\C}_{\e}(\c)=\frac{1}{\prod_{s: e_s=0}C_s}\sum_{\un-\e\leq \c' \leq \C \odot (\un-\e)} \sum_{\ell=1}^{N_{\c+\c'}}f(Y_{\ell,\c+\c'})-\frac{1}{\Pi_C}\sum_{\un\leq \j\leq \C}\sum_{\ell\geq 1}^{N_{\j}}f(Y_{\ell,\j}).$$
	
	We have 
	$$\frac{1}{\prod_{i:e_i=1}C_i}\sum_{\e\leq \c\leq \C\odot \e}\left(a^{\C}_{\e}(\c)\right)^2\convAS \Cov\left(\sum_{\ell=1}^{N_{\un}}f(Y_{\ell,\un}),\sum_{\ell=1}^{N_{\deu-\e}}f(Y_{\ell,\deu-\e})\right),$$
	$$\text{and }\frac{1}{\prod_{i:e_i=1}C_i}\sum_{\e\leq \c\leq \C\odot \e}\left(a^{\C}_{\e}(\c)\right)^2\mathds{1}\left\{|a^{\C}_{\e}(\c)|\geq \left(\Pi_{i:e_i=1}C_i\right)^{1/2}\varepsilon\right\}\convAS 0, \text{ for every }\varepsilon>0.$$

	\label{lem:boot}
\end{lem}

\textbf{Proof: }
Let $\mu_C(\widetilde{f})$ the short-cut for $\frac{1}{\Pi_C}\sum_{\un\leq \j\leq \C}\sum_{\ell\geq 1}^{N_{\j}}f(Y_{\ell,\j})$.

For every $\c$ such that $\c\wedge \un =\e$, let $$b(\C\odot(\un-\e),\c)=a_{\e}^{\C}(\c)+\mu_C(\widetilde{f})=\frac{1}{\prod_{s:e_s=0}C_s}\sum_{\un-\e\leq \c' \leq \C \odot (\un-\e)} \sum_{\ell=1}^{N_{\c+\c'}}f(Y_{\ell,\c+\c'}),$$
$$\text{ and }A=\frac{1}{\prod_{i: e_i=1} C_i}\sum_{\e\leq \c\leq \C\odot \e}\left(b(\C\odot(\un-\e),\c)\right)^2.$$

By definition of $a_{\e}^{\C}(\c)$ and $A$
\begin{align*}\frac{1}{\prod_{i:e_i=1}C_i}\sum_{\e\leq \c\leq \C\odot \e}\left(a^{\C}_{\e}(\c)\right)^2
&=A-\left(\mu_C(\widetilde{f})\right)^2.\end{align*}

The representation of $\left\{N_{\j},\vec{Y}_{\j}\right\}_{\j\geq\un}$ in terms of the $U$'s implies that $\left\{N_{\j},\vec{Y}_{\j}\right\}_{\j\geq\un}$ forms a dissociated, separately exchangeable array. %(see \cite{kallenberg05}, page 339 or \cite{Aldous83}, page 125 for a definition of dissociation). 
Lemma 7.35 in \cite{kallenberg05} is therefore applicable and ensures
\begin{align*}\mu_C(\widetilde{f})&\convAS \E\left(\sum_{\ell=1}^{N_{\un}}f(Y_{\ell,\un})\right).
\end{align*}

We now focus on the limit of $A$, which is less straightforward to obtain. First, note that we can rewrite $A$ as
\begin{align*}
&A=\frac{1}{|\mathcal{B}_e|}\sum_{(\j,\j')\in \mathcal{B}_e}\sum_{\ell=1}^{N_{\j}}f(Y_{\ell,\j})\sum_{\ell=1}^{N_{\j'}}f(Y_{\ell,\j'})=\\
&\frac{1}{\prod_{i: e_i=1} C_i}\frac{1}{\prod_{s: e_s=0}C_s^2}\sum_{\e\leq \c\leq \C\odot \e}\sum_{\un-\e\leq \c' \leq \C \odot (\un-\e)} \sum_{\un-\e\leq \c'' \leq \C \odot (\un-\e)}\sum_{\ell=1}^{N_{\c+\c'}}f(Y_{\ell,\c+\c'})\sum_{\ell=1}^{N_{\c+\c''}}f(Y_{\ell,\c+\c''}).\end{align*}
A $k$ dimensional jointly exchangeable array, is an array such that Condition 1 in Assumption \ref{as:dgp} holds for $\pi_1=\pi_2=...=\pi_k$.

Note that $$\left\{\sum_{\ell=1}^{N_{\c+\c'}}f(Y_{\ell,\c+\c'})\times \sum_{\ell=1}^{N_{\c+\c''}}f(Y_{\ell,\c+\c''})\right\}_{\c,\c',\c'': \c\wedge \un=\e, \c'\wedge \un=\c''\wedge \un=\un-\e}$$ is a jointly exchangeable array indexed by the non-zero components of $\c,\c',\c''$ of dimension $l=2\sum_{i=1}^k (1-e_i)+\sum_{i=1}^k e_i=2k-\sum_{i=1}^k e_i$. Moreover, this array is dissociated. Lemma 7.35 in \cite{kallenberg05} is again applicable. As a result, $A$ admits an almost sure limit that takes the form
$$\lim_{\underline{C}\rightarrow \infty} \E\left[\frac{1}{|\mathcal{B}_e|}\sum_{(\j,\j')\in \mathcal{B}_e}\sum_{\ell=1}^{N_{\j}}f(Y_{\ell,\j})\sum_{\ell=1}^{N_{\j'}}f(Y_{\ell,\j'})\right].$$

Using $\lim_{\underline{C}\rightarrow \infty}\frac{|\mathcal{A}_e|}{|\mathcal{B}_e|}=1$ and the Cauchy-Schwarz inequality
\begin{align*}
&\limsup_{\underline{C}\rightarrow \infty}\left|\E\left[\frac{1}{|\mathcal{B}_e|}\sum_{(\j,\j')\in \mathcal{B}_e\backslash \mathcal{A}_e}\sum_{\ell=1}^{N_{\j}}f(Y_{\ell,\j})\sum_{\ell=1}^{N_{\j'}}f(Y_{\ell,\j'})\right]\right|\\
&\leq \limsup_{\underline{C}\rightarrow \infty} \frac{1}{|\mathcal{B}_e|} \sum_{(\j,\j')\in \mathcal{B}_e\backslash \mathcal{A}_e} \sqrt{\E\left(\left(\sum_{\ell=1}^{N_{\j}}f(Y_{\ell,\j})\right)^2\right)} \sqrt{\E\left(\left(\sum_{\ell=1}^{N_{\j'}}f(Y_{\ell,\j'})\right)^2\right)}\\
&\leq \limsup_{\underline{C}\rightarrow \infty} \frac{|\mathcal{B}_e|-|\mathcal{A}_e|}{|\mathcal{B}_e|} \E\left(\left(\sum_{\ell=1}^{N_{\un}}f(Y_{\ell,\un})\right)^2\right)\\
&\leq 0.
\end{align*}
On the other hand
\begin{align*}
&\E\left[\frac{1}{|\mathcal{B}_e|}\sum_{(\j,\j')\in \mathcal{A}_e}\sum_{\ell=1}^{N_{\j}}f(Y_{\ell,\j})\sum_{\ell=1}^{N_{\j'}}f(Y_{\ell,\j'})\right]\\
&=\frac{1}{|\mathcal{B}_e|}\sum_{(\j,\j')\in \mathcal{A}_e}\E\left[\sum_{\ell=1}^{N_{\j}}f(Y_{\ell,\j})\sum_{\ell=1}^{N_{\j'}}f(Y_{\ell,\j'})\right]\\
&=\frac{|\mathcal{A}_e|}{|\mathcal{B}_e|}\E\left[\sum_{\ell=1}^{N_{\un}}f(Y_{\ell,\un})\sum_{\ell=1}^{N_{\deu-\e}}f(Y_{\ell,\deu-\e})\right]\\
&\rightarrow \E\left[\sum_{\ell=1}^{N_{\un}}f(Y_{\ell,\un})\sum_{\ell=1}^{N_{\deu-\e}}f(Y_{\ell,\deu-\e})\right].
\end{align*}

Then
\begin{align*}\frac{1}{\prod_{i:e_i=1}C_i}\sum_{\e\leq \c\leq \C\odot \e}\left(a^{\C}_{\e}(\c)\right)^2\convAS&\E\left[\sum_{\ell=1}^{N_{\un}}f(Y_{\ell,\un})\sum_{\ell=1}^{N_{\deu-\e}}f(Y_{\ell,\deu-\e})\right]-\E\left[\sum_{\ell=1}^{N_{\un}}f(Y_{\ell,\un})\right]^2\\
&= \Cov\left(\sum_{\ell=1}^{N_{\un}}f(Y_{\ell,\un}),\sum_{\ell=1}^{N_{\deu}}f(Y_{\ell,\deu})\right).
\end{align*}

For every $\varepsilon>0$, the inequalities $\mathds{1}_{|a+b|\geq \varepsilon}\leq \mathds{1}_{|a|\geq \varepsilon/2}+\mathds{1}_{|b|\geq \varepsilon/2}$, $(a+b)^2\leq 2a^2+2b^2$ and the monotonicity of $b\mapsto \mathds{1}_{b\geq \varepsilon}$ together ensure
\begin{align*}&\frac{1}{\prod_{i:e_i=1}C_i}\sum_{\e\leq \c\leq \C\odot \e}\left(a^{\C}_{\e}(\c)\right)^2\mathds{1}\left\{|a^{\C}_{\e}(\c)|\geq \left(\prod_{i:e_i=1}\sqrt{C_i}\right)\varepsilon\right\}\\
&\leq \frac{1}{\prod_{i:e_i=1}C_i}\sum_{\e\leq \c\leq \C\odot \e}\left(a^{\C}_{\e}(\c)\right)^2\mathds{1}\left\{|b(\C\odot(\un-\e),\c)|\geq \left(\prod_{i:e_i=1}\sqrt{C_i}\right)\varepsilon/2\right\}\\
&+\mathds{1}\left\{|\mu_C(\widetilde{f})|\geq \left(\prod_{i:e_i=1}\sqrt{C_i}\right)\varepsilon/2\right\}\frac{1}{\prod_{i:e_i=1}C_i}\sum_{\e\leq \c\leq \C\odot \e}\left(a^{\C}_{\e}(\c)\right)^2\\
&\leq 2\frac{1}{\prod_{i:e_i=1}C_i}\sum_{\e\leq \c\leq \C\odot \e}\left(b(\C\odot(\un-\e),\c)\right)^2\mathds{1}\left\{|b(\C\odot(\un-\e),\c)|\geq \left(\prod_{i:e_i=1}\sqrt{C_i}\right)\varepsilon/2\right\}\\
&+2(\mu_C(\widetilde{f}))^2\frac{1}{\prod_{i:e_i=1}C_i}\sum_{\e\leq \c\leq \C\odot \e}\mathds{1}\left\{|b(\C\odot(\un-\e),\c)|\geq \left(\prod_{i:e_i=1}\sqrt{C_i}\right)\varepsilon/2\right\}\\
&+\mathds{1}\left\{|\mu_C(\widetilde{f})|\geq \left(\prod_{i:e_i=1}\sqrt{C_i}\right)\varepsilon/2\right\}\frac{1}{\prod_{i:e_i=1}C_i}\sum_{\e\leq \c\leq \C\odot \e}\left(a^{\C}_{\e}(\c)\right)^2\\
&\leq \left(2A+2(\mu_C(\widetilde{f}))^2\right) \mathds{1}\left\{\max_{\e\leq \c\leq \C\odot\e}|b(\C\odot(\un-\e),\c)|\geq \left(\prod_{i:e_i=1}\sqrt{C_i}\right)\varepsilon/2\right\}\\
&+\mathds{1}\left\{|\mu_C(\widetilde{f})|\geq \left(\prod_{i:e_i=1}\sqrt{C_i}\right)\varepsilon/2\right\}\frac{1}{\prod_{i:e_i=1}C_i}\sum_{\e\leq \c\leq \C\odot \e}\left(a^{\C}_{\e}(\c)\right)^2.
\end{align*}
We have already shown that $A=O_{as}(1)$, $\mu_C(\widetilde{f})=O_{as}(1)$ and $\frac{1}{\prod_{i:e_i=1}C_i}\sum_{\e\leq \c\leq \C\odot \e}\left(a^{\C}_{\e}(\c)\right)^2=O_{as}(1)$. Then
\begin{align}\label{eq:lem_boot_1}
&\frac{1}{\prod_{i:e_i=1}C_i}\sum_{\e\leq \c\leq \C\odot \e}\left(a^{\C}_{\e}(\c)\right)^2\mathds{1}\left\{|a^{\C}_{\e}(\c)|\geq \left(\prod_{i:e_i=1}\sqrt{C_i}\right)\varepsilon\right\}\nonumber \\
&\leq O_{as}(1)\mathds{1}\left\{\max_{\e\leq \c\leq \C\odot\e}|b(\C\odot(\un-\e),\c)|\geq \left(\prod_{i:e_i=1}\sqrt{C_i}\right)\varepsilon/2\right\}+o_{as}(1)O_{as}(1)\nonumber \\
&\leq O_{as}(1)\mathds{1}\left\{\max_{\e\leq \c\leq \C\odot\e}\frac{(b(\C\odot(\un-\e),\c))^2}{\prod_{i:e_i=1}C_i}\geq \varepsilon/2\right\}+o_{as}(1).
\end{align}

Let us show that $\max_{\e\leq \c\leq \C\odot\e}\frac{(b(\C\odot(\un-\e),\c))^2}{\prod_{i:e_i=1}\C_i}\convAS 0$ when $\underline{C}\rightarrow \infty$. This is sufficient to get the result.

\begin{align*}
\frac{\max_{(\C-\un)\odot \e < \c \leq \C\odot \e }(b(\C\odot(\un-\e),\c))^2}{\prod_{i:e_i=1}C_i}\leq&
\frac{1}{\prod_{i:e_i=1}C_i}\sum_{(\C-\un)\odot \e < \c \leq \C\odot \e }(b(\C\odot(\un-\e),\c))^2\\&=\frac{1}{\prod_{i:e_i=1}C_i}\sum_{ \e \leq \c \leq \C\odot \e }(b(\C\odot(\un-\e),\c))^2\\
&-\frac{\prod_{i:e_i=1}(C_i-1)}{\prod_{i:e_i=1}C_i}\frac{1}{\prod_{i:e_i=1}(C_i-1)}\sum_{ \e \leq \c \leq (\C-\un)\odot \e }(b(\C\odot(\un-\e),\c))^2
\end{align*}
We have shown that $\frac{1}{\prod_{i:e_i=1}C_i}\sum_{ \e \leq \c \leq \C\odot \e }(b(\C\odot(\un-\e),\c))^2$ converges almost surely to $$\E\left[\sum_{\ell=1}^{N_{\un}}f(Y_{\ell,\un})\sum_{\ell=1}^{N_{\deu-\e}}f(Y_{\ell,\deu-\e})\right].$$ This is also the case for $\frac{1}{\prod_{i:e_i=1}C_i-1}\sum_{ \e \leq \c \leq (\C-\un)\odot \e }(b(\C\odot(\un-\e),\c))^2$. Because $\lim_{\underline{C}\rightarrow \infty}\frac{\prod_{i:e_i=1}C_i-1}{\prod_{i:e_i=1}C_i}=1$, we deduce that
$$\frac{\max_{(\C-\un)\odot \e < \c \leq \C\odot \e }(b(\C\odot(\un-\e),\c))^2}{\prod_{i:e_i=1}C_i}\convAS 0.$$

Fix $\eta$ arbitrarily small. There exists $\underline{C}^{\ast}$ such that for any $\C\geq \underline{C}^{\ast}\times \un$, we have
$$\frac{\max_{(\C-\un)\odot \e < \c \leq \C\odot \e }(b(\C\odot(\un-\e),\c))^2}{\prod_{i:e_i=1}C_i}\leq \frac{\eta}{2}.$$
We have
\begin{align*}
\frac{\max_{\e \leq \c \leq \C\odot \e }(b(\C\odot(\un-\e),\c))^2}{\prod_{i:e_i=1}C_i}=& \max_{\un<\widetilde{\C}\leq \C}\frac{\max_{(\widetilde{\C}-\un)\odot \e < \c \leq \widetilde{\C}\odot \e }(b(\C\odot(\un-\e),\c))^2}{\prod_{i:e_i=1}C_i}\\
&\leq \frac{\max_{\un<\widetilde{\C}\leq \underline{C}^{\ast}\times \un }\max_{(\widetilde{\C}-\un)\odot \e < \c \leq \widetilde{\C}\odot \e }(b(\C\odot(\un-\e),\c))^2}{\prod_{i:e_i=1}C_i} \\
&+\max_{\underline{C}^{\ast}\times \un<\widetilde{\C}\leq \C}\frac{\max_{(\widetilde{\C}-\un)\odot \e < \c \leq \widetilde{\C}\odot \e }(b(\C\odot(\un-\e),\c))^2}{\prod_{i:e_i=1}C_i}\\
&\leq \frac{\max_{\un<\widetilde{\C}\leq \underline{C}^{\ast}\times \un }\max_{(\widetilde{\C}-\un)\odot \e < \c \leq \widetilde{\C}\odot \e }(b(\C\odot(\un-\e),\c))^2}{\prod_{i:e_i=1}C_i} \\
&+\max_{\underline{C}^{\ast}\times \un<\widetilde{\C}\leq \C }\frac{\max_{(\widetilde{\C}-\un)\odot \e < \c \leq \widetilde{\C}\odot \e }(b(\C\odot(\un-\e),\c))^2}{\prod_{i:e_i=1}\widetilde{C}_i}
\end{align*}
For $\C$ and $\widetilde{\C}$ such that $\underline{C}^{\ast}\times \un<\widetilde{\C}\leq \C$, let $\overline{\C}=\widetilde{\C}\odot \e+\C\odot(\un-\e)$. We have $\overline{\C}\odot \e=\widetilde{\C}\odot \e$, $(\overline{\C}-\un)\odot \e=(\widetilde{\C}-\un)\odot \e$, $\overline{\C}\odot (\un-\e)=\C\odot (\un-\e)$, $\prod_{i:e_i=1}\widetilde{C}_i=\prod_{i:e_i=1}\overline{C}_i$ and $\overline{\C}>\underline{C}^{\ast}\times \un$.
\begin{align*}\max_{\underline{C}^{\ast}\times \un<\widetilde{\C}\leq \C }\frac{\max_{(\widetilde{\C}-\un)\odot \e < \c \leq \widetilde{\C}\odot \e }(b(\C\odot(\un-\e),\c))^2}{\prod_{i:e_i=1}\widetilde{C}_i}&=\max_{\underline{C}^{\ast}\times \un<\widetilde{\C}\leq \C }\frac{\max_{(\overline{\C}-\un)\odot \e < \c \leq \overline{\C}\odot \e }(b(\overline{\C}\odot(\un-\e),\c))^2}{\prod_{i:e_i=1}\overline{C}_i}\\
&\leq \frac{\eta}{2}\end{align*}

On the other hand, for $\underline{C}$ sufficiently large, we have
$$\frac{\max_{\un<\widetilde{\C}\leq \underline{C}^{\ast}\times \un }\max_{(\widetilde{\C}-\un)\odot \e < \c \leq \widetilde{\C}\odot \e }(b(\C\odot(\un-\e),\c))^2}{\prod_{i:e_i=1}C_i}\leq \frac{\eta}{2},$$
and next (still for $\underline{C}$ sufficiently large)
$$\frac{\max_{\e \leq \c \leq \C\odot \e }(b(\C\odot(\un-\e),\c))^2}{\prod_{i:e_i=1}C_i}\leq \eta.$$
Because $\eta$ is arbitrarily small, this means that
\begin{align}\label{eq:lem_boot_2}
	&\frac{\max_{\e \leq \c \leq \C\odot \e }(b(\C\odot(\un-\e),\c))^2}{\prod_{i:e_i=1}C_i}\convAS 0.
\end{align}

Combine (\ref{eq:lem_boot_1}) and (\ref{eq:lem_boot_2}) to see that
$$\frac{1}{\prod_{i:e_i=1}C_i}\sum_{\e\leq \c\leq \C\odot \e}\left(a^{\C}_{\e}(\c)\right)^2\mathds{1}\left\{|a^{\C}_{\e}(\c)|\geq \left(\Pi_{i:e_i=1}C_i\right)^{1/2}\varepsilon\right\}\convAS 0, \text{ for every }\varepsilon>0.$$

\end{document}